\documentclass[aps,11pt,prd,groupedaddress,nofootinbib,notitlepage,eqsecnum,preprintnumbers]{revtex4-2}
\usepackage[utf8]{inputenc}
\usepackage{hyperref}
\usepackage{xcolor}
\usepackage{graphicx}
\usepackage{amsmath,amssymb}
\usepackage{bm}
\usepackage{comment}
\usepackage[shortlabels]{enumitem}
\usepackage{overpic}
\usepackage{ulem}
\usepackage{makecell}
\usepackage{tabularx}
\usepackage{booktabs}
\usepackage{diagbox}
\usepackage{pifont}

\begin{document}
\title{Enhanced induced gravitational waves in Horndeski gravity}

\author{\textsc{Guillem Dom\`enech$^{a,b}$}}
    \email{{guillem.domenech}@{itp.uni-hannover.de}}

\author{\textsc{Alexander Ganz}$^{a}$}
\email{alexander.ganz@itp.uni-hannover.de}

\affiliation{$^a$ Institute for Theoretical Physics, Leibniz University Hannover, Appelstraße 2, 30167 Hannover, Germany.}
\affiliation{$^b$ Max-Planck-Institut für Gravitationsphysik, Albert-Einstein-Institut, 30167 Hannover, Germany}

\graphicspath{{./figures/}}

\begin{abstract}
We study secondary gravitational wave production in Horndenski gravity, when the scalar field dominates the very early universe. We find that higher derivative interactions easily dominate the source term on subhorizon scales and significantly enhance the amplitude of induced GWs. We analytically derive, for the first time, the Horndeski induced GW spectrum for a general class of power-law solutions. The main effect of modifications of gravity are stronger resonances and a growth of tensor fluctuations on small scales. The maximum attainable amplitude of the induced GW spectrum is bounded by the possible backreaction of higher derivatives on curvature fluctuations, thereby shutting down the source term to induced GWs. We argue that the maximum attainable amplitude depends linearly on the primordial curvature spectrum ($\Omega_{\rm GW}\propto {\cal P}_\zeta$), as opposed to the standard case where it depends quadratically. Resonances may further enhance the maximum amplitude by a factor $(k/{\cal H}_t)^2$ or  $(k/{\cal H}_t)$ respectively for sharp and broad peaks (including a scale invariant) primordial spectrum,  where ${\cal H}_t$ is the comoving horizon at the time when standard gravity is recovered.  Remarkably, in the scale invariant case, the Horndeski-induced GW spectrum grows as $k^3$. This opens up the interesting possibility that induced GWs might be observable despite no enhancement of the primordial curvature spectrum. Our formalism can be generalized to a wider class of solutions and to more general scalar-tensor theories, such as DHOST and spatially covariant gravity. In the appendices we provide discussions on the gauge issue and disformal transformations of induced GWs.
\end{abstract}

\maketitle

\section{Introduction}

There is a big effort to investigate general modifications of General Relativity (GR) as the theory of gravity (see, e.g., Refs.~\cite{DeFelice:2010aj,Koyama:2015vza,Nojiri:2017ncd,Langlois:2018dxi,Kobayashi:2019hrl,Shankaranarayanan:2022wbx} for various reviews). One of the motivations to do so is to understand and explain the origins of the current accelerated expansion of the universe. But, down to a more fundamental level, one also wonders about the nature of gravity at very high energies, where effects of new fields and new interactions are expected to show up. Curiously, Cosmic Microwave Background (CMB) observations  \cite{Akrami:2018odb} point towards a rather flat inflaton potential during cosmic inflation  \cite{Starobinsky:1979ty,Sato:1980yn,Guth:1980zm}. This is, in fact, the general effect of non-minimal couplings to the Ricci curvature in the large field limit in the context of scalar-tensor theories of gravity \cite{Fujii:2003pa}. Nice examples of this are Higgs Inflation \cite{Bezrukov:2007ep} (see, e.g., Ref.~\cite{Rubio:2018ogq} for a review) and Starobinsky’s inflation \cite{Starobinsky:1980te}. Now that we have access to Gravitational Wave (GW) observations, we have a new way test non-minimal couplings to gravity.

However, non-minimal couplings to the Ricci curvature are but a small subset of non-minimal couplings to gravity. One interesting possibility are higher derivative couplings to gravity, which may appear in higher dimensional theories after dimensional reduction, including the Gauss-Bonnet term, and in the decoupling limit of massive gravity \cite{Dvali:2000hr,Luty:2003vm,deRham:2010eu,Goon:2011qf,Goon:2011uw,Trodden:2011xh,VanAcoleyen:2011mj,vandeBruck:2018jlz,deRham:2011by,Heisenberg:2014kea}. They also emerge when considering generalized metric transformations that include derivatives of a scalar field \cite{Zumalacarregui:2013pma,Bettoni:2013diz}, the so-called disformal transformations \cite{Bekenstein:1992pj}. Here we take a general approach to such theories by considering Horndeski gravity \cite{Horndeski:1974wa}, also known as Generalized Galileon \cite{Deffayet:2009wt,Deffayet:2009mn,Deffayet:2011gz,Kobayashi:2011nu} (see Ref.~\cite{Kobayashi:2019hrl} for a nice review). These theories are derived by requiring that the system of gravity plus a scalar field only has three degrees of freedom, despite the presence of higher derivatives. Even more general theories are the so-called beyond Hordenski \cite{Gleyzes:2014dya}, Degenerate Higher Order Scalar Tensor (DHOST) theories \cite{Langlois:2015cwa,Langlois:2018dxi}, U-DHOST \cite{DeFelice:2018ewo} and spatially covariant theories of gravity \cite{Gao:2014fra,Gao:2014soa} (see also Refs.~\cite{Gleyzes:2013ooa,Fujita:2015ymn,Crisostomi:2016tcp,BenAchour:2016cay,BenAchour:2016fzp,Takahashi:2017zgr,BenAchour:2020wiw,Babichev:2021bim} for their related Effective Field Theories (EFTs)).

How can we test Horndeski theories in the very early universe? If a Horndeski scalar was responsible for inflation, we may look for a blue tilted tensor spectrum \cite{Kobayashi:2010cm,Cai:2014uka} and curvature and tensor bispectrum \cite{DeFelice:2011uc,Gao:2011vs,Renaux-Petel:2011zgy,Gao:2012ib,DeFelice:2013ar,Tahara:2017wud} on CMB scales. Alternatively, it is also possible that curvature and tensor fluctuations are enhanced on the very small scales \cite{Hirano:2016gmv,Mylova:2018yap}. But, there is another interesting possibility in the case that a Horndeski scalar was present after inflation and before Big Bang Nucleosynthesis (BBN).\footnote{To be more precise that standard model of particle physics and cosmology is recovered before the temperature of the universe is $T\sim 4\,{\rm MeV}$ \cite{Kawasaki:1999na,Kawasaki:2000en,Hannestad:2004px,Hasegawa:2019jsa}} Any fluctuations generated during inflation act as a secondary source of GWs after inflation ends \cite{Tomita,Matarrese:1992rp,Matarrese:1993zf,Matarrese:1997ay}.\footnote{Note that Horndeski modifications could also lead to resonances in the linear equations for GWs \cite{Lin:2015nda,Kuroyanagi:2017kfx,Cai:2020ovp,Ye:2023xyr,Cai:2023ykr}.} These are the so-called induced GWs \cite{Ananda:2006af,Baumann:2007zm,Assadullahi:2009nf,Alabidi:2013lya}, and are a crucial test of the Primordial Black Hole (PBH) scenario \cite{Saito:2008jc,Saito:2009jt,Bugaev:2009zh,Bugaev:2009kq,Bugaev:2010bb} (see Refs.~\cite{Khlopov:2008qy,Sasaki:2018dmp,Green:2020jor,Carr:2020gox,Escriva:2022duf} for reviews on PBHs). In fact, induced GWs are one possible explanation \cite{Dandoy:2023jot,Franciolini:2023pbf,Franciolini:2023wjm,Inomata:2023zup,Cai:2023dls,Wang:2023ost,Liu:2023ymk,Unal:2023srk,Figueroa:2023zhu,Yi:2023mbm,Zhu:2023faa,Firouzjahi:2023lzg,Li:2023qua,You:2023rmn,Balaji:2023ehk,HosseiniMansoori:2023mqh,Zhao:2023joc,Liu:2023pau,Yi:2023tdk,Bhaumik:2023wmw,Choudhury:2023hfm,Yi:2023npi,Harigaya:2023pmw,Basilakos:2023xof,Jin:2023wri,Cannizzaro:2023mgc,Zhang:2023nrs,Liu:2023hpw,Choudhury:2023fwk,Tagliazucchi:2023dai,Basilakos:2023jvp,Inomata:2023drn,Li:2023xtl,Domenech:2023dxx,Gangopadhyay:2023qjr,Cyr:2023pgw,Lozanov:2023rcd,Madge:2023dxc,Domenech:2024rks} for the tentative PTA signal of a GW background \cite{EPTA:2023fyk,EPTA:2023sfo,EPTA:2023xxk,Zic:2023gta,Reardon:2023gzh,Reardon:2023zen,NANOGrav:2023hde,NANOGrav:2023gor,NANOGrav:2023hvm,InternationalPulsarTimingArray:2023mzf,Xu:2023wog}. For a detailed review of induced GWs see Ref.~\cite{Domenech:2021ztg} and for a pedagogical approach see Ref.~\cite{Domenech:2024kmh}. Other helpful reviews can be found in Refs.~\cite{Yuan:2021qgz,LISACosmologyWorkingGroup:2023njw,Domenech:2023jve,Domenech:2024cjn}. It is expected that modifications of gravity will significantly impact the production of induced GWs. Basically, as a secondary effect, the source of induced GWs in modified gravity will have much more terms than in GR, some of which will include higher spatial derivatives.

In this work, we investigate, for the first time, the effects of Horndeski gravity on the generation of induced GWs in the very early universe. For analytical viability, due to the large functional freedom, we focus on general power-law solutions in which the scale factor as well as the scalar field evolve as a power-law of time. We can then compute the induced GW spectrum without specifying the explicit form of the functions, since all coefficients are constant on the power-law solution. We also require that GR is recovered before BBN. This is important to ensure the standard big bang cosmology and compatibility with late universe GW observations \cite{LIGOScientific:2017zic,Ezquiaga:2017ekz} and the matter power spectrum \cite{Bellini:2015wfa}. We assume, for simplicity, an instantaneous transition to GR, though we also comment on possible effects of a gradual one. 

Despite the simplifying assumptions, we find that the Horndeski-scalar induced GWs have a rich phenomenology and several new features (which are reflected by the length of the paper). Because of the length, we provide a short summary of the key results below. We follow the standard notation in Horndeski gravity \cite{Kobayashi:2019hrl} to denote with $G_3$ the coupling to $\Box\phi$, with $G_4$ the coupling to the Ricci curvature $R$ and with $G_5$ the derivative coupling to the Einstein tensor $G^{\mu\nu}\nabla_\mu\nabla_\nu\phi$. With this notation, we proceed to the summary:
\begin{itemize}
    \item \textbf{Effects of higher derivatives on the generation of induced GWs:} Higher derivative interactions enhance the amplitude of the induced GW spectrum.
    \begin{enumerate}
        \item The $G_4$ and $G_5$ functions lead to an enhancement of the induced GW spectrum proportional to the ratio of the horizon scale to the wavelength, namely $(k/{\cal H})^2$, with respect to the GR prediction. For resonant modes the enhancement grows as $(k/{\cal H})^4$.
        \item The $G_3$ function enhances resonant modes only,  with respect to GR, with a growth of the induced GW spectrum proportional to $(k/{\cal H})^2$.
        \item Higher derivative terms also enhance  curvature fluctuations. Requiring no backreaction (which roughly imposes that ${\rm Horndeski\,functions}\times(k/{\cal H})\times P_{\zeta}^{1/2}<1$) leads to a maximum attainable amplitude of the induced GW spectrum of the order of $\Omega^{\rm max}_{\rm GW}\propto {\cal P}_\zeta$. In comparison, the GR prediction always has that $\Omega_{\rm GW}\propto {\cal P}_\zeta^2$.
    \end{enumerate}
    \item \textbf{New features of Horndeski-induced GW spectrum:} assuming an instantaneous transition to GR, we find distinct predictions also shown in Tab.~\ref{tab:table0}.

\begin{table}
\def\arraystretch{2}
\setlength\tabcolsep{1.5mm}
\begin{tabular}{|c|c|c|c|}
\toprule[2.0pt]\addlinespace[0mm]
\makecell{\textbf{Horndeski} \\\textbf{function} } & \makecell{\textbf{Sharply peaked}\\\textbf{spectrum at $\mathbf{k_{\rm p}}$}  }  & \makecell{\textbf{Broadly peaked}\\\textbf{spectrum at $\mathbf{k_{\rm p}}$}  } & \makecell{\textbf{Flat spectrum} \\ \textbf{with cut at $\mathbf{k_{\rm uv}}$}   }\\
\hline
\makecell{ \makecell{\textbf{$\mathbf{G_4}$ \& $\mathbf{G_5}$} }} &$\Omega^{\rm res}_{\rm GW}\sim (k_p/k_t)^4{\cal P}_\zeta^2$& $\Omega^{\rm res}_{\rm GW}\sim (k_p/k_t)^3{\cal P}_\zeta^2$&  $\Omega_{\rm GW}(k)\sim (k/k_t)^3{\cal P}_\zeta^2$ \\
\hline
\makecell{\makecell{\textbf{$\mathbf{G_3}$ only}   }} &$\Omega^{\rm res}_{\rm GW}\sim (k_p/k_t)^2{\cal P}_\zeta^2$& $\Omega^{\rm res}_{\rm GW}\sim (k_p/k_t){\cal P}_\zeta^2$& $\Omega_{\rm GW}(k)\sim (k/k_t){\cal P}_\zeta^2$ \\
\toprule[2.0pt]
\end{tabular}
\caption{Summary table of new features of the Horndeski-induced GW spectrum. We consider a sharply peaked, broadly peaked and flat primordial spectrum of curvature fluctuations, respectively in the second, third and fourth columns. The peaked spectra peak at $k_p$ and the flat spectrum is cut at $k_{\rm uv}$ to avoid higher derivative non-linearitires. $k_t$ is the comoving horizon size at the time of the transition to GR. In the rows, we separate the cases of non-trivial $G_4$ and $G_5$ functions and the case with $G_3$ function only. The difference comes from the additional number of derivatives in the source terms of induced GWs. For the peaked spectrum we show the estimate for the amplitude at the resonant peak in the induced GW spectrum. For the flat spectrum, we give an analytical estimate of the induced GW spectrum for $k>k_t$.\label{tab:table0}}
\end{table}
    \begin{enumerate}
        \addtocounter{enumi}{3}
        \item \underline{Sharply peaked primordial spectrum at $k_p$:} We find that the amplitude of the resonant peak of the induced GW spectrum is enhanced by a factor $(k_p/k_t)^4$, for $G_4$ and $G_5$, and $(k_p/k_t)^2$ for $G_3$. Taking into account the backreaction bound, the maximum attainable amplitude at the peak is then given by $\Omega^{\rm max}_{\rm GW}\sim (k_p/k_t)^2{\cal P}_\zeta$ and $\Omega^{\rm max}_{\rm GW}\sim {\cal P}_\zeta$, respectively for $G_4$/$G_5$ and $G_3$. For broad peaks, one adds a factor $(k_t/k_p)$ suppression.
        \item \underline{Flat primordial spectrum cut off at $k_{\rm uv}$:} The resulting Horndeski-induced GW spectrum is not scale invariant. In fact, it goes as $k^3$ and $k$ respectively for the $G_{4}$/$G_5$ and $G_{3}$, reflecting the subhorizon growth of the source term due to higher derivatives.
    \end{enumerate}
\end{itemize}

\begin{figure}
    \centering
    \includegraphics[width=0.8\columnwidth]{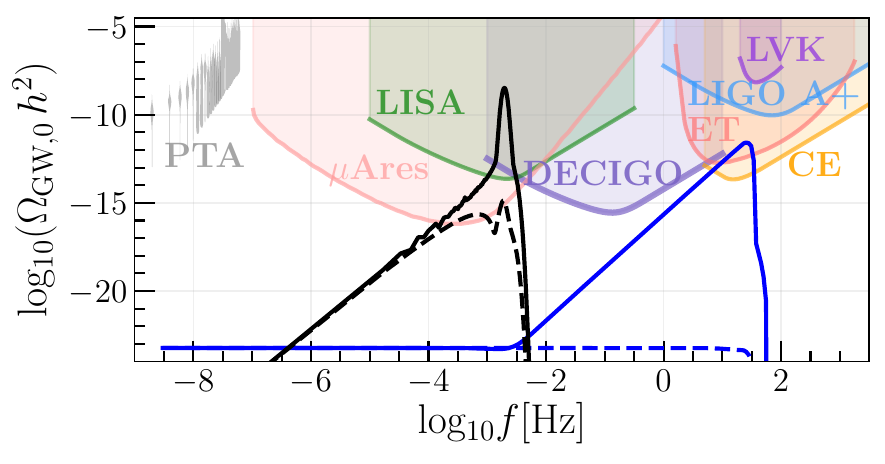}
    \caption{Induced GW spectrum today for a log-normal (with the peak at $k_p$) and a scale invariant (with a high-k cut at $k_{\rm uv}$) primordial spectrum of curvature fluctuations, respectively shown in black and blue lines. Solid lines indicate the Horndeski-scalar induced GWs, while dashed lines show the corresponding GR limit. We fix the amplitude of the power spectrum to $A_\zeta =10^{-5}$ and $A_\zeta=10^{-9}$ for the log-normal and scale invariant cases, respectively. We also chose
    $\tilde \Gamma=10^{-3}$ (roughly the amplitude of the non-trivial $G_{4}$),  and $k_p/k_t =10^4$ and $k_{\rm uv}/k_t =10^5$ respectively for the peaked and flat primoridal spectra (see section \ref{sec:horndeskiconcrete} for more details). For the log-normal distribution we fixed $f_p = k_p/(2\pi)\approx 1.7\times 10^{-3}\,\mathrm{Hz}$ and for the scale-invariant spectrum $f_{\rm uv} =k_{\rm uv}/(2\pi)\approx 30\,\mathrm{Hz}$. For illustration purposes, we show the power-law integrated sensitivity curves as described in Ref.~\cite{Thrane:2013oya,Schmitz:2020syl}
    for LIGO A+ \cite{A+}, Einstein Telescope (ET) (15km arms triangular configuration \cite{Branchesi:2023mws}), Cosmic Explorer (CE) \cite{ce}, DECIGO \cite{Yagi:2011wg,Kawamura:2020pcg}, LISA \cite{Barke:2014lsa} and $\mu$-Ares \cite{Sesana:2019vho} experiments. The purple shaded region shows the upper bounds from the LVK collaboration \cite{KAGRA:2021kbb}. The gray violins show the posterior distributions of the NANOGrav 15yr data \cite{NANOGrav:2023hde,NANOGrav:2023hvm}. One application to the PTA data could be to move the peak of the solid blue line near the nHz frequency range. We briefly discuss this possibility at the end of Sec.~\ref{sec:horndeskiconcrete}.}
    \label{fig:enter-label}
\end{figure}

We show two examples of Horndeski-induced GWs in Fig.~\ref{fig:enter-label}. We consider a peaked and a scale invariant primordial spectrum of curvature fluctuations. It is interesting that we have the possibility that induced GWs may be observable by future GW detectors such as $\mu$-Ares \cite{Sesana:2019vho}, LISA, Taiji \cite{Barke:2014lsa,Ruan:2018tsw}, TianQin \cite{Gong:2021gvw},  DECIGO \cite{Yagi:2011wg,Kawamura:2020pcg}, Einstein Telescope (ET) \cite{Maggiore:2019uih}, Cosmic Explorer (CE) \cite{ce}, Voyager \cite{A+,voyager} and LIGO A+, despite no (or little) enhancement of ${\cal P}_\zeta$ during inflation (see, e.g., Refs.~\cite{Ozsoy:2023ryl,Kristiano:2024ngc} for recent reviews). Naively, we do not expect any significant PBH formation in that case, though a detailed study is needed to draw any definitive conclusions (specially because of the effect of higher derivative terms). We discuss possible applications to PTAs \cite{EPTA:2023fyk,EPTA:2023sfo,EPTA:2023xxk,Zic:2023gta,Reardon:2023gzh,Reardon:2023zen,NANOGrav:2023hde,NANOGrav:2023gor,NANOGrav:2023hvm,InternationalPulsarTimingArray:2023mzf,Xu:2023wog} at the end of Sec.~\ref{sec:horndeskiconcrete}. Lastly, we note that our work will mainly focus on the effects from $G_4$ and $G_5$ with a brief discussion on the case of $G_3$ only. The main reason is that the effects of $G_4$ and $G_5$ are more pronounced as they introduce more spatial derivatives to the source term of induced GWs than $G_3$. We plan to study the effects in a general $G_3$ theory in subsequent works, as this term has the physical interpretation of an imperfect fluid \cite{Pujolas:2011he}, whereas the physical interpretation behind $G_4$ and $G_5$ in terms of a fluid is not yet clear. This paper also introduces the basic formalism to generalize the calculation of induced GWs in more general solutions and theories, such as DHOST \cite{Langlois:2015cwa,Langlois:2018dxi}, U-DHOST \cite{DeFelice:2018ewo} and spatially covariant theories of gravity \cite{Gao:2014fra,Gao:2014soa}. It should be noted that, in a complementary direction, Refs.~\cite{Hu:2024hzo,Feng:2024yic,Zhang:2024vfw} investigate the induced GW spectrum in parity violating modifications of GR, in scalar-tensor theories and in teleparallel gravity.

This paper is organized as follows. In Sec.~\ref{sec:powerlaw}, we introduce the power-law solutions when the Horndeski scalar dominates the very early universe. In Sec.~\ref{sec:preliminaries}, we provide a physically intuitive picture of the generation of induced GWs in Horndeski gravity. This allows us to qualitatively explain the origin of the unique features of Horndeski-scalar induced GWs. We then proceed to the general calculation in Sec.~\ref{sec:horndeskigeneral}, where we derive the explicit form of the Kernel. We also discuss the impact of the Horndeski modifications to the induced GW spectrum assuming that the evaluation at the end of the Horndeski phase provides a good approximation. In Sec.~\ref{sec:horndeskiconcrete}, we consider a minimal case with a concrete model for the transition to GR. In our minimal scenario, modifications to GR only appear at second order in cosmological perturbation theory. We end our paper with several discussions and future direction in Sec.~\ref{sec:discussions}. We provide many details of the calculations in the appendices.

\section{Power-law solutions in Horndeski theory \label{sec:powerlaw}}

Our starting point is Horndeski theory, also known as generalized Galileon, for which the general action reads \cite{Kobayashi:2019hrl}
\begin{align}\label{eq:fulllagrangian}
    S = \int d^4x\,\sqrt{-g}\,\Big[& G_2-G_3 \Box \phi+ G_4 \mathcal{R} + G_{4,X}  \left((\Box \phi)^2-(\nabla_\mu\nabla_\nu\phi)^2 \right)\nonumber\\&+G_5 G^{\mu\nu}\nabla_\mu\nabla_\nu\phi-\frac{1}{6}G_{5,X}\left((\Box \phi)^3-3\Box\phi(\nabla_\mu\nabla_\nu\phi)^2+2(\nabla_\mu\nabla_\nu\phi)^3\right)\Big]\,,
\end{align}
where $R$ is the Ricci scalar, $G^{\mu\nu}$ the Einstein tensor, $\phi$ a scalar field, $X= - \frac{1}{2} \nabla_\mu\phi\nabla^\mu\phi$ and $G_{i}(\phi,X)$ with $i=\{2,3,4,5\}$ are arbitrary functions of the scalar field and its first derivative. Here and throughout the paper we define $G_{i,X}\equiv \partial G_i/\partial X$. The same notation applies to derivatives with respect to $\phi$. For the metric we consider a perturbed flat Friedmann–Lemaître–Robertson–Walker (FLRW) metric.

\subsection{Background solutions}

Before going into the details of the perturbations, let us explain the type of background solutions that we consider. For analytical clarity and viability, we are interested in those solutions that lead to constant dimensionless coefficients in the equations for perturbations. This is a great simplification when dealing with the induced GWs, similar to the analytical solutions found in Refs.~\cite{Domenech:2019quo,Domenech:2020kqm} for a constant equation of state of the universe. It will also be helpful when comparing with the existing literature on induced GWs. Here we find such solutions as follows. We first compute the Horndeski action \eqref{eq:fulllagrangian} in the FLRW background metric, where the line element reads
\begin{align}
ds^2=-N^2dt^2+a^2d\mathbf{x}^2\,,
\end{align}
and we keep explicitly the lapse $N$ in order to derive the first Friedmann equation, or equivalently the Hamiltonian constraint. After some integration by parts (see Sec.~3 of Ref.~\cite{Gleyzes:2013ooa} for some simplifying tricks), we obtain that
\begin{align}\label{eq:bgaction}
S_{\rm bg}=\int d^3x dt N a^3\Big\{G_2+&(\sqrt{2X}f_{3\phi}-2X G_{3,\phi})+\frac{3H}{N}(f_3-2\sqrt{2X}G_{4,\phi})\nonumber\\&-\frac{6H^2}{N^2}(G_4-2XG_{4,X}+XG_{5,\phi})+\frac{H^3}{N^3}(2X)^{3/2}G_{5,X}\Big\}\,,
\end{align}
where $a$ is the scale factor, $H=\dot a/a$, with $\dot a= da/dt$ and we defined
\begin{align}
f_3=\int dX \sqrt{2X}{G_{3,X}}\,.
\end{align}

For a constant equation of state of the universe, the scale factor is a power-law of time, say 
\begin{align}\label{eq:scalefactorandH}
a=a_*(t/t_*)^p\quad{\rm  and}\quad H=p/t\,,
\end{align}
where $t_*$ is an arbitrary pivot time and $1/p=3(1+w)/2$ where $w$ is the “equation of state” of the universe. For a radiation dominated universe we have that $w=1/3$ and $p=1/2$. We must then choose the functions $G_i$’s such that the action Eq.~\eqref{eq:bgaction} is compatible with the power-law ansatz. To simplify the task we work in a field redefinition where
\begin{align}\label{eq:phiunitary}
\phi=\phi_*(t/t_*)\,,
\end{align}
in the particular solution. Note that if we started with a concrete model, this field redefinition would shuffle the definitions of $G_i$’s. However, it does not matter for our purposes, since we are considering general solutions and $\phi$ does not couple directly to matter. Furthermore, without loss of generality, we assume that $G_4$ is constant in the particular solution. We could consider a $G_4$ which also scales as a power-law of time but such effect could be absorbed by a conformal transformation. Then, since $X=X_*={\rm constant}$, we conclude that
\begin{align}\label{eq:Gisolutions1}
G_2(\phi,X)&=\Lambda_2^4\left(\frac{\phi}{\phi_*}\right)^{-2}a_2(Y)\quad,\quad G_3(\phi,X)=\Lambda_3 \left(\frac{\phi}{\phi_*}\right)^{-1}a_3(Y)\,,\\
G_4(\phi,X)&=\Lambda_4^2\,a_4(Y)\quad,\quad G_5(\phi,X)=\frac{1}{\Lambda_5}\left(\frac{\phi}{\phi_*}\right)a_5(Y)\,,\label{eq:Gisolutions2}
\end{align}
so that all terms in Eq.~\eqref{eq:bgaction} scale as $t^{-2}$. 
 In Eqs.~\eqref{eq:Gisolutions1} and \eqref{eq:Gisolutions2}, we defined for convenience the normalized variable
\begin{align}
Y=\frac{X}{X_*}\,,
\end{align}
which is $Y=1$ in the particular solution.\footnote{We keep the notation $Y$ and not $X$ since $Y$ is normalized and it can be generalized to arbitrary time dependence of $\phi$ (i.e. $\phi\propto t^q$) by defining $Y\to Y=X\phi^{2\frac{1-q}{q}}$ such that it is constant in the power-law solution \cite{Amendola:2018ltt}.} We also introduced $a_i$’s which are arbitrary functions of $Y$ only and are constants in the particular solution. $\Lambda_i$’s denote the energy scale of each function. Note that with a simple manipulation of Eqs.~\eqref{eq:scalefactorandH} and \eqref{eq:phiunitary}, we see that such particular solutions satisfy
\begin{align}\label{eq:scalingrelation}
\frac{\dot\phi}{\phi H}=\frac{1}{p}\,.
\end{align}
This type of solutions fall inside the so-called “tracker” and “scaling” solutions where $\dot \varphi \propto H^\alpha$ (after a field redefinition given by $\dot \varphi\sim \dot\phi/\phi$) \cite{Amendola:1999qq,Amendola:2006qi,Gomes:2013ema,Gomes:2015dhl,Amendola:2017xhl,Amendola:2018ltt,Frusciante:2018tvu,Frusciante:2018aew,Vaskonen:2020lbd}. In the context of de Sitter these solutions are also called self-accelerating \cite{Chow:2009fm,Silva:2009km,Deffayet:2010qz,Kimura:2010di,DeFelice:2010pv,DeFelice:2010nf,Crisostomi:2017pjs}. Normally, these solutions are called “scaling” solutions instead, if one considers an additional matter field and further requires that the total equation of state is constant \cite{Amendola:1999qq}. Here we will simply call them \textit{power-law solutions}. Note that in general such power-law solutions are attractors \cite{Amendola:2018ltt}.

The first and second Friedmann equations follow, respectively, from taking the variation of \eqref{eq:bgaction} with respect to $N$ and $a$. In the power-law solutions they yield (see also Refs.~\cite{Kobayashi:2011nu,Kobayashi:2019hrl} for the explicit expression of Friedmann equations)
\begin{align}
\lambda_2^4 p^2
   \left(2
   a_{2,Y}-a_2\right)&+\lambda_3 \left(6
   p a_{3,Y}+a_{3}\right)-3
   \lambda_4^2 p^2 \left(a_{4}-4
   \left(a_{4,Y}+a_{4,YY}\right)\right
   )\nonumber\\&\qquad\qquad\qquad\qquad+\lambda_5 \left(4 p a_{5,YY}+2
   (5 p-3) a_{5,Y}-9 a_{5}\right)=0\label{eq:fried1}\,,\\
   \lambda_2^4 p^2a_{2} +\lambda_3a_{3}
   &+\lambda_4^2 p (3
   p-2) \left(a_{4}-2
   a_{4,Y}\right)\nonumber\\&\qquad\qquad\qquad\qquad+\lambda_5 \left(2
   (1-2 p) a_{5,Y}+a_{5}
   \left(3-\frac{2}{p}\right)\right)=0\label{eq:fried2}\,,
\end{align}
where we defined
\begin{align}
\Lambda_2^2=M_{\rm pl} H_* \lambda_2^2\quad,\quad \Lambda_3=\frac{M_{\rm pl}^2}{\phi_*}\lambda_3\quad,\quad \Lambda_4^2=\frac{M_{\rm pl}^2}{2}\lambda_4^2\quad,\quad \Lambda_5=\frac{H_*^2\phi_*}{M_{\rm pl}^2\lambda_5}\,,
\end{align}
such that $\lambda_i$’s are dimensionless. Eqs.~\eqref{eq:fried1} and \eqref{eq:fried2} should be understood as algebraic relations between the functions $a_i$’s once evaluated in the power-law solution.

It should also be noted that the general equations of motion from Eq.~\eqref{eq:bgaction} have a particular structure, because $\phi$ always appears multiplied by $H$. This means that the action can be written as $S=\int d^3x dt N a^3H^2 f(Q,P,N)$ where $Q\equiv \phi H$, $P\equiv \dot\phi/(\phi H)$ and $f(Q,P,N)$ is a general function of $Q$, $P$ and $N$. Because of this structure the equations of motion reduce to
\begin{align}\label{eq:generalsolutions}
\frac{dP}{d{\cal N}}=F(P)\,,
\end{align}
where $d{\cal N}=d\ln a$ are the number of e-folds. To get to \eqref{eq:generalsolutions} one needs to use that the second Friedmann equation relates $d\ln{\cal H}/d{\cal N}$ with $Q$ and $P$ and then that the Hamiltonian constraint gives a relation between $Q$ and $P$. Though Eq.~\eqref{eq:generalsolutions} is just a formal simplification as it is impossible to solve $Q(P)$ in general from the Hamiltonian constraint,  it illustrates a couple of important features of the system. First, we see that the dynamics do not depend on the precise value of $Q$. There is also the symmetry $Q\to {\rm constant}\times Q$, which leaves $P$ invariant. This shares some similarities with shift symmetric models. Related to the shift symmetry invariance, we will later find that the scalar field perturbations are massless. Second, the perturbed background equations around the particular power-law solution (say taking $Q=Q_{p.l.}+\delta Q$ and $P=P_{p.l.}+\delta P$, where the subscript “$p.l.$” refers to evaluation at the power-law particular solution) in general reads
\begin{align}\label{eq:deltaP}
\frac{d\delta P}{d{\cal N}}=\frac{1-3p}{p}\delta P\,,
\end{align}
where we used Eqs.~\eqref{eq:fried1} and \eqref{eq:fried2} for simplifications. Integrating twice Eq.~\eqref{eq:deltaP} and using the definition of $P$ (see paragraph above Eq.~\eqref{eq:generalsolutions}), we have that
\begin{align}\label{eq:attractorsolution}
\phi=C_1+C_2 e^{\frac{1-3p}{p}{\cal N}}\,,
\end{align}
where $C_1$ and $C_2$ are constants of integration.  This implies that the expanding power-law solutions, where in our definition ${\cal N}$ increases with time, are always attractors for $p>1/3$ and $p<0$ (or equivalently $w<1$) and that the attractor is always approached exponentially fast. Contracting power-law solutions are attractors for $p<1/3$, that is $w>1$. This is the same behavior found in Refs.~\cite{Creminelli:2004jg,Khoury:2009my} in the context of a canonical, minimally coupled, scalar field in an exponential potential. We would also like to mention that this solutions could also be used for exact power-law solutions in Horndeski inflation.

\subsubsection{Examples of a minimally coupled scalar field}
\label{subsubsec:Canonical_scalar_field}
To provide context on the Horndeski power-law solutions, we show that they also contain the case of a canonical scalar field with an exponential potential \cite{Lucchin:1984yf,Russo:2004ym,Andrianov:2011fg}. We recover this case by setting $a_4=1$, $a_3=a_5=0$ and
\begin{align}
a_2(Y)=c_1+c_2 Y\,.
\end{align}
Eqs.~\eqref{eq:fried1} and \eqref{eq:fried2} then yield 
\begin{align}
c_1= \frac{(1-3 p)\lambda_4^2}{p\lambda_2^4
   }\quad,\quad c_2=\frac{\lambda_4^2}{p\lambda_2^4 }\,.
\end{align}
And, after a field redefinition given by
\begin{align}
\phi=e^{\frac{\lambda\varphi}{2M_{\rm pl}}}\,,
\end{align}
we find that
\begin{align}
G_2(\phi,X)=-\frac{1}{2}\partial_\mu\varphi\partial^\nu\varphi-V_*e^{-\frac{\lambda(\varphi-\varphi_*)}{M_{\rm pl}}}\,,
\label{eq:Canonical_scalar}
\end{align}
where $V_*=H_*^2 M_{\rm pl}^2 (3-1/p)$, $\lambda^2=2/p$ and we have absorbed the factor $\lambda_4^2$ into $V_*$ and $\lambda$ for better comparison with Ref.~\cite{Lucchin:1984yf}.

\subsubsection{Adiabatic perfect fluid}
\label{subsubsec:Adiabatic_perfect_fluid}
Another useful example is an adiabatic perfect fluid with an equation of state $\omega = 2/(3p)-1$ modeled by a scalar field. We can recover this case by setting
\begin{align}
    a_2(Y)= \frac{\lambda_4^2 (2-3p)}{\lambda_2^4 p} Y^{\frac{1}{2-3p}}\,. 
\end{align}
Performing a field redefinition given by
\begin{align}
    \phi = \frac{\phi_*}{t_*} \left( 2^{\frac{1}{2-3p}} p (2-3p) (1-3p)^{\frac{2}{-2+3p}} \lambda_4^2  \right)^{\frac{2-3p}{2-6p}} \varphi^{\frac{1}{3p-1}}\,,
\end{align}
we obtain that
\begin{align}
    G_2 = M_{\rm pl}^2 \left( - \frac{1}{2} \partial_\mu \varphi \partial^\mu \varphi \right)^{\frac{1}{2-3p}}\,.  
\end{align}

\section{Preliminaries -- linear perturbations and \\ GWs during the Horndeski phase \label{sec:preliminaries}}

We now consider perturbations on top of the flat FLRW metric. For convenience we work in the ADM decomposition and in the uniform-$\phi$ slicing (i.e.  $\delta\phi=0$), where the general line element then reads
\begin{align}
ds^2=-N^2dt^2+ a^2e^{2\zeta}(e^h)_{ij}(dx^i+N^idt)(dx^j+N^jdt)\,,
\end{align}
and $N=1+\delta N$, $N_i=\partial_i\beta$. For simplicity, we neglected vector perturbations and fixed the spatial gauge to remove the traceless scalar component of the spatial metric. For our field definition of $\phi$ (i.e. $\phi\propto t$), such gauge choice is commonly referred to as unitary gauge. The reason for our choice of gauge is twofold. First, it greatly simplifies the perturbative expansion as the functions $G_i$ only depend on the perturbations through $\delta N$. Second, we are interested in transitioning to GR and the standard hot Big Bang cosmology. We will model the transition such that GR is recovered instantaneously after $\phi$ reaches a given value, say at $\phi=\phi_t$ where “$t$” stands for transition. Thus, the uniform-$\phi$ slicing is most suitable to deal with such transition, where the matching hypersurface is well-defined \cite{Padilla:2012ze,Nishi:2014bsa}. Lastly, one must be careful when computing the spectral density of induced GWs due to the known gauge dependence \cite{Hwang:2017oxa,Gong:2019mui,Tomikawa:2019tvi,Sipp:2022kmb,DeLuca:2019ufz,Inomata:2019yww,Domenech:2020xin}. This is why the transition to GR is also important, where the gauge issue is better understood, see e.g. Refs.~\cite{Domenech:2017ems,Domenech:2020xin}. We proceed to briefly review the results of linear perturbations and qualitatively discuss the effects of the higher derivative couplings on the generation of induced GWs. 

\subsection{Linear perturbations}

The second order action in general Horndeski theory reads \cite{Kobayashi:2019hrl} (for the original paper see Ref.~\cite{Kobayashi:2011nu})
\begin{align}\label{eq:secondorderaction}
S_{2}=\int dtd^3 x \,\,\frac{a^3}{8} \left({\cal G}_t  \dot h_{ij} \dot h_{ij} +a^{-2}{\cal F}_t h_{ij}\Delta h_{ij} \right)+ a^3 \left({\cal G}_s  {\dot \zeta}^2 +a^{-2}{\cal F}_s\zeta\Delta \zeta \right)\,,
\end{align}
where, in the power-law solutions, we have that
\begin{align}
M_{\rm pl}^{-2}{\cal G}_t&=\lambda_4^2 \left(a_4-2
   a_{4Y}\right)+{\lambda_5}{p^{-2}}\left(a_5-2 p
   a_{5Y}\right)\,,\\
M_{\rm pl}^{-2}{\cal F}_t&=\lambda_4^2 a_4+{\lambda_5}{p^{-2}}a_5\,.
\end{align}
The explicit expressions for ${\cal G}_s$ and ${\cal F}_s$ in the power-law solutions can be found in App.~\ref{app:explicitperturbations}. The general expressions can be found in Ref.~\cite{Kobayashi:2019hrl}. The relevant point here is that both ${\cal G}_s$  and ${\cal F}_s$ are constant in the power-law solution. In the GR limit we have that ${\cal F}_t={\cal G}_t=1$ and ${\cal F}_s =\epsilon=1/p$. The propagation speed of tensor and scalar fluctuations are respectively defined by
\begin{align}
c_t^2={\cal F}_t/{\cal G}_t\quad{\rm and}\quad c_s^2={\cal F}_s/{\cal G}_s\,,
\end{align}
and are also constant in the power-law solutions.\footnote{Our framework provides a concrete realization of the model considered in Ref.~\cite{Chen:2024fir}, where they considered scalar induced GWs with a constant $c_t\neq1$.} We note that one could do a disformal transformation to go to the GW Einstein frame where ${\cal G}_t={\cal F}_t=c_t=1$ \cite{Creminelli:2014wna}. For the moment, we do not pursue this direction and refer the interested reader to App.~\ref{app:Newtondisformal}.  From now on and for simplicity, we work in units where $M_{\rm pl}=1$ otherwise stated.

The advantage of working with power-law solutions is that 
there exist exact solutions to the linear equations of motion of the tensor and scalar perturbations. In conformal time, defined by $d\tau=dt/a$, they read
\begin{align}\label{eq:eomzetaandh}
\frac{1}{a^2{\cal G}_t}\frac{\partial}{\partial\tau}\left(a^2 {\cal G}_t h'_{ij}\right)-c_t^2\Delta h_{ij}=0\quad,\quad
\frac{1}{a^2{\cal G}_s}\frac{\partial}{\partial\tau}\left(a^2{{\cal G}_s}\zeta'\right) -{ c_s^2}\Delta \zeta=0 \,.
\end{align}
The solutions for Fourier modes\footnote{Our convention for the Fourier expansion for the tensor mode is $h_{ij}=\int \frac{d^3k}{(2\pi)^3}h_{\mathbf{k},\lambda} e_{ij}^\lambda(\mathbf{k}) e^{i\mathbf{k}\cdot\mathbf{x}}$ where $e_{ij}^\lambda(\mathbf{k})$ are the normalized polarization tensors and $\lambda$ is the polarization. For concrete calculations we use the Left and Right circular polarization. The same Fourier expansion applies to the curvature fluctuations, albeit without polarization.} are then given by
\begin{align}\label{eq:linearsolution}
\zeta_{\mathbf k}&=A_1{\cal J}_{\nu}( c_s x)+A_2\,{\cal Y}_\nu( c_s x)\quad{\rm with}\quad \nu=1/2+b\quad{\rm and}\quad x=k\tau\,,
\end{align}
and where we defined for later convenience
\begin{align}
b=\frac{1-3w}{1+3w}=\frac{1-2p}{p-1}\,,
\end{align}
and
\begin{align}
{\cal J}_{\nu}(x)=x^{-\nu}J_{\nu}(x)\quad{\rm and}\quad {\cal Y}_{\nu}(x)=x^{-\nu}Y_{\nu}(x) \,,
\end{align}
where $J_\nu$ and $Y_\nu$ are the Bessel functions of the first and second kind of order $\nu$.
The same solutions apply to $h_{{\mathbf k},\lambda}$ simply replacing $c_s$ by $c_t$. For initial conditions  after inflation on superhorizon scales one has that $A_2=0$ and 
\begin{align}\label{eq:A1}
A_1=\zeta^{\rm p}_{\mathbf k}\times 2^{1/2+b}\Gamma[3/2+b]\,,
\end{align}
where $\zeta^{\rm p}_{\mathbf k}$ is the initial condition set by inflation. The superscript ${\rm p}$ refers to its primordial value. We will assume that $\zeta^{\rm p}_{\mathbf k}$ and $h^{\rm p}_{\mathbf k}$ are random Gaussian variables.

\subsection{Main features of secondary GWs}

Before going into the exact calculations, it is instructive to qualitatively understand the generation of induced GWs in a simplified limit. This will also help us to clarify the main differences between the generation of induced GW in Horndeski and GR. To do that, we neglect all metric perturbations, except for $h_{ij}$, and look at the effect of scalar field fluctuations $\delta\phi$. This is loosely the flat slicing but neglecting the lapse and shift. Although, strictly speaking, this is not a very accurate limit due to the mixing of gravitational and scalar degrees of freedom, it will provide us with good intuition. To simplify the calculations, we look at the leading interactions in terms of derivatives of $G_i$’s, without loss of generality. Expanding the action \eqref{eq:fulllagrangian} in that limit, we find that\footnote{For the $G_{4,X}$ term we  used that
\begin{align}
(\Delta \delta\phi)^2-\partial_i\partial_j\delta\phi\partial_i\partial_j\delta\phi=\partial_i\partial_j(\partial_i\delta\phi\partial_j\delta\phi)-\Delta(\partial_i\delta\phi\partial_i\delta\phi)\,.
\end{align}
For $G_{5,X}$ we find that the highest order interaction term from $G_{5X}G^{\mu\nu}\nabla_\mu X\nabla_\nu\phi$ which is given by $\dot h^{ij}\Delta(\partial_i\delta\phi\partial_j\delta\phi)$ (where we used that $R_{ij}\supset-\frac{1}{2}\Delta h_{ij}$) adds up to a total derivative when we consider the terms in $(\Box\phi)^3-3\Box\phi(\nabla\nabla\phi)^2+2(\nabla\nabla\phi)^3$. This can be found by considering only the highest derivative interactions of $\delta\phi$. For instance, $\Box\phi(\nabla\nabla\phi)^2\sim -a^{-4}\dot\phi \dot h_{ij}\partial_i\partial_j\delta\phi\Delta\delta\phi$ and $(\nabla\nabla\phi)^3\sim -\frac{3}{2}a^{-4}\dot\phi \dot h_{ij}\partial_i\partial_k\delta\phi\partial_j\partial_k\delta\phi$.}
\begin{align}\label{eq:hdeltaphideltaphi}
S_{h\delta\phi\delta\phi}\sim\int dt d^3x\, a\,\Bigg\{&\frac{1}{2}G_{2,X} h^{ij}\partial_i\delta\phi\partial_j\delta\phi+\dot\phi G_{3,X}h^{ij}\partial_i\delta\dot\phi\partial_j\delta\phi\nonumber\\&-\frac{1}{2a^2}G_{4,X}h^{ij}\Delta(\partial_i\delta\phi\partial_j\delta\phi)+\frac{3H\dot\phi G_{5,X}}{2a^2} \dot h^{ij}\partial_i\delta\dot\phi\partial_j\delta\phi+...\Bigg\}\,,
\end{align}
where the first term comes from $G_2\sim G_{2X}\delta X$, the second from $G_{3X}\nabla_\mu X\nabla^\mu\phi$ (after integration by parts), the fourth from $G_{4X}((\Box\phi)^2-(\nabla\nabla\phi)^2$) and the fifth from $G_{5X}\Box\phi(\nabla\nabla\phi)^2$.\footnote{Note that the qualitative interactions correspond to the terms in the action in the Newton gauge Eq.~\eqref{eq:simplifiednewton} that contain the shift vector $\beta$. This is to be expected since in the uniform-$\phi$ and the Newton gauge are related by $\beta_u\sim \delta\phi_N/\dot\phi$.} The equations of motion for induced tensor modes, after taking the variation of Eqs.~\eqref{eq:secondorderaction} and \eqref{eq:hdeltaphideltaphi} with respect to $h_{ij}$, are given by
\begin{align}\label{eq:inducedtensorequation}
\ddot h_{ij}+3{\cal H}\dot h_{ij}-\frac{c_t^2}{a^2}\Delta h_{ij}=\widehat{TT}^{ab}_{ij}S_{ab}\,,
\end{align}
where $\widehat{TT}^{ab}_{ij}$ is the transverse-traceless projector (for the explicit expression see e.g. Ref.~\cite{Domenech:2021ztg}) and the source term $S_{ij}$ reads
\begin{align}\label{eq:inducedtensorequationsource}
S_{ij}=\frac{2}{a^2{\cal G}_t}\Big(G_{2,X}& \partial_i\delta\phi\partial_j\delta\phi+2\dot\phi G_{3,X}\partial_i\delta\dot\phi\partial_j\delta\phi-\frac{G_{4,X}}{a^2}\Delta(\partial_i\delta\phi\partial_j\delta\phi)-\frac{\dot\phi H G_{5,X}}{a^2}{\partial_t}(\partial_i\delta\dot\phi\partial_j\delta\phi)\Big)\,.
\end{align}

To find the time dependence of $\delta\phi$ in the power-law solutions, we can use (after a gauge transformation from the uniform-$\phi$ gauge to the flat gauge) that 
\begin{align}
\zeta=\frac{H}{\dot\phi}\delta\phi=p\frac{\delta\phi}{\phi}\,,
\end{align}
where in the last step we used the power-law relation \eqref{eq:scalingrelation}. Then, dropping irrelevant numerical factors, we see that in the power-law solution the source term in conformal time $\tau$ is given by
\begin{align}
a^2S_{ij}\sim  a_{2,Y} \partial_i\frac{\delta\phi}{\phi}\partial_j\frac{\delta\phi}{\phi}&+\frac{a_{3,Y}}{aH}\partial_i\frac{\delta\phi'}{\phi}\partial_j\frac{\delta\phi}{\phi}\nonumber\\&-\frac{a_{4,Y}}{a^2H^2}\Delta\left(\partial_i\frac{\delta\phi}{\phi}\partial_j\frac{\delta\phi}{\phi}\right)-\frac{a_{5,Y}}{a^2H^2}\partial_\tau\left(\partial_i\frac{\delta\phi'}{\phi}\partial_j\frac{\delta\phi}{\phi}\right)\,,
\end{align}
where we also used Eq.~\eqref{eq:scalingrelation} to replace $\dot\phi$. It then follows that the equations of motion for the induced tensor modes in conformal time read
\begin{align}\label{eq:hkconformal}
h''_{\mathbf{k},\lambda}+2{\cal H}h'_{\mathbf{k},\lambda}+c_t^2k^2h_{\mathbf{k},\lambda}={\cal S}_\mathbf{k}\,,
\end{align}
with
\begin{align}\label{eq:Sksimple}
{\cal S}_\mathbf{k}\sim \int &\frac{d^3q}{(2\pi)^3}\varepsilon_{ij}(k)q_iq_j\frac{\delta\phi^{\rm p}_{\mathbf{q}}}{\phi}\frac{\delta\phi^{\rm p}_{\mathbf{k-q}}}{\phi}\nonumber\\&\times\Big\{a_{2,Y}{\cal J}_{\nu}(c_sq\tau){\cal J}_{\nu}(c_s|\mathbf{k}-\mathbf{q}|\tau)+a_{3,Y}\frac{\partial {\cal J}_{\nu}(c_sq\tau)}{\partial\ln (c_sq\tau)}{\cal J}_{\nu}(c_s|\mathbf{k}-\mathbf{q}|\tau)\nonumber\\&+a_{4,Y}{k^2\tau^2}{\cal J}_{\nu}(c_sq\tau){\cal J}_{\nu}(c_s|\mathbf{k}-\mathbf{q}|\tau)-a_{5,Y}\frac{\partial}{\partial\ln(k\tau)}\left(\frac{\partial {\cal J}_{\nu}(c_sq\tau)}{\partial\ln (c_sq\tau)}{\cal J}_{\nu}(c_s|\mathbf{k}-\mathbf{q}|\tau)\right)\Big\}\,,
\end{align}
where we used the linear solution for $\zeta$, Eq.~\eqref{eq:linearsolution}, and dropped again irrelevant numerical factors.

The integral \eqref{eq:Sksimple} cannot be done in general but since we are interested in the main characteristics we do the following approximations. First, we replace $\delta\phi^{\rm p}_{\mathbf{q}}$ by its root mean square value and consider only fluctuations on a single scale, say $k_p$. This means that for all practical purposes we take
\begin{align}
\frac{\delta\phi^{\rm p}_{\mathbf{q}}}{\phi}\sim A_{\delta\phi}\delta(\ln(q/k_p))\,.
\end{align}
Then, the source term \eqref{eq:Sksimple} is roughly given by
\begin{align}\label{eq:diracdeltasksimple}
{\cal S}_\mathbf{k}\sim &k_p^2A_{\delta\phi}^2\left(1-\frac{k^2}{4k_p^2}\right)\Theta(2k_p-k){\cal J}_{\nu}(c_sk_p\tau)\nonumber\\&\times\left\{a_{2,Y}{\cal J}_{\nu}(c_sk_p\tau)+a_{3,Y}\frac{\partial {\cal J}_{\nu}(c_sk_p\tau)}{\partial\ln (c_sq\tau)}+a_{4,Y}{k^2\tau^2}{\cal J}_{\nu}(c_sk_p\tau)+a_{5,Y}\frac{\partial^2 {\cal J}_{\nu}(c_sk_p\tau)}{\partial\ln (c_sq\tau)^2}\right\}\,,
\end{align}
where we neglected the integral over the azimuthal angle, otherwise it vanishes.\footnote{This simply means that $\langle h_{ij}\rangle = 0$, as expected from random variables with zero mean.} The Heaviside $\Theta$ comes from momentum conservation and implies $k<2k_p$. The solution to Eq.~\eqref{eq:hkconformal} can be found via the Green’s function method, which yields
\begin{align}\label{eq:greenssolutions}
h_{\mathbf{k},\lambda}=\int_0^\tau d\tau G(\tau,\tilde\tau){\cal S}_\mathbf{k}(\tilde \tau)\,,
\end{align}
where the Green’s function is given by
\begin{align}\label{eq:greensfunction}
kG(\tau,\tilde\tau)=\frac{\pi}{2}{(c_tk\tilde \tau)^{1+2\nu}}\left({\cal J}_{\nu}(c_tk\tilde \tau){\cal Y}_{\nu}(c_t k\tau)-{\cal Y}_{\nu}(c_tk\tilde \tau){\cal J}_{\nu}(c_t k\tau)\right)\,.
\end{align}
We are now ready to qualitatively explore the behavior of induced tensor modes in Horndeski.

\subsubsection{Superhorizon scalar modes}

Let us first study the growth of induced tensor modes when the scalar modes are superhorizon, that is when $c_sk_p\tau\ll1$. In that limit, the source term is given by
\begin{align}\label{eq:ssuperhorizon}
{\cal S}_\mathbf{k}\sim {k_p^2A_{\delta\phi}^2}\times\left\{a_{2,Y}+(v_p^2a_{3,Y}+a_{4,Y}+v_p^2a_{5,Y})x^2\right\}\,,
\end{align}
where we introduced the notation
\begin{align}\label{eq:xdefinition}
x=k\tau \quad,\quad x_p=k_p\tau\quad,\quad v_p=k_p/k\,,
\end{align}
and we dropped numerical prefactors and we took the leading order term in $x$ for each $a_i$. Neglecting $k^2$ term in the left hand side of Eq.~\eqref{eq:hkconformal}, it is easy to see that induced tensor modes grow  on superhozion scales as
\begin{align}\label{eq:roughsolutionsuperh}
h_{\mathbf{k},\lambda}(c_tk\ll {\cal H})\sim k_p^2A_{\delta\phi}^2\left(a_{2,Y}x^2+(v_p^2a_{3,Y}+a_{4,Y}+v_p^2a_{5,Y})x^4\right)\,.
\end{align}
This means that in the limit where $x,x_p\ll1$, there is no relevant difference with respect to the GR case. However, modifications become more important as the tensor mode approaches the time of horizon crossing. Note that, although we may have $v_p\gg1$ in Eq.~\eqref{eq:roughsolutionsuperh}, we are always within the perturbative expansion as $v_p x=x_p\ll1$.

\subsubsection{Subhorizon scalar modes}

After the scalar mode enters the sound horizon, i.e. when $c_sx_p\gg1$, the source term is approximately given by
\begin{align}\label{eq:ssubhorizon}
{\cal S}_\mathbf{k}\sim \frac{k_p^2A_{\delta\phi}^2}{(c_sx_p)^{2(1+b)}}\times\cos(2c_sx_p)\left(a_{2,Y}+a_{3,Y}v_p x+a_{4,Y}x^2+a_{5,Y}v^2_\star x^2\right)\,,
\end{align}
where again we only consider the leading order term in $x$ for each $a_i$ and we dropped numerical prefactors and phases. Note that the source term oscillates but the terms containing $a_{3,Y}$, $a_{4,Y}$ and $a_{5,Y}$ do not decay as usual. From the Green’s method solution, Eq.~\eqref{eq:greenssolutions}, we roughly have that
\begin{align}\label{eq:subhgreensrough}
h_{\mathbf{k},\lambda}(c_sx_p\gg 1)\sim\frac{k_p^2A_{\delta\phi}^2}{x^{1+b}(c_sv_p)^{2(1+b)}} &\int \frac{d\tilde x}{\tilde x^{1+b}} \sin(c_t(x-\tilde x))\nonumber\\&\times\sin(2c_sx_p)\left(a_{2,Y}+a_{3,Y}v_p x+a_{4,Y}x^2+a_{5,Y}v^2_\star x^2\right)\,.
\end{align}
Let us study Eq.~\eqref{eq:subhgreensrough} in more detail in the limits of superhorizon and subhorizon tensor modes, separately.

First, we see that, if we consider superhorizon tensor modes where $c_tx\ll1$, the solution of induced tensor modes is roughly given by
\begin{align}\label{eq:hksuperhsubhsolution}
h_{\mathbf{k},\lambda}(c_tk\ll {\cal H})\sim k_*^2A_{\delta\phi}^2\left({\rm constant}+a_{2,Y}x^{-2b}+a_{3,Y}x^{1-2b}+a_{4,Y}x^{2-2b}+a_{5,Y}x^{2-2b}\right)\,,
\end{align}
where we neglected the oscillations. Note how, unless $b$ is large enough (e.g. $b>3/2$ for $a_{5,Y}$), induced tensor modes keep growing even after the scalar mode enters the horizon. This is because, while the scalar mode decays, the source term grows due to higher derivative terms. This also suggests that if the last term in the right hand side of Eq.~\eqref{eq:hksuperhsubhsolution} dominates, the IR tail of the induced GW spectrum will have a transient slope of $k^{4-2|b|}$ instead of the standard $k^{2-2|b|}$. Note that the above discussion assumes that one can match the solution of the induced tensor modes during Horndeski to the free tensor modes in GR in an instantaneous transition. Nevertheless, it gives the right scaling for the near infrared tail of the spectrum, as we shall show in more detail in Secs.~\ref{sec:horndeskigeneral} and \ref{sec:horndeskiconcrete}.

Second, from Eq.~\eqref{eq:subhgreensrough} we can also see that there is a resonance when $c_t=2c_sv_p$, which corresponds to the case when the frequency of the homogeneous tensor mode ($c_tk$) is equal to twice the frequency of the source term ($2c_sk_p$). We then expect that the induced GW spectrum has a peak at
\begin{align}
k_{\rm res}=\frac{2c_sk_p}{c_t}\,.
\end{align}
Note that $c_s$ and $c_t$ appear as a ratio, since what matters is the relative propagation speed. For example, in the GW Einstein frame where $c_t=1$, the resonant peak is in the same expected position as in GR, where $k^{\rm GR}_{\rm res}={2c_sk_p}$.
However, the resonance in general Horndeski is in principle more pronounced as it grows as
\begin{align}\label{eq:resonantgrowth}
h_{\mathbf{k},\lambda}(k=k_{\rm res}\gg {\cal H})\sim a_{2,Y}x^{-b}+a_{3,Y}x^{1-b}+a_{4,Y}x^{2-b}+a_{5,Y}x^{2-b}.
\end{align}
We see that even in the case of radiation domination when $b=0$, there is a $x^2$ growth in addition to the standard $\ln x$ one. 

To understand the oscillatory behavior of $h_{\mathbf{k},\lambda}$ as well, let us focus  for simplicity on the case $b=0$.  Upon integration of Eq.~\eqref{eq:subhgreensrough}, we find that subhorizon induced tensor modes roughly behave as
\begin{align}\label{eq:roughsubtensor}
h_{\mathbf{k},\lambda}(c_tk\gg {\cal H})\sim\,& a_{2,Y}\frac{\sin(c_t x)}{c_t x}+a_{3,Y}\frac{\sin(2c_s v_p x)}{(2c_sv_p- c_t)x}\nonumber\\&+(a_{4,Y}+a_{5,Y})\left(\frac{\sin(2c_s v_p x)}{(2c_sv_p- c_t)}+\frac{\cos(2c_s v_p x)}{(2c_sv_p- c_t)^2x}\right)\,.
\end{align}
From Eq.~\eqref{eq:roughsubtensor}, we see that the term proportional to $a_{2,Y}$ behaves as a standard gravitational wave, i.e. it oscillates with frequency proportional to $c_t$ and decays as $1/x\sim 1/a$. The term with $a_{3,Y}$ decays in the expected way but the frequency is proportional to $2c_sv_p$. This could be attributed to kinetic gravity braiding, whereby the scalar field oscillations are imprinted on the tensor oscillations. Lastly, the terms with $a_{4,Y}$ and $a_{5,Y}$ oscillate like the scalar field squared but do not decay as standard tensor modes. The relative growth of these terms is due to the higher derivative interactions which constantly source tensor modes. In passing, we also check the abovementioned resonance/divergence for $2c_sv_p=c_t$. We see that while $a_{4,Y}$ and $a_{5,Y}$ have a divergence that goes as $(2c_sv_p-c_t)^{-2}$ in the $a_{3,Y}$ term it is $(2c_sv_p-c_t)^{-1}$. We will check this in the exact calculation. Note that there is also a resonant logarithmic term proportional to $a_{2,Y}$, which we neglect as it is the standard resonance of induced GWs in GR.

Lastly, it is important to note that the integral \eqref{eq:subhgreensrough} does not converge for $b<1$ ($w>0$) because of the higher derivative interactions. This either means that the Horndeski phase cannot last arbitrarily long or that the perturbative expansion for higher derivative interactions breaks down. Let us argue below that it is the latter case.

\subsubsection{Backreaction of higher derivatives}

So far we have considered the effect of scalar field fluctuations and higher derivative terms on the induced tensor modes. However, a similar equation to Eq.~\eqref{eq:inducedtensorequation} with a source like Eq.~\eqref{eq:inducedtensorequationsource} also applies to induced scalar field fluctuations, obviously without the tensor indices. Roughly speaking, we will encounter that
\begin{align}\label{eq:inducedscalarequation}
\delta\ddot\phi+3{\cal H}\delta\dot\phi-\frac{c_s^2}{a^2}\Delta \delta\phi=S_{\delta\phi^2}\,,
\end{align}
where $S_{\delta\phi^2}$ is very crudely proportional to the trace of Eq.~\eqref{eq:inducedtensorequationsource}. If higher derivative terms like those coming from $a_{4,Y}$ and $a_{5,Y}$ dominate in the tensor induced source \eqref{eq:inducedtensorequationsource},\footnote{We will explicitly see later in Sec.~\ref{sec:horndeskigeneral} that $a_{3,Y}$ is not as dangerous.} they will dominate the scalar induced source term in Eq.~\eqref{eq:inducedscalarequation}. Eventually $S_{\delta\phi^2}$ will be larger than $\Delta \delta\phi$. This is nothing extraordinary in the context of Horndeski theory. For example, one encounters a similar situation in the so-called Vainshtein mechanism \cite{Vainshtein:1972sx}, where higher derivative interactions are responsible for screening the fifth force due to scalar fluctuations inside massive bodies \cite{Kimura:2011dc,Koyama:2013paa,Babichev:2013usa,Kobayashi:2019hrl}.\footnote{Without entering in the higher derivative non-linear regime, higher derivative terms also affect the matter bispectrum significantly \cite{Bellini:2015wfa}.}

Unfortunately, although the general third order action for the curvature perturbation can be found in Refs.~\cite{Gao:2012ib}, the perturbative expansion in terms of derivatives breaks down. So, to solve the system, one needs to take into account the tower of all higher derivative interactions. As far as we are concerned, this is only possible in the quasi-static limit. In that limit, the Vainshtein mechanism in Horndeski has been studied in a cosmological context in Ref.~\cite{Kimura:2011dc}. Using their results, we may speculate the end point of the scalar field fluctuations after higher derivatives become important. Looking at Eqs.~(28)-(30) of Ref.~\cite{Kimura:2011dc}, we see that depending on the coefficients there can be static solutions to $\delta\phi$ even in the absence of matter. If so, the scalar field fluctuations develop a radial profile as $\delta\phi\propto r^2$, where $r$ is the radial coordinate. Otherwise, the only solution is the trivial solution where $\delta\phi={\rm constant}$. Thus, although it is only speculation, it is likely that effect of higher derivatives is to smooth out scalar field fluctuations on the very small scales, assuming that the theory is non-linearly stable. 

In this work, we place a cut-off when higher order derivatives terms in the source become as important as the linear terms. Namely, we stop our induced GW calculations whenever
\begin{align}\label{eq:roughbackreaction}
\frac{k^2}{c_s^2{\cal H}^2}\delta\phi\times 
{\rm max}\left(a_{4,Y},a_{5,Y},a_{4,YY},...\right)\sim 1\,,
\end{align}
where “...” means higher derivatives of $a_{4}$ and $a_{5}$ with respect to $Y$.\footnote{In the case of $G_3$, backreation might come from terms proportional to $\delta\phi'^2\Delta\delta\phi$. Eq.\eqref{eq:roughbackreaction} would then have similar powers of $k$ in the subhorizon limit also for $G_3$.} We will provide a more concrete expression when computing the general kernel in Sec.~\ref{sec:horndeskigeneral}.
We expect this to yield a conservative estimate for the production of induced GWs during the Horndeski phase, since it is likely that more GWs are produced regardless of the evolution of scalar modes after entering the non-linear higher derivative regime.  Although not the topic of our paper, we mention that the cut-off is also consistent with an effective field theory approach to Horndeski theories.\footnote{Note that during inflation higher derivatives interactions in Hordenski are not per se a problem. In the infinite past, when $-k\tau\gg1$, the $i\epsilon$ prescription in the in-in formalism ensures that the scalar field fluctuations are in the vacuum of the free theory. Afterwards, the exponential expansion makes sure that the higher derivatives never dominate. We thank S.~Renaux-Petel for clarifying this point.} We briefly discuss the GW energy density during the Horndeski phase in App.~\ref{app:GWenergydensity}

\section{Induced GWs during the Horndeski phase \label{sec:horndeskigeneral}}

We proceed with the detailed calculation of the induced GW spectrum generated during the Horndeski phase. We will see at the end of this section that our exact calculations match the expectations discussed in Sec.~\ref{sec:preliminaries}. We start with the third order action in the unitary gauge that contains scalar-scalar-tensor interactions, which is given by \cite{Gao:2012ib}
\begin{align}\label{eq:sshaction}
    {\cal L}_{ssh} &=  \frac{a}{2} \dot h_{ij} \partial_{i}\partial_j\beta \left( \frac{\Gamma \mathcal{G}_t}{\Theta} \dot \zeta - 3 \mathcal{G}_t \zeta \right) - a h_{ij} \left(  \mathcal{F}_t \partial_i\zeta \partial_j\zeta + \frac{2 \mathcal{G}_t ^2}{\Theta} \partial_i\zeta  \partial_j\dot\zeta - \frac{\mathcal{G}_T}{4 a^2} \Delta (\partial_i\beta \partial_j\beta) \right)\nonumber\\&
    +a\mu\left(\frac{{\cal G}_t}{\Theta}\dot h_{ij}\partial_i\dot\zeta\partial_j\zeta-\dot h_{ij}\partial_i\dot\zeta\partial_j\beta-\frac{{\cal G}_t}{a^2\Theta}h_{ij}\Delta(\partial_i\dot\zeta\partial_j\beta)+\frac{1}{2a^2}\dot h_{ij}\Delta(\partial_i\beta\partial_j\beta)\right)\,,
\end{align}
where $\beta$ is the scalar component of the shift and reads
\begin{align}
\beta=&\frac{1}{a\mathcal{ G}_t}\left(a^3\mathcal{ G}_s\Delta^{-1}\dot\zeta
-\frac{a\mathcal{ G}_t^2}{\Theta}\zeta\right)\,,\label{eq:shiftsolution}
\end{align}
with $\Delta^{-1}$ formally the inverse Laplace operator. Since the focus of the paper is on the power-law solutions only, we rewrite the action in terms of variables which are constant in the power-law solution. We denote these variables with a tilde and are defined by
\begin{align}
\beta=a\tilde \beta\quad,\quad \Theta=H \tilde \Theta \quad{\rm and}\quad \mu=\tilde\mu/H\,.
\end{align}
Then, the third order action in conformal time is given by
\begin{align}\label{eq:actionconformal2}
    a^{-2}{\cal L}_{ssh} &=   - h_{ij} \left(  \mathcal{F}_t \partial_i\zeta \partial_j\zeta + \frac{2 \mathcal{G}_t^2}{\tilde\Theta{\cal H}} \partial_i\zeta  \partial_j\zeta' - \frac{\mathcal{G}_T}{4 } \Delta (\partial_i\tilde\beta \partial_j\tilde\beta) +\frac{\tilde\mu}{{\cal H}}\frac{{\cal G}_t}{{\cal H}\tilde\Theta}\Delta(\partial_i\zeta'\partial_j\tilde\beta)\right)\nonumber\\
    &+\frac{1}{{\cal H}}h_{ij}'\left(\frac{\tilde\mu{\cal G}_t}{{\cal H}\tilde\Theta} \partial_i\zeta'\partial_j\zeta- \left(\tilde\mu+\frac{1}{2}   \frac{\Gamma \mathcal{G}_t}{\tilde\Theta}\right)\partial_i\zeta'\partial_j\tilde\beta+\frac{\tilde\mu}{2} \Delta(\partial_i\tilde\beta\partial_j\tilde\beta) +\frac{3}{2}{\cal H}   \mathcal{G}_t \partial_i\tilde\beta\partial_{j}\zeta \right)\,.
\end{align}

Taking the variation of Eq.~\eqref{eq:actionconformal2} with respect to $h_{ij}$ we arrive at the equations of motion of the induced tensor modes, namely
\begin{align}\label{eq:inducedgwsrealspace}
h''_{ij}+&2{\cal H}h_{ij}'-c_t^2\Delta h_{ij}\nonumber\\&=-\frac{4}{{\cal G}_t}\widehat{TT}^{ab}_{ij}\Bigg[ \mathcal{F}_t \partial_i\zeta \partial_j\zeta + \frac{2 \mathcal{G}_t^2}{\tilde\Theta{\cal H}} \partial_i\zeta  \partial_j\zeta' - \frac{\mathcal{G}_t}{4 } \Delta (\partial_i\tilde\beta \partial_j\tilde\beta) +\frac{\tilde\mu}{{\cal H}}\frac{{\cal G}_t}{{\cal H}\tilde\Theta}\Delta(\partial_i\zeta'\partial_j\tilde\beta)\nonumber\\
    &+\frac{1}{a^2}\frac{\partial}{\partial\tau}\left(\frac{a^2}{{\cal H}}\left(\frac{\tilde\mu{\cal G}_t}{{\cal H}\tilde\Theta} \partial_i\zeta'\partial_j\zeta- \left(\tilde\mu+\frac{1}{2}   \frac{\Gamma \mathcal{G}_t}{\tilde\Theta}\right)\partial_i\zeta'\partial_j\tilde\beta+\frac{\tilde\mu}{2} \Delta(\partial_i\tilde\beta\partial_j\tilde\beta) +\frac{3}{2}{\cal H}   \mathcal{G}_t \partial_i\tilde\beta\partial_{j}\zeta \right)\right)\Bigg]\,.
\end{align}
Using the linear solution for $\zeta$ given by Eqs.~\eqref{eq:linearsolution} and \eqref{eq:A1}, we write Eq.~\eqref{eq:inducedgwsrealspace} in Fourier modes as
\begin{align}
h''_{\mathbf{k},\lambda}+&2{\cal H}h_{\mathbf{k},\lambda}-c_t^2\Delta h_{\mathbf{k},\lambda}={\cal S}_{\mathbf k,\lambda}\,,
\end{align}
where
\begin{align}
S_{\mathbf k,\lambda}=-{4}\int \frac{d^3q}{(2\pi)^3}\varepsilon_{ij,\lambda}(\mathbf{k})q_iq_j\times f(q,|\mathbf{k}-\mathbf{q}|,\tau)\zeta^{\rm p}_{\mathbf{q}}\zeta^{\rm p}_{\mathbf{k-q}}\,.
\end{align}
The function $f(q,|\mathbf{k}-\mathbf{q}|,\tau)$ contains the transfer functions of $\zeta$ and, in the power-law solution \eqref{eq:linearsolution}, it explicitly reads
\begin{align}\label{eq:generalf}
f(v,u,x)=&\frac{2^{2b} \Gamma^2[b+\tfrac{3}{2}]}{{\cal G}_t}\Big[{\cal F}_t{\cal J}_\nu(c_svx){\cal J}_\nu(c_sux)-\frac{2{\cal G}_t^2(c_sux)^2}{\tilde \Theta(1+b)}{\cal J}_\nu(c_svx){\cal J}_{\nu+1}(c_sux)\nonumber\\&+\frac{{\cal G}_tx^2}{4}{\cal B}_\nu(c_svx){\cal B}_\nu(c_sux)-\frac{\tilde \mu{\cal G}_t(c_sux^2)^2}{\tilde\Theta(1+b)}{\cal B}_\nu(c_svx){\cal J}_{\nu+1}(c_sux)\nonumber\\
    &-\frac{x^{-2(1+b)}}{(1+b)}\frac{\partial}{\partial x}\Big({x^{3+2b}}\Big(\frac{\tilde\mu{\cal G}_t(c_sux)^2}{\tilde\Theta(1+b)} {\cal J}_\nu(c_svx){\cal J}_{\nu+1}(c_sux)+\frac{\tilde\mu x^2}{2} {\cal B}_\nu(c_svx){\cal B}_{\nu}(c_sux)\nonumber\\&+ \left(\tilde\mu+\frac{1}{2}   \frac{\Gamma \mathcal{G}_t}{\tilde\Theta}\right)\frac{(c_sux)^2}{(1+b)}{\cal B}_\nu(c_svx){\cal J}_{\nu+1}(c_sux) +\frac{3(1+b)}{2}   \mathcal{G}_t {\cal B}_\nu(c_svx){\cal J}_{\nu}(c_sux) \Big)\Big)\nonumber\\&+u\leftrightarrow v\Big]\,,
\end{align}
where, for compactness and later convenience, we have defined
\begin{align}
v=\frac{q}{k}\quad,\quad u=\frac{|\mathbf{k}-\mathbf{q}|}{k}\,,
\end{align}
and
\begin{align}\label{eq:calB}
   {\cal B}_\nu(c_sx)=-\frac{1}{\tau}\frac{\tilde\beta}{2^{1/2+b}\Gamma[3/2+b]}= \frac{{\cal G}_t}{\tilde \Theta(1+b)}{\cal J}_\nu(c_sx)-\frac{{\cal F}_s}{{\cal G}_t}{\cal J}_{\nu+1}(c_sx)\,.
\end{align}
We remind the reader that $x=k\tau$, see Eq.~\eqref{eq:xdefinition}. We also used that $d{\cal J}_\nu/dx=-x{\cal J}_{\nu+1}$.

We now use the Green's method solution, given in Eq.~\eqref{eq:greenssolutions}, to compute the two-point correlator of $h_{\mathbf{k},\lambda}$, namely
\begin{align}
    \langle h_{\mathbf{k},\lambda}h_{\mathbf{k}',\lambda'}&\rangle= (2\pi)^3\delta^{(3)}(\mathbf{k}+\mathbf{k'})\nonumber\\&\times\frac{2}{k^2}{4^2}\int \frac{d^3q}{(2\pi)^3}\varepsilon_{ij,\lambda}(k)q_iq_j\varepsilon^*_{ml,\lambda'}(k)q_mq_lI^2(u,v,x)\frac{(2\pi^2)^2}{q^3|\mathbf{k}-\mathbf{q}|^3}{\cal P}_\zeta(vk){\cal P}_\zeta(uk)\,,
\end{align}
where, as standard practice in the calculations of induced GWs, we introduced the kernel function
\begin{align}\label{eq:I}
    I(v,u,x)=\int_0^x dx_1 kG(x,x_1)f(v,u,x_1)\,,
\end{align}
and $G(x,x_1)$ is the Green's function given in Eq.~\eqref{eq:greensfunction}. The dimensionless power-spectrum\footnote{The dimensionless power-spectrum is defined by
\begin{align}
    \langle h_{\mathbf{k},\lambda}h_{\mathbf{k}',\lambda'}\rangle=\frac{2\pi^2}{k^3}{\cal P}_{h,\lambda}(k)\times (2\pi)^3\delta_{\lambda\lambda'}\delta^{(3)}(\mathbf{k}+\mathbf{k'})\,.
\end{align}} for the induced tensor modes can then be written as
\begin{align}\label{eq:Phktau}
    {\cal P}_{h}(k,\tau)&=\sum_\lambda {\cal P}_{h,\lambda}(k,\tau)\nonumber\\&=8\int_0^\infty dv\int_{|1-v|}^{1+v}du\left(\frac{4v^2-(1-u^2+v^2)^2}{4uv}\right)^2 I^2(v,u,x){\cal P}_\zeta(vk){\cal P}_\zeta(uk)\,.
\end{align}
Equipped with Eq.~\eqref{eq:Phktau}, everything boils down to the calculation of the kernel function $I(v,u,x)$ \eqref{eq:I}. Unfortunately, we have not found general analytical formulas for $I(v,u,x)$ for a general constant equation of state $w$ (or equivalently $b$). One of the reasons is that known expressions are only valid in the limit where $x\to\infty$, but as we have seen in Sec.~\ref{sec:preliminaries} some terms in the integrals do not converge. Since our main focus is the effects of modifications of gravity, we restrict ourselves to the case where the Horndeski scalar behaves as a radiation-like fluid at the background level, i.e. $w=1/3$ and $b=0$. In this case, the integration can be performed analytically. Another option is to consider a dust-like fluid with $w=0$ and $b=1$. We do not consider this case as to minimize the effect of the transition to standard GR. We leave a study for general $w$ for future work.

\subsection{Radiation-like Horndeski phase}
For simplicity and analytical viability we focus on the case when $b=0$, where the Bessel functions reduce to trigonometric functions. Then, the source term \eqref{eq:generalf} reduces, after using trigonometric identities, to
\begin{align}\label{eq:fsimplifiedincis}
f(v,u,x)=&\left(c^-_{0}+\frac{c^-_{2}}{x^2}+\frac{c^-_{4}}{x^4}\right)\cos(c_s(u-v)x)+\left(c^+_{0}+\frac{c^+_{2}}{x^2}+\frac{c^+_{4}}{x^4}\right)\cos(c_s(u+v)x)\nonumber\\&
+\left(\frac{c^-_{1}}{x}+\frac{c^-_{3}}{x^3}\right)\sin(c_s(u-v)x)+\left(\frac{c^+_{1}}{x}+\frac{c^+_{3}}{x^3}\right)\sin(c_s(u+v)x)\,,
\end{align}
where the coefficients $c_i^{\pm}$'s are constant in time, but depend on the internal momenta. Their explicit expressions are given in App.~\ref{app:explicit}. To compare with the preliminary discussion of Sec.~\ref{sec:preliminaries}, we check the asymptotic behavior of $f(v,u,x)$ on superhorizon ($ux,vx\ll1$) and subhorizon ($ux,vx\gg1$) scales. First, on superhorizon scales we have that
\begin{align}\label{eq:fsuperhorizon}
f(v,u,x\ll 1)=&\frac{3{\cal F}_s}{2{\cal G}_t} + c_t^2 -\frac{9 {\cal G}_t}{ 2 \tilde\Theta }+{\cal O}(x^2)=\frac{1}{2} \left(\frac{{\cal G}_t (3-6 p)}{\tilde\Theta  p}-c_t^2\right)+{\cal O}(x^2)\,,
\end{align}
where in the last step we used that in the power-law solutions
\begin{align}\label{eq:Fexplicit}
{\cal F}_s= \left(1 + \frac{1}{p}\right)\frac{{\cal G}_t^2}{\tilde\Theta}- {\cal F}_t\,.
\end{align}
As expected, we find a constant source in Eq.~\eqref{eq:ssuperhorizon} on superhorizon scales.\footnote{It is interesting to note that for $p\neq 1/2$ there could be no constant source on superhorizon scales if $c_t^2={\cal F}_t/{\cal G}_t=\frac{{\cal G}_t (3-6 p)}{\tilde\Theta  p}$. However, it is not clear to us what would be the main effect of the absence of a constant source on superhorizon scales. For instance, the source term in the isocurvature induced GWs \cite{Domenech:2021and} is not constant on superhorizon scales and the resulting spectrum is similar to the standard adiabatic case.} Second, on subhorizon scales we find that
\begin{align}\label{eq:fsubhorizon}
f(v,u,x\gg 1)=&c^-_{0}\cos(c_s(u-v)x)+c^+_{0}\cos(c_s(u+v)x)\nonumber\\&
+\frac{c^-_{1}}{x}\sin(c_s(u-v)x)+\frac{c^+_{1}}{x}\sin(c_s(u+v)x)\,.
\end{align}
These oscillatory terms are related to the higher derivative interactions and are similar to those we found in Eq.~\eqref{eq:ssubhorizon}. We note, though, that the oscillatory terms of Eq.~\eqref{eq:fsubhorizon} are not exclusively due to higher derivative interactions. Even in GR we have have that $c^-_{0}, c^+_{0}\neq 0$  in the unitary gauge. The origin of such oscillatory behavior in GR is the term $\Delta (\partial_i\beta \partial_j\beta)$ in Eq.~\eqref{eq:sshaction}, which seemingly has higher derivatives. Nevertheless, in GR this term is removed by a gauge transformation to, e.g., the Newton gauge. Unfortunately, we found no gauge transformation that eliminates the higher derivative terms in a general Horndeski theory. So, even in the Newton or flat gauge, the higher derivative terms are physical in Horndeski gravity. This is why we must keep track of the induced GWs in the unitary gauge until we transition to GR.  We discuss the form of the action under gauge transformations  in more detail in App.~\ref{app:thirdorder}.

We proceed to do the integrals. We first split Eq.~\eqref{eq:I} into two integrals, namely we compute
\begin{align}
I(v,u,x)=\frac{\sin (c_t x)}{c_t x} I_c(v,u,x)-\frac{\cos (c_t x)}{c_t x} I_s(v,u,x)\,,
\end{align}
where
\begin{align}\label{eq:Idefinition}
I_{c/s}(v,u,x)=\int_0^x d\tilde x \,\tilde x \left\{\begin{aligned}
\cos(c_t x)\\
\sin(c_t x)
\end{aligned}\right\}f(v,u,\tilde x)\,.
\end{align}
Due to their length, we provide the explicit expressions of the integrals in App.~\ref{app:explicit}.
 Here we study analytically the $x\gg1$ limit of the kernel, which should be a good approximation for modes which enter the horizon much before the transition to GR. Note that for the numerical calculations we will use the exact analytical solution for finite values of $x$. In the $x\gg 1$ limit, the kernel \eqref{eq:I} can be written as
\begin{align}\label{eq:explicitIxgg1}
I(v,u,x\gg1)&\approx-\frac{\sin (c_t x)}{c_t x} I_{c,0}(v,u)-\frac{\cos (c_t x)}{c_t x} I_{s,0}(v,u)\nonumber\\&+\frac{c_0^- }{2  c_s^2  u  v (1 - y_t)}\cos (c_s x (u-v))-\frac{c_0^+ }{2  c_s^2  u  v (1 + y_t)}\cos (c_s x (u+v))\nonumber\\&+\frac{1}{2(c_s^2uv(1-y_t))^2}
\left(c_0^-+\frac{c_1^-}{c_s(u-v)}
   c_s^2 u v (1-y_t)\right)\frac{\sin (c_s (u-v)x)}{c_s(u-v)x}\nonumber\\&+\frac{1}{2(c_s^2uv(1+y_t))^2}
\left(c_0^+-\frac{c_1^+}{c_s(u+v)}
   c_s^2 u v (1+y_t)\right)\frac{\sin (c_s (u+v)x)}{c_s(u+v)x}\,,
\end{align}
where
\begin{align}\label{eq:Ic0}
I_{c,0}(v,u)=&\frac{{c_0^+ (1-y_t)}-{c_0^-}{(1+y_t)}}{2c_s^2uv(1-y_t^2)}+\frac{c_t^2\left(c_0^-(1+y_t)^2+c_0^+(1-y_t)^2\right)}{2c_s^4u^2v^2(1-y_t^2)^2}+c_4^-
   c_s^2 u v\nonumber\\&+\frac{c_1^- (u-v)}{2  c_s u  v (1 - y_t)}-\frac{c_1^+ (u+v)}{2  c_s  u  v (1 + y_t)}+\frac{1}{2}\left(c_2^{-}-c_4^{-}c_s^2uv(1-y_t)\right)\ln\left|\frac{1-y_t}{1+y_t}\right|\,,\\
I_{s,0}(v,u)
=&-\frac{\pi}{2}\left(c_2^--c_4^-  c_s^2  u  v (1 - y_t)\right){\rm sign}(1+y_t)\Theta(1-y_t^2)\,,\label{eq:Is0}
\end{align}
and for compactness we defined
\begin{align}\label{eq:yt}
y_t=\frac{c_s^2(u^2+v^2)-c_t^2}{2c_s^2uv}\,.
\end{align}
Let us analyze some interesting features and properties of the Kernel Eq.~\eqref{eq:explicitIxgg1}. We will separately treat the the first line of Eq.~\eqref{eq:explicitIxgg1} from the remaining three. This is because the first line is $x$ independent and takes a similar form as the standard kernel for induced GWs in radiation domination (see, e.g., Refs.~\cite{Kohri:2018awv,Domenech:2021ztg}), while the other lines are absent in the GR (after a suitable gauge transformation).

\subsubsection{Main features of the “standard” kernel}
Let us discuss in some detail Eqs.~\eqref{eq:Ic0} and \eqref{eq:Is0}. Let us start by pointing out the similarities with the GR case. First, looking at the limit where $u\sim v\gg1$ and $y_t\sim 1$, which corresponds to the low-frequency limit of the induced GW spectrum, we find from $I_{c,0}$ \eqref{eq:Ic0} that
\begin{align}\label{eq:Ic0uvlimit}
I_{c,0}(v\sim u\gg1)\approx &\frac{{\cal G}_t^3 (\Gamma +7 \tilde\Theta )+4 {\cal G}_t^2 \tilde\Theta  \tilde\mu-{\cal G}_t \tilde\Theta  (\Gamma  {\cal F}_s+2 {\cal F}_t \tilde\Theta )-2 {\cal F}_s \tilde\Theta ^2
   \mu  }{4 c_s^2
   {\cal G}_t^2 \tilde\Theta ^2 }v^{-2}\ln\frac{2c_sv}{c_t}\,.
\end{align}
Note that there is also a $v^{-2}$ term but we considered only $v^{-2}\ln\frac{2c_sv}{c_t}$ as the dominant contribution.
Such asymptotic form of the Kernel leads to the $k^3\ln^2k$ (or $k^2\ln^2k$ for very sharp peaks) scaling of the IR tail of the induced GW spectrum in radiation domination \cite{Cai:2018dig,Cai:2019cdl,Yuan:2019wwo}. Second, we see that, for $c_s<c_t$, there is a logarithmic divergence associated to the resonance when $c_s(u+v)=c_t$ (or, equivalently, $y_t=-1$) from the last term in $I_{c,0}$ \eqref{eq:Ic0}. For $c_s>c_t$ there is no resonance.

However, looking more closely to Eqs.~\eqref{eq:Ic0} and \eqref{eq:Is0} we also see that it may be possible to have a mid-infrared slope with $v^{-4}$ scaling instead of $v^{-2}$ and that the logarithmic resonance might not be the dominant contribution to the divergent term. First, consider again the large momenta limit (that is $u\sim v\gg1$ and $y_t\to 1$) and look at the next leading order term which is proportional to $v^{-4}\ln\frac{2c_sv}{c_t}$. Let us call $v_{\ddagger}$ the value of $v$ where the terms $v^{-4}$ and $v^{-2}$ have equal contribution to the kernel. Explicitly, this is given by
\begin{align}\label{eq:v4}
v_\ddagger^2=\left|\frac{\Gamma  c_t^2 {\cal F}_s {\cal G}_t^2 \tilde\Theta +{\cal F}_s \tilde\Theta  \tilde\mu  \left(2
   c_t^2 {\cal G}_t \tilde\Theta -2 {\cal F}_s \tilde\Theta +8 {\cal G}_t^2\right)+{\cal F}_s
   {\cal G}_t \tilde\Theta  \left(3 c_t^2 {\cal G}_t \tilde\Theta +{\cal F}_s \tilde\Theta -2
   {\cal G}_t^2\right)}{2 \Gamma  c_s^2 {\cal G}_t \left({\cal G}_t^3-{\cal F}_s {\cal G}_t \tilde\Theta \right)+4
   c_s^2 {\cal G}_t \tilde\Theta  \tilde\mu  \left(2 {\cal G}_t^2-{\cal F}_s \tilde\Theta \right)+2
   c_s^2 {\cal G}_t^2 \tilde\Theta  \left(7 {\cal G}_t^2-2 {\cal F}_t \tilde\Theta \right)}\right|\,.
\end{align}
For $v<v_\ddagger$ the $v^{-4}$ terms might dominate. Thus, whenever $v_\ddagger\gg 1$, i.e. the change in slope is far from the peak, we will see a running in the near-IR proportional to $k^5\ln^2k$ (or $k^4\ln^2k$ for the very sharp peak case). Eq.~\eqref{eq:v4} might not be very helpful in the current form but we can explore some particular cases. For example, in GR we have that $v_\ddagger=1/c_s$, which is close to the location of the resonant peak. Thus, in GR we never find any $k^4\ln^2k$ regime. However, there is enough freedom in Horndeski to achieve $v_\ddagger\gg 1$. For instance, we can find values of the parameters for which the denominator in Eq.~\eqref{eq:v4} vanishes,\footnote{Note that one could ask if there is a choice of the parameters such that all the coefficients in front of all the terms $v^{-2}$ in the total Kernel \eqref{eq:explicitIxgg1} vanish. This yields 4 conditions, which could be in principle satisfied by a very particular choice of the Horndeski functions. Unfortunately, this is not an easy task to do analytically. Numerically, we found no real solutions to the system when $\mu=0$. For $\mu=1$ we found one solution which has ${\cal F}_s,{\cal F}_t,{\cal G}_t>0$ when $\Gamma\approx 0.36, {\cal F}_t\approx 1.82, \tilde\Theta\approx  0.13, {\cal G}_t\approx 0.92$. This is, however, a very fine tuned situation. And, since we will recover GR nevertheless, the universal infrared scaling \cite{Cai:2019cdl} will be recovered for those modes that enter the horizon after the transition.} so that $v_\ddagger\to\infty$. Solving for $\Gamma$ we find that  $v_\ddagger\to\infty$ when 
\begin{align}
\Gamma\to\frac{\tilde\Theta  \left(2 {\cal F}_s \tilde\Theta  \tilde\mu +2 {\cal F}_t  {\cal G}_t \tilde\Theta -7
   {\cal G}_t^3-4 {\cal G}_t^2 \tilde\mu \right)}{{\cal G}_t({\cal G}_t^2-{\cal F}_s ) \tilde\Theta}\,.
\end{align}
In the case where $\Gamma$ is the only effect different from GR (by, e.g., having $G_{4,X}=G_{5}=G_3=0$ but $G_{4,XX}\neq0$) we find that $\Gamma\to5$ will enhance the visibility of the $k^4\ln^2k$ term. We will check this later when calculating a concrete induced GW spectrum evaluated today. 

The most interesting new feature, though, is that there are divergent terms in Eq.~\eqref{eq:Ic0} which scale as $(1-y_t^2)^{-2}$ and $(1-y_t^2)^{-1}$. Thus, in the resonant limit, which is $c_s(u+v)=c_t$ and $y_t=-1$, we have a stronger divergence than in GR. If we look at the coefficients near the resonance, we find that
\begin{align}\label{eq:Ic0div}
I_{c,0}(y_t\approx -1)\approx&\frac{D_2}{16 c_s \tilde\Theta ^2 v (c_t-c_s v) (c_t-c_s (u+v))^2}
   \nonumber\\&+\frac{D_1}{16 c_s^2 {\cal G}_t^2 \tilde\Theta ^2 v^2 (c_t-c_s v)^2 (c_t-c_s
   (u+v))}+...
\end{align}
where “...” indicate the standard logarithmic divergence and other non-divergent terms. In Eq.~\eqref{eq:Ic0div}, we defined 
\begin{align}
D_2={\cal G}_t\left(c_t^2 \Gamma- {\cal G}_t\right) -2
   \tilde\mu  \left({\cal G}_t-4 c_t^2 \tilde\Theta \right)\,,
\end{align}
and we will shortly present the explicit expression for $D_1$ in a particular case. But, first, we would like to note that $D_2=0$ whenever $\tilde\mu=0$ and $\Gamma={\cal G}_t={\cal F}_t$ (and $c_t^2=1$). This case corresponds to power-law solutions with $G_{4,X}=G_{4,XX}=0$ and $G_5=0$. Thus, the $(c_t-c_s (u+v))^{-2}$ divergence is due to the presence of a non-trivial $G_4(X)$ and $G_5(X)$ functions. The explicit expression of the $D_1$ coefficient is too long in general. However, it is interesting to write it down for $\mu=0$ and $\Gamma={\cal G}_t={\cal F}_t$, where it is given by
\begin{align}
D_1=4 c_s {\cal G}_t^3 v (1 - c_s v) ({\cal G}_t - \tilde\Theta)\,.
\end{align}
Interestingly, the condition that $C_1^-=0$, which in this limit implies ${\cal G}_t=\tilde\Theta$ corresponds to the condition of no kinetic gravity braiding \cite{Deffayet:2010qz,Bellini:2014fua}. Thus, in the presence of a non-trivial $G_3(X)$ the resonant peak of the induced GW spectrum is also enhanced with respect to GR. Note that near the resonance we have that $u\sim (c_t-c_sv)/c_s$ and, since $u>0$ we have that $c_t>c_sv$. In the exact limit when $u\to0$ we also have that $v\to 1$. This means that the limit $c_t=c_s$ should be treated separatedly. However, in this case, the resonance lies exactly at the limit of the integration plane where there is the suppression due to the fact that the scalar modes become exactly parallel to the tensor mode and no more production takes place. Thus, the terms proportional to $(c_t-c_sv)$ in Eq.~\eqref{eq:Ic0div} are not per se problematic, though might change the power-law scaling of the high-frequency tail (see, e.g., Ref.~\cite{Balaji:2022dbi} in the case of GR and $c_s=1$).

\subsubsection{New terms and possible gauge artifacts}

Let us turn to the new terms unique to the Horndeski phase, which are contained in the last three lines of Eq.~\eqref{eq:explicitIxgg1}. However, we must be careful that: $(i)$ there are no gauge artifacts and $(ii)$ they survive the transition to GR. We look at the point $(i)$ below and at point $(ii)$ in a concrete example of transition in Sec.~\ref{sec:horndeskiconcrete}. Before going into the details, let us also clarify that it is known that the uniform-$\phi$ slicing is not the best choice of coordinates to compute the (so far) gauge dependent energy density of induced GWs \cite{Domenech:2020xin}. Nevertheless, we know that in GR, choices such as the Newton, Flat, synchronous and constant Hubble gauges yield the same prediction for GWs deep inside the horizon \cite{Domenech:2020xin}. Thus, we can check in which cases the new terms unique to the Horndeski (last three lines of Eq.~\eqref{eq:explicitIxgg1}) are also present in those gauges. For simplicity, we choose to transform to the Newton gauge. 

The tensor modes in the Newton gauge are related to those in the uniform-$\phi$ gauge at second order via (see Apps.~\ref{app:Newtondisformal} and \ref{app:gauge})
\begin{align}
h_{ij}^N= h^u_{ij}-a^{-2}\partial_i\beta\partial_j\beta\,,
\end{align}
where the superscripts “$N$” and “$u$” refer to Newton and uniform-$\phi$ gauge, respectively.\footnote{In addition to gauge transformation, one could also work with disformal transformations. We provide some examples in App.~\ref{app:Newtondisformal}. Also see Ref.~\cite{Cai:2023ykr,BenAchour:2024tqt} for a related discussion on disformal transformations and GWs.}
Working in Fourier space, we see that such transformation amounts to a shift of the kernel given by
\begin{align}\label{eq:kernelchange}
I^N(v,u,x)=I^u(v,u,x)-\frac{\pi}{8}x^2{\cal B}_\nu(c_svx){\cal B}_\nu(c_sux)\,,
\end{align}
where $I^u(v,u,x)$ and ${\cal B}_\nu(x)$ are given in Eqs.~\eqref{eq:explicitIxgg1} and \eqref{eq:calB}, respectively. After some algebra, the kernel in the Newton gauge, at leading order in $x\gg1$, reads
\begin{align}\label{eq:explicitIxgg1N}
I^N(v,u,x\gg1)&\approx-\frac{\sin (c_t x)}{c_t x} I_{c,0}(v,u)-\frac{\cos (c_t x)}{c_t x} I_{s,0}(v,u)\nonumber\\&+C_0^-\cos (c_s x (u-v))+C_1^-\frac{\sin (c_s (u-v)x)}{c_s(u-v)x}\nonumber\\&+C_0^+\cos (c_s x (u+v))+C_1^+\frac{\sin (c_s (u+v)x)}{c_s(u+v)x}\,.
\end{align}
We provide a detailed explanation of the coefficients $C_0^\pm$ and $C_1^\pm$, shortly. We note, though, that since $C_0^+=C_0^-(v\to -v)$ and $C_1^+=C_1^-(v\to -v)$ we will only discuss the features of $C_0^-$ and $C_1^-$. 

In the case of $C_0^-$, its general explicit expression is informative. It reads
\begin{align}\label{eq:C0minus}
C_0^-=\frac{1}{4 c_s^4 u^2 v^2 {\cal G}^2_t(y_t-1) \tilde\Theta^2}\Bigg({\cal G}_t^3(c_t^2-1) &+c_s^2 {\cal G}_t^2 (u-v)^2(\Gamma-{\cal G}_t)\nonumber\\&+\tilde\mu  \left(4 c_s^2 {\cal G}_t \tilde\Theta  (u-v)^2+4 {\cal F}_t \tilde\Theta
   -2 {\cal G}_t^2\right)\Bigg)\,.
\end{align}
An important feature of $C_0^-$ is that it vanishes when ${\cal G}_t={\cal F}_t=\Gamma$ (which also implies $c_t=1$) and $\tilde\mu=0$, which again corresponds to power-law solutions with $G_{4,X}=G_{4,XX}=0$ and $G_5=0$. This means that in such cases the cosine terms in the Kernel \eqref{eq:explicitIxgg1} are gauge artifacts. This is to be expected as there are no higher derivative interactions of the scalar field. We expect that this conclusion also holds for $G_4=f(\phi)$, i.e. standard non-minimally coupled scalar-tensor theories like Brans-Dicke. It is also interesting to note that depending on the coefficients, e.g. when $c_t=1$, $C_0^{\pm}$ could roughly depend on $v^{-2}$, which together with the $x$ term may lead to an intermediate $k^6$ scaling, at least for a Dirac delta. We will see one numerical example later.

The general expression for $C_1^-$ is rather long and uninformative. The important point is that it does not vanish in the case when ${\cal G}_t={\cal F}_t=\Gamma$ and $\tilde\mu=0$, like $C_0^-$ does. In that limit, it is given instead by
\begin{align}\label{eq:C1minus}
C_1^-=\frac{ ({\cal G}_t-\tilde\Theta) (u-v)^2}{ c_s^2 u^2 v^2 \tilde\Theta ^2\left(y_t-1\right)}\,,
\end{align}
where we also used Eq.~\eqref{eq:Fexplicit}.
Interestingly, the condition that $C_1^-=0$ corresponds again to the case of no kinetic gravity braiding \cite{Deffayet:2010qz,Bellini:2014fua}. In other words, the sine terms in \eqref{eq:explicitIxgg1} are gauge artifacts only for $G_{3,X}=0$. Thus, we can associate the sine terms in the kernel \eqref{eq:explicitIxgg1} to the presence of kinetic gravity braiding. 

We also find that the new oscillatory terms have a similar behavior to the “standard” Kernel given by  Eqs.~\eqref{eq:Ic0} and \eqref{eq:Is0}. For instance, we find that in the large momenta limit (i.e. $u\sim v\gg 1$), the cosine terms scale as $v^{-2}$. We also find a divergence proportional to $(1-y_t^2)^{-2}$ coming from the sine terms and a divergence proportional to $(1-y_t^2)^{-1}$ coming from both the sine and cosine. The exact form of the coefficients is not relevant to the present discussion as they satisfy the same limits as those discussed previously. Namely, the resonance proportional to $(1-y_t^2)^{-2}$ vanish when ${\cal G}_t={\cal F}_t=\Gamma$ and $\tilde\mu=0$ and the one proportional to $(1-y_t^2)^{-1}$ when $\tilde\Theta={\cal G}_t$. 

However, there is a very important difference with the “standard” Kernel which we anticipated in Sec.~\ref{sec:preliminaries} in Eq.~\eqref{eq:resonantgrowth}. Namely that, for the exact resonant mode, the amplification is proportional to $x^2$ and $x$, instead of $\ln x$. We can see this effect by rewriting the Kernel in the standard form, that is
\begin{align}\label{eq:standardIxgg1N}
I^N(v,u,x\gg1)&\approx\frac{\sin (c_t x)}{c_t x} I_{c}(v,u,x)-\frac{\cos (c_t x)}{c_t x} I_{s}(v,u,x)\,,
\end{align}
where now $I_{c}(v,u,x)$ and $I_{s}(v,u,x)$ contain terms such as $\cos((c_t-c_s(u+v))x)$ and $\sin((c_t-c_s(u+v))x)$. For the explicit expressions we refer the reader to App.~\ref{app:explicit}. The important point is that in the resonant limit where $c_t\approx c_s(u+v)$, the function $I_{c}(v,u,x)$ goes as
\begin{align}\label{eq:Icresonance}
I_{c}(c_t\approx& c_s(u+v),x) \nonumber\\&\approx x^2\frac{c_t{\cal G}_t^3 \left(c_t^2-1\right)+(\Gamma-{\cal G}_t)  c_s^2 c_t {\cal G}_t^2 (u+v)^2+c_t \tilde\mu 
   \left(4 c_s^2 {\cal G}_t \tilde\Theta  (u+v)^2-4 {\cal F}_s \tilde\Theta +10
   {\cal G}_t^2\right)}{16 c_s^2 {\cal G}_t^2 \tilde\Theta ^2 u v }\,.
\end{align}
The $x^2$ is absent whenever $c_t=1$, $\mu=0$ and ${\cal G}_t=\Gamma$, as expected. The $I_s$ term is more involved and we only show it in the limit when ${\cal G}_t={\cal F}_t=\Gamma=1$ and $\mu=0$, which reads
\begin{align}\label{eq:Isresonance}
I_{s}(c_t\approx& c_s(u+v),x)\approx  x \frac{(1-\tilde\Theta)(u+v)}{4c_s u v\tilde\Theta ^2}\,.
\end{align}
This means that in the case where $G_4$ and $G_5$ modifications are absent, the growth due to $G_3$ term is proportional to $x$.

In passing, we would like to note that, as we also found in Sec.~\ref{sec:preliminaries}, the fact that the frequency of the oscillations in the last two lines of Eq.~\eqref{eq:explicitIxgg1N} of the induced tensor modes follow that of the scalar field, i.e. they oscillate with a frequency proportional to $c_s(u\pm v)$ instead of $c_t$ indicates that they come from a direct mixing between the scalar field and gravity. Then, there comes the interesting question on whether these terms should be regarded as physical GWs (since they do not propagate with speed equal to $c_t$) and whether one can remove them by some kind of field redefinition \cite{Dalang:2019rke,Dalang:2020eaj,Garoffolo:2020vtd}. Unfortunately, we found no field redefinition to remove these terms in general. But, in our understanding, these terms would be seen as a “strain” on a GW detector. One might discern them from the standard GWs by their different propagation speed. In fact, we will see that once we recover GR, such $\cos(c_s(u\pm v)x)$ and $\sin(c_s(u\pm v)x)$ terms get imprinted at the transition in the induced GW spectrum as fossils of the higher derivative interactions. However, the oscillations on the induced GW spectrum are only pronounced for a very sharp peak in the primordial curvature spectrum.

\subsection{Approximate energy density of induced GWs after Horndeski phase}

Before studying the details of the transition from the Horndeski phase to standard GR in a concrete model, which we will do in the section \ref{sec:horndeskiconcrete},  it is interesting to estimate the induced GW spectrum in the general case. To do that, we consider an instantaneous transition at time $\tau=\tau_t$ with a continuous kernel, namely that $I^N_-(\tau_t)=I^N_{+}(\tau_t)$ and $(I_-^N)^\prime(\tau_t)=(I^{N}_+)^\prime(\tau_t)$, where $+$ and $-$ means after and before the transition. For $I^{N}_+$ we take free GWs in a radiation dominate universe, namely
\begin{align}\label{eq:INGRapprox}
    I^N_{+} \approx  B_1 \frac{\cos x}{x} + B_2 \frac{\sin x }{x}\,.
\end{align}
The coefficients $B_1$ and $B_2$ are fixed by the matching conditions.\footnote{Alternatively we could take the coefficients from Eq.~\eqref{eq:standardIxgg1N}, which for $c_t=1$ directly match $B_1$ and $B_2$ in Eq.~\eqref{eq:INGRapprox}.} We do not provide the explicit expressions as they are rather lengthy but we show their numerical behavior. Our simplified spectrum dismissed corrections due to the transition but, nevertheless, should be a good order of magnitude estimate for the amplitude and the spectral shape. Once in standard GR, we can evaluate the induced GW spectrum via
\begin{align}
    \Omega_{\rm GW}(\tau_t) = \frac{k^2}{12 \mathcal{H}^2} \overline{\mathcal{P}_h(k,\tau)} \Big \vert_{\tau=\tau_t}~\,,
\end{align}
where $\mathcal{P}_h(k,\tau)$ is given by Eq.~\eqref{eq:Phktau}, the overline indicates oscillation average and we used that $\overline{I^2}\to \tfrac{1}{2x^2}(B_1^2+B_2^2)$. Let us study two cases separately, namely a Dirac delta and a scale invariant primordial spectrum of curvature fluctuations. Note that our approximation is valid for $k\gg k_t$ where $k_t=a_tH_t$ is the scale that enters the Hubble radius at the transition.

\subsection{Dirac delta primordial spectrum}
\begin{figure}
    \centering
    \includegraphics[scale=0.55]{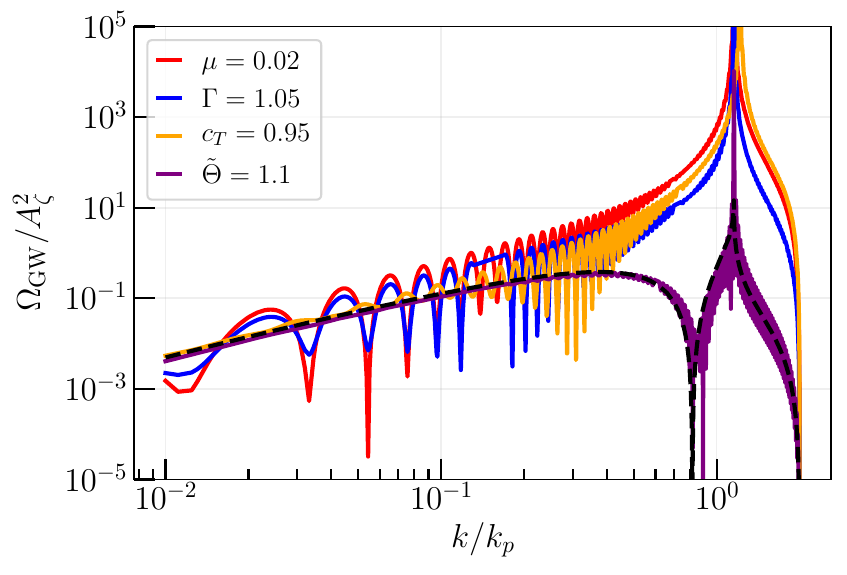}
    \includegraphics[scale=0.55]{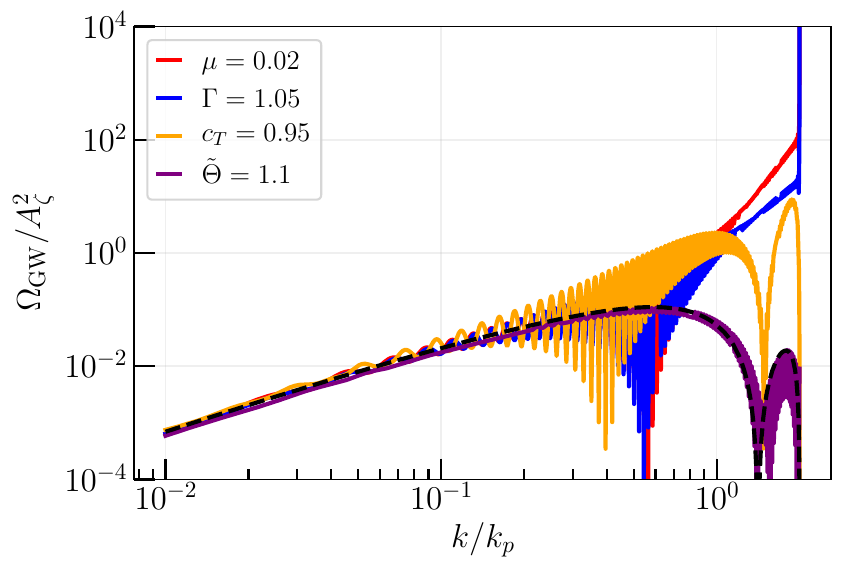}
    \caption{Induced GW spectrum in general Horndeski for a Dirac delta primordial spectrum \ref{eq:diracdeltapzeta}. On the left panel we show the case when $c_t>c_s$ fixing $c_s=1/\sqrt{3}$. On the right panel we show $c_t\leq c_s$ fixing $c_s=1$. By default we choose the GR limit (black dashed line) and then turn on one parameter at a time, these are the colored lines. In purple, we set $\tilde\Theta=1.1$. In orage, we fix $c_t=0.95$. And in blue and line we show the effect of $\Gamma=1.05$ and $\tilde\mu=0.02$, respectively. Note that some terms multiplying $\tilde\mu$ have a relative factor $v^{-2}$ in front, with respect to $\Gamma$. For $c_t<c_s$, these terms become more important as one approaches the cut-off. We note that the red line (with $\tilde\mu\neq0$) grows faster, as $k^6$, than the blue line (with $1-\Gamma\neq0$), which goes as $k^4$.}
    \label{fig:Omega_NT}
\end{figure}

We consider first Dirac Delta power spectrum, which allows for analytical calculations. In this case, the spectrum of curvature fluctuations is given by
\begin{align}\label{eq:diracdeltapzeta}
    \mathcal{P}_{\mathcal{\zeta}} = A_\zeta \delta(\ln({k}/{k_p}) )\,,
\end{align}
where $k_{p}$ is the peak scale. Then we have that
\begin{align}
    \Omega_{\rm GW} = \frac{A_{\zeta}^2 }{3{\cal G}_t^2}\frac{k_p^2}{k^2}\left( 1 - \frac{k^2}{4 k_p^2}  \right)^2 \left(B_1^2+B_2^2\right)\, \Theta(2 k_p -k)\,,
\end{align}
where the Heaviside theta comes from momentum conservation. In Fig. \ref{fig:Omega_NT} we plot the resulting gravitational wave energy density 
for a fixed transition time $k_p \tau_t=300$ and varied one of the parameters $\mu$, $\Gamma$, $\tilde \Theta$ or $\mathcal{F}_T$ and fixed the other parameters to the standard values of GR.
We can see that if we only consider cubic Horndeski $\tilde \Theta \neq 1$ the presence of the sine terms in the kernel lead to oscillations on top of the standard spectrum and the resonance gets enhanced. Varying $\mathcal{F}_T$, $\Gamma$ or $\mu$ corresponds to the higher derivative terms which leads to even more pronounced oscillations and a sharper resonance peak. Further, as discussed before, we get an intermediate power-law scaling which differs from the universal $k^2 \ln^2 k $.  Note, that by varying $\mu$ or $\Gamma $ we still get a resonance peak for $c_s=c_t=1$ at $k=2 k_p$ which is absent in standard GR since the resonance scales only logarithmic and the projection $\epsilon_{ij} q^i q^j$ vanishes quadratically at this point. However, since for the higher derivative terms the resonance is quartic the resonance is still present. Note that, though it seems that in this case the GW spectrum abruptly ends at $k=2k_p$, it is an artifact of the Dirac delta spectrum. The decay is smooth once we consider a primordial spectrum with a finite width. For the case with $c_t < c_s$ the resonance is absent despite the presence of the higher derivative terms because it is not allowed by momentum conservation.

\subsection{Scale invariant primordial spectrum \label{sec:scaleinvariantgeneral}}

It is also interesting to estimate the GW spectrum in the case of a scale invariant primordial spectrum. For instance, this is the expectation if one extrapolates the CMB measurements \cite{Planck:2018vyg,Akrami:2018odb} down to very small scales. However, due to the fact that high-$k$ modes might backreact due to the higher derivative terms, we consider that the scale invariant spectrum has a cut-off at $k=k_{\rm uv}$, namely
\begin{align}\label{eq:scaleinvariant}
    \mathcal{P}_{\mathcal{\zeta}} = A_\zeta \Theta(k_{\rm uv}-k)\,.
\end{align}

In what follows, let us focus on the case where $c_t>c_s$, which is the most interesting one because of the resonance. For a scale invariant spectrum, we have that all $k$-modes (with $k_{t}<k<k_{\rm uv}$) will eventually resonate. For non-trivial $G_4$ and $G_5$ functions we know that once the modes start to resonate their amplitude grows as $x^2$, see e.g. Eq.~\eqref{eq:Icresonance}. This means that, when evaluated at $\tau=\tau_t$, each $k$ mode will have grown approximately by a factor $(k/k_t)^2$. However, the resonance is only a small portion of the integration plane. To estimate the GW spectrum from the resonance, we follow the procedure described in Ref.~\cite{Inomata:2019ivs}.\footnote{Although Ref.~\cite{Inomata:2019ivs} deals with an instantaneous transition from matter to radiation domination, the behavior of the Kernel is similar to the one of higher derivative terms due to Horndeski gravity. The interpretation is different, though. While in Ref.~\cite{Inomata:2019ivs} the resonant induced GW are generated right after the transition, in our case the resonance is present before the transition.} We introduce the variables $z=(c_t-c_s(u+v))x_t$, which vanishes at the resonance, and $s=u-v$. The Jacobian of the transformation yields a factor $1/(2c_sx_t)$. We then take the leading order terms in $I_c$ from Eq.~\eqref{eq:Icresonance} in the limit $z\sim 0$. These terms go as $\sin(z)/z$. Then, we evaluate the integrand at $z=0$ except for the $\sin^2(z)/z^2$ term which we integrate.\footnote{To be slightly more precise, we use that $\int_{-\infty}^{\infty}\sin^2(z)/z^2=\pi$ but one could also evaluate this term at $z=0$ and get only $O(1)$ differences.} After this procedure, we obtain that
\begin{align}\label{eq:resonancegeneral}
\Omega^{\rm res}_{\rm GW}\sim &\frac{\pi A_\zeta^2  \left(c_s^2-c_t^2\right)^2 }{96 c_s {\cal G}_t^4 \tilde\Theta^4}\left(\frac{k}{k_t}\right)^3\nonumber\\&\quad\quad\times\left({\cal G}_t^2 \left({\cal G}_t-\Gamma  c_t^2\right) +\tilde\mu  \left(4 {\cal F}_s \tilde\Theta-4 c_t^2 {\cal G}_t \tilde\Theta  -10
   {\cal G}_t^2\right)\right)^2
\int_{-s_0(k)}^{s_0(k)}\frac{\left(1-s^2\right)^2}{\left(c_t^2-c_s^2 s^2\right)^4}\,,
\end{align}
where
\begin{align}\label{eq:s0}
s_0(k)=\left\{
\begin{aligned}
&1\qquad &v_{\rm uv}>\tfrac{c_t+c_s}{2c_s}\\
&2v_{\rm uv}-\frac{c_t}{c_s}\qquad & \tfrac{c_t}{2c_s}<v_{\rm uv}<\tfrac{c_t+c_s}{2c_s}\\
&0\qquad &v_{\rm uv}<\tfrac{c_t}{2c_s}
\end{aligned}
\right.\,,
\end{align}
where $v_{\rm uv}=k_{\rm uv}/k$. The spectrum has a cut-off at $k={2c_s}k_{\rm uv}/{c_t}$. In the case of $G_3$ only the spectrum will grow as $(k/k_{\rm uv})$ instead, with an amplitude proportional to Eq.~\eqref{eq:Isresonance}. It is interesting to note that, in contrast with Ref.~\eqref{eq:Icresonance}, there is no divergence in the Kernel in the limit $z=0$, but the finite contribution is enhanced by a factor $x_t^2$ with respect to the others. We conclude that the presence of higher derivative terms also enhance the scale invariant spectrum and has a typical slope of $k^3$.

\subsection{Backreaction of higher derivatives \label{sec:backreaction}}
To estimate the backreaction to the curvature fluctuations from higher derivatives, we consider the terms in the third order action of curvature fluctuations given in Ref.~\cite{Gao:2012ib} that contain the highest order of derivatives. These terms are
\begin{align}\label{eq:relevantLss}
{\cal L}_{sss}\supset&-\left(\frac{\Gamma}{2a}\left(\frac{{\cal G}_t}{\Theta}\right)^3+\frac{3\mu}{a}\left(\frac{{\cal G}_t}{\Theta}\right)^2\right)\times \dot\zeta\left(\partial_i\partial_j\zeta\partial_i\partial_j\zeta-\Delta\zeta\Delta\zeta\right)\nonumber\\&
+\frac{3}{2a}{\cal G}_t\left(\frac{{\cal G}_t}{\Theta}\right)^2\zeta\left(\partial_i\partial_j\zeta\partial_i\partial_j\zeta-\Delta\zeta\Delta\zeta\right)-\frac{2}{a}{\cal G}_t\left(\frac{{\cal G}_t}{\Theta}\right)^2\Delta\zeta\partial_i\zeta\partial_i\zeta\,,
\end{align}
where we already used the expressions for the lapse and shift, that is $\alpha$ and $\beta$, provided in Ref.~\cite{Gao:2012ib}. We did not include the terms proportional to ${\cal G}_t$ only as they can be removed by a gauge transformation. This is also the reason why we subtracted ${\cal G}_t$ from the full $\Gamma$ in Eq.~\eqref{eq:relevantLss}. After integration by parts in Eq.~\eqref{eq:relevantLss}, we find that
\begin{align}
{\cal L}_{sss}\supset\frac{1}{2}\left\{\frac{d}{dt}\left(\frac{\Gamma}{2a}\left(\frac{{\cal G}_t}{\Theta}\right)^3+\frac{3\mu}{a}\left(\frac{{\cal G}_t}{\Theta}\right)^2\right)+\frac{{\cal G}_t}{4a}\left(\frac{{\cal G}_t}{\Theta}\right)^2\right\}\times \Delta\zeta\partial_i\zeta\partial_i\zeta\,.
\end{align}
To remove possible gauge artifacts we do a gauge transformation to the Newton gauge via Eq.~\eqref{eq:gauge-uniform-newton}. After the transformation, the action in the power-law solutions and in conformal time reads
\begin{align}
{\cal L}_{sss}\supset\frac{a^2}{2}\left(\frac{{\cal G}_t}{\tilde\Theta}\right)^2\left\{\left(\frac{3}{p}-1\right)\left(\frac{\Gamma}{2}\frac{{\cal G}_t}{\tilde\Theta}+{3\tilde\mu}\right)+\frac{1}{2}{\cal G}_t-\frac{3}{2}{\cal F}_s\right\}\times \frac{1}{{\cal H}^2}\Delta\zeta\partial_i\zeta\partial_i\zeta\,.
\end{align}
One can check that the above Lagrangian vanishes in the GR limit.
Thus, in order to avoid backreaction of higher derivative interactions, we require that
\begin{align}\label{eq:backreactionbound}
\left|\frac{1}{2}\left(\frac{{\cal G}_t}{\tilde\Theta}\right)^2\left\{\left(\frac{3}{p}-1\right)\left(\frac{\Gamma}{2}\frac{{\cal G}_t}{\tilde\Theta}+{3\tilde\mu}\right)+\frac{1}{2}{\cal G}_t-\frac{3}{2}{\cal F}_s\right\}\times \frac{\Delta\zeta}{{\cal F}_s{\cal H}^2}\right|<1\,,
\end{align}
so that the third order action is always smaller than the second order action \eqref{eq:secondorderaction}. In Fourier space and in radiation domination, and assuming that the transition to GR takes place at $\tau=\tau_{t}$, we interpret the backreaction bound \eqref{eq:backreactionbound} as an upper bound on the coefficients, that is
\begin{align}\label{eq:backreactionbound2}
\frac{1}{2c_s^2{\cal F}_s}\left(\frac{{\cal G}_t}{\tilde\Theta}\right)^2\left|\frac{5\Gamma}{2}\frac{{\cal G}_t}{\tilde\Theta}+{15\tilde\mu}+\frac{1}{2}{\cal G}_t-\frac{3}{2}{\cal F}_s\right|< \frac{1}{c_sq\tau_{t}\times {{\cal P}_\zeta^{1/2}}(q)}\,,
\end{align}
where we took the root mean square of $\zeta$ in terms of its primordial dimensionless power spectrum ${\cal P}_\zeta(q)$. We also used Eqs.~\eqref{eq:linearsolution} and \eqref{eq:A1} and we neglected the oscillations as we are interested in the limit $c_sq\tau_{ t}\gg1$. Eq.~\eqref{eq:backreactionbound2} provides a rough estimation of the backreaction bound for sharp primordial spectrum, which is our main interest. Note that the physical meaning of Eq.~\eqref{eq:backreactionbound2} is that the amplification factor of the induced GWs is bounded by how much gradients of the curvature fluctuations can grow. This is the reason why the bound depends on $P_\zeta$. In the case of a theory with $G_3$ only, the third order action from Ref.~\cite{Gao:2012ib} contain terms proportional to $G_{3,XX}\zeta'^2\Delta\zeta$. If such term also backreacts, the backreation bound should be similar to Eq.~\eqref{eq:backreactionbound2}, though it is not completely clear to us up to what extent the $G_3$ term can backreact.

It is interesting to note, though, that even considering the bound \eqref{eq:backreactionbound2} we have that the dominant contribution to the induced GW spectrum, from Eqs.~\eqref{eq:explicitIxgg1N} and \eqref{eq:C0minus}, very roughly scales as
\begin{align}
\Omega_{\rm GW}(\tau_{t})\propto (\Gamma-{\cal G}_t)^2x_t^2{\cal P}_\zeta^2=\left((\Gamma-{\cal G}_t)x_t{\cal P}_\zeta^{1/2}\right)^2\times {\cal P}_\zeta\,.
\end{align}
By saturating the backreaction bound \eqref{eq:backreactionbound2}, we find that there is an upper bound on the maximum induced GWs generated during the Horndeski phase, which is crudely given by
\begin{align}\label{eq:upperbound}
\Omega_{\rm GW}(\tau_{t})\lesssim  {\cal P}_\zeta\,.
\end{align}
Thus, if we compare it to the GR prediction, which roughly is $\Omega^{\rm GR}_{\rm GW}\propto{\cal P}_\zeta^2$, the induced GWs during the Horndeski phase can be enhanced by a factor ${\cal P}_\zeta^{-1}$. Note that saturating this bound means that the backreaction takes place around the time of the transition. If the backreaction takes place earlier, the GW spectrum should be evaluated at the time of the backreation as we do not expect any more enhancement of the GW spectrum after that time.

If we now include the effect of the resonance, which is to enhance the spectrum by an additional factor $x_t$, we have that an the upper bound on the resonant peak of the induced GW spectrum is given by
\begin{align}\label{eq:upperbound2}
\Omega^{\rm peak}_{\rm GW}(\tau_{t})\lesssim  x_t^2{\cal P}_\zeta\,.
\end{align}
This means that the resonant peak can have a significant amplitude even when ${\cal P}_\zeta$ is small, if $x_t$ is large enough. For example, if we take ${\cal P}_\zeta\sim 10^{-9}$ and we include the redshift factor of $\sim 10^{-5}$ we have that for $x_t\sim 10^2$ (evaluated at the scale corresponding to the peak) the amplitude of the resonant peak is $\Omega_{\rm GW}^{\rm peak}\sim 10^{-10}$ which is observable by future GW detectors such as $\mu$-Ares, LISA, Taiji, TianQin,  DECIGO, ET and CE. For broad peaks, like the scale invariant case, we have instead that $\Omega^{\rm peak}_{\rm GW}(\tau_{t})\lesssim  x_t{\cal P}_\zeta$.

As clarification, note that, though we focused our discussion on the factor $\Gamma-{\cal G}_t$, for simplicity and because it vanishes in the GR limit, the same applies to $\tilde\mu$. We also note that Eq.~\eqref{eq:upperbound} should be understood as a rough order of magnitude estimate, where the prefactor of the upper bound \eqref{eq:upperbound} depends on the details of the Horndeski functions. We will derive more precise estimates in the concrete model we study below.

\section{Transition to standard gravity \\ and induced GW spectrum today\label{sec:horndeskiconcrete}}

In Secs.~\ref{sec:preliminaries} and \ref{sec:horndeskigeneral} we have found that, during the Horndeski phase, induced GWs appear to mix with scalar fluctuations through the higher derivative terms, leading to a time-dependence of the tensor modes which is different from what we often associate to GWs. Namely, there are terms that do not correspond to plane waves propagating with speed equal to $c_t$. Some of these terms also appear in the GR limit in the unitary gauge, which can be removed by a suitable gauge transformation to, e.g., the Newton gauge. But, there remains some doubt of whether one should call the remaining term GWs or whether there is a (complicated) field redefinition that could absorb them. Although an interesting question by itself, these ambiguities completely disappear if we follow the GW production until the universe transitions to GR. 

Similar (but not quite) to the case of an early matter dominated universe, the transition to the standard radiation dominated universe could be  important \cite{Inomata:2019ivs,Inomata:2019zqy}. In our case, we do not expect an enhancement of the GW production after the transition, like in the case of a sudden transition from matter to radiation \cite{Inomata:2019ivs}, basically because the curvature perturbation during the Hordenski dominated phase decays as usual (by construction of the power-law solutions) as long as $c_s\neq 0$. However, in general, we expect corrections  due to the transition of the same order than the amplitude generated during Hordenski phase. For instance, an instantaneous transition to GR would lead to several Dirac delta-like terms in the background and perturbation equations, which would modify the matching conditions from the Horndeski phase to GR. The source term of the induced GW will also have in general such Dirac delta-like terms. If the transition is gradual, the overall enhancement of the induced GW spectrum will decrease (simply because of decreasing coefficients in Eq.~\eqref{eq:inducedgwsrealspace}) and the oscillations due to the scalar mixing will be smeared out. Thus, the precise form of the induced GW spectrum today might depend on the details of the transition. We do not expect a big change in terms of an order of magnitude estimate, though.

In this section, we construct the simplest Horndeski model with a transition to GR that has the same evolution for the background and linear perturbations as in GR but a non-trivial source to the induced GWs. We then evaluate the induced GW spectral density once GR is recovered in the standard way, namely \cite{Inomata:2016rbd,Domenech:2021ztg}
\begin{align}\label{eq:GWstoday}.
\Omega_{\rm GW,0}h^2&=1.62\times 10^{-5}\left(\frac{\Omega_{\rm rad,0}h^2}{4.18\times 10^{-5}}\right)\left(\frac{g_{\rho}(T_c)}{106.75}\right)\left(\frac{g_{s}(T_c)}{106.75}\right)^{-4/3}\Omega_{\rm GW}(\tau_c)\,,
\end{align}
where $\tau_c\geq\tau_t$ is a time when the GWs are deep enough inside the horizon such that the energy density ratio is constant. For $k>k_t$, where $k_t=a_tH_t$, one should take the equality, i.e. $\tau_c=\tau_t$. In Eq.~\eqref{eq:GWstoday}, $g_{\rho}(T)$ and $g_{s}(T)$ are the effective relativistic degrees of freedom of the radiation fluid at a temperature $T$ (see, e.g., Refs.~\cite{Inomata:2016rbd,Saikawa:2018rcs}). We proceed to look into the details of the transition and to compute $\Omega_{\rm GW}(\tau_c)$.

\subsection{Transition to standard gravity}
Our purpose is to build a Horndeski model which transitions to GR in the simplest possible way. For analytical viability, we assume a sudden transition to GR parametrized by some kind of Heaviside-$\Theta$ function. We also assume that the system transitions from a Horndeski power-law solution to a GR power-law solution (examples of which can be found in Secs.~\ref{subsubsec:Canonical_scalar_field} and \ref{subsubsec:Adiabatic_perfect_fluid}). Since $\dot\phi$ is constant in our power-law solutions, the only option is to consider the Heaviside-$\Theta$ as function of $\phi$, say $\Theta(\phi_t-\phi)$ where $\phi_t=\phi(t_t)=\phi_*(t_t/t_*)$. To fully recover GR, we multiply all Horndeski functions in Eqs.~\eqref{eq:Gisolutions1} and \eqref{eq:Gisolutions2} by $\Theta(\phi_t-\phi)$. Furthermore, aiming for the transition to be as smooth as possible, we add the terms needed for the GR power-law solution already in the Horndeski phase. With all these requirements, we write
\begin{align}\label{eq:ansatztransition}
  & G_2= G_2^{\rm GR}(\phi,Y) + \Lambda_2^4 \left( \frac{\phi_*}{\phi} \right)^2 \tilde a_2(Y) \Theta(\phi_t - \phi)  , \quad & G_3=\Lambda_3 \left( \frac{\phi_*}{\phi} \right) a_3(Y) \Theta( \phi_t - \phi), \\
   &  G_4 = \frac{1}{2} + \Lambda_4^2\, \tilde a_4\left( Y \right) \Theta(\phi_t - \phi), \qquad & G_5 = \Lambda_5^{-1} \left( \frac{\phi}{\phi_*} \right) a_5(Y) \Theta(\phi_t - \phi)\label{eq:ansatztransition2}
\end{align}
where we set $M_{\rm pl}=1$ and $G_2^{\rm GR}$ is the standard K-essence term discussed in subsections \ref{subsubsec:Canonical_scalar_field} and \ref{subsubsec:Adiabatic_perfect_fluid}.\footnote{To compare with Eq.~\eqref{eq:Gisolutions2}, we have that $a_4(Y)=1/(2\Lambda_4^2)+\tilde a_4(Y)$.} As is clear from Eqs.~\eqref{eq:ansatztransition} and \eqref{eq:ansatztransition2}, all Horndeski terms vanish for $\phi>\phi_t$. Note that we also take into account the term $G_2^{\rm GR}$ in our analysis before the transition. For the numerical investigations, and to explore the effects of a gradual transition, we will model the step function via 
\begin{align}\label{eq:thetasmooth}
    \Theta(x) = \lim_{\xi\rightarrow \infty }\frac{1+ \tanh(\xi x)}{2}\,,
\end{align}
when necessary. 

In order to match the power-law solution in Horndeski to standard GR we use the generalized junction conditions discussed in Ref.~\cite{Padilla:2012ze,Nishi:2014bsa}. In general, derivatives of the step function will provide localized sources on the matching hypersurface at $\phi=\phi_t$ (or $q=\phi_t-\phi=0$ in the notation of \cite{Padilla:2012ze,Nishi:2014bsa}) leading to a discontinuous evolution for $\dot \phi$ and $H$ as well as for the linear perturbations. However, there are cases where the transition could be smooth for background quantities and linear perturbations, that is $\zeta$ and $h_{ij}$. We find that the simplest ansatz that leads to a smooth background is given by $\tilde a_{2}= a_{3}= a_{3,Y}=\tilde a_{4}=\tilde a_{4,Y}= a_{5}= a_{5,Y}=0$ evaluated at the power-law solution, i.e. at $Y=1$. Note that these conditions imply $c_t({Y=1})=1$. In addition we must also require that
\begin{align}
a_{5,YY}({Y=1})=-\frac{2 \lambda_4^2 p }{\lambda_5}\tilde a_{4,YY}({Y=1})\,.
\end{align}
The above condition can be understood from the cancellation of the variation with respect to $X$ of the terms containing $G_{4,X}$ and $G_{5,X}$ in the background action \eqref{eq:bgaction}. Recall the $a_2(Y=1)$ and $a_{2,Y}(Y=1)$, in the $G_2^{\rm GR}$ term, satisfy Fridemann equations \eqref{eq:fried1} and \eqref{eq:fried2}. Lastly, we require a smooth transition of the curvature perturbation, which yields
\begin{align}\label{eq:a3yycondition}
a_{3,YY}({Y=1})=-\frac{1}{3 \lambda_3}\left(p \lambda_2^4 \tilde a_{2,YY}+3 \lambda_4^2 \left((2-p) \tilde a_{4,YY}+2p
   \tilde a_{4,YYY} \right)+2 \lambda_5a_{5,YYY}\right)\,,
\end{align}
where the right hand side is also evaluated at $Y=1$. Eq.~\eqref{eq:a3yycondition}
which comes from requiring that $\dot {\cal G}_s$ has no $\delta(\phi_t-\phi)$ (we will see later that it has an inevitable $\dot\delta(\phi_t-\phi)$ term, though it turns out to be mostly harmless for our purposes). We find that Horndenski functions given by
\begin{align}\label{eq:ansatz1}
    \tilde a_2(Y) &= - \frac{4\lambda_4^2}{\lambda_2^4 }(Y-1)\quad,\quad a_3(Y) =  \frac{3 \lambda_4^2}{2\lambda_3 } (Y-1)^2 \\
    \tilde a_4(Y) &=  (Y-1)^2\quad,\quad a_5 (Y)= - \frac{\lambda_4^2}{\lambda_5} (Y-1)^2\,,\label{eq:ansatz2}
\end{align}
satisfy all these conditions in the Horndeski radiation-like dominated universe. Note that, with our choice, the background evolution before and after the transition are driven by $G_{2}^{\rm GR}$. Also note that since $\tilde a_{4,YY}\neq 0$ we maintain a non-trivial higher order cubic interaction. Also, note that $\lambda_4^2$ is our only free parameter, as the other $\lambda_i$’s drop out in $G_i$’s. 

\begin{figure}
    \centering
    \includegraphics[scale=0.6]{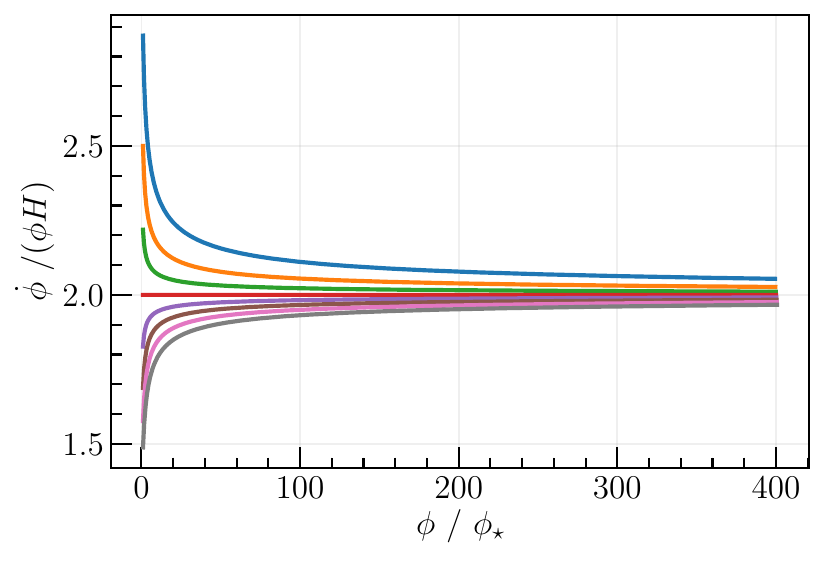}
    \caption{Phase space of $\phi$ during the transition regime for a gradual transition $\xi=5\times 10^{-3}$, see Eq.~\eqref{eq:thetasmooth}. Note how the scalar field approaches the power-law solution, that is given by constant line at $\dot\phi/(\phi H)=2$. We set the transition at $\phi_t=200 \phi_\star$. See how it does not change the overall behavior.}
    \label{fig:Scaling_solution_BG}
\end{figure}

For the above parameter choice the transition does not impact the background evolution as the power-law solution is the same before and after the transition and independent of the higher derivative terms. Recall that we required that the localized sources vanish on the matching hypersurface. In Fig. \ref{fig:Scaling_solution_BG} we plot the background evolution confirming the attractor behavior of the power-law solution given by Eq.~\eqref{eq:attractorsolution}. We also see that the attractor behavior is not impacted by the transition, even for initial conditions outside of the attractor.

\subsubsection{Linear perturbations at the transition}

Let us study now linear perturbations. Using the ansatz given in Eq.~\eqref{eq:ansatz1} and \eqref{eq:ansatz2}, which allows for a smooth background transition, we find that the coefficients for the linear perturbations read
\begin{align}
    \label{eq:Choice_a_i}
    \mathcal{G}_t = 1\,, \qquad \mathcal{F}_t=  1\,, \qquad \mathcal{G}_s = \frac{\epsilon}{c_s^2} - 12  {\lambda_4^2}\frac{\dot \phi}{H} \delta(\phi_t - \phi)\,, \qquad \mathcal{F}_s = \epsilon\,,
\end{align}
where $\epsilon=1/p=2$ for radiation domination. Eq.~\eqref{eq:Choice_a_i} yields equations of motion for $\zeta$ and $h_{ij}$, i.e. Eq.~\eqref{eq:eomzetaandh}, which are exactly like the ones in GR for an adiabatic perfect fluid or for a scalar field.

Since there is no jump in ${\cal G}_t$ and ${\cal F}_t$, we recover the usual matching conditions of GR for the tensor modes, namely
\begin{align}
    [h_{ij}]_-^+=0, \qquad [\dot h_{ij}]_-^+=0,
\end{align}
where $+$ and $-$ respectively indicate evaluation after and before the transition. This result is convenient as the Green’s function for the tensor modes is unchanged throughout at the transition. Thus, to study the effect of the transition on the induced GWs we shall only look at the source term $f(v,u,x)$ given by Eq.~\eqref{eq:generalf}.

The matching of curvature perturbations is not as trivial. For instance, from Eq.~\eqref{eq:Choice_a_i} and the matching conditions of Ref.~\cite{Nishi:2014bsa}, we obtain that $[\zeta ]_-^+=0$. One would then expect that the Dirac delta in ${\cal G}_s$ \eqref{eq:Choice_a_i} leads to a localized source at the matching hypersurfaces. However, after integrating once the linear equations of motion, we have that
\begin{align}\label{eq:conditionstrange}
    & \int_{t_t-\varepsilon}^{t_t + \varepsilon} {\rm d} t \frac{{\rm d} }{{\rm d} t} \left( \frac{\epsilon}{c_s^2} a^3 \dot \zeta \right) = 12  \lambda_4^2  \int_{t_t-\varepsilon}^{t_t + \varepsilon} {\rm d}t  \frac{{\rm d}}{{\rm d} t} \left( \frac{a^3 \dot \phi}{H}\dot \zeta \delta(\phi_t - \phi)\right)=0\,,
\end{align}
which due to being a total derivative does not yield much inside as it vanishes trivially. It indicates, however, that there might no jumps in $\zeta'$ and $\zeta''$. To see this, it is more helpful to look at the equations of motion for $\zeta$ \eqref{eq:eomzetaandh} directly, namely
\begin{align}
  \frac{1}{a^2{\cal G}_s}\frac{\partial}{\partial\tau}\left(a^2{{\cal G}_s}\zeta'\right)+\frac{\epsilon}{{\cal G}_s}k^2 \zeta=0 \,.
\end{align}
For sudden transitions we have that ${\cal G}_s\gg 1$ near the transition time. For scales such that ${\cal G}_s\gg \epsilon k^2$ we can ignore the gradient terms and take the $k\to0$ solution to $\zeta$ which is given by
\begin{align}
\zeta'\approx\frac{\rm constant}{a^2{{\cal G}_s}}\,.
\end{align}
We can see that, before the transition, $\zeta$ follows the standard evolution, very close to $\tau_t$ then ${\cal G}_s$ blows up and $\zeta'$ quickly decreases. After the transition, $\zeta$ again behaves as usual. For $\zeta''$ we have, at leading order in ${\cal G}_s$, that
\begin{align}\label{eq:zetaprimeprime}
    \zeta''\approx -\zeta'\frac{{\cal G}_s'}{{\cal G}_s}\,.
\end{align}
Since ${\cal G}_s$ first increases and then decreases we have that, around the transition time, $\zeta''$ first decreases and then increases. After that it follows the standard evolution. Note that for a sudden but finite width transition, we have that ${\cal G}_s'(\tau=\tau_t)=0$ simply because ${\cal G}_s$ peaks at the transition time. If we take the exact instantaneous transition limit, the effect of an Dirac delta in ${\cal G}_s$ seems to be to completely freeze the curvature perturbation, i.e. $\zeta'=\zeta''=0$. So, in that sense, the transition in the exact instantaneous limit might not well-defined. In what follows, we consider a sudden but finite width transition with the approximation that the curvature perturbation is not modified by the transition, i.e. we take that $[\dot\zeta ]_-^+=0$. We expect this to be a good approximation, as numerical calculations confirm very minor changes to $\zeta$. We will also consider small values for $\lambda_4$ in Eq.~\eqref{eq:Choice_a_i}, thus reducing the impact of the transition.

\begin{figure}
    \centering
     \includegraphics[scale=0.55]{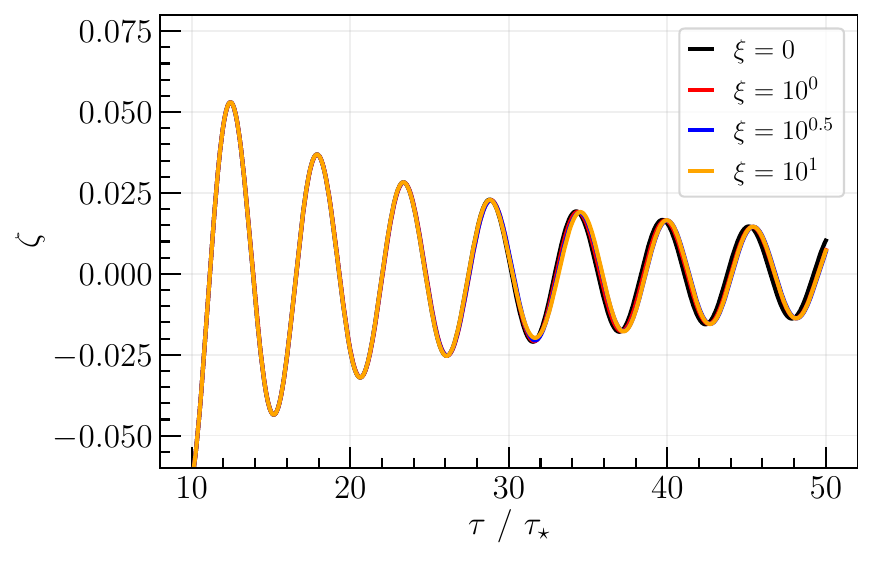}
    \includegraphics[scale=0.55]{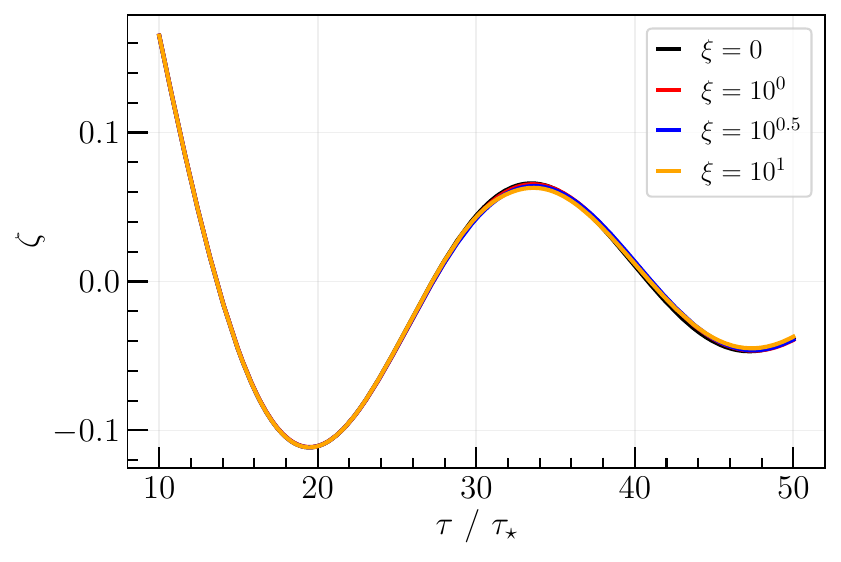}
    \caption{Evolution of the curvature perturbation for three different transition widths, namely $\xi=10^{0}$ (red), $\xi=10^{0.5}$ (blue)  and $\xi=10^{1}$ (orange), compared to the GR evolution without transition $\xi =0$ (black), for $k/k_*=2$ (left figure) and $k/k_*=4\times10^{-1}$ (right figure). We set the transition at $\tau_t/\tau_*=10^{1.5}$ and the initial conditions are set at $\tau_*=1$ where the equations of motion are not impacted by the transition. Note how the transition leads to a phase shift and a slight damping, both of which depend on $k$. }
    \label{fig:Lin_Evolution}
\end{figure}

To confirm the above discussion, we show the numerical solution for the curvature perturbation for two different values of $k/k_*=2$ (left hand side) and $k/k_*=4 \times 10^{-1}$ (right hand side), and for three different transition widths $\xi$ \eqref{eq:thetasmooth} in comparison to the GR evolution without transition $\xi=0$, in Fig.~\ref{fig:Lin_Evolution}. To make the impact more visible we fix $\lambda_4=0.1$ in the figure. We find that when the transition is gradual (though not necessarily slow in terms of a Hubble time), there appears a phase shift to the curvature perturbation, which depends on the value of $k$. The $k$-dependence can be understood by noting that the effect of the transition is a change in the friction term and sound speed in the vicinity of the transition point.  As the transition becomes more sudden, the impact on the evolution of the curvature perturbation becomes more pronounced, with its amplitude related to the size of the modified gravity terms, i.e. $\lambda_4$. For smaller values, which we will later consider for the induced GW spectrum the impact becomes non-visible. Thus, we only take an instantaneous transition to GR for simplicity as a good enough estimate.

\subsubsection{Source of the induced GWs through the transition}

\begin{figure}
    \centering
    \includegraphics[scale=0.55]{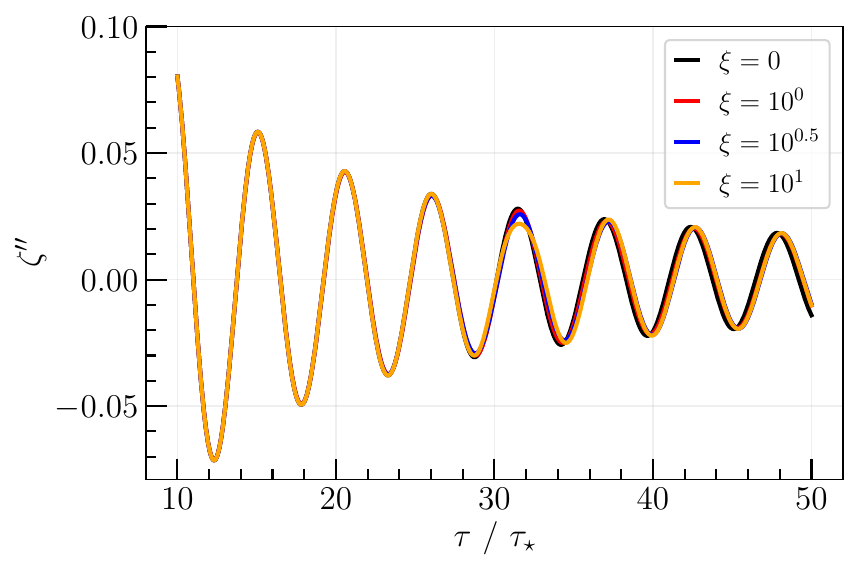}
    \includegraphics[scale=0.55]{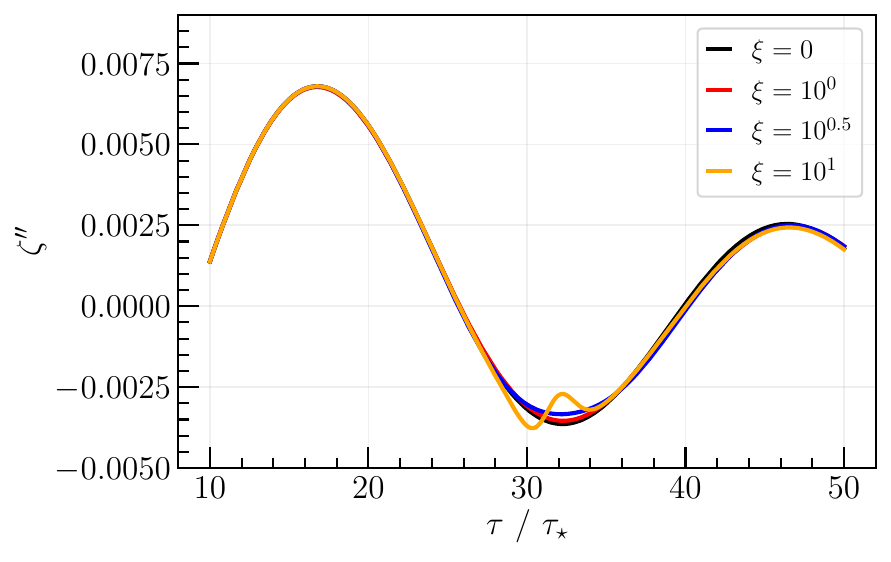}
    \caption{Evolution of the second derivative of the curvature perturbation for three different transition widths, namely $\xi=10^{0}$ (red), $\xi=10^{0.5}$ (blue)  and $\xi=10^{1}$ (orange), compared to the GR evolution without transition $\xi =0$ (black), for $k/k_*=2$ (left figure) and $k/k_*=4\times10^{-1}$ (right figure). We set the transition at $\tau_t/\tau_*=10^{1.5}$ and the initial conditions are set at $\tau_*=1$ where the equations of motion are not impacted by the transition. }
    \label{fig:Second_Derivative_Evolution}
\end{figure}

We now explore the effect on the source term to the induced GWs. The coefficients relevant to the induced GWs for our choice given by Eq.~\eqref{eq:Choice_a_i} read
\begin{align}\label{eq:gammatilde}
    \mu=0\quad{\rm and}\quad \Gamma=1+  16 \lambda_4^2 \Theta( \phi_t - \phi)=1+\tilde \Gamma \,\Theta(\phi_t -\phi),
\end{align}
where we introduced $\tilde \Gamma= 16 \lambda_4^2$ for later convenience as it parametrizes all deviations from GR. Plugging Eqs.~\eqref{eq:Choice_a_i} and \eqref{eq:gammatilde} in the source \eqref{eq:generalf}, we find that $f(v,u,x)$ is modified via 
\begin{align}\label{eq:change_of_f}
    f(v,u,x)  \to & f(v,u,x) +\Delta f(v,u,x_t)\tilde\Gamma\delta(x_t-x) ,
\end{align}
where in the right hand side of Eq.~\eqref{eq:change_of_f} the function $f(v,u,x)$ is given by Eq.~\eqref{eq:fsimplifiedincis} and we defined
\begin{align}
\Delta f(v,u,x_t)=&\left({u^2-u v+v^2}+\frac{u^2+v^2}{c_s^2 u v x_t^2}\right)\frac{\cos(c_s(u-v)x_t)}{c_s^2 u^2 v^2 x_t}\nonumber\\&
-(u-v)\left(1-\frac{2 \left(u^2+v^2\right)}{c_s^2 u^2 v^2 x_t^2}\right)\frac{\sin(c_s(u-v)x_t)}{2c_suv}+(v\to -v).
\end{align}
Note that $(v\to -v)$ means to take the whole expression and replace $v$ for $-v$ even in the symmetric terms. Thus, taking into account the transition adds an additional localized source on the matching hypersurface. It should be noted that there could also be a localized source from $\zeta''$ or $\tilde\beta'$. It turns out that this terms do not yield any Delta function because of the matching condition \eqref{eq:conditionstrange}. We also checked numerically, which we show in Fig. \ref{fig:Second_Derivative_Evolution}, that $\zeta^{\prime\prime}$ does not jump at the transition (see also the discussion below Eq.~\eqref{eq:zetaprimeprime}). Therefore, as a good approximation we can consider a continuous evolution.

We write the Kernel in the standard GR regime and in the Newton gauge as
\begin{align}
    I^N(v,u,x)= &  \frac{\sin x}{x} \left( -I^{\rm GR}_c(v,u)  +  \tilde \Gamma \times {\cal I}_c^{\tilde \Gamma}(v,u,x_t) \right)  -  \frac{\cos x}{x} \left(  I^{\rm GR}_s(v,u) + \tilde \Gamma \times {\cal I}_s^{\tilde \Gamma}(v,u,x_t)   \right)\,,
\end{align}
where we did the gauge transformation to the Newton gauge using Eq.~\eqref{eq:kernelchange} (in the GR limit) and we defined $I^{\rm GR}_{c/s}$ as the Kernel in GR and included in ${\cal I}^{\tilde \Gamma}_{c/s}$ all terms proportional to $\tilde\Gamma$. It should be noted that since $c_t=1$ and $h_{ij}$ is continuous we use Eq.~\eqref{eq:Idefinition} to compute $I(v,u,x)$ for any value of $x$. This means that to compute $I^{\rm GR}_{c/s}(v,u)$ we can take the limit $x\gg 1$ as these particular terms are not affected by the transition. Instead, the term ${\cal I}_{c/s}^{\tilde \Gamma}(v,u,x_t)$ stops at $x_t$ because there is no more “Horndeski” source after $x=x_t$.

For completeness, we provide the explicit expressions for $I^{\rm GR}_{c/s}$ (which follow from Eq.~\eqref{eq:Ic0} and \eqref{eq:Is0} in the GR limit), namely
\begin{align}\label{eq:IcGR}
I^{\rm GR}_{c}(u,v)&=-\frac{y_t}{c_s^2uv}\left(1-\frac{1}{2}y_t\,\ln\left|\frac{1+y_t}{1-y_t}\right|\right)\,,\\
I^{\rm GR}_{s}(u,v)&=\frac{\pi y_t^2}{2 c_s^2 u v}\,{\rm sign}(1+y_t)\Theta(1-y_t^2)\,,\label{eq:IsGR}
\end{align}
with $y_t$ given by Eq.~\eqref{eq:yt} when $c_t=1$. We note that Eqs.~\eqref{eq:IcGR} and \eqref{eq:IsGR} agree with Refs.~\cite{Kohri:2018awv,Domenech:2019quo,Domenech:2020kqm,Domenech:2021ztg}. The new, Horndeski, terms read
\begin{align}
{\cal I}^{\tilde\Gamma}_{c}(u,v,x_t)&=x_t\cos(x_t)\Delta f(v,u,x_t)+I_c^{\tilde\Gamma}(u,v,x_t)-I_{c,0}^{\tilde\Gamma}(u,v)\,,\\
{\cal I}^{\tilde\Gamma}_{s}(u,v,x_t)&=x_t\sin(x_t)\Delta f(v,u,x_t)+I_s^{\tilde\Gamma}(u,v,x_t)\,,
\end{align}
where $I_{c/s}^{\tilde\Gamma}$ are given by Eqs.~\eqref{eq:generalIc} and \eqref{eq:generalIs} after subtracting Eqs.~\eqref{eq:IcGR} and \eqref{eq:IsGR} evaluated at $x_t$. Their explicit expression is given by
\begin{align}
I_c^{\tilde\Gamma}(u,v,x_t)&=\frac{1}{16u^3v^3}\left\{{\left(\frac{u^2+v^2}{c_s^4 x_t^2}-\frac{u^2 v^2 (u-v)^2}{y_1^2}-\frac{2 u v (u-v) \left(u^2+v^2\right)}{c_s y_1}\right) \cos (y_1x_t)}\right.\nonumber\\&+{\left(\frac{u^2+v^2}{c_s^4 x_t^2}-\frac{u^2 v^2 (u-v)^2}{y_4^2}+\frac{2 u v (u-v) \left(u^2+v^2\right)}{c_s y_4}\right) \cos ( y_4x_t)}\nonumber\\&-\left({u^2 v^2 x^2 (u-v)^2}+\frac{\left(u^2+v^2\right)}{c_s^4}
   y_1y_4\right) \left(\frac{\sin ( y_1x_t)}{y_1x_t}+\frac{\sin (
   y_4x_t)}{y_4x}\right)\nonumber\\&\frac{\left(u^2+v^2+c_s^2 (u^4+v^4)\right)}{c_s^4} \left(\text{Ci}(y_1x_t)+\text{Ci}(y_4x_t)\right)\Bigg\}+(v\to-v)\,,
\end{align} 
\begin{align}
I_{c,0}^{\tilde\Gamma}(u,v)=&\frac{u^2+v^2+c_s^2 (u^4 +v^4)}{32 c_s^4 u^3 v^3}\ln\left|\frac{1+y_t}{1-y_t}\right|+\frac{u^2+v^2}{8 c_s^2 u^2 v^2}\nonumber\\&+
\frac{(u-v)^2}{4 u v\left(1-c_s^2 (u-v)^2\right)}\left(\frac{ \left(1+c_s^2 (u-v)^2\right)}{2 \left(1-c_s^2
   (u-v)^2\right)}+\frac{\left(u^2+v^2\right)}{u v }\right)+(v\to-v)\,,
\end{align}
and
\begin{align}
I_s^{\tilde\Gamma}(u,v,x_t)&=\frac{1}{16u^3v^3}\Bigg\{{\left(\frac{u^2+v^2}{c_s^4 x^2}-\frac{u^2 v^2 (u-v)^2}{y_1^2}-\frac{2 u v (u-v) \left(u^2+v^2\right)}{c_sy_1}\right) \sin (y_1x_t)}\nonumber\\&+{\left(\frac{u^2+v^2}{c_s^4 x^2}-\frac{u^2 v^2 (u-v)^2}{y_4^2}+\frac{2 u v (u-v) \left(u^2+v^2\right)}{c_s y_4}\right) \sin (y_4x_t)}\nonumber\\&+{\left(\frac{\left(u^2+v^2\right) y_1y_4}{c_s^4}-{u^2 v^2
   x_t^2 (u-v)^2}\right) \frac{\cos (y_1x_t)}{
   y_1x_t}}\nonumber\\&+{\left(\frac{\left(u^2+v^2\right) y_1y_4}{c_s^4}+{u^2 v^2 x_t^2 (u-v)^2}\right)\frac{\cos (y_4x_t)}{y_4x_t}}\nonumber\\&+{\frac{\left(u^2+v^2+c_s^2 (u^4+ v^4)\right)}{c_s^4} \left(\text{Si}(y_1x_t)+\text{Si}(y_4x_t)\right)}\Bigg\}+(v\to-v)\,,
\end{align}
where, for compactness, we used the definitions on App.~\ref{app:explicit} for $y_1=1-c_s(u-v)$ and $y_4=1+c_s(u-v)$. 

Lastly, and most important for the calculations of the induced GW spectrum, the averaged Kernel squared is given by
\begin{align}
\overline{(I^N(u,v,x))^2}=\frac{1}{2x^2}\left\{\left(I^{\rm GR}_c(v,u)  -  \tilde \Gamma \times {\cal I}_c^{\tilde \Gamma}(v,u,x_t) \right)^2+\left(I^{\rm GR}_s(v,u)  +  \tilde \Gamma \times {\cal I}_s^{\tilde \Gamma}(v,u,x_t) \right)^2\right\}
 \end{align}
 We then compute the induced GW spectrum at $\tau=\tau_c$ as
 \begin{align}\label{eq:omegataucparticular}
    \Omega_{\rm GW}(\tau_c)=\frac{1}{3}&\int_0^\infty dv\int_{|1-v|}^{1+v}du\left(\frac{4v^2-(1-u^2+v^2)^2}{4uv}\right)^2 {\cal P}_\zeta(vk){\cal P}_\zeta(uk)\nonumber\\&
    \times \left\{\left(I^{\rm GR}_c(v,u)  -  \tilde \Gamma \times {\cal I}_c^{\tilde \Gamma}(v,u,x_t) \right)^2+\left(I^{\rm GR}_s(v,u)  +  \tilde \Gamma \times {\cal I}_s^{\tilde \Gamma}(v,u,x_t) \right)^2\right\}\,.
\end{align}

\subsection{Induced Gravitational Wave Spectrum}
We proceed now to study the induced GW spectrum given by Eq.~\eqref{eq:omegataucparticular} in three particular cases: a sharp peak modeled by a Dirac Delta, a log-normal peak and a flat (scale invariant) spectrum. This will provide us with a general understanding of the impact of the higher derivative terms on the induced GW spectrum.

\subsubsection{Dirac Delta power spectrum}
We consider first the simplest case of a very sharp peak in the curvature power spectrum modeled by a Dirac Delta, given by Eq.~\eqref{eq:diracdeltapzeta}. Plugging Eq.~\eqref{eq:diracdeltapzeta} in Eq.~\eqref{eq:omegataucparticular}, and using that $u=v=v_p=k_p/k$, leads us to
\begin{align}\label{eq:omegagwsdirac}
    \Omega_{\rm GWs}(\tau_c) = \frac{2A_{\zeta}^2}{3} \frac{k_p^2}{k^2}\left( 1 - \frac{k^2}{4 k_p^2}  \right)^2 \overline{(I^{N}(k_p/k,k_p/k,x_t))^2}\, \Theta(2 k_p -k)\,.
\end{align}
For the far infrared tail $x_t \ll 1$ we note that the contribution from the Horndeski gravity terms are suppressed by a power of $\mathcal{O}(\tilde \Gamma x_t^2)$ in comparison to the ones from standard GR. Therefore, far in the infrared we recover asymptotically the results of induced GWs in standard GR. This is to be expected as modes with $k\ll k_t$ entered the horizon after the Hordenski phase ended. Thus, we recover the standard behavior of the infrared tail for induced GWs \cite{Cai:2019cdl}, namely the $k^2\ln^2k$ scaling for the Dirac delta \cite{Cai:2018dig,Yuan:2019wwo}.

The most interesting regime, and where we get new effects, is when the Dirac delta peak enters the horizon before the transition to standard GR, that is $k_p \tau_t > 1$. Let us then focus on the induced GWs with $x_t=k/k_t>1$. First, we find oscillations proportional to $\sin^2(1\pm 2c_sv)$, the amplitude of which grows as $x_t^2$. Indeed, expanding Eq.~\eqref{eq:omegagwsdirac} for $x_t \gg 1$ we obtain
\begin{align}\label{eq:Icdiracdelta}
     I_c^{\tilde \Gamma}\left(v_p,v_p,x_t\right)   \approx &\, x_t \left(\frac{(1-4 c_s v_p) \sin ((1-2 c_s v_p)x_t)}{8 c_s v_p (1-2 c_s
   v_p)}+\frac{(1+4 c_s v_p) \sin ((1+2 c_s v_p)x_t)}{8 c_s v_p(1+2 c_s v_p)}\right)\,,
\end{align}
and
\begin{align}\label{eq:Isdiracdelta}
     I_s^{\tilde \Gamma}\left(k_p/k,k_p/k,x_t\right)   \approx &\, x_t \left(\frac{(1-4 c_s v) \cos ((1-2 c_s v_p)x_t)}{8 c_s v (1-2 c_s
   v)}-\frac{(1+4 c_s v) \cos ((1+2 c_s v_p)x_t)}{8 c_s v(1+2 c_s v_p)}\right)\,.
\end{align}
As is clear from Eq.~\eqref{eq:Icdiracdelta} and \eqref{eq:Isdiracdelta}, the amplitude and the frequency of the oscillations grow with $k$. Thus, in the Dirac delta case, the oscillations due to the mixing of tensor and scalar modes, is visible in the induced GW spectrum. However, we will later see that a finite width of the primordial spectrum will smear out the oscillations.

\begin{figure}
    \centering
    \includegraphics[scale=0.55]{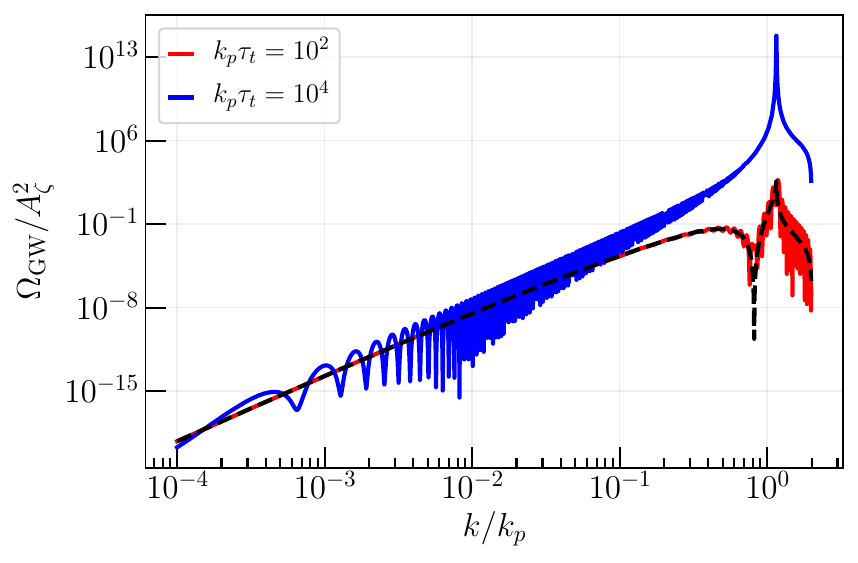}
    \includegraphics[scale=0.55]{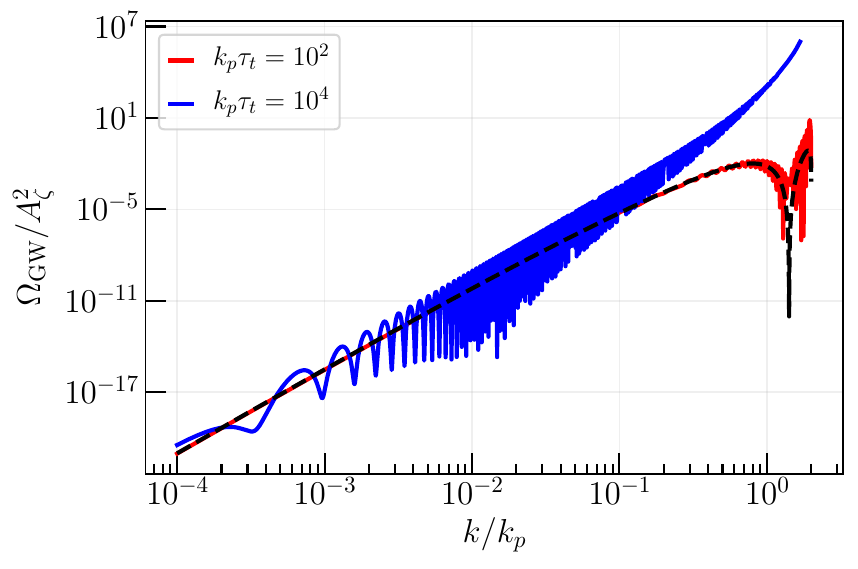}
    \caption{The induced GW spectrum for a Dirac delta primordial spectrum \eqref{eq:diracdeltapzeta} for two different parameter combinations of $k_p \tau_t$ and $\tilde \Gamma$, compared to the prediction in GR. In red and blue lines we respectively show the induced GW spectrum for $k_p \tau_t=10^2$ and $k_p \tau_t=10^4$, both for a fixed and $\tilde \Gamma=10^{-3}$. Dashed black lines show the GR prediction. In the left and right figures, we respectively show the cases for $c_s^2=1/3$ and $c_s^2=1$. We can see that the resonance peak gets enhanced as one increases $k_p \tau_t$ and $\tilde \Gamma$. We also see the oscillations due to the higher derivative mixing of scalar and tensor modes for $k>k_t$. The amplitude of the induced GW spectrum around the peak is proportional to $\tilde \Gamma^2$. We require that the backreaction bound \eqref{eq:backreactionbound2} holds.}
    \label{fig:Omega_GW_Dirac_Delta}
\end{figure}

We show in Fig.~\ref{fig:Omega_GW_Dirac_Delta} the Horndeski-scalar induced GW spectrum $\Omega_{\rm GW}$ (normalized by $A_\zeta^2$) compared to the GR prediction for $c_s^2=1/3$ and $c_s=1$, respectively on the left and right panels. We can see that if the modes entered the horizon way before the transition $k_p \tau_t \gg 1$ we obtain sizeable oscillations on top of the GR spectrum. We also see that the resonance peak is enhanced with respect to GR, its amplitude scaling as $(\tilde \Gamma x_t^2)^2$. The upper bounds given in Eqs.~\eqref{eq:upperbound} and \eqref{eq:upperbound2} from the backreaction of higher derivative terms in the curvature perturbation apply in this case as well (not considering the divergent peak). We also confirm that the deep infrared tail is the same as in GR, as expected.
Furthermore, we observe that in intermediate scales, when the contribution from higher derivative terms dominate the induced GW spectrum, the spectrum does not grow as $k^2$ but instead as $k^4$. This is consistent with our discussion around Eq.~\eqref{eq:hksuperhsubhsolution} in Sec.~\ref{sec:preliminaries}. Lastly, it is interesting to note that we get a resonant peak even for $c_s=1$. Such resonant peak is not present in GR because the resonance is logarithmic and killed by the projection $\epsilon_{ij} q^i q^j$ when $k=2k_p$. However, in Horndeski gravity the resonance diverges with a higher power and can, therefore, compensate by the suppression from the projection. Nevertheless, this is an artifact of the Dirac delta spectrum. We will see later that a finite-width primordial spectrum has a smooth decay of the induced GW spectrum.

\subsubsection{Log-normal peak}

\begin{figure}
    \centering
    \includegraphics[scale=0.55]{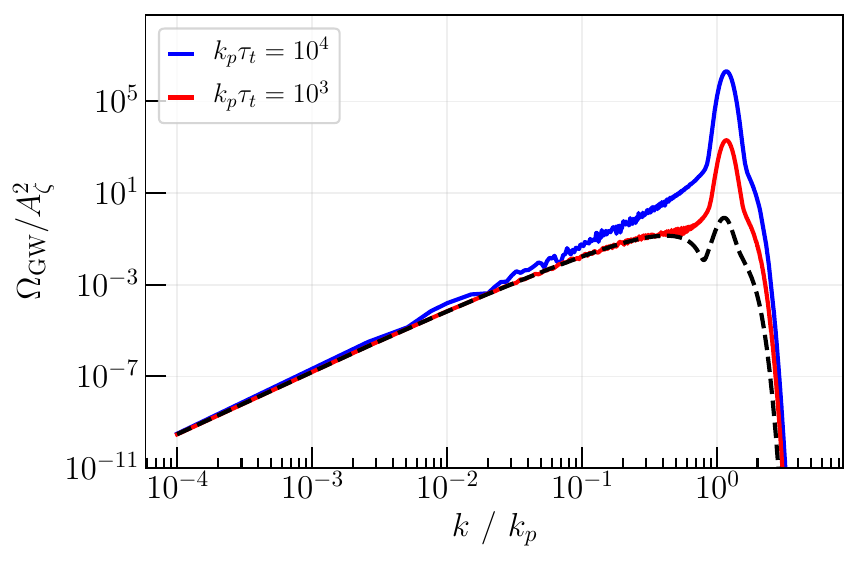}
    \includegraphics[scale=0.55]{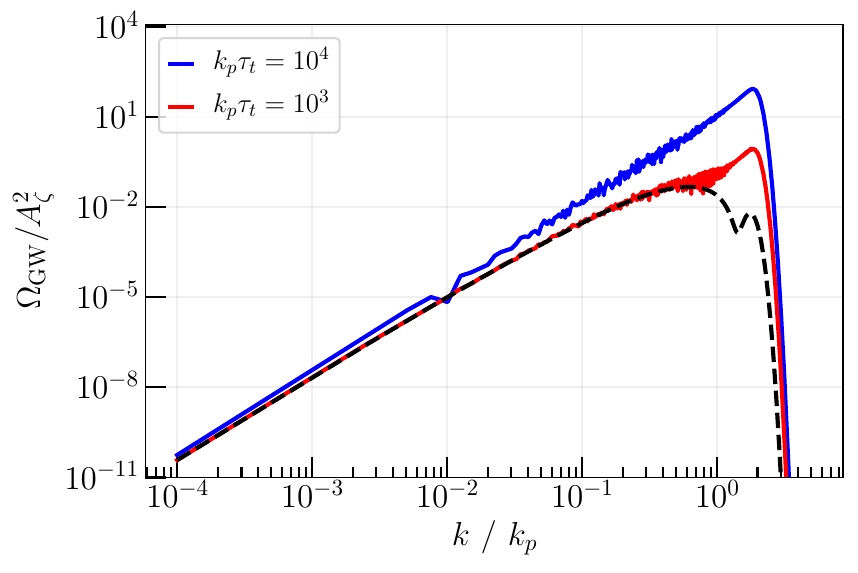}
    \caption{Induced GW spectrum for a log-normal peak in the primordial spectrum \eqref{eq:lognormal} with $\Delta =0.1$. Black-dashed lines show the GR prediction, while the blue and red solid lines show the Horndeski prediction respectively for $k_p \tau_t=10^3$ and $k_p \tau_t=10^4$ for a fixed $\tilde \Gamma=10^{-3}$. On the left and right panels we respectively show the cases with $c_s^2=1/3$ and $c_s^2=1$. Note that the overall amplitude of the peak is proportional $\tilde \Gamma^2$. We also require that the backreaction bound \eqref{eq:backreactionbound2} from higher derivative interactions to the curvature perturbation is satisfied.}
    \label{fig:Omega_GW_LogNormal}
\end{figure}

To investigate the effects of a finite-width peak in the primordial curvature power spectrum, we consider a log-normal peak given by
\begin{align}\label{eq:lognormal}
    P_\zeta = \frac{A_\zeta}{\sqrt{2 \pi \Delta^2}} e^{-\frac{\ln^2( k/k_p)}{2 \Delta^2}}\,.
\end{align}
In this case, we computed the induced GW spectrum  numerically. Nevertheless, one can confirm analytically that the universal infra-red slope proportional to $k^3\ln^2k$ is recovered \cite{Cai:2018dig,Yuan:2019wwo}. We show in Fig.~\ref{fig:Omega_GW_LogNormal} the induced GW spectrum for $\Delta =0.1$ and $\tilde \Gamma=10^{-3}$ for $c_s=1/\sqrt{3}$ (left panel) and $c_s=1$ (right panel). The first thing we note is that the oscillations in the induced GW spectrum are washed out in comparison to the Dirac delta case. However, we see that the enhancement due to the resonance is still much more pronounced that in GR. We also see that the induced GW spectrum then quickly decays for $k>2k_p$, even for $c_s=1$. The amplitude of the peak in the induced GW spectrum does not scale exactly as in the Dirac delta case, because the spectrum has a finite width. Instead, we find a similar situation than in the scale invariant case of Sec.~\ref{sec:scaleinvariantgeneral}. Namely, only a narrow window of momenta around the resonance contributes giving an additional factor of $1/x_t$. Thus, for not so sharp peak peak, the amplitude of the resonant peak is a factor $\tilde \Gamma^2 x_t^3$ larger than GR. We numerically find that
\begin{align}
\Omega_{\rm GW}^{\Delta,\rm peak}\approx 2 A_\zeta^2\tilde \Gamma^2 \left(\frac{k_p}{k_t}\right)^3\,.
\end{align}
It then follows from Eq.~\eqref{eq:upperbound2} that the peak of the induced GW spectrum is bounded from above. Namely, we have that $\Omega_{\rm GW}^{\Delta,\rm peak}<A_\zeta \times \left({k_p}/{k_t}\right)$.

\subsubsection{Scale Invariant primordial spectrum}

As a last example, we consider a scale invariant primordial spectrum given by Eq.~\eqref{eq:scaleinvariant}. This is well-motivated from CMB observations if slow-roll inflation lasted until the end of inflation. Note that we must consider a high-$k$, or “uv”, cut-off to the scale invariant spectrum. Otherwise, high $k$ curvature fluctuations will eventually enter the regime where higher derivative terms backreact. To avoid this regime, we require that the cut-off at $k_{\rm uv}$ always satisfied the backreaction bound \eqref{eq:backreactionbound2}.

\begin{figure}
    \centering
    \includegraphics[scale=0.7]{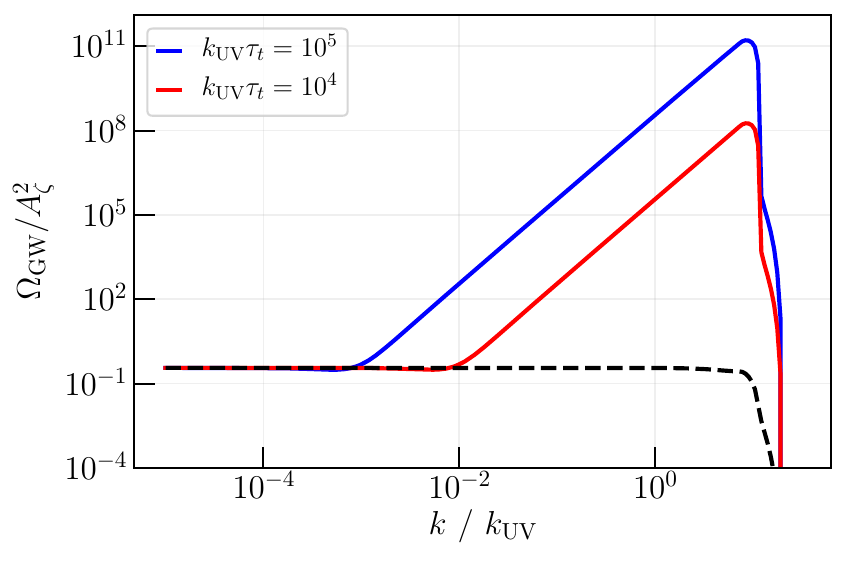}
    \caption{Induced GW spectrum from a scale invariant primordial spectrum with a cut-off at $k=k_{\rm uv}$ \eqref{eq:scaleinvariant} for $c_s=1/\sqrt{3}$. The black-dashed line shows the prediction in GR, which has an amplitude $\sim 0.82$ \cite{Kohri:2018awv}. The red and blue solid lines respectively show the induced GW spectrum for $k_{\rm UV}\tau_t=10^4$ and $k_{\rm UV}\tau_t=10^5$ for $\tilde \Gamma=10^{-3}$. As in previous examples, we satisfy the backreation bound \eqref{eq:backreactionbound2}. Note that the induced GW spectrum scales as $k^3$ for $k>k_t$.}
    \label{fig:Omega_GW_Scale_Invariance}
\end{figure}

We find that, although the exact shape of the GW spectrum has to be calculated numerically, it is well approximated by the resonant contribution to the momentum integral in Eq.~\eqref{eq:omegataucparticular} when the higher derivative terms dominate. Thus, to provide an analytical estimate, we select the leading order terms in $x_t$ and carry out the integration only around the resonance in the Kernel, as explained in Ref.~\cite{Inomata:2019ivs}. For a more detail discussion between the differences with Ref.~\cite{Inomata:2019ivs}, see the paragraph above Eq.~\eqref{eq:resonancegeneral}. The induced GW spectrum from the resonant contribution reads
\begin{align}\label{eq:omegaresflat}
    \Omega_{\rm GW}^{\rm res}(k)\approx \frac{A_\zeta^2}{3}\tilde\Gamma^2x_t^4\int_0^\infty dv&\int_{|1-v|}^{1+v}du\left(\frac{4v^2-(1-u^2+v^2)^2}{4uv}\right)^2\nonumber\\&\times\frac{(u+v)^4\sin^2 ((1-c_s (u+v))x_t)}{4^4 u^2 v^2 (1-c_s(u+v))^2x_t^2} \Theta(v_{\rm uv}-v)\Theta(v_{\rm uv}-u)\,,
\end{align}
where $v_{\rm uv}=k_{\rm uv}/k$. We continue introducing the variable $z=(1-c_s (u+v))x_t$ and $s=u-v$ and evaluate at the resonant point, i.e. $z=0$, except for the $\sin z/z$. Then we find that
\begin{align}\label{eq:estimateflat}
\Omega^{\rm res}_{\rm GW}\sim &\frac{\pi   \left(1-c_s^2\right)^2 }{96c_s}A_\zeta^2\tilde\Gamma^2\left(\frac{k}{k_t}\right)^3
\int_{-s_0(k)}^{s_0(k)}\frac{\left(1-s^2\right)^2}{\left(1-c_s^2 s^2\right)^4}\,,
\end{align}
where $s_0(k)$ is given by Eq.~\eqref{eq:s0} in the limit when $c_t=1$. This agrees with our earlier estimate Eq.~\eqref{eq:resonancegeneral} in the general case, which indicates that the resonance is not sensitive to the transition as it is developed during the Horndeski phase. We checked that our analytical estimated \eqref{eq:estimateflat} agrees with the numerical integration up to ${\cal O}(1)$ factors, which is good enough for our purposes. We see that the spectrum grows as $k^3$ and cuts off at $k\sim 2c_sk_{\rm p}$. We show in Fig.~\ref{fig:Omega_GW_Scale_Invariance} the numerical results for the induced GW spectrum from a scale invariant primordial spectrum with a cut-off.

It is interesting to discuss the maximum amplitude that the induced GW spectrum could have without the backreaction of higher derivatives \eqref{eq:backreactionbound2}. Assuming that the cutoff scale $k_{\rm uv}$ is below the scale for which the non-linearities become important, we can impose an upper limit on the transition time $\tau_t$. Using the limit $\tilde \Gamma (k_{\rm uv} \tau_t) P_\zeta^{1/2} < 1$ 
In this case we can estimate the maximal amplitude of $\Omega_{GW}$ at the cutoff frequency as
\begin{align}
    \Omega_{\rm{GW}, 0}(k_{\rm uv}) h^2 \sim  1.62\times 10^{-5}\tilde \Gamma^2 {\cal P}_\zeta^2 (k_{\rm uv} \tau_t)^3 < 1.62\times 10^{-5} (k_{\rm uv}/k_t) {\cal P}_\zeta\,.
\end{align}
Interestingly, if we take the CMB normalization, roughly $P_\zeta\sim 10^{-9}$, we have that for $k_{\rm uv}>10^{n}k_t \sim 10^{n} k_t$, the maximum amplitude of the induced GW spectrum today can be roughly as large as $\Omega_{\rm{GW}, 0}(k_{\rm uv}) h^2 \lesssim 10^{n-14}$ for $n>2$ the signal is observable in GW detectors such as LISA, ET and CE. It is also interesting to note that the $k^3$ slope is consistent with the recent results from PTAs. However, the main issue is that one would need $n\sim 6$ to have an amplitude of $10^{-8}$, which implies that if we fix $f_{\rm uv}\sim 10^{-7}\,{\rm Hz}$ the transition frequency is around $f_{\rm t}\sim 10^{-13}\,{\rm Hz}$. This means that the transition must happen well after BBN, thus ruling out the explanation. Thus, to explain the PTA results we still need an enhancement of the primordial spectrum or to consider that the Horndeski scalar is a subdominant component of the universe. We leave the latter option for future work. But, for the former, if we fix $f_{\rm t}\sim 10^{-10}\,{\rm Hz}$ and $f_{\rm uv}\sim 10^{-7}\,{\rm Hz}$, we need ${\cal P}_\zeta\sim 10^{-5}$ to achieve $\Omega_{\rm GW,0}h^2\sim 10^{-7}$. Such value of ${\cal P}_\zeta$ is much lower than the one needed for induced GWs to explain the PTA results, which is about ${\cal P}_\zeta\sim 10^{-1}$.

\section{Discussion and conclusions \label{sec:discussions}}

GWs offer an ideal way to probe modifications of gravity. In this work, we studied the production of induced GWs, after inflation, during a period in the very early universe dominated by a Horndeski scalar field. This could be possible, e.g., within Horndeski inflation \cite{Kobayashi:2010cm}, if the Horndeski-scalar responsible for inflation does not decay at the end of inflation but “rolls” faster (like in Quintessential inflation scenarios \cite{Peebles:1998qn,Hossain:2014xha}). Or, perhaps, inflation ends because a Horndeski-scalar dominates. Interestingly, we have found that the generation of induced GWs is enhanced during such Horndeski phase.

The Horndeski phase translates into stronger resonances in the induced GW spectrum as well as an amplification of the power on small scales. We provide the general expression for the Kernel in Eq.~\eqref{eq:explicitIxgg1}, though most of the explicit expression of the coefficients can be found in App.~\ref{app:explicit}. Focusing on the effects of $G_4$ and $G_5$, we studied the induced GW spectrum for a sharply peaked and a scale invariant primordial curvature spectrum, ${\cal P}_\zeta$. We find that, for sharp peaks in the primordial spectrum, the peak amplitude of the GW background scales as $\Omega_{\rm GW}\propto (k_p/k_t)^4 {\cal P}_\zeta^2$ where $k_p$ and $k_t$ are respectively the scale of the peak and the scale of the transition to GR. Broad peaks in the primordial spectrum yield one power less of $(k_p/k_t)$ in the induced GW spectrum. For the scale invariant case, we find that the induced GW spectrum is well approximated by the resonant contribution \eqref{eq:resonancegeneral}. In this case we find that $\Omega_{\rm GW}\propto (k/k_t)^3 {\cal P}_\zeta^2$. Lastly, requiring that no backreaction from higher derivative terms takes place (which basically bounds  how much gradients of the curvature fluctuations can grow and roughly imposes that $(k_p/k_t)P^{-1/2}_\zeta<1$; see Sec.~\ref{sec:backreaction} for me details), we find that the maximum peak amplitude of the induced GW spectrum is bounded by $\Omega^{\rm peak}_{\rm GW}< (k_p/k_t)^2 {\cal P}_\zeta$ and $\Omega^{\rm peak}_{\rm GW}< (k_p/k_t) {\cal P}_\zeta$ respectively for the sharply peaked and the broadly peak (including scale invariant) primordial spectrum. We also confirmed our analytical calculations in a concrete Horndeski model with a smooth transition to GR in Sec.~\ref{sec:horndeskiconcrete}.

One important question is whether one can tell apart the induced GW spectrum generated during the Horndeski-like or the standard radiation dominated phase, beside other possible degeneracies with other cosmic sources of the GW background (see, e.g., Refs.~\cite{,Guzzetti:2016mkm,Caprini:2018mtu}). In principle, it may be possible to design a primordial curvature spectrum such that the resulting spectrum of induced GWs during the standard radiation dominated phase (and even considering a general equation of state as in Refs.~\cite{Domenech:2019quo,Domenech:2020kqm}) resembles the Horndeski-scalar induced GW spectrum we derived in this work. We would like to emphasize, though, that this issue is general to cosmic GW sources. Nevertheless, an attractive feature of the Horndeski-scalar induced GWs is that one may render the induced GW spectrum observable by future GW detectors such as $\mu$-Ares, LISA, Taiji, TianQin,  DECIGO, ET, CE, Voyager, without requiring an enhancement of ${\cal P}_\zeta$ during inflation. In this case, it is likely that the Horndeski-scalar induced GW signal does not have a PBH counterpart, though one should check the formation of PBHs in the Horndeski-phase to confirm this.

We have focused on power-law solutions; a requirement that considerably fixes the functional freedom of the $G_i$ functions to Eqs.~\eqref{eq:Gisolutions1} and \eqref{eq:Gisolutions2}. Though we do not expect the main characteristic features of the Horndeski-scalar induced GWs to be affected by a different class of solutions, the amplitude and the spectral slope of the GW spectrum might increase/decrease depending on the time-dependence of the coefficients. It would be interesting to repeat our calculations in other well-motivated models, including gradual transitions to GR. It would also be interesting to consider scaling solutions \cite{Amendola:1999qq,Amendola:2006qi,Gomes:2013ema,Gomes:2015dhl,Amendola:2017xhl,Amendola:2018ltt,Frusciante:2018tvu,Frusciante:2018aew,Vaskonen:2020lbd} together with a radiation fluid and explore how the amplitude of the induced GW spectrum depends on the energy density fraction of the Horndeski-scalar field. In addition, one could explore then production of induced GWs inside the non-linear regime, when higher derivative terms backreact, as occurs in the Vainshtein mechanism \cite{Vainshtein:1972sx,Kimura:2011dc,Koyama:2013paa,Babichev:2013usa,Kobayashi:2019hrl}. As future work, we plan to extend our analysis to more general theories under an EFT approach \cite{Gleyzes:2013ooa,Fujita:2015ymn,Crisostomi:2016tcp,BenAchour:2016cay,BenAchour:2016fzp,Takahashi:2017zgr,BenAchour:2020wiw,Babichev:2021bim}.

Lastly, we note that recently there has been a lot of interest in the opposite regime, namely scalar fluctuations induced by GWs. While in our work we showed that higher derivative terms modify and enhance the induced GWs, Refs.~\cite{Creminelli:2019kjy,Creminelli:2019nok} study instabilities in the scalar sector and resonant decays of GWs into the scalar field. It would be interesting to understand this effect in the very early universe in our formalism, perhaps extending our calculations to the inclusion of primordial tensor modes and scalar-tensor mixings \cite{Chang:2022vlv,Yu:2023lmo,Bari:2023rcw,Picard:2023sbz} as well as tensor-induced density fluctuations as in Ref.~\cite{Bari:2021xvf}.

\begin{acknowledgments}
We would like to thank S.~Renaux-Petel, M.~Sasaki, A.~Vikman, M.~Yamaguchi, and M.~Zumalacárregui for helpful comments and discussions. This research is supported by the DFG under the Emmy-Noether program, project number 496592360, and by the JSPS KAKENHI grant No. JP24K00624.
\end{acknowledgments}

\appendix 

\section{Explicit formulas for perturbations \label{app:explicitperturbations}}

In this appendix, for the readers convenience, we list all the coefficients that appear in the perturbative expansion in the second and third order actions, as derived by Ref.~\cite{Gao:2012ib}. First, the coefficients for the second order action \eqref{eq:secondorderaction} for tensor modes are given by
\begin{align}
\mathcal{ F}_t&=2\left(G_4
-X\left( \ddot\phi G_{5X}+G_{5\phi}\right)\right),
\\
\mathcal{ G}_t&=2\left(G_4-2 XG_{4X}
-X\left(H\dot\phi G_{5X} -G_{5\phi}\right)\right).
\end{align}
The ones for the second order action \eqref{eq:secondorderaction} for scalar modes are
\begin{align}
\mathcal{ F}_s=&\frac{1}{a}\frac{d}{dt}\left(\frac{a}{\Theta}\mathcal{ G}_t^2\right)
-\mathcal{ F}_t,
\\
\mathcal{ G}_s=&\frac{\Sigma }{\Theta^2}\mathcal{ G}_t^2+3\mathcal{ G}_t. 
\end{align}
In the equations above, $\Sigma$ and $\Theta$ are defined by
\begin{align}
\Sigma=&XG_{2,X}+2X^2G_{2,XX}+12H\dot\phi XG_{3X}
+6H\dot\phi X^2G_{3XX}
-2XG_{3\phi}-2X^2G_{3\phi X}-6H^2G_4\nonumber\\&
+6\left[H^2\left(7XG_{4X}+16X^2G_{4XX}+4X^3G_{4XXX}\right)
-H\dot\phi\left(G_{4\phi}+5XG_{4\phi X}+2X^2G_{4\phi XX}\right)
\right]\nonumber\\&
+30H^3\dot\phi XG_{5X}+26H^3\dot\phi X^2G_{5XX}
+4H^3\dot\phi X^3G_{5XXX} \nonumber\\&
-6H^2X\left(6G_{5\phi}
+9XG_{5\phi X}+2 X^2G_{5\phi XX}\right),
\\
\Theta=&-\dot\phi XG_{3X}+
2HG_4-8HXG_{4X}
-8HX^2G_{4XX}+\dot\phi G_{4\phi}+2X\dot\phi G_{4\phi X}\nonumber\\&
-H^2\dot\phi\left(5XG_{5X}+2X^2G_{5XX}\right)
+2HX\left(3G_{5\phi}+2XG_{5\phi X}\right).
\end{align}

In the third order action for scalar-scalar-tensor interactions we have that
\begin{align}
\mu=\dot\phi XG_{5X}\,,
\end{align}
and
\begin{align}
\Gamma=& 2G_4-8XG_{4X}-8X^2G_{4XX}
-2H\dot\phi\left(5XG_{5X}+2X^2G_{5XX}\right)
+2X\left(3G_{5\phi}+2XG_{5\phi X}\right).
\end{align}
\subsection{Coefficients in the power-law solutions}
Now we write the explicit expression of the coefficients when evaluated on the power-law solutions.
First, for ${\cal G}_t$ and ${\cal F}_t$, we find that
\begin{align}
M_{\rm pl}^{-2}{\cal G}_t&=\lambda_4^2 \left(a_4-2
   a_{4Y}\right)+{\lambda_5}{p^{-2}}\left(a_5-2 p
   a_{5Y}\right)\,,\\
M_{\rm pl}^{-2}{\cal F}_t&=\lambda_4^2 a_4+{\lambda_5}{p^{-2}}a_5\,.
\end{align}
Second, ${\cal G}_s$ and ${\cal F}_s$ are given by
\begin{align}
M_{\rm pl}^{-2}{\cal G}_s&=\tilde\Sigma\frac{{\cal G}_t^2}{\tilde \Theta^2 M_{\rm pl}^2}+3{\cal G}_t\,,\\
M_{\rm pl}^{-2}{\cal F}_s&=\left(1+\frac{1}{p}\right)\frac{{\cal G}_t^2}{\tilde \Theta M_{\rm pl}^2}-{\cal F}_t\,,
\end{align}
where we have defined
\begin{align}
\tilde \Theta&=\frac{\Theta}{HM_{\rm pl}^2}\nonumber\\&
=-\frac{\lambda_3
   }{p}a_{3,Y}+\lambda_4^2
   \left(a_4-4
   \left(a_{4,Y}+a_{4,YY}\right)\right
   )+\frac{\lambda_5 }{p^2}\left(3
   a_5+(2-5 p) a_{5,Y}-2 p
   a_{5,YY}\right)\,,
\end{align}
and
\begin{align}
\tilde\Sigma&=\frac{\Sigma}{H^2 M_{\rm pl}^2}\nonumber\\&
=2\lambda_2^4
   a_{2,YY}+\frac{\lambda_3}{p^2}\left((9 p+1)
   a_{3,Y}+6
   p a_{3,YY}\right)\nonumber\\&+\frac{\lambda_4^2}{p}\left(a_4 (1-3
   p)+2 (9 p-1) a_{4,Y}+6 p(7p
   a_{4,YY}+2
   a_{4,YYY})\right)\nonumber\\&+\frac{\lambda_5 }{p^3}\left(
   a_5 (1-15
   p)+(27 p-25)
   pa_{5,Y}+6 (4 p-1)
   pa_{5,YY}+4p^2 a_{5,YYY}\right)\,.
\end{align}
Note that we used Eqs.~\eqref{eq:fried1} and \eqref{eq:fried2} to eliminate $a_2$ and $a_{2,Y}$ in the expressions for $\Theta$ and $\Sigma$.

Lastly, the $\Gamma$ and $\mu$ coefficients are given by
\begin{align}
M_{\rm pl}^{-2}\Gamma = \lambda_4^2 \left(a_4-4
   \left(a_{4,Y}+a_{4,YY}\right)\right
   )+\frac{\lambda_5 }{p^2}\left(3
   a_{5}+(2-10 p) a_{5,Y}-4 p
   a_{5,YY}\right)\,,
\end{align}
and
\begin{align}
\tilde\mu=\frac{\mu\,H}{M_{\rm pl}^2}=\frac{\lambda_5 
   }{p}a_{5,Y}\,.
\end{align}

\section{The GW energy density during the Horndeski phase \label{app:GWenergydensity}}

Before proceeding to the explicit calculation of the induced GWs, it is interesting to look at the definition of GW energy density in Hordenski theory. To do that without entering into the messy equations of motion (see, e.g., Ref.~\cite{Gao:2011qe} for the explicit non-linear expression), we look into the third order action in the uniform-$\phi$ gauge and take the terms that contain the lapse function and two tensor modes. These are given by \cite{Gao:2012ib}
\begin{align}
    {\cal L}_{shh}^u =-\frac{a^3\alpha}{8}\left(\Gamma\dot h_{ij}\dot h_{ij}+\frac{{\cal G}_t}{a^2}\partial_k h_{ij}\partial_k h_{ij}+\frac{4\mu}{a^2}\dot h_{ij}\Delta h_{ij}\right)+...\,,
\end{align}
where “...” refers to scalar-tensor-tensor terms that do not contain the lapse and
\begin{align}
\Gamma&=\lambda_4^2 \left(a_4-4
   \left(a_{4,Y}+a_{4,YY}\right)\right)+\frac{\lambda_5 }{p^2}\left(3 a_{5}+(2-10 p) a_{5,Y}-4 p
   a_{5,YY}\right)\,,\\
\mu\,H&=\frac{\lambda_5}{p}a_{5,Y}\,.
\end{align}
For the explicit expression of $\Gamma$ and $\mu$ in terms of $G_i$’s see Ref.~\cite{Gao:2012ib}, which for completeness we also include in App.~\ref{app:explicitperturbations}.

We define the energy density of GWs as the $00$-component of Einstein equations, which are related to the variation with respect to the lapse. We find at second order in cosmological perturbation theory that
\begin{align}
\rho_{\rm GW}=-\frac{1}{a^3}\left\langle\frac{\delta {\cal L}_{shh}}{\delta g^{00}}\right\rangle=-\frac{1}{2a^3}\left\langle\frac{\delta {\cal L}_{shh}}{\delta \alpha}\right\rangle=\frac{1}{8}\left\langle\Gamma\dot h_{ij}\dot h_{ij}+\frac{{\cal G}_t}{a^2}\partial_k h_{ij}\partial_k h_{ij}+\frac{4\mu}{a^2}\dot h_{ij}\Delta h_{ij}\right\rangle\,,
\end{align}
where the brackets mean some kind of time or spatial average. Note that in GR we have $\Gamma={\cal G}_t=1$ and $\mu=0$. To have an idea of the impact of GW energy density on the background, we write the first Friedmann equation, in the power-law solutions, as
\begin{align}
3H^2\times \left(\Gamma + \frac{20}{3}\mu H+ \frac{8\lambda_5}{3p}8 a_{5,YY}\right)=\rho_\phi+\rho_{\rm GWs}\,,
\end{align}
where $\rho_\phi$ include the rest of the Hordenski terms which are not proportional to $H^2$ or $H^3$. Thus, during the Horndeski phase we have that the energy density ratio is given by
\begin{align}\label{eq:spectraldensityinHordenski}
\Omega_{\rm GWs}=\frac{\left\langle\Gamma\dot h_{ij}\dot h_{ij}+\frac{{\cal G}_t}{a^2}\partial_k h_{ij}\partial_k h_{ij}+\frac{4\mu}{a^2}\dot h_{ij}\Delta h_{ij}\right\rangle}{24H^2\left(\Gamma + \frac{20}{3}\mu H+ \frac{8\lambda_5}{3p}8 a_{5,YY}\right)}\,.
\end{align}
We can use Eq.~\eqref{eq:spectraldensityinHordenski}, e.g., to investigate for example possible backreaction of GWs during the Horndeski phase. We will not pursue this direction in this work as one should include an effective energy density due to the curvature fluctuations as well, due to the higher derivative terms. This is out of the scope of this paper.

Nevertheless, it is interesting to note that the term proportional to $\mu$ in the numerator of Eq.~\eqref{eq:spectraldensityinHordenski} has a distinct scaling as the other two. In fact, using the linear GW equation, the term proportional to $\mu$ scales with an additional power of $k/{\cal H}$ with respect to the standard $\dot h_{ij}\dot h_{ij}$ term. On subhorizon scales, where $k\gg {\cal H}$ this term can easily dominate unless $\mu$ is very small. In the case were $\mu=0$, the energy density of GWs in Horndeski, in the subhorizon limit and using the linear GW solution, is the one in GR but rescaled by a factor $(1+{\cal G}_t/\Gamma)$. If ${\cal G}_t\gg \Gamma$, which for $G_5=0$ corresponds to $2XG_{4,XX}>G_{4,X}$, the energy density ratio of GWs will be enhanced during the Horndeski phase. The situation will be different in the case of induced GWs since, from Eq.~\eqref{eq:roughsubtensor}, they do not behave as free GWs. We will come back to this when computing the detailed induced GW spectrum during the Horndeski phase. Let us advance, though, that we will require that the Horndeski phase ends and GR is recovered. After the transition, the energy density of GWs takes the standard form. Thus, any modification to the energy density of GWs during the Horndeski phase is not important for the prediction of the GW spectrum today. Only the modification to the source term makes a difference.

\section{Explicit formulas for the source and kernel \label{app:explicit}}

In this appendix, we write down the explicit expression of the coefficients in the source function $f(v,u,x)$ \eqref{eq:fsimplifiedincis} using the parameters of \ref{app:explicitperturbations}. We have that
\begin{align}
c_0^-=\frac{1}{
   8uv\tilde\Theta ^2 }\Bigg(\frac{{\cal G}_t^3-10{\cal G}_t^2\tilde\mu+4{\cal F}_s\tilde\Theta\tilde\mu}{c_s^2{\cal G}_t}-(u-v)^2 (\Gamma  {\cal G}_t+4 \tilde\Theta  \tilde\mu )\Bigg)\,,
\end{align}
\begin{align}
c_1^-=\frac{(u-v) }{ 8  c_s  u^2  v^2{\cal G}_t^2    \tilde\Theta }\Bigg(\frac{{\cal F}_s ({\cal G}_t^3 - 4 {\cal G}_t^2 \tilde\mu + 2 {\cal F}_s \tilde\Theta\tilde\mu)}{c_s^2{\cal G}_t}+2{\cal G}_t^3uv-{\cal F}_s(u^2+v^2)\left({\cal G}_t\Gamma+2\tilde\Theta\tilde\mu\right)  \Bigg)\,,
\end{align}
\begin{align}
c_2^-=\frac{1}{2c_s^2uv{\cal G}_t}\left(2c_s^4u^2v^2c_4^--c_1^-c_s(u-v)-c_1^+c_s(u+v)+\frac{3{\cal F}_s+2{\cal F}_t}{2}-\frac{{\cal G}_t^2(\Gamma+9\tilde\Theta)}{2\tilde\Theta^2}-\frac{2{\cal G}_t\tilde\mu}{\tilde\Theta}\right)\,,
\end{align}
and
\begin{align}
c_4^-=\frac{{\cal F}_s }{8  c_s^4  u^3  v^3  {\cal G}_t^2 \tilde\Theta }\Bigg(\frac{{\cal F}_s \tilde\Theta ({\cal G}_t + 2 \tilde\mu)}{c_s^2{\cal G}_t}-(u^2 + v^2) ({\cal G}_t (\Gamma+ 3 \tilde\Theta) + 
   2 \tilde\Theta\tilde\mu)\Bigg)\,.
\end{align}
The remaining coefficients follow from the following relations,
\begin{align}
{c_{0}^+}+c_{0}^-=\frac{(\Gamma  {\cal G}_t+4 \tilde\Theta  \tilde\mu) }{2 \tilde\Theta ^2}\quad,\quad {c_{2}^+}+{c_{2}^-}\pm 2c_4^{\pm}c_s^2uv=0\quad,\quad {c_{4}^+}+{c_{4}^-}=0\,,
\end{align}
\begin{align}
\frac{c_{1}^+}{v+u}+\frac{c_{1}^-}{v-u}=-\frac{{\cal G}_t}{2c_suv\tilde\Theta}\quad,\quad c_3^{\pm}=c_s(u\pm v)c_4^{\pm}\,.
\end{align}
Note that, because of symmetry, we have that ${c_{i}^-}(v\to-v)=c_i^{+}$. Since the kernel is symmetric in $u$ and $v$, the same applies to a flip in $u$’s sign.

Turning to the explicit expressions for the integrals in $I(v,u,x)$ \eqref{eq:Idefinition}, we find that
\begin{align}\label{eq:generalIc}
I_c(v,u,x)&=\frac{1}{4}\Bigg\{\cos (x y_1) \left(
   \frac{2c_0^-}{y_1^2}+\frac{2c_1^-}{y_1}-\frac{c_4^-}{x^2}\right)+\sin (x y_1) \left(\frac{2c_0^- x}{y_1}+\frac{2
   c_3^-+c_4^- y_1}{x}\right)\nonumber\\&+\cos (x y_2) \left(\frac{2c_0^+}{
   y_2^2}-\frac{2c_1^+}{
   y_2}-\frac{c_4^+}{x^2}\right)+\sin (x y_2) \left(\frac{2c_0^+
   x}{y_2}+\frac{c_4^+ y_2-2 c_3^+}{ x}\right)\nonumber\\&+\cos (x
   y_3) 
   \left(\frac{2c_0^+}{y_3^2}+\frac{2c_1^+}{y_3}-\frac{c_4^+}{ x^2}\right)+\sin (x
   y_3) \left(\frac{2c_0^+ x}{ y_3}+\frac{2 c_3^++c_4^+
   y_3}{x}\right)\nonumber\\&+\cos (x y_4) \left(\frac{2c_0^-}{
   y_4^2}-\frac{2c_1^- }{
   y_4}-\frac{c_4^-}{x^2}\right)+\sin (x y_4) \left(\frac{2c_0^-
   x}{ y_4}+\frac{c_4^- y_4-2 c_3^-}{ x}\right)\nonumber\\&+\text{Ci}(x |y_1|) (2 c_2^--y_1 (2
   c_3^-+c_4^- y_1))+ \text{Ci}(x
  |y_2|) (2 c_2^++y_2 (2 c_3^+-c_4^+ y_2))\nonumber\\&+
   \text{Ci}(x |y_3|) (2 c_2^+-y_3 (2 c_3^++c_4^+ y_3))+\text{Ci}(x |y_4|) (2
   c_2^-+y_4 (2 c_3^--c_4^- y_4))\Bigg\}\nonumber\\&
   -I_{c,0}(v,u)\,,
\end{align}
where
\begin{align}\label{eq:Ic0appendix}
I_{c,0}(v,u)=&\frac{c_0^- \left(c_t^2+c_s^2
   (u-v)^2\right)}{\left(c_t^2-c_s^2
   (u-v)^2\right)^2}+\frac{c_0^+ \left(c_t^2+c_s^2
   (u+v)^2\right)}{\left(c_t^2-c_s^2
   (u+v)^2\right)^2}+\frac{c_1^- c_s (u-v)}{c_t^2-c_s^2
   (u-v)^2}+\frac{c_1^+ c_s (u+v)}{c_t^2-c_s^2 (u+v)^2}\nonumber\\&+c_4^-
   c_s^2 u v+\frac{1}{4}\left(2c_2^{-}-c_4^{-}(c_t^2-c_s^2(u-v)^2\right)\ln\left|\frac{c_t^2-c_s^2(u-v)^2}{c_t^2-c_s^2(u+v)^2}\right|\,,
\end{align}
and
\begin{align}\label{eq:generalIs}
I_s(v,u,x)&=\frac{1}{4}\Big\{-\cos (x y_1) \left(\frac{2c_0^-
   x}{ y_1}+\frac{c_4^- y_1+2 c_3^-}{ x}\right)+\sin (x  y_1)
   \left(\frac{2c_0^-}{y_1^2}+\frac{2c_1^-}{y_1}-\frac{c_4^-}{ x^2}\right)\nonumber\\&
  -\cos (x y_2)\left(\frac{2c_0^+ x}{ y_2}+\frac{ c_4^+ y_2-2
   c_3^+}{ x}\right)+\sin (x y_2) \left(\frac{ 2c_0^+ }{  y_2^2}-\frac{2c_1^+}{  y_2}-\frac{c_4^+}{ x^2}\right)\nonumber\\&
   -\cos (x y_3) \left(\frac{2 c_0^+ x}{y_3}+\frac{c_4^+
   y_3+2 c_3^+}{x}\right)+ \sin (x y_3) \left(\frac{2 c_0^+}{y_3^2}+\frac{2
   c_1^+}{y_3}-\frac{c_4^+}{x^2}\right)\nonumber\\&
   -\cos (x y_4)\left(\frac{2 c_0^- x}{ y_4}+\frac{c_4^- y_4-2
   c_3^-}{x}\right)+\sin (x y_4)\left(\frac{2 c_0^-}{ y_4^2}-\frac{2 c_1^-}{ y_4}-\frac{c_4^-
   }{ x^2}\right)\nonumber\\&
   +\text{Si}(x y_1) \left(2 c_2^--2 c_3^- y_1-c_4^-
  y_1^2\right)+ \text{Si}(x y_2) (2 c_2^++ 2y_2 c_3^+-c_4^+
   y_2^2)\nonumber\\&+\text{Si}(x y_3) \left(2 c_2^+-2 c_3^+ y_3-c_4^+
   y_3^2\right)+\text{Si}(x y_4) (2 c_2^-+ 2y_4 c_3^--c_4^-
   y_4^2)\Big\}\,.
\end{align}
In the above equations we have defined for compactness
\begin{align}
y_1=c_t-c_s(u-v)\quad,\quad y_2=c_t+c_s(u+v)\quad,\quad y_3=c_t-c_s(u+v)\quad,\quad y_4=c_t+c_s(u-v)\,.
\end{align}

We also provide the explicit expression in the superhorizon and subhorizon limit. For superhorizon modes, we have that
\begin{align}
I_c(v,u,x\ll1)=\frac{1}{2} (c_0^{-} + c_0^{+}+ 
   c_s (c_1^{-} (u - v) + c_1^{+}(u + v) + 2 c_s u v (c_2^{-} - c_4^{-} c_s^2 u v))) x^2\,,
\end{align}
and
\begin{align}
I_s(v,u,x\ll1)=\frac{1}{3}c_t (c_0^{-} + c_0^{+} + 
   c_s (c_1^{-} (u - v) + c_1^{+} (u + v) + 2 c_s u v (c_2^{-} - c_4^{-} c_s^2 u v))) x^3\,,
\end{align}
For subhorizon modes, the total kernel reads
\begin{align}\label{eq:largexappendix}
I(v,u,x)&=\frac{\sin (c_t x)}{c_t x} I_{c,0}(v,u)-\frac{\cos (c_t x)}{c_t x} I_{s,0}(v,u)\nonumber\\&+\frac{c_0^- }{2  c_s^2  u  v (1 - y_t)}\cos (c_s x (u-v))-\frac{c_0^+ }{2  c_s^2  u  v (1 + y_t)}\cos (c_s x (u+v))\nonumber\\&+\frac{1}{2(c_s^2uv(1-y_t))^2}
\left(c_0^-+\frac{c_1^-}{c_s(u-v)}
   c_s^2 u v (1-y_t)\right)\frac{\sin (c_s (u-v)x)}{c_s(u-v)x}\nonumber\\&+\frac{1}{2(c_s^2uv(1+y_t))^2}
\left(c_0^+-\frac{c_1^+}{c_s(u+v)}
   c_s^2 u v (1+y_t)\right)\frac{\sin (c_s (u+v)x)}{c_s(u+v)x}\,,
\end{align}
where we defined
\begin{align}\label{eq:Is0appendix}
I_{s,0}(v,u)&=-\frac{\pi}{4}\left(2c_2^--c_4^-(c_t^2-c_s^2(u-v)^2)\right){\rm sign}(1+y_t)\Theta(1-y_t^2)\nonumber\\&
=-\frac{\pi}{2}\left(c_2^--c_4^-  c_s^2  u  v (1 - y_t)\right){\rm sign}(1+y_t)\Theta(1-y_t^2)\,,
\end{align}
with $(y_t^2-1)/(2c_s^2uv)^2=y_1y_2y_3y_4$, or equivalently
\begin{align}
y_t=\frac{c_s^2(u^2+v^2)-c_t^2}{2c_s^2uv}\,.
\end{align}
In deriving Eq.~\eqref{eq:Is0appendix}, we used that
\begin{align}
\lim_{x\to\infty}\big(\text{Si}(x y_1) -\text{Si}(x y_2) -\text{Si}(x y_3) &+\text{Si}(x y_4)\big) =\pi{\rm sign}(y_1y_4)\Theta(-y_1y_2y_3y_4)\nonumber\\&
=\pi{\rm sign}(-y_2y_3)\Theta(-y_1y_2y_3y_4)=\pi\,{\rm sign}(1+y_t)\Theta(1-y_t^2)\,.
\end{align}
Since we have that $|u-v|<1$ and $u+v>1$, it follows that $c_t^2>y_1y_4>c_t^2-c_s^2$ and $y_2y_3<c_t^2-c_s^2$. If we consider the case $c_t>c_s$ then $y_1y_4>0$ and we have that $\Theta(1-y_t^2)=\Theta(c_s(u+v)-c_t)$ as in the standard case. We also present Eq.~\eqref{eq:Ic0appendix} in terms of variable $y_t$, namely
\begin{align}
I_{c,0}(v,u)=&-\frac{c_0^- }{2c_s^2uv(1-y_t)}\left(1-\frac{c_t^2}{c_s^2uv(1-y_t)}\right)+\frac{c_0^+ }{2c_s^2uv(1+y_t)}\left(1+\frac{c_t^2}{c_s^2uv(1+y_t)}\right)\nonumber\\&+\frac{c_1^- c_s (u-v)}{2  c_s^2  u  v (1 - y_t)}-\frac{c_1^+ c_s (u+v)}{2  c_s^2  u  v (1 + y_t)}\nonumber\\&+c_4^-
   c_s^2 u v+\frac{1}{2}\left(c_2^{-}-c_4^{-}c_s^2uv(1-y_t)\right)\ln\left|\frac{1-y_t}{1+y_t}\right|\,.
\end{align}

Lastly, when computing the change in the Kernel after going to the Newton gauge, that is Eq.~\eqref{eq:kernelchange}, we used that
\begin{align}
\tilde\beta_q\tilde\beta_{|\mathbf{k-q}|}=\zeta^{\rm p}_q\zeta^{\rm p}_{|\mathbf{k-q}|}\times A_1^2\,x^2{\cal B}_\nu(c_svx){\cal B}_\nu(c_sux)\,,
\end{align}
and
\begin{align}
A_1^2x^2{\cal B}_\nu(c_svx){\cal B}_\nu(c_sux)&=\frac{{\cal G}_t^2\cos (c_s x (u-v))}{2 c_s^2 u v\tilde\Theta^2}-\frac{{\cal G}_t^2\cos (c_s x (u+v))}{2c_s^2 u v\tilde\Theta^2}\nonumber\\&+\frac{{\cal F}_s(u-v) \sin (c_s x (u-v))}{c_s^3 u^2 v^2 \tilde\Theta x}+\frac{{\cal F}_s(u+v) \sin (c_s
   x (u+v))}{c_s^3 u^2 v^2 \tilde\Theta x}\,,
\end{align}
which exactly cancels the problematic terms in Eq.~\eqref{eq:largexappendix}.

\section{Second order action in the Newton gauge}
As an interesting exercise, and because we did not find it in the literature, we derive the second order action for scalar perturbations in the Newton gauge in Horndeski. We start from the second order action in the uniform-$\phi$ gauge, namely
\begin{align}
 {\cal L}_{ss}^u = a^3 \left({\cal G}_s \dot \zeta^2 +{\cal F}_s\zeta\Delta \zeta \right)\,.
\end{align}
Then we compute the Hamiltonian, which is given by
\begin{align}
 {\cal H}_{ss}^u =\pi_\zeta\dot\zeta- {\cal L}_{ss}^u = \frac{\pi_\zeta^2}{4a^3{\cal G}_s}-a{\cal F}_s\zeta\Delta \zeta\,.
\end{align}
From the gauge transformation (see App.~\ref{app:gauge}) we have that
\begin{align}\label{eq:psitozeta}
\Psi=\zeta+\beta H\,,
\end{align}
where
\begin{align}
\beta=\frac{1}{2a{\cal G}_t}\Delta^{-1}\pi_\zeta-\frac{{\cal G}_t}{\Theta}\zeta\,.
\end{align}

We compute the transformation to the second order action by means of a canonical transformation, see e.g. Ref.~\cite{Domenech:2017ems} for the GR case, given by
\begin{align}
\pi_\zeta=\frac{2a{\cal G}_t}{H}\Delta\Psi\quad,\quad \zeta=\frac{1}{a{\cal F}_s}\frac{\partial }{\partial t}\left(\frac{a{\cal G}_t}{H}\right)\Psi-\frac{H}{2a{\cal G}_t}\Delta^{-1}\pi_\Psi\,.
\end{align}
Then, we find that
\begin{align}
{\cal H}_{ss}^N=-\frac{H^2 {\cal F}_s}{{4a\cal G}_t^2}\pi_\Psi\Delta^{-1}\pi_\Psi+\frac{1}{aH^2}\frac{{\cal G}^2_t}{{\cal G}_s}\Delta\Psi\Delta\Psi-\frac{a{\cal G}_t}{H}\frac{\partial}{\partial t}\left(\frac{1}{a{\cal F}_s}\frac{\partial}{\partial t}\left(\frac{a{\cal G}_t}{H}\right)\right)\Psi\Delta\Psi\,.
\end{align}
The Lagrangian then reads
\begin{align}
{\cal L}_{ss}^N=\pi_\Psi\dot\Psi-{\cal H}_{ss}^N=-\frac{a {\cal G}^2_t}{{\cal F}_sH^2}\dot \Psi\Delta\dot\Psi-\frac{1}{aH^2}\frac{{\cal G}^2_t}{{\cal G}_s}\Delta\Psi\Delta\Psi+\frac{a{\cal G}_t}{H}\frac{\partial}{\partial t}\left(\frac{1}{a{\cal F}_s}\frac{\partial}{\partial t}\left(\frac{a{\cal G}_t}{H}\right)\right)\Psi\Delta\Psi\,.
\end{align}

\section{Third order action in Newton gauge \label{app:thirdorder}}

In this appendix, we present the change of the third order action under a gauge transformation from the uniform-$\phi$ slicing to the Newton gauge. We also discuss how other field redefinitions simplify the third order action and we provide an interpretation for such simplifications. For the transformation we consider the third order action \eqref{eq:sshaction} as well as the second order action \eqref{eq:secondorderaction}.

\subsection{From uniform-$\phi$ to Newton gauge}
As shown in App.~\ref{app:gauge}, the field redefinition that follows from the gauge transformation from the uniform-$\phi$ to Newton gauge is given by
\begin{align}\label{eq:transformationfirst}
h_{ij}&\to h_{ij}-\frac{1}{a^2}\left(\partial_i \beta\partial_j \beta\right)^{TT}\,,\\
\zeta&\to \zeta+\frac{1}{4}\Delta^{-1}\left(\dot h_{ij}\partial_i\partial_j \beta\right)^{TT}+\alpha_1\Delta^{-1}\left(h_{ij}\partial_i\partial_j \left(\zeta+\frac{\Theta}{{\cal G}_t}\beta\right)\right)\,,
\end{align}
where we only wrote down the relevant terms for the scalar-scalar-tensor third order action. Note that we added an additional term to $\zeta$ proportional to $h_{ij}$ because it greatly simplifies the action in the Newton gauge in GR \cite{Domenech:2017ems}. We may fix the value of $\alpha_1$ to simplify some terms. In GR the $\alpha_1$ parameter in Eq.~\eqref{eq:transformationfirst} is equal to $1$. Here we leave it as a free parameter. Plugging Eq.~\eqref{eq:transformationfirst} into \eqref{eq:secondorderaction}, and after several integrations by parts, we find that
\begin{align}\label{eq:actionewtonfirst}
    {\cal L}_{ssh} &=  a{\cal F}_t (2\alpha_1-1) h_{ij}\partial_i\zeta\partial_j\zeta\nonumber\\&+2a\alpha_1\left(c_t^2\Theta-{\cal G}_t \frac{\partial\ln\alpha_1}{\partial t}\right) h_{ij}\partial_i\beta\partial_j\zeta+2a{\cal G}_t\left(\frac{H{\cal G}_t}{\Theta}\alpha_1-\frac{3+c_t^2}{4}\right)h_{ij}\partial_i\beta\partial_j\dot\zeta\nonumber\\&+\left(\frac{2\Theta}{{\cal G}_t}\frac{\partial}{\partial t}\left(a{\cal G}_t\right)-\frac{\partial}{\partial t}\left(a\tilde C_\beta\right)\right)h_{ij}\partial_i\beta\partial_j\beta+\frac{a{\cal G}_t}{\Theta}(\alpha_1-1)h_{ij}\partial_i\zeta\partial_j\dot\zeta\nonumber\\&
    +2a\left(\Theta\alpha_1-{\tilde C_\beta}\right)h_{ij}\partial_i\dot\beta\partial_j\beta+\frac{a{\cal G}_t}{2\Theta}\left({{\cal G}_t}-\Gamma\right)\dot h_{ij}\partial_i\beta\partial_j\dot\zeta+\frac{{\cal G}_t}{4a}(1-c_t^2)h_{ij}\Delta(\partial_i\beta\partial_j\beta)\nonumber\\
    &+\mu\left(\frac{{\cal G}_t}{\Theta}\dot h_{ij}\partial_i\dot\zeta\partial_j\zeta-\dot h_{ij}\partial_i\dot\zeta\partial_j\beta-\frac{{\cal G}_t}{a^2\Theta}h_{ij}\Delta(\partial_i\dot\zeta\partial_j\beta)+\frac{1}{2a^2}\dot h_{ij}\Delta(\partial_i\beta\partial_j\beta)\right)\,,
\end{align}
where for compactness we have defined
\begin{align}
\tilde C_\beta={{\cal G}_tH}\left(1+\frac{1}{2}\frac{\partial \ln{\cal G}_t}{Hdt}\right)\,.
\end{align}
As we shall see later, this action also recovers the action in the Newton gauge in the GR limit for $\alpha_1=1$ (see App.~\ref{app:GRlimit}).

\subsection{Disformed Newton gauge}
The third order action can also be simplified from the following field redefinition
\begin{align}\label{eq:transformationtonewton}
h_{ij}&\to h_{ij}-\frac{1}{a^2c_t^2}\left(\partial_i \beta\partial_j \beta\right)^{TT}\,,\\
\zeta&\to \zeta+\frac{1}{4c_t^2}\Delta^{-1}\left(\dot h_{ij}\partial_i\partial_j \beta\right)^{TT}+\Delta^{-1}\left(h_{ij}\partial_i\partial_j \left(\zeta+\frac{\Theta}{{\cal G}_t}\beta\right)\right)\,.
\end{align}
It should be noted that such field redefinition has the same GR limit as Eq.~\eqref{eq:transformationfirst}. Note that for simplicity we have set $\alpha_1=1$ in \eqref{eq:transformationfirst}. In general, as we show in App.~\ref{app:Newtondisformal}, it corresponds to the Newton gauge of the GW Einstein frame after a disformal transformation.\footnote{This is also clear by using that the disformal transformation to the GW Einstein frame, where $c_t=1$, is given by
$g_{\mu\nu}\to g_{\mu\nu}+(1-c_t^2)\delta^0_\mu\delta_\nu^0$. This changes the shift by $\beta\to c_t \beta$. Therefore the original tensor modes in the uniform-$\phi$ slicing in terms of the disformed Newton slicing read exactly as Eq.~\eqref{eq:transformationtonewton}. The same applies to the redefinition of the curvature perturbation $\zeta$. } After several integrations by parts we find that
\begin{align}\label{eq:simplifiednewton}
    {\cal L}_{ssh} &=  a{\cal F}_t h_{ij}\partial_i\zeta\partial_j\zeta+2ac_t^2\Theta h_{ij}\partial_i\beta\partial_j\zeta+\left(\frac{2\Theta}{{\cal G}_t}\frac{\partial}{\partial t}\left(a{\cal G}_t\right)-\frac{\partial}{\partial t}\left(aC_\beta\right)\right)h_{ij}\partial_i\beta\partial_j\beta\nonumber\\&
    +2a\left(\Theta-{C_\beta}\right)h_{ij}\partial_i\dot\beta\partial_j\beta+\frac{a{\cal G}_t}{2\Theta}\left(\frac{{\cal G}_t}{c_t^2}-\Gamma\right)\dot h_{ij}\partial_i\beta\partial_j\dot\zeta\nonumber\\
    &+a\mu\left(\frac{{\cal G}_t}{\Theta}\dot h_{ij}\partial_i\dot\zeta\partial_j\zeta-\dot h_{ij}\partial_i\dot\zeta\partial_j\beta-\frac{{\cal G}_t}{a^2\Theta}h_{ij}\Delta(\partial_i\dot\zeta\partial_j\beta)+\frac{1}{2a^2}\dot h_{ij}\Delta(\partial_i\beta\partial_j\beta)\right)\,,
\end{align}
where for compactness we have defined
\begin{align}
C_\beta=\frac{{\cal G}_tH}{c_t^2}\left(1+\frac{1}{2}\frac{\partial \ln(c_t{\cal G}_t)}{Hdt}\right)\,.
\end{align}
This action also recovers the action in the Newton gauge in the GR limit (see App.~\ref{app:GRlimit}). The advantage of this action is that we remove most of the $\dot h_{ij}$ terms, which simplifies the calculation of the source term for the induced GWs. Unfortunately, we found no field redefinition that completely eliminates all four derivative terms in the action. 

\subsection{GR limit with canonical scalar field\label{app:GRlimit}}
We end this appendix by showing that the third order actions \eqref{eq:actionewtonfirst} and \eqref{eq:simplifiednewton} reduce to the results of Ref.~\cite{Domenech:2017ems} in GR and a canonical scalar field. In that limit, both actions reduce, after more integration by parts, to
\begin{align}
    {\cal L}_{ssh}^N &=  a\left(1+\epsilon\right)H^2h_{ij}\partial_i\beta^u\partial_j\beta^u+ah_{ij}\partial_i\zeta\partial_j\zeta+2aHh_{ij}\partial_i\zeta\partial_j\beta^u\\
    &=ah_{ij}\partial_i\Psi\partial_j\Psi+a\epsilon H^2h_{ij}\partial_i\beta^u\partial_j\beta^u\,,
\end{align}
where in the last line we use that $\beta^u=\delta\phi_N/\dot\phi$ as well as Eq.~\eqref{eq:psitozeta}.

\section{Gauge transformations \label{app:gauge}}
For self-consistency, we also derive the relevant gauge transformation rules at first and second order in cosmological perturbation theory. As a starting point, we consider the active approach to gauge transformations \cite{Malik_2009}, which gives
\begin{align}
{\tilde g}_{\mu\nu}={\rm e}^{{\cal L}_\xi}{g}_{\mu\nu}={ g}_{\mu\nu}+{\cal L}_\xi{ g}_{\mu\nu}+\frac{1}{2}{\cal L}^2_\xi{ g}_{\mu\nu}+...\,.
\end{align}
The same formula applies to any tensor. For instance, we have that \cite{Domenech:2023fuz}
\begin{align}
&{\cal L}_\xi A=\xi^\alpha\partial_\alpha A\,,\\
&{\cal L}_\xi B_\mu=\xi^\alpha\partial_\alpha B_\mu+B_\alpha\partial_\mu \xi^\alpha \,,\\
&{\cal L}_\xi C_{\mu\nu}=\xi^\alpha\partial_\alpha C_{\mu\nu}+2C_{\alpha(\mu}\partial_{\nu)}\xi^{\alpha}\,,
\end{align}
and
\begin{align}
&{\cal L}^2_\xi A=\xi^\beta\partial_\beta\left(\xi^\alpha\partial_\alpha A\right)\,,\\
&{\cal L}^2_\xi B_\mu=\xi^\beta\partial_\beta\left(\xi^\alpha\partial_\alpha B_\mu+B_\alpha\partial_\mu \xi^\alpha\right)+\xi^\alpha\partial_\alpha u_\beta\partial_\mu\xi^\beta+B_\alpha\partial_\beta \xi^\alpha\partial_\mu \xi^\beta\,, \\
&{\cal L}^2_\xi C_{\mu\nu}=\xi^\beta\partial_\beta\left(\xi^\alpha\partial_\alpha C_{\mu\nu}+2C_{\alpha(\mu}\partial_{\nu)}\xi^{\alpha}\right)+2\xi^\alpha \partial_\alpha C_{\beta(\mu}\partial_{\nu)}\xi^\beta\nonumber\\&\qquad\qquad\qquad\qquad\qquad\qquad\qquad\qquad\quad+2C_{\alpha\beta}\partial_{(\mu}\xi^\alpha\partial_{\nu)}\xi^\beta+2\partial_\beta\xi^\alpha C_{\alpha(\mu}\partial_{\nu)}\xi^\beta\,,
\end{align}
where $A$ is an arbitrary scalar, $B$ is an arbitrary vector and $C$ is an arbitrary tensor.

We take the metric given by the ADM decomposition, namely
\begin{align}
ds^2=-N^2dt^2+H_{ij}(dx^i+N^idt)(dx^j+N^jdt)\,,
\end{align}
and consider the perturbations $N=1+\delta N$, $N_i=\partial_i\beta$ and
\begin{align}\label{eq:exponentialansatz}
H_{ij}=a^2e^{2\psi}(e^h)_{ij}\,,
\end{align}
where $\psi$ is the curvature perturbation and $h_{ij}$ the tensor perturbation, i.e. the transverse-traceless component. We fixed the spatial gauge to remove any traceless scalar component in the $(e^h)_{ij}$ part and we neglected vector modes. This is because we will be mainly interested in temporal gauge transformations. Then, using the above formulas we find, assuming $\xi^\alpha=(T,L^i)$ where $L^i$ is a second order quantity (see later why), that
\begin{align}
\delta\tilde g_{00}&=\delta g_{00}-2T\delta N-2\dot T-4\delta N\dot T-T\ddot T-2\dot T^2\,\\
\delta\tilde g_{0i}&=\delta g_{0i}+T\partial_i\dot\beta-(1+2\delta N)\partial_i T+\partial_i\beta \dot T-\frac{1}{2}T\partial_i\dot T-2\dot T\partial_i T+a^2\dot L^i \label{eq:g0igauge}\\
\delta\tilde g_{ij}&=\delta g_{ij}+2HTa^2(1+2\psi+h_{ij})+Ta^2(2\dot\psi+\dot h_{ij})+2\partial_{(i}\beta\partial_{j)}T\nonumber\\&\qquad\qquad\qquad\qquad\qquad\qquad\qquad+\frac{1}{2}T\frac{\partial }{\partial t}\left(2a^2HT\right)-\partial_iT\partial_jT+2a^2\partial_{(i} L_{j)}\,,
\end{align}
where the brackets denote normalized symmetrization of indices. For the scalar field one has
\begin{align}
\delta\tilde\phi=\delta\phi+T(\dot\phi+\dot\delta\phi)+\frac{1}{2}T\frac{\partial}{\partial t}(T\dot\phi)\,.
\end{align}
Let us now write down the transformations for first and second order separately.

\subsection{Linear order}

At linear level we recover the well-known transformation rules, namely
\begin{align}
\delta\tilde N= \delta N +\dot T\quad,\quad \tilde\beta=\beta-T\quad,\quad \tilde\psi=\psi + HT \quad,\quad \delta \tilde\phi=\delta\phi+\dot \phi T\quad,\quad \tilde h_{ij}= h_{ij}\,.
\end{align}

The relation from the uniform-$\phi$ slicing to the Newton slicing at linear order is given by
\begin{align}
\beta^u=-T\quad,\quad \delta\phi^N=-\dot\phi T\,.
\end{align}
Since we know the solution of $\beta^u$ we may write that $\delta\phi^N=\dot\phi\beta^u$.

\subsection{Second order}

Now we look at second order transformations and focus on the curvature and tensor perturbations. We note that the exponential ansatz \eqref{eq:exponentialansatz} is not preserved under the time reparametrization, as noted in Ref.~\cite{Maldacena:2002vr}. Thus, we use $L^i$ to recover it. The suitable parameter $L^i$ is found by solving
\begin{align}
\left(\Delta^{-1}\partial_i\partial_j-\frac{1}{3}\delta_{ij}\right)\delta\tilde g_{ij}=0\,,
\end{align}
where repeated lower indices are contracted by $\delta_{ij}$. The remaining trace part is absorbed by $\psi$ via
\begin{align}
\psi\to \psi+\alpha\,,
\end{align}
where
\begin{align}
4\alpha=\left(\delta_{ij}-\Delta^{-1}\partial_i\partial_j\right)\delta\tilde g_{ij}\,.
\end{align}
Note that no $L^i$ appears in the above equation for $\alpha$.

We are interested in the scalar-scalar terms in the gauge transformation of tensor perturbations an in the scalar-tensor terms in the gauge transformation of scalar modes. These are given by
\begin{align}\label{eq:gaugetransformsecond}
\tilde h_{ij}&=h_{ij}+a^{-2}\left(2\partial_{(i}\beta\partial_{j)}T-\partial_i T\partial_j T\right)^{TT}+...\\
\tilde \psi&=\psi-\frac{1}{4}\Delta^{-1}\left(\dot h_{ij}\partial_i\partial_j T\right)-\frac{1}{4}\left(\partial_iT\partial_jT-\Delta^{-1}\partial_i\partial_j(\partial_iT\partial_jT)\right)+...\,,
\end{align}
where the “...” refer to the other terms which we neglect for simplicity. From the above formulas we find the relation between the curvature and tensor perturbations in the uniform-$\phi$ slicing ($\delta\phi^u=0$) and the Newton gauge ($\beta^N=0$). The subscript $u$ and $N$ respectively denotes evaluation in the uniform-$\phi$ and Newton slicings. From the linear gauge transformation of Eq.~\eqref{eq:g0igauge} we find that $T=-\beta_u$ and, therefore, it follows that
\begin{align}\label{eq:gauge-uniform-newton}
h^u_{ij}&=h^N_{ij}-a^{-2}\left(\partial_i \beta^u\partial_j \beta^u\right)^{TT}+...\\
\zeta&=\Psi+\frac{1}{4}\Delta^{-1}\left(\dot h_{ij}\partial_i\partial_j \beta^u\right)-\frac{1}{4}\left(\partial_i\beta\partial_j\beta-\Delta^{-1}\partial_i\partial_j(\partial_i\beta\partial_j\beta)\right)+...\,,
\end{align}
where we called $\zeta=\psi^u$ and $\Psi=\psi^N$. At linear order we have that $\Phi=\zeta+H\beta^u$, where $H=\dot a/a$ is the Hubble parameter. We also have that $\delta\phi_N=\dot\phi\beta^u$.

\section{Field redefinitions, Disformal transformations and Newton gauge \label{app:Newtondisformal}}

In general, it is interesting to understand the interplay between gauge transformations and disformal transformations. For instance, Ref.~\cite{Domenech:2015hka} proves the non-linear equivalence between disformal frames in the uniform-$\phi$ slicing (see also Ref.~\cite{Minamitsuji:2014waa,Motohashi:2015pra} for the analysis at linear level and Ref.~\cite{BenAchour:2024tqt} for a coordinate independent analysis). However, once we depart from the uniform-$\phi$ slicing, the “equivalence” becomes more involved. For instance, there will be a mixing between tensors and scalar fluctuations in different frames at second order in cosmological perturbation theory. Such mixing is important for induced GWs.

Consider the disformal transformation given by
\begin{align}
g_{\mu\nu}=\tilde g_{\mu\nu}+D(\phi)\partial_\mu\phi\partial_\nu\phi\,.
\end{align}
At the background level such transformation leads to
\begin{align}
 N_0^2 = \tilde N^2-D(t)\left(\frac{d\phi}{d\tilde t}\right)^2\,,
\end{align}
which can be absorbed by a redefinition of time \cite{Domenech:2015hka}, namely
\begin{align}
d t= d\tilde t\sqrt{1-D\left(\frac{d\phi}{d\tilde t}\right)}\,.
\end{align}
We also have that $a=\tilde a$. For linear perturbations, we find that
\begin{align}
\delta  N=\delta \tilde N-D\left(\frac{d\phi}{d\tilde t}\right)^2\delta\dot\phi\quad,\quad
\beta =\tilde \beta+D\frac{d\phi}{d\tilde t}\delta\phi\quad,\quad
\psi=\tilde\psi\quad,\quad
h_{ij}=\tilde h_{ij}\,.
\end{align}

If we focus only on the tensor perturbations, we have at second order  that \cite{Cai:2023ykr}
\begin{align}
h_{ij}=\tilde h_{ij}+\tilde a^{-2}D\left(\partial_i\delta\phi\partial_j\delta\phi\right)^{TT}\,.
\end{align}
In the Newton gauge we have that $\delta\phi\neq 0$ and therefore $h_{ij}=\tilde h_{ij}$ are in principle not the same, though we need to correct by the fact that a disformal transformation leads to a non-zero shift. To correct for this we do a temporal gauge transformation as in App.~\ref{app:gauge}. The transformation of the shift is then
\begin{align}
\beta&=\tilde{\tilde \beta}-T+D\frac{d\phi}{d\tilde t}\delta\phi+D\left(\frac{d\phi}{d\tilde t}\right)^2T\,.
\end{align}
Thus the condition that $\beta^{N}=\tilde{\tilde \beta}^{\tilde{\tilde N}}=0$ yields
\begin{align}\label{eq:Tdisformal}
T=\frac{D\left({d\phi}/{d\tilde t}\right)^2}{1-D\left({d\phi}/{d\tilde t}\right)^2}\frac{\delta\phi^{\tilde{\tilde N}}}{{d\phi}/{d\tilde t}}\,.
\end{align}
Note that we denoted with a subscript $\tilde{\tilde N}$ the Newton gauge in the tilde metric that matches the definition of Newton gauge in the metric. This difference is important since the scalar field perturbation also changes under the time reparametrization. For example, we now have that
\begin{align}
\delta\phi^N=\delta\phi^{\tilde N}=\frac{\delta\phi^{\tilde{\tilde N}}}{1-D(d\phi/d\tilde t)^2}.
\end{align}

At second order, we have from \eqref{eq:gaugetransformsecond} in their corresponding metric that
\begin{align}
 h_{ij}= \tilde {\tilde h}_{ij}+\tilde a^{-2}\left(2\partial_{(i}\beta\partial_{j)}T-\partial_i T\partial_j T\right)^{TT}+\tilde a^{-2}D\left(\partial_i\left(\delta\phi+ \frac{d\phi}{d\tilde t} T\right)\partial_j\left(\delta\phi+ \frac{d\phi}{d\tilde t} T\right)\right)^{TT}\,.
\end{align}
Using the solution \eqref{eq:Tdisformal} we have that
\begin{align}\label{eq:hnmess}
 h^{ N}_{ij}=\tilde {\tilde h}^{\tilde{\tilde N}}_{ij}+\tilde a^{-2}\frac{D\left({d\phi}/{d\tilde t}\right)^2}{1-D\left({d\phi}/{d\tilde t}\right)^2}\left(\partial_i\frac{\delta\phi^{ \tilde{\tilde N}}}{{d\phi}/{d\tilde t}}\partial_j\frac{\delta\phi^{ \tilde{\tilde N}}}{{d\phi}/{d\tilde t}}\right)^{TT}\,.
\end{align}

The same follows using the equivalence in the uniform-$\phi$ gauge. For instance, we can take that
\begin{align}
\tilde h^u_{ij}&=\tilde {\tilde h}^{\tilde{\tilde N}}_{ij}- \tilde a^{-2}\left(\partial_i \frac{\delta\phi^{\tilde{\tilde N}}}{d\phi/d\tilde t}\partial_j \frac{\delta\phi^{\tilde{\tilde N}}}{d\phi/d\tilde t}\right)^{TT}
\quad {\rm and}\quad 
h^u_{ij}&=h^N_{ij}-a^{-2}\left(\partial_i \frac{\delta\phi^N}{d\phi/d t}\partial_j \frac{\delta\phi^N}{d\phi/d t}\right)^{TT}\,.
\end{align}
Then, using that $\tilde h^u_{ij}= h^u_{ij}$ and the relation between $d\tilde t$ and $dt$, we recover Eq.~\eqref{eq:hnmess}.

\subsection{ GW Einstein frame}

Let us now require that the disformal transformation sets $c_t=1$ for the metric without tildes. For example, if we look at the second order action in the barred frame, we have
\begin{align}
S_{tt}&=\int d^3x d t \,\frac{ a^3}{8} \left( {\cal G}_t \frac{\partial}{\partial  t} h_{ij} \frac{\partial}{\partial  t}  h_{ij} +a^{-2} {\cal F}_t  h_{ij}\Delta  h_{ij} \right)\\&
=\int d^3x d \tilde t \,\frac{ \tilde a^3 {\cal G}_t}{8\sqrt{1-D(d\phi/d\tilde t)^2}} \left( \frac{\partial}{\partial  \tilde t} h_{ij} \frac{\partial}{\partial  \tilde t} h_{ij} +\tilde a^{-2} c_t^2(1-D(d\phi/d\tilde t)^2) h_{ij}\Delta h_{ij} \right)\,.
\end{align}
We then choose that $ c_t^2(1-D(d\phi/d\tilde t)^2)=\tilde c_t^2=1$. We will remove the remaining pre-factor by a conformal transformation later. With the condition of GW Einstein frame, Eq.~\eqref{eq:Tdisformal}  then becomes
\begin{align}\label{eq:newtontonewton}
 h^{N}_{ij}=\tilde {\tilde h}^{\tilde{\tilde N}}_{ij}+\left( c_t^2-1\right)\tilde a^{-2}\left(\partial_i\frac{\delta\phi^{\tilde{\tilde N}}}{d\phi/d \tilde t}\partial_j\frac{\delta\phi^{\tilde{\tilde N}}}{d\phi/d \tilde t}\right)^{TT}=\tilde {\tilde h}^{\tilde{\tilde N}}_{ij}+\left(1-c_t^{-2}\right) a^{-2}\left(\partial_i\frac{\delta\phi^{N}}{d\phi/d  t}\partial_j\frac{\delta\phi^{N}}{d\phi/d  t}\right)^{TT}\,.
\end{align}

Lastly, to obtain Eq.~\eqref{eq:transformationtonewton} we have to use a conformal transformation to go to the GW Einstein frame. Specifically, we have that
\begin{align}
 \tilde g_{\mu\nu}= c_t^{-1} \bar  g_{\mu\nu}\,,
\end{align}
The effect of the conformal transformation is to further redefine time and scale factor, namely $d\tilde t = c_t^{-1/2} d\bar t$ and $\tilde a = c_t^{-1/2} \bar a$.  Then, the action in the barred frame is
\begin{align}
S_{tt}&
=\int d^3x d \tilde t\, \frac{ c_t\tilde a^3 {\cal G}_t}{8} \left( \frac{\partial}{\partial  \tilde t} h_{ij} \frac{\partial}{\partial \tilde  t} h_{ij} +\tilde a^{-2}  h_{ij}\Delta h_{ij} \right)\\&=\int d^3x d \bar t \,\frac{ \bar a^3 {\cal G}_t}{8} \left( \frac{\partial}{\partial\bar  t} h_{ij} \frac{\partial}{\partial\bar  t} h_{ij} +\bar a^{-2}  h_{ij}\Delta h_{ij} \right)\,.
\end{align}
We then relate the tensor modes in the original frame in the uniform-$\phi$ slicing $h_{ij}^u$ to the tensor modes in the GW Eintein frame in the Newton slicing by
\begin{align}
h^u_{ij}={\bar h}^{\bar N}_{ij}-\frac{1}{a^{2}c_t^2}\left(\partial_i \frac{\delta\phi}{d\phi/d t}\partial_j \frac{\delta\phi}{d\phi/d  t}\right)^{TT}
\end{align}
where we used that ${\bar h}^{\bar N}_{ij}=\tilde {\tilde h}^{\tilde N}_{ij}$ as a conformal transformation does not affect the tensor perturbations. We also used that $\tilde ad\phi/d\tilde t=\bar a d\phi/d\bar t$. The last step is to use the gauge transformation rule between the Newton gauge and the uniform-$\phi$ gauge, which leads us to
\begin{align}
h^u_{ij}={\bar h}^{\bar N}_{ij}-\frac{1}{a^{2}c_t^2}\left(\partial_i \beta^u\partial_j \beta^u\right)^{TT}\,.
\end{align}
Note that with a further conformal transformation we could remove ${\cal G}_t$ completely but the above transformation rule for the tensor fluctuations would remain unchanged. Lastly, we note that the effect of going to the GW Einstein frame on curvature fluctuations is the replacement of $c_s^2$ by $\bar{c}_s^2={{c}_s^2}/{c_t^2}$.

\bibliography{refgwscalar.bib}

\begin{thebibliography}{234}%
\makeatletter
\providecommand \@ifxundefined [1]{%
 \@ifx{#1\undefined}
}%
\providecommand \@ifnum [1]{%
 \ifnum #1\expandafter \@firstoftwo
 \else \expandafter \@secondoftwo
 \fi
}%
\providecommand \@ifx [1]{%
 \ifx #1\expandafter \@firstoftwo
 \else \expandafter \@secondoftwo
 \fi
}%
\providecommand \natexlab [1]{#1}%
\providecommand \enquote  [1]{``#1''}%
\providecommand \bibnamefont  [1]{#1}%
\providecommand \bibfnamefont [1]{#1}%
\providecommand \citenamefont [1]{#1}%
\providecommand \href@noop [0]{\@secondoftwo}%
\providecommand \href [0]{\begingroup \@sanitize@url \@href}%
\providecommand \@href[1]{\@@startlink{#1}\@@href}%
\providecommand \@@href[1]{\endgroup#1\@@endlink}%
\providecommand \@sanitize@url [0]{\catcode `\\12\catcode `\$12\catcode
  `\&12\catcode `\#12\catcode `\^12\catcode `\_12\catcode `\%12\relax}%
\providecommand \@@startlink[1]{}%
\providecommand \@@endlink[0]{}%
\providecommand \url  [0]{\begingroup\@sanitize@url \@url }%
\providecommand \@url [1]{\endgroup\@href {#1}{\urlprefix }}%
\providecommand \urlprefix  [0]{URL }%
\providecommand \Eprint [0]{\href }%
\providecommand \doibase [0]{https://doi.org/}%
\providecommand \selectlanguage [0]{\@gobble}%
\providecommand \bibinfo  [0]{\@secondoftwo}%
\providecommand \bibfield  [0]{\@secondoftwo}%
\providecommand \translation [1]{[#1]}%
\providecommand \BibitemOpen [0]{}%
\providecommand \bibitemStop [0]{}%
\providecommand \bibitemNoStop [0]{.\EOS\space}%
\providecommand \EOS [0]{\spacefactor3000\relax}%
\providecommand \BibitemShut  [1]{\csname bibitem#1\endcsname}%
\let\auto@bib@innerbib\@empty
\bibitem [{\citenamefont {De~Felice}\ and\ \citenamefont
  {Tsujikawa}(2010{\natexlab{a}})}]{DeFelice:2010aj}%
  \BibitemOpen
  \bibfield  {author} {\bibinfo {author} {\bibfnamefont {A.}~\bibnamefont
  {De~Felice}}\ and\ \bibinfo {author} {\bibfnamefont {S.}~\bibnamefont
  {Tsujikawa}},\ }\bibfield  {title} {\bibinfo {title} {{f(R) theories}},\
  }\href {https://doi.org/10.12942/lrr-2010-3} {\bibfield  {journal} {\bibinfo
  {journal} {Living Rev. Rel.}\ }\textbf {\bibinfo {volume} {13}},\ \bibinfo
  {pages} {3} (\bibinfo {year} {2010}{\natexlab{a}})},\ \Eprint
  {https://arxiv.org/abs/1002.4928} {arXiv:1002.4928 [gr-qc]} \BibitemShut
  {NoStop}%
\bibitem [{\citenamefont {Koyama}(2016)}]{Koyama:2015vza}%
  \BibitemOpen
  \bibfield  {author} {\bibinfo {author} {\bibfnamefont {K.}~\bibnamefont
  {Koyama}},\ }\bibfield  {title} {\bibinfo {title} {{Cosmological Tests of
  Modified Gravity}},\ }\href {https://doi.org/10.1088/0034-4885/79/4/046902}
  {\bibfield  {journal} {\bibinfo  {journal} {Rept. Prog. Phys.}\ }\textbf
  {\bibinfo {volume} {79}},\ \bibinfo {pages} {046902} (\bibinfo {year}
  {2016})},\ \Eprint {https://arxiv.org/abs/1504.04623} {arXiv:1504.04623
  [astro-ph.CO]} \BibitemShut {NoStop}%
\bibitem [{\citenamefont {Nojiri}\ \emph {et~al.}(2017)\citenamefont {Nojiri},
  \citenamefont {Odintsov},\ and\ \citenamefont {Oikonomou}}]{Nojiri:2017ncd}%
  \BibitemOpen
  \bibfield  {author} {\bibinfo {author} {\bibfnamefont {S.}~\bibnamefont
  {Nojiri}}, \bibinfo {author} {\bibfnamefont {S.~D.}\ \bibnamefont
  {Odintsov}},\ and\ \bibinfo {author} {\bibfnamefont {V.~K.}\ \bibnamefont
  {Oikonomou}},\ }\bibfield  {title} {\bibinfo {title} {{Modified Gravity
  Theories on a Nutshell: Inflation, Bounce and Late-time Evolution}},\ }\href
  {https://doi.org/10.1016/j.physrep.2017.06.001} {\bibfield  {journal}
  {\bibinfo  {journal} {Phys. Rept.}\ }\textbf {\bibinfo {volume} {692}},\
  \bibinfo {pages} {1} (\bibinfo {year} {2017})},\ \Eprint
  {https://arxiv.org/abs/1705.11098} {arXiv:1705.11098 [gr-qc]} \BibitemShut
  {NoStop}%
\bibitem [{\citenamefont {Langlois}(2019)}]{Langlois:2018dxi}%
  \BibitemOpen
  \bibfield  {author} {\bibinfo {author} {\bibfnamefont {D.}~\bibnamefont
  {Langlois}},\ }\bibfield  {title} {\bibinfo {title} {{Dark energy and
  modified gravity in degenerate higher-order scalar\textendash{}tensor (DHOST)
  theories: A review}},\ }\href {https://doi.org/10.1142/S0218271819420069}
  {\bibfield  {journal} {\bibinfo  {journal} {Int. J. Mod. Phys. D}\ }\textbf
  {\bibinfo {volume} {28}},\ \bibinfo {pages} {1942006} (\bibinfo {year}
  {2019})},\ \Eprint {https://arxiv.org/abs/1811.06271} {arXiv:1811.06271
  [gr-qc]} \BibitemShut {NoStop}%
\bibitem [{\citenamefont {Kobayashi}(2019)}]{Kobayashi:2019hrl}%
  \BibitemOpen
  \bibfield  {author} {\bibinfo {author} {\bibfnamefont {T.}~\bibnamefont
  {Kobayashi}},\ }\bibfield  {title} {\bibinfo {title} {{Horndeski theory and
  beyond: a review}},\ }\href {https://doi.org/10.1088/1361-6633/ab2429}
  {\bibfield  {journal} {\bibinfo  {journal} {Rept. Prog. Phys.}\ }\textbf
  {\bibinfo {volume} {82}},\ \bibinfo {pages} {086901} (\bibinfo {year}
  {2019})},\ \Eprint {https://arxiv.org/abs/1901.07183} {arXiv:1901.07183
  [gr-qc]} \BibitemShut {NoStop}%
\bibitem [{\citenamefont {Shankaranarayanan}\ and\ \citenamefont
  {Johnson}(2022)}]{Shankaranarayanan:2022wbx}%
  \BibitemOpen
  \bibfield  {author} {\bibinfo {author} {\bibfnamefont {S.}~\bibnamefont
  {Shankaranarayanan}}\ and\ \bibinfo {author} {\bibfnamefont {J.~P.}\
  \bibnamefont {Johnson}},\ }\bibfield  {title} {\bibinfo {title} {{Modified
  theories of gravity: Why, how and what?}},\ }\href
  {https://doi.org/10.1007/s10714-022-02927-2} {\bibfield  {journal} {\bibinfo
  {journal} {Gen. Rel. Grav.}\ }\textbf {\bibinfo {volume} {54}},\ \bibinfo
  {pages} {44} (\bibinfo {year} {2022})},\ \Eprint
  {https://arxiv.org/abs/2204.06533} {arXiv:2204.06533 [gr-qc]} \BibitemShut
  {NoStop}%
\bibitem [{\citenamefont {Akrami}\ \emph {et~al.}(2020)\citenamefont {Akrami}
  \emph {et~al.}}]{Akrami:2018odb}%
  \BibitemOpen
  \bibfield  {author} {\bibinfo {author} {\bibfnamefont {Y.}~\bibnamefont
  {Akrami}} \emph {et~al.} (\bibinfo {collaboration} {Planck}),\ }\bibfield
  {title} {\bibinfo {title} {{Planck 2018 results. X. Constraints on
  inflation}},\ }\href {https://doi.org/10.1051/0004-6361/201833887} {\bibfield
   {journal} {\bibinfo  {journal} {Astron. Astrophys.}\ }\textbf {\bibinfo
  {volume} {641}},\ \bibinfo {pages} {A10} (\bibinfo {year} {2020})},\ \Eprint
  {https://arxiv.org/abs/1807.06211} {arXiv:1807.06211 [astro-ph.CO]}
  \BibitemShut {NoStop}%
\bibitem [{\citenamefont {Starobinsky}(1979)}]{Starobinsky:1979ty}%
  \BibitemOpen
  \bibfield  {author} {\bibinfo {author} {\bibfnamefont {A.~A.}\ \bibnamefont
  {Starobinsky}},\ }\bibfield  {title} {\bibinfo {title} {{Spectrum of relict
  gravitational radiation and the early state of the universe}},\ }\href@noop
  {} {\bibfield  {journal} {\bibinfo  {journal} {JETP Lett.}\ }\textbf
  {\bibinfo {volume} {30}},\ \bibinfo {pages} {682} (\bibinfo {year}
  {1979})}\BibitemShut {NoStop}%
\bibitem [{\citenamefont {Sato}(1981)}]{Sato:1980yn}%
  \BibitemOpen
  \bibfield  {author} {\bibinfo {author} {\bibfnamefont {K.}~\bibnamefont
  {Sato}},\ }\bibfield  {title} {\bibinfo {title} {{First Order Phase
  Transition of a Vacuum and Expansion of the Universe}},\ }\href@noop {}
  {\bibfield  {journal} {\bibinfo  {journal} {Mon. Not. Roy. Astron. Soc.}\
  }\textbf {\bibinfo {volume} {195}},\ \bibinfo {pages} {467} (\bibinfo {year}
  {1981})}\BibitemShut {NoStop}%
\bibitem [{\citenamefont {Guth}(1981)}]{Guth:1980zm}%
  \BibitemOpen
  \bibfield  {author} {\bibinfo {author} {\bibfnamefont {A.~H.}\ \bibnamefont
  {Guth}},\ }\bibfield  {title} {\bibinfo {title} {{The Inflationary Universe:
  A Possible Solution to the Horizon and Flatness Problems}},\ }\href
  {https://doi.org/10.1103/PhysRevD.23.347} {\bibfield  {journal} {\bibinfo
  {journal} {Phys. Rev. D}\ }\textbf {\bibinfo {volume} {23}},\ \bibinfo
  {pages} {347} (\bibinfo {year} {1981})}\BibitemShut {NoStop}%
\bibitem [{\citenamefont {Fujii}\ and\ \citenamefont
  {Maeda}(2007)}]{Fujii:2003pa}%
  \BibitemOpen
  \bibfield  {author} {\bibinfo {author} {\bibfnamefont {Y.}~\bibnamefont
  {Fujii}}\ and\ \bibinfo {author} {\bibfnamefont {K.}~\bibnamefont {Maeda}},\
  }\href {https://doi.org/10.1017/CBO9780511535093} {\emph {\bibinfo {title}
  {{The scalar-tensor theory of gravitation}}}},\ Cambridge Monographs on
  Mathematical Physics\ (\bibinfo  {publisher} {Cambridge University Press},\
  \bibinfo {year} {2007})\BibitemShut {NoStop}%
\bibitem [{\citenamefont {Bezrukov}\ and\ \citenamefont
  {Shaposhnikov}(2008)}]{Bezrukov:2007ep}%
  \BibitemOpen
  \bibfield  {author} {\bibinfo {author} {\bibfnamefont {F.~L.}\ \bibnamefont
  {Bezrukov}}\ and\ \bibinfo {author} {\bibfnamefont {M.}~\bibnamefont
  {Shaposhnikov}},\ }\bibfield  {title} {\bibinfo {title} {{The Standard Model
  Higgs boson as the inflaton}},\ }\href
  {https://doi.org/10.1016/j.physletb.2007.11.072} {\bibfield  {journal}
  {\bibinfo  {journal} {Phys. Lett. B}\ }\textbf {\bibinfo {volume} {659}},\
  \bibinfo {pages} {703} (\bibinfo {year} {2008})},\ \Eprint
  {https://arxiv.org/abs/0710.3755} {arXiv:0710.3755 [hep-th]} \BibitemShut
  {NoStop}%
\bibitem [{\citenamefont {Rubio}(2019)}]{Rubio:2018ogq}%
  \BibitemOpen
  \bibfield  {author} {\bibinfo {author} {\bibfnamefont {J.}~\bibnamefont
  {Rubio}},\ }\bibfield  {title} {\bibinfo {title} {{Higgs inflation}},\ }\href
  {https://doi.org/10.3389/fspas.2018.00050} {\bibfield  {journal} {\bibinfo
  {journal} {Front. Astron. Space Sci.}\ }\textbf {\bibinfo {volume} {5}},\
  \bibinfo {pages} {50} (\bibinfo {year} {2019})},\ \Eprint
  {https://arxiv.org/abs/1807.02376} {arXiv:1807.02376 [hep-ph]} \BibitemShut
  {NoStop}%
\bibitem [{\citenamefont {Starobinsky}(1980)}]{Starobinsky:1980te}%
  \BibitemOpen
  \bibfield  {author} {\bibinfo {author} {\bibfnamefont {A.~A.}\ \bibnamefont
  {Starobinsky}},\ }\bibfield  {title} {\bibinfo {title} {{A New Type of
  Isotropic Cosmological Models Without Singularity}},\ }\href
  {https://doi.org/10.1016/0370-2693(80)90670-X} {\bibfield  {journal}
  {\bibinfo  {journal} {Phys. Lett. B}\ }\textbf {\bibinfo {volume} {91}},\
  \bibinfo {pages} {99} (\bibinfo {year} {1980})}\BibitemShut {NoStop}%
\bibitem [{\citenamefont {Dvali}\ \emph {et~al.}(2000)\citenamefont {Dvali},
  \citenamefont {Gabadadze},\ and\ \citenamefont {Porrati}}]{Dvali:2000hr}%
  \BibitemOpen
  \bibfield  {author} {\bibinfo {author} {\bibfnamefont {G.~R.}\ \bibnamefont
  {Dvali}}, \bibinfo {author} {\bibfnamefont {G.}~\bibnamefont {Gabadadze}},\
  and\ \bibinfo {author} {\bibfnamefont {M.}~\bibnamefont {Porrati}},\
  }\bibfield  {title} {\bibinfo {title} {{4-D gravity on a brane in 5-D
  Minkowski space}},\ }\href {https://doi.org/10.1016/S0370-2693(00)00669-9}
  {\bibfield  {journal} {\bibinfo  {journal} {Phys. Lett. B}\ }\textbf
  {\bibinfo {volume} {485}},\ \bibinfo {pages} {208} (\bibinfo {year}
  {2000})},\ \Eprint {https://arxiv.org/abs/hep-th/0005016}
  {arXiv:hep-th/0005016} \BibitemShut {NoStop}%
\bibitem [{\citenamefont {Luty}\ \emph {et~al.}(2003)\citenamefont {Luty},
  \citenamefont {Porrati},\ and\ \citenamefont {Rattazzi}}]{Luty:2003vm}%
  \BibitemOpen
  \bibfield  {author} {\bibinfo {author} {\bibfnamefont {M.~A.}\ \bibnamefont
  {Luty}}, \bibinfo {author} {\bibfnamefont {M.}~\bibnamefont {Porrati}},\ and\
  \bibinfo {author} {\bibfnamefont {R.}~\bibnamefont {Rattazzi}},\ }\bibfield
  {title} {\bibinfo {title} {{Strong interactions and stability in the DGP
  model}},\ }\href {https://doi.org/10.1088/1126-6708/2003/09/029} {\bibfield
  {journal} {\bibinfo  {journal} {JHEP}\ }\textbf {\bibinfo {volume} {09}},\
  \bibinfo {pages} {029}},\ \Eprint {https://arxiv.org/abs/hep-th/0303116}
  {arXiv:hep-th/0303116} \BibitemShut {NoStop}%
\bibitem [{\citenamefont {de~Rham}\ and\ \citenamefont
  {Tolley}(2010)}]{deRham:2010eu}%
  \BibitemOpen
  \bibfield  {author} {\bibinfo {author} {\bibfnamefont {C.}~\bibnamefont
  {de~Rham}}\ and\ \bibinfo {author} {\bibfnamefont {A.~J.}\ \bibnamefont
  {Tolley}},\ }\bibfield  {title} {\bibinfo {title} {{DBI and the Galileon
  reunited}},\ }\href {https://doi.org/10.1088/1475-7516/2010/05/015}
  {\bibfield  {journal} {\bibinfo  {journal} {JCAP}\ }\textbf {\bibinfo
  {volume} {05}},\ \bibinfo {pages} {015}},\ \Eprint
  {https://arxiv.org/abs/1003.5917} {arXiv:1003.5917 [hep-th]} \BibitemShut
  {NoStop}%
\bibitem [{\citenamefont {Goon}\ \emph
  {et~al.}(2011{\natexlab{a}})\citenamefont {Goon}, \citenamefont
  {Hinterbichler},\ and\ \citenamefont {Trodden}}]{Goon:2011qf}%
  \BibitemOpen
  \bibfield  {author} {\bibinfo {author} {\bibfnamefont {G.}~\bibnamefont
  {Goon}}, \bibinfo {author} {\bibfnamefont {K.}~\bibnamefont
  {Hinterbichler}},\ and\ \bibinfo {author} {\bibfnamefont {M.}~\bibnamefont
  {Trodden}},\ }\bibfield  {title} {\bibinfo {title} {{Symmetries for Galileons
  and DBI scalars on curved space}},\ }\href
  {https://doi.org/10.1088/1475-7516/2011/07/017} {\bibfield  {journal}
  {\bibinfo  {journal} {JCAP}\ }\textbf {\bibinfo {volume} {07}},\ \bibinfo
  {pages} {017}},\ \Eprint {https://arxiv.org/abs/1103.5745} {arXiv:1103.5745
  [hep-th]} \BibitemShut {NoStop}%
\bibitem [{\citenamefont {Goon}\ \emph
  {et~al.}(2011{\natexlab{b}})\citenamefont {Goon}, \citenamefont
  {Hinterbichler},\ and\ \citenamefont {Trodden}}]{Goon:2011uw}%
  \BibitemOpen
  \bibfield  {author} {\bibinfo {author} {\bibfnamefont {G.}~\bibnamefont
  {Goon}}, \bibinfo {author} {\bibfnamefont {K.}~\bibnamefont
  {Hinterbichler}},\ and\ \bibinfo {author} {\bibfnamefont {M.}~\bibnamefont
  {Trodden}},\ }\bibfield  {title} {\bibinfo {title} {{A New Class of Effective
  Field Theories from Embedded Branes}},\ }\href
  {https://doi.org/10.1103/PhysRevLett.106.231102} {\bibfield  {journal}
  {\bibinfo  {journal} {Phys. Rev. Lett.}\ }\textbf {\bibinfo {volume} {106}},\
  \bibinfo {pages} {231102} (\bibinfo {year} {2011}{\natexlab{b}})},\ \Eprint
  {https://arxiv.org/abs/1103.6029} {arXiv:1103.6029 [hep-th]} \BibitemShut
  {NoStop}%
\bibitem [{\citenamefont {Trodden}\ and\ \citenamefont
  {Hinterbichler}(2011)}]{Trodden:2011xh}%
  \BibitemOpen
  \bibfield  {author} {\bibinfo {author} {\bibfnamefont {M.}~\bibnamefont
  {Trodden}}\ and\ \bibinfo {author} {\bibfnamefont {K.}~\bibnamefont
  {Hinterbichler}},\ }\bibfield  {title} {\bibinfo {title} {{Generalizing
  Galileons}},\ }\href {https://doi.org/10.1088/0264-9381/28/20/204003}
  {\bibfield  {journal} {\bibinfo  {journal} {Class. Quant. Grav.}\ }\textbf
  {\bibinfo {volume} {28}},\ \bibinfo {pages} {204003} (\bibinfo {year}
  {2011})},\ \Eprint {https://arxiv.org/abs/1104.2088} {arXiv:1104.2088
  [hep-th]} \BibitemShut {NoStop}%
\bibitem [{\citenamefont {Van~Acoleyen}\ and\ \citenamefont
  {Van~Doorsselaere}(2011)}]{VanAcoleyen:2011mj}%
  \BibitemOpen
  \bibfield  {author} {\bibinfo {author} {\bibfnamefont {K.}~\bibnamefont
  {Van~Acoleyen}}\ and\ \bibinfo {author} {\bibfnamefont {J.}~\bibnamefont
  {Van~Doorsselaere}},\ }\bibfield  {title} {\bibinfo {title} {{Galileons from
  Lovelock actions}},\ }\href {https://doi.org/10.1103/PhysRevD.83.084025}
  {\bibfield  {journal} {\bibinfo  {journal} {Phys. Rev. D}\ }\textbf {\bibinfo
  {volume} {83}},\ \bibinfo {pages} {084025} (\bibinfo {year} {2011})},\
  \Eprint {https://arxiv.org/abs/1102.0487} {arXiv:1102.0487 [gr-qc]}
  \BibitemShut {NoStop}%
\bibitem [{\citenamefont {van~de Bruck}\ and\ \citenamefont
  {Longden}(2019)}]{vandeBruck:2018jlz}%
  \BibitemOpen
  \bibfield  {author} {\bibinfo {author} {\bibfnamefont {C.}~\bibnamefont
  {van~de Bruck}}\ and\ \bibinfo {author} {\bibfnamefont {C.}~\bibnamefont
  {Longden}},\ }\bibfield  {title} {\bibinfo {title} {{Einstein-Gauss-Bonnet
  gravity with extra dimensions}},\ }\href
  {https://doi.org/10.3390/galaxies7010039} {\bibfield  {journal} {\bibinfo
  {journal} {Galaxies}\ }\textbf {\bibinfo {volume} {7}},\ \bibinfo {pages}
  {39} (\bibinfo {year} {2019})},\ \Eprint {https://arxiv.org/abs/1809.00920}
  {arXiv:1809.00920 [gr-qc]} \BibitemShut {NoStop}%
\bibitem [{\citenamefont {de~Rham}\ and\ \citenamefont
  {Heisenberg}(2011)}]{deRham:2011by}%
  \BibitemOpen
  \bibfield  {author} {\bibinfo {author} {\bibfnamefont {C.}~\bibnamefont
  {de~Rham}}\ and\ \bibinfo {author} {\bibfnamefont {L.}~\bibnamefont
  {Heisenberg}},\ }\bibfield  {title} {\bibinfo {title} {{Cosmology of the
  Galileon from Massive Gravity}},\ }\href
  {https://doi.org/10.1103/PhysRevD.84.043503} {\bibfield  {journal} {\bibinfo
  {journal} {Phys. Rev. D}\ }\textbf {\bibinfo {volume} {84}},\ \bibinfo
  {pages} {043503} (\bibinfo {year} {2011})},\ \Eprint
  {https://arxiv.org/abs/1106.3312} {arXiv:1106.3312 [hep-th]} \BibitemShut
  {NoStop}%
\bibitem [{\citenamefont {Heisenberg}\ \emph {et~al.}(2014)\citenamefont
  {Heisenberg}, \citenamefont {Kimura},\ and\ \citenamefont
  {Yamamoto}}]{Heisenberg:2014kea}%
  \BibitemOpen
  \bibfield  {author} {\bibinfo {author} {\bibfnamefont {L.}~\bibnamefont
  {Heisenberg}}, \bibinfo {author} {\bibfnamefont {R.}~\bibnamefont {Kimura}},\
  and\ \bibinfo {author} {\bibfnamefont {K.}~\bibnamefont {Yamamoto}},\
  }\bibfield  {title} {\bibinfo {title} {{Cosmology of the proxy theory to
  massive gravity}},\ }\href {https://doi.org/10.1103/PhysRevD.89.103008}
  {\bibfield  {journal} {\bibinfo  {journal} {Phys. Rev. D}\ }\textbf {\bibinfo
  {volume} {89}},\ \bibinfo {pages} {103008} (\bibinfo {year} {2014})},\
  \Eprint {https://arxiv.org/abs/1403.2049} {arXiv:1403.2049 [hep-th]}
  \BibitemShut {NoStop}%
\bibitem [{\citenamefont {Zumalac\'arregui}\ and\ \citenamefont
  {Garc\'\i{}a-Bellido}(2014)}]{Zumalacarregui:2013pma}%
  \BibitemOpen
  \bibfield  {author} {\bibinfo {author} {\bibfnamefont {M.}~\bibnamefont
  {Zumalac\'arregui}}\ and\ \bibinfo {author} {\bibfnamefont {J.}~\bibnamefont
  {Garc\'\i{}a-Bellido}},\ }\bibfield  {title} {\bibinfo {title} {{Transforming
  gravity: from derivative couplings to matter to second-order scalar-tensor
  theories beyond the Horndeski Lagrangian}},\ }\href
  {https://doi.org/10.1103/PhysRevD.89.064046} {\bibfield  {journal} {\bibinfo
  {journal} {Phys. Rev. D}\ }\textbf {\bibinfo {volume} {89}},\ \bibinfo
  {pages} {064046} (\bibinfo {year} {2014})},\ \Eprint
  {https://arxiv.org/abs/1308.4685} {arXiv:1308.4685 [gr-qc]} \BibitemShut
  {NoStop}%
\bibitem [{\citenamefont {Bettoni}\ and\ \citenamefont
  {Liberati}(2013)}]{Bettoni:2013diz}%
  \BibitemOpen
  \bibfield  {author} {\bibinfo {author} {\bibfnamefont {D.}~\bibnamefont
  {Bettoni}}\ and\ \bibinfo {author} {\bibfnamefont {S.}~\bibnamefont
  {Liberati}},\ }\bibfield  {title} {\bibinfo {title} {{Disformal invariance of
  second order scalar-tensor theories: Framing the Horndeski action}},\ }\href
  {https://doi.org/10.1103/PhysRevD.88.084020} {\bibfield  {journal} {\bibinfo
  {journal} {Phys. Rev. D}\ }\textbf {\bibinfo {volume} {88}},\ \bibinfo
  {pages} {084020} (\bibinfo {year} {2013})},\ \Eprint
  {https://arxiv.org/abs/1306.6724} {arXiv:1306.6724 [gr-qc]} \BibitemShut
  {NoStop}%
\bibitem [{\citenamefont {Bekenstein}(1993)}]{Bekenstein:1992pj}%
  \BibitemOpen
  \bibfield  {author} {\bibinfo {author} {\bibfnamefont {J.~D.}\ \bibnamefont
  {Bekenstein}},\ }\bibfield  {title} {\bibinfo {title} {{The Relation between
  physical and gravitational geometry}},\ }\href
  {https://doi.org/10.1103/PhysRevD.48.3641} {\bibfield  {journal} {\bibinfo
  {journal} {Phys. Rev. D}\ }\textbf {\bibinfo {volume} {48}},\ \bibinfo
  {pages} {3641} (\bibinfo {year} {1993})},\ \Eprint
  {https://arxiv.org/abs/gr-qc/9211017} {arXiv:gr-qc/9211017} \BibitemShut
  {NoStop}%
\bibitem [{\citenamefont {Horndeski}(1974)}]{Horndeski:1974wa}%
  \BibitemOpen
  \bibfield  {author} {\bibinfo {author} {\bibfnamefont {G.~W.}\ \bibnamefont
  {Horndeski}},\ }\bibfield  {title} {\bibinfo {title} {{Second-order
  scalar-tensor field equations in a four-dimensional space}},\ }\href
  {https://doi.org/10.1007/BF01807638} {\bibfield  {journal} {\bibinfo
  {journal} {Int. J. Theor. Phys.}\ }\textbf {\bibinfo {volume} {10}},\
  \bibinfo {pages} {363} (\bibinfo {year} {1974})}\BibitemShut {NoStop}%
\bibitem [{\citenamefont {Deffayet}\ \emph
  {et~al.}(2009{\natexlab{a}})\citenamefont {Deffayet}, \citenamefont
  {Esposito-Farese},\ and\ \citenamefont {Vikman}}]{Deffayet:2009wt}%
  \BibitemOpen
  \bibfield  {author} {\bibinfo {author} {\bibfnamefont {C.}~\bibnamefont
  {Deffayet}}, \bibinfo {author} {\bibfnamefont {G.}~\bibnamefont
  {Esposito-Farese}},\ and\ \bibinfo {author} {\bibfnamefont {A.}~\bibnamefont
  {Vikman}},\ }\bibfield  {title} {\bibinfo {title} {{Covariant Galileon}},\
  }\href {https://doi.org/10.1103/PhysRevD.79.084003} {\bibfield  {journal}
  {\bibinfo  {journal} {Phys. Rev. D}\ }\textbf {\bibinfo {volume} {79}},\
  \bibinfo {pages} {084003} (\bibinfo {year} {2009}{\natexlab{a}})},\ \Eprint
  {https://arxiv.org/abs/0901.1314} {arXiv:0901.1314 [hep-th]} \BibitemShut
  {NoStop}%
\bibitem [{\citenamefont {Deffayet}\ \emph
  {et~al.}(2009{\natexlab{b}})\citenamefont {Deffayet}, \citenamefont {Deser},\
  and\ \citenamefont {Esposito-Farese}}]{Deffayet:2009mn}%
  \BibitemOpen
  \bibfield  {author} {\bibinfo {author} {\bibfnamefont {C.}~\bibnamefont
  {Deffayet}}, \bibinfo {author} {\bibfnamefont {S.}~\bibnamefont {Deser}},\
  and\ \bibinfo {author} {\bibfnamefont {G.}~\bibnamefont {Esposito-Farese}},\
  }\bibfield  {title} {\bibinfo {title} {{Generalized Galileons: All scalar
  models whose curved background extensions maintain second-order field
  equations and stress-tensors}},\ }\href
  {https://doi.org/10.1103/PhysRevD.80.064015} {\bibfield  {journal} {\bibinfo
  {journal} {Phys. Rev. D}\ }\textbf {\bibinfo {volume} {80}},\ \bibinfo
  {pages} {064015} (\bibinfo {year} {2009}{\natexlab{b}})},\ \Eprint
  {https://arxiv.org/abs/0906.1967} {arXiv:0906.1967 [gr-qc]} \BibitemShut
  {NoStop}%
\bibitem [{\citenamefont {Deffayet}\ \emph {et~al.}(2011)\citenamefont
  {Deffayet}, \citenamefont {Gao}, \citenamefont {Steer},\ and\ \citenamefont
  {Zahariade}}]{Deffayet:2011gz}%
  \BibitemOpen
  \bibfield  {author} {\bibinfo {author} {\bibfnamefont {C.}~\bibnamefont
  {Deffayet}}, \bibinfo {author} {\bibfnamefont {X.}~\bibnamefont {Gao}},
  \bibinfo {author} {\bibfnamefont {D.~A.}\ \bibnamefont {Steer}},\ and\
  \bibinfo {author} {\bibfnamefont {G.}~\bibnamefont {Zahariade}},\ }\bibfield
  {title} {\bibinfo {title} {{From k-essence to generalised Galileons}},\
  }\href {https://doi.org/10.1103/PhysRevD.84.064039} {\bibfield  {journal}
  {\bibinfo  {journal} {Phys. Rev. D}\ }\textbf {\bibinfo {volume} {84}},\
  \bibinfo {pages} {064039} (\bibinfo {year} {2011})},\ \Eprint
  {https://arxiv.org/abs/1103.3260} {arXiv:1103.3260 [hep-th]} \BibitemShut
  {NoStop}%
\bibitem [{\citenamefont {Kobayashi}\ \emph {et~al.}(2011)\citenamefont
  {Kobayashi}, \citenamefont {Yamaguchi},\ and\ \citenamefont
  {Yokoyama}}]{Kobayashi:2011nu}%
  \BibitemOpen
  \bibfield  {author} {\bibinfo {author} {\bibfnamefont {T.}~\bibnamefont
  {Kobayashi}}, \bibinfo {author} {\bibfnamefont {M.}~\bibnamefont
  {Yamaguchi}},\ and\ \bibinfo {author} {\bibfnamefont {J.}~\bibnamefont
  {Yokoyama}},\ }\bibfield  {title} {\bibinfo {title} {{Generalized
  G-inflation: Inflation with the most general second-order field equations}},\
  }\href {https://doi.org/10.1143/PTP.126.511} {\bibfield  {journal} {\bibinfo
  {journal} {Prog. Theor. Phys.}\ }\textbf {\bibinfo {volume} {126}},\ \bibinfo
  {pages} {511} (\bibinfo {year} {2011})},\ \Eprint
  {https://arxiv.org/abs/1105.5723} {arXiv:1105.5723 [hep-th]} \BibitemShut
  {NoStop}%
\bibitem [{\citenamefont {Gleyzes}\ \emph {et~al.}(2015)\citenamefont
  {Gleyzes}, \citenamefont {Langlois}, \citenamefont {Piazza},\ and\
  \citenamefont {Vernizzi}}]{Gleyzes:2014dya}%
  \BibitemOpen
  \bibfield  {author} {\bibinfo {author} {\bibfnamefont {J.}~\bibnamefont
  {Gleyzes}}, \bibinfo {author} {\bibfnamefont {D.}~\bibnamefont {Langlois}},
  \bibinfo {author} {\bibfnamefont {F.}~\bibnamefont {Piazza}},\ and\ \bibinfo
  {author} {\bibfnamefont {F.}~\bibnamefont {Vernizzi}},\ }\bibfield  {title}
  {\bibinfo {title} {{Healthy theories beyond Horndeski}},\ }\href
  {https://doi.org/10.1103/PhysRevLett.114.211101} {\bibfield  {journal}
  {\bibinfo  {journal} {Phys. Rev. Lett.}\ }\textbf {\bibinfo {volume} {114}},\
  \bibinfo {pages} {211101} (\bibinfo {year} {2015})},\ \Eprint
  {https://arxiv.org/abs/1404.6495} {arXiv:1404.6495 [hep-th]} \BibitemShut
  {NoStop}%
\bibitem [{\citenamefont {Langlois}\ and\ \citenamefont
  {Noui}(2016)}]{Langlois:2015cwa}%
  \BibitemOpen
  \bibfield  {author} {\bibinfo {author} {\bibfnamefont {D.}~\bibnamefont
  {Langlois}}\ and\ \bibinfo {author} {\bibfnamefont {K.}~\bibnamefont
  {Noui}},\ }\bibfield  {title} {\bibinfo {title} {{Degenerate higher
  derivative theories beyond Horndeski: evading the Ostrogradski
  instability}},\ }\href {https://doi.org/10.1088/1475-7516/2016/02/034}
  {\bibfield  {journal} {\bibinfo  {journal} {JCAP}\ }\textbf {\bibinfo
  {volume} {02}},\ \bibinfo {pages} {034}},\ \Eprint
  {https://arxiv.org/abs/1510.06930} {arXiv:1510.06930 [gr-qc]} \BibitemShut
  {NoStop}%
\bibitem [{\citenamefont {De~Felice}\ \emph {et~al.}(2018)\citenamefont
  {De~Felice}, \citenamefont {Langlois}, \citenamefont {Mukohyama},
  \citenamefont {Noui},\ and\ \citenamefont {Wang}}]{DeFelice:2018ewo}%
  \BibitemOpen
  \bibfield  {author} {\bibinfo {author} {\bibfnamefont {A.}~\bibnamefont
  {De~Felice}}, \bibinfo {author} {\bibfnamefont {D.}~\bibnamefont {Langlois}},
  \bibinfo {author} {\bibfnamefont {S.}~\bibnamefont {Mukohyama}}, \bibinfo
  {author} {\bibfnamefont {K.}~\bibnamefont {Noui}},\ and\ \bibinfo {author}
  {\bibfnamefont {A.}~\bibnamefont {Wang}},\ }\bibfield  {title} {\bibinfo
  {title} {{Generalized instantaneous modes in higher-order scalar-tensor
  theories}},\ }\href {https://doi.org/10.1103/PhysRevD.98.084024} {\bibfield
  {journal} {\bibinfo  {journal} {Phys. Rev. D}\ }\textbf {\bibinfo {volume}
  {98}},\ \bibinfo {pages} {084024} (\bibinfo {year} {2018})},\ \Eprint
  {https://arxiv.org/abs/1803.06241} {arXiv:1803.06241 [hep-th]} \BibitemShut
  {NoStop}%
\bibitem [{\citenamefont {Gao}(2014{\natexlab{a}})}]{Gao:2014fra}%
  \BibitemOpen
  \bibfield  {author} {\bibinfo {author} {\bibfnamefont {X.}~\bibnamefont
  {Gao}},\ }\bibfield  {title} {\bibinfo {title} {{Hamiltonian analysis of
  spatially covariant gravity}},\ }\href
  {https://doi.org/10.1103/PhysRevD.90.104033} {\bibfield  {journal} {\bibinfo
  {journal} {Phys. Rev. D}\ }\textbf {\bibinfo {volume} {90}},\ \bibinfo
  {pages} {104033} (\bibinfo {year} {2014}{\natexlab{a}})},\ \Eprint
  {https://arxiv.org/abs/1409.6708} {arXiv:1409.6708 [gr-qc]} \BibitemShut
  {NoStop}%
\bibitem [{\citenamefont {Gao}(2014{\natexlab{b}})}]{Gao:2014soa}%
  \BibitemOpen
  \bibfield  {author} {\bibinfo {author} {\bibfnamefont {X.}~\bibnamefont
  {Gao}},\ }\bibfield  {title} {\bibinfo {title} {{Unifying framework for
  scalar-tensor theories of gravity}},\ }\href
  {https://doi.org/10.1103/PhysRevD.90.081501} {\bibfield  {journal} {\bibinfo
  {journal} {Phys. Rev. D}\ }\textbf {\bibinfo {volume} {90}},\ \bibinfo
  {pages} {081501} (\bibinfo {year} {2014}{\natexlab{b}})},\ \Eprint
  {https://arxiv.org/abs/1406.0822} {arXiv:1406.0822 [gr-qc]} \BibitemShut
  {NoStop}%
\bibitem [{\citenamefont {Gleyzes}\ \emph {et~al.}(2013)\citenamefont
  {Gleyzes}, \citenamefont {Langlois}, \citenamefont {Piazza},\ and\
  \citenamefont {Vernizzi}}]{Gleyzes:2013ooa}%
  \BibitemOpen
  \bibfield  {author} {\bibinfo {author} {\bibfnamefont {J.}~\bibnamefont
  {Gleyzes}}, \bibinfo {author} {\bibfnamefont {D.}~\bibnamefont {Langlois}},
  \bibinfo {author} {\bibfnamefont {F.}~\bibnamefont {Piazza}},\ and\ \bibinfo
  {author} {\bibfnamefont {F.}~\bibnamefont {Vernizzi}},\ }\bibfield  {title}
  {\bibinfo {title} {{Essential Building Blocks of Dark Energy}},\ }\href
  {https://doi.org/10.1088/1475-7516/2013/08/025} {\bibfield  {journal}
  {\bibinfo  {journal} {JCAP}\ }\textbf {\bibinfo {volume} {08}},\ \bibinfo
  {pages} {025}},\ \Eprint {https://arxiv.org/abs/1304.4840} {arXiv:1304.4840
  [hep-th]} \BibitemShut {NoStop}%
\bibitem [{\citenamefont {Fujita}\ \emph {et~al.}(2016)\citenamefont {Fujita},
  \citenamefont {Gao},\ and\ \citenamefont {Yokoyama}}]{Fujita:2015ymn}%
  \BibitemOpen
  \bibfield  {author} {\bibinfo {author} {\bibfnamefont {T.}~\bibnamefont
  {Fujita}}, \bibinfo {author} {\bibfnamefont {X.}~\bibnamefont {Gao}},\ and\
  \bibinfo {author} {\bibfnamefont {J.}~\bibnamefont {Yokoyama}},\ }\bibfield
  {title} {\bibinfo {title} {{Spatially covariant theories of gravity:
  disformal transformation, cosmological perturbations and the Einstein
  frame}},\ }\href {https://doi.org/10.1088/1475-7516/2016/02/014} {\bibfield
  {journal} {\bibinfo  {journal} {JCAP}\ }\textbf {\bibinfo {volume} {02}},\
  \bibinfo {pages} {014}},\ \Eprint {https://arxiv.org/abs/1511.04324}
  {arXiv:1511.04324 [gr-qc]} \BibitemShut {NoStop}%
\bibitem [{\citenamefont {Crisostomi}\ \emph {et~al.}(2016)\citenamefont
  {Crisostomi}, \citenamefont {Hull}, \citenamefont {Koyama},\ and\
  \citenamefont {Tasinato}}]{Crisostomi:2016tcp}%
  \BibitemOpen
  \bibfield  {author} {\bibinfo {author} {\bibfnamefont {M.}~\bibnamefont
  {Crisostomi}}, \bibinfo {author} {\bibfnamefont {M.}~\bibnamefont {Hull}},
  \bibinfo {author} {\bibfnamefont {K.}~\bibnamefont {Koyama}},\ and\ \bibinfo
  {author} {\bibfnamefont {G.}~\bibnamefont {Tasinato}},\ }\bibfield  {title}
  {\bibinfo {title} {{Horndeski: beyond, or not beyond?}},\ }\href
  {https://doi.org/10.1088/1475-7516/2016/03/038} {\bibfield  {journal}
  {\bibinfo  {journal} {JCAP}\ }\textbf {\bibinfo {volume} {03}},\ \bibinfo
  {pages} {038}},\ \Eprint {https://arxiv.org/abs/1601.04658} {arXiv:1601.04658
  [hep-th]} \BibitemShut {NoStop}%
\bibitem [{\citenamefont {Ben~Achour}\ \emph
  {et~al.}(2016{\natexlab{a}})\citenamefont {Ben~Achour}, \citenamefont
  {Langlois},\ and\ \citenamefont {Noui}}]{BenAchour:2016cay}%
  \BibitemOpen
  \bibfield  {author} {\bibinfo {author} {\bibfnamefont {J.}~\bibnamefont
  {Ben~Achour}}, \bibinfo {author} {\bibfnamefont {D.}~\bibnamefont
  {Langlois}},\ and\ \bibinfo {author} {\bibfnamefont {K.}~\bibnamefont
  {Noui}},\ }\bibfield  {title} {\bibinfo {title} {{Degenerate higher order
  scalar-tensor theories beyond Horndeski and disformal transformations}},\
  }\href {https://doi.org/10.1103/PhysRevD.93.124005} {\bibfield  {journal}
  {\bibinfo  {journal} {Phys. Rev. D}\ }\textbf {\bibinfo {volume} {93}},\
  \bibinfo {pages} {124005} (\bibinfo {year} {2016}{\natexlab{a}})},\ \Eprint
  {https://arxiv.org/abs/1602.08398} {arXiv:1602.08398 [gr-qc]} \BibitemShut
  {NoStop}%
\bibitem [{\citenamefont {Ben~Achour}\ \emph
  {et~al.}(2016{\natexlab{b}})\citenamefont {Ben~Achour}, \citenamefont
  {Crisostomi}, \citenamefont {Koyama}, \citenamefont {Langlois}, \citenamefont
  {Noui},\ and\ \citenamefont {Tasinato}}]{BenAchour:2016fzp}%
  \BibitemOpen
  \bibfield  {author} {\bibinfo {author} {\bibfnamefont {J.}~\bibnamefont
  {Ben~Achour}}, \bibinfo {author} {\bibfnamefont {M.}~\bibnamefont
  {Crisostomi}}, \bibinfo {author} {\bibfnamefont {K.}~\bibnamefont {Koyama}},
  \bibinfo {author} {\bibfnamefont {D.}~\bibnamefont {Langlois}}, \bibinfo
  {author} {\bibfnamefont {K.}~\bibnamefont {Noui}},\ and\ \bibinfo {author}
  {\bibfnamefont {G.}~\bibnamefont {Tasinato}},\ }\bibfield  {title} {\bibinfo
  {title} {{Degenerate higher order scalar-tensor theories beyond Horndeski up
  to cubic order}},\ }\href {https://doi.org/10.1007/JHEP12(2016)100}
  {\bibfield  {journal} {\bibinfo  {journal} {JHEP}\ }\textbf {\bibinfo
  {volume} {12}},\ \bibinfo {pages} {100}},\ \Eprint
  {https://arxiv.org/abs/1608.08135} {arXiv:1608.08135 [hep-th]} \BibitemShut
  {NoStop}%
\bibitem [{\citenamefont {Takahashi}\ \emph {et~al.}(2017)\citenamefont
  {Takahashi}, \citenamefont {Motohashi}, \citenamefont {Suyama},\ and\
  \citenamefont {Kobayashi}}]{Takahashi:2017zgr}%
  \BibitemOpen
  \bibfield  {author} {\bibinfo {author} {\bibfnamefont {K.}~\bibnamefont
  {Takahashi}}, \bibinfo {author} {\bibfnamefont {H.}~\bibnamefont
  {Motohashi}}, \bibinfo {author} {\bibfnamefont {T.}~\bibnamefont {Suyama}},\
  and\ \bibinfo {author} {\bibfnamefont {T.}~\bibnamefont {Kobayashi}},\
  }\bibfield  {title} {\bibinfo {title} {{General invertible transformation and
  physical degrees of freedom}},\ }\href
  {https://doi.org/10.1103/PhysRevD.95.084053} {\bibfield  {journal} {\bibinfo
  {journal} {Phys. Rev.}\ }\textbf {\bibinfo {volume} {D95}},\ \bibinfo {pages}
  {084053} (\bibinfo {year} {2017})},\ \Eprint
  {https://arxiv.org/abs/1702.01849} {arXiv:1702.01849 [gr-qc]} \BibitemShut
  {NoStop}%
\bibitem [{\citenamefont {Ben~Achour}\ \emph {et~al.}(2020)\citenamefont
  {Ben~Achour}, \citenamefont {Liu},\ and\ \citenamefont
  {Mukohyama}}]{BenAchour:2020wiw}%
  \BibitemOpen
  \bibfield  {author} {\bibinfo {author} {\bibfnamefont {J.}~\bibnamefont
  {Ben~Achour}}, \bibinfo {author} {\bibfnamefont {H.}~\bibnamefont {Liu}},\
  and\ \bibinfo {author} {\bibfnamefont {S.}~\bibnamefont {Mukohyama}},\
  }\bibfield  {title} {\bibinfo {title} {{Hairy black holes in DHOST theories:
  Exploring disformal transformation as a solution-generating method}},\ }\href
  {https://doi.org/10.1088/1475-7516/2020/02/023} {\bibfield  {journal}
  {\bibinfo  {journal} {JCAP}\ }\textbf {\bibinfo {volume} {02}},\ \bibinfo
  {pages} {023}},\ \Eprint {https://arxiv.org/abs/1910.11017} {arXiv:1910.11017
  [gr-qc]} \BibitemShut {NoStop}%
\bibitem [{\citenamefont {Babichev}\ \emph {et~al.}(2022)\citenamefont
  {Babichev}, \citenamefont {Izumi}, \citenamefont {Tanahashi},\ and\
  \citenamefont {Yamaguchi}}]{Babichev:2021bim}%
  \BibitemOpen
  \bibfield  {author} {\bibinfo {author} {\bibfnamefont {E.}~\bibnamefont
  {Babichev}}, \bibinfo {author} {\bibfnamefont {K.}~\bibnamefont {Izumi}},
  \bibinfo {author} {\bibfnamefont {N.}~\bibnamefont {Tanahashi}},\ and\
  \bibinfo {author} {\bibfnamefont {M.}~\bibnamefont {Yamaguchi}},\ }\bibfield
  {title} {\bibinfo {title} {{Invertibility conditions for field
  transformations with derivatives: Toward extensions of disformal
  transformation with higher derivatives}},\ }\href
  {https://doi.org/10.1093/ptep/ptab151} {\bibfield  {journal} {\bibinfo
  {journal} {PTEP}\ }\textbf {\bibinfo {volume} {2022}},\ \bibinfo {pages}
  {013A01} (\bibinfo {year} {2022})},\ \Eprint
  {https://arxiv.org/abs/2109.00912} {arXiv:2109.00912 [hep-th]} \BibitemShut
  {NoStop}%
\bibitem [{\citenamefont {Kobayashi}\ \emph {et~al.}(2010)\citenamefont
  {Kobayashi}, \citenamefont {Yamaguchi},\ and\ \citenamefont
  {Yokoyama}}]{Kobayashi:2010cm}%
  \BibitemOpen
  \bibfield  {author} {\bibinfo {author} {\bibfnamefont {T.}~\bibnamefont
  {Kobayashi}}, \bibinfo {author} {\bibfnamefont {M.}~\bibnamefont
  {Yamaguchi}},\ and\ \bibinfo {author} {\bibfnamefont {J.}~\bibnamefont
  {Yokoyama}},\ }\bibfield  {title} {\bibinfo {title} {{G-inflation: Inflation
  driven by the Galileon field}},\ }\href
  {https://doi.org/10.1103/PhysRevLett.105.231302} {\bibfield  {journal}
  {\bibinfo  {journal} {Phys. Rev. Lett.}\ }\textbf {\bibinfo {volume} {105}},\
  \bibinfo {pages} {231302} (\bibinfo {year} {2010})},\ \Eprint
  {https://arxiv.org/abs/1008.0603} {arXiv:1008.0603 [hep-th]} \BibitemShut
  {NoStop}%
\bibitem [{\citenamefont {Cai}\ \emph {et~al.}(2015)\citenamefont {Cai},
  \citenamefont {Gong}, \citenamefont {Pi}, \citenamefont {Saridakis},\ and\
  \citenamefont {Wu}}]{Cai:2014uka}%
  \BibitemOpen
  \bibfield  {author} {\bibinfo {author} {\bibfnamefont {Y.-F.}\ \bibnamefont
  {Cai}}, \bibinfo {author} {\bibfnamefont {J.-O.}\ \bibnamefont {Gong}},
  \bibinfo {author} {\bibfnamefont {S.}~\bibnamefont {Pi}}, \bibinfo {author}
  {\bibfnamefont {E.~N.}\ \bibnamefont {Saridakis}},\ and\ \bibinfo {author}
  {\bibfnamefont {S.-Y.}\ \bibnamefont {Wu}},\ }\bibfield  {title} {\bibinfo
  {title} {{On the possibility of blue tensor spectrum within single field
  inflation}},\ }\href {https://doi.org/10.1016/j.nuclphysb.2015.09.025}
  {\bibfield  {journal} {\bibinfo  {journal} {Nucl. Phys. B}\ }\textbf
  {\bibinfo {volume} {900}},\ \bibinfo {pages} {517} (\bibinfo {year}
  {2015})},\ \Eprint {https://arxiv.org/abs/1412.7241} {arXiv:1412.7241
  [hep-th]} \BibitemShut {NoStop}%
\bibitem [{\citenamefont {De~Felice}\ and\ \citenamefont
  {Tsujikawa}(2011{\natexlab{a}})}]{DeFelice:2011uc}%
  \BibitemOpen
  \bibfield  {author} {\bibinfo {author} {\bibfnamefont {A.}~\bibnamefont
  {De~Felice}}\ and\ \bibinfo {author} {\bibfnamefont {S.}~\bibnamefont
  {Tsujikawa}},\ }\bibfield  {title} {\bibinfo {title} {{Inflationary
  non-Gaussianities in the most general second-order scalar-tensor theories}},\
  }\href {https://doi.org/10.1103/PhysRevD.84.083504} {\bibfield  {journal}
  {\bibinfo  {journal} {Phys. Rev. D}\ }\textbf {\bibinfo {volume} {84}},\
  \bibinfo {pages} {083504} (\bibinfo {year} {2011}{\natexlab{a}})},\ \Eprint
  {https://arxiv.org/abs/1107.3917} {arXiv:1107.3917 [gr-qc]} \BibitemShut
  {NoStop}%
\bibitem [{\citenamefont {Gao}\ \emph {et~al.}(2011)\citenamefont {Gao},
  \citenamefont {Kobayashi}, \citenamefont {Yamaguchi},\ and\ \citenamefont
  {Yokoyama}}]{Gao:2011vs}%
  \BibitemOpen
  \bibfield  {author} {\bibinfo {author} {\bibfnamefont {X.}~\bibnamefont
  {Gao}}, \bibinfo {author} {\bibfnamefont {T.}~\bibnamefont {Kobayashi}},
  \bibinfo {author} {\bibfnamefont {M.}~\bibnamefont {Yamaguchi}},\ and\
  \bibinfo {author} {\bibfnamefont {J.}~\bibnamefont {Yokoyama}},\ }\bibfield
  {title} {\bibinfo {title} {{Primordial non-Gaussianities of gravitational
  waves in the most general single-field inflation model}},\ }\href
  {https://doi.org/10.1103/PhysRevLett.107.211301} {\bibfield  {journal}
  {\bibinfo  {journal} {Phys. Rev. Lett.}\ }\textbf {\bibinfo {volume} {107}},\
  \bibinfo {pages} {211301} (\bibinfo {year} {2011})},\ \Eprint
  {https://arxiv.org/abs/1108.3513} {arXiv:1108.3513 [astro-ph.CO]}
  \BibitemShut {NoStop}%
\bibitem [{\citenamefont {Renaux-Petel}(2012)}]{Renaux-Petel:2011zgy}%
  \BibitemOpen
  \bibfield  {author} {\bibinfo {author} {\bibfnamefont {S.}~\bibnamefont
  {Renaux-Petel}},\ }\bibfield  {title} {\bibinfo {title} {{On the redundancy
  of operators and the bispectrum in the most general second-order
  scalar-tensor theory}},\ }\href
  {https://doi.org/10.1088/1475-7516/2012/02/020} {\bibfield  {journal}
  {\bibinfo  {journal} {JCAP}\ }\textbf {\bibinfo {volume} {02}},\ \bibinfo
  {pages} {020}},\ \Eprint {https://arxiv.org/abs/1107.5020} {arXiv:1107.5020
  [astro-ph.CO]} \BibitemShut {NoStop}%
\bibitem [{\citenamefont {Gao}\ \emph {et~al.}(2013)\citenamefont {Gao},
  \citenamefont {Kobayashi}, \citenamefont {Shiraishi}, \citenamefont
  {Yamaguchi}, \citenamefont {Yokoyama},\ and\ \citenamefont
  {Yokoyama}}]{Gao:2012ib}%
  \BibitemOpen
  \bibfield  {author} {\bibinfo {author} {\bibfnamefont {X.}~\bibnamefont
  {Gao}}, \bibinfo {author} {\bibfnamefont {T.}~\bibnamefont {Kobayashi}},
  \bibinfo {author} {\bibfnamefont {M.}~\bibnamefont {Shiraishi}}, \bibinfo
  {author} {\bibfnamefont {M.}~\bibnamefont {Yamaguchi}}, \bibinfo {author}
  {\bibfnamefont {J.}~\bibnamefont {Yokoyama}},\ and\ \bibinfo {author}
  {\bibfnamefont {S.}~\bibnamefont {Yokoyama}},\ }\bibfield  {title} {\bibinfo
  {title} {{Full bispectra from primordial scalar and tensor perturbations in
  the most general single-field inflation model}},\ }\href
  {https://doi.org/10.1093/ptep/ptt031} {\bibfield  {journal} {\bibinfo
  {journal} {PTEP}\ }\textbf {\bibinfo {volume} {2013}},\ \bibinfo {pages}
  {053E03} (\bibinfo {year} {2013})},\ \Eprint
  {https://arxiv.org/abs/1207.0588} {arXiv:1207.0588 [astro-ph.CO]}
  \BibitemShut {NoStop}%
\bibitem [{\citenamefont {De~Felice}\ and\ \citenamefont
  {Tsujikawa}(2013)}]{DeFelice:2013ar}%
  \BibitemOpen
  \bibfield  {author} {\bibinfo {author} {\bibfnamefont {A.}~\bibnamefont
  {De~Felice}}\ and\ \bibinfo {author} {\bibfnamefont {S.}~\bibnamefont
  {Tsujikawa}},\ }\bibfield  {title} {\bibinfo {title} {{Shapes of primordial
  non-Gaussianities in the Horndeski's most general scalar-tensor theories}},\
  }\href {https://doi.org/10.1088/1475-7516/2013/03/030} {\bibfield  {journal}
  {\bibinfo  {journal} {JCAP}\ }\textbf {\bibinfo {volume} {03}},\ \bibinfo
  {pages} {030}},\ \Eprint {https://arxiv.org/abs/1301.5721} {arXiv:1301.5721
  [hep-th]} \BibitemShut {NoStop}%
\bibitem [{\citenamefont {Tahara}\ and\ \citenamefont
  {Yokoyama}(2018)}]{Tahara:2017wud}%
  \BibitemOpen
  \bibfield  {author} {\bibinfo {author} {\bibfnamefont {H.~W.~H.}\
  \bibnamefont {Tahara}}\ and\ \bibinfo {author} {\bibfnamefont
  {J.}~\bibnamefont {Yokoyama}},\ }\bibfield  {title} {\bibinfo {title} {{CMB
  B-mode auto-bispectrum produced by primordial gravitational waves}},\ }\href
  {https://doi.org/10.1093/ptep/ptx185} {\bibfield  {journal} {\bibinfo
  {journal} {PTEP}\ }\textbf {\bibinfo {volume} {2018}},\ \bibinfo {pages}
  {013E03} (\bibinfo {year} {2018})},\ \Eprint
  {https://arxiv.org/abs/1704.08904} {arXiv:1704.08904 [astro-ph.CO]}
  \BibitemShut {NoStop}%
\bibitem [{\citenamefont {Hirano}\ \emph {et~al.}(2016)\citenamefont {Hirano},
  \citenamefont {Kobayashi},\ and\ \citenamefont {Yokoyama}}]{Hirano:2016gmv}%
  \BibitemOpen
  \bibfield  {author} {\bibinfo {author} {\bibfnamefont {S.}~\bibnamefont
  {Hirano}}, \bibinfo {author} {\bibfnamefont {T.}~\bibnamefont {Kobayashi}},\
  and\ \bibinfo {author} {\bibfnamefont {S.}~\bibnamefont {Yokoyama}},\
  }\bibfield  {title} {\bibinfo {title} {{Ultra slow-roll G-inflation}},\
  }\href {https://doi.org/10.1103/PhysRevD.94.103515} {\bibfield  {journal}
  {\bibinfo  {journal} {Phys. Rev. D}\ }\textbf {\bibinfo {volume} {94}},\
  \bibinfo {pages} {103515} (\bibinfo {year} {2016})},\ \Eprint
  {https://arxiv.org/abs/1604.00141} {arXiv:1604.00141 [astro-ph.CO]}
  \BibitemShut {NoStop}%
\bibitem [{\citenamefont {Mylova}\ \emph {et~al.}(2018)\citenamefont {Mylova},
  \citenamefont {\"Ozsoy}, \citenamefont {Parameswaran}, \citenamefont
  {Tasinato},\ and\ \citenamefont {Zavala}}]{Mylova:2018yap}%
  \BibitemOpen
  \bibfield  {author} {\bibinfo {author} {\bibfnamefont {M.}~\bibnamefont
  {Mylova}}, \bibinfo {author} {\bibfnamefont {O.}~\bibnamefont {\"Ozsoy}},
  \bibinfo {author} {\bibfnamefont {S.}~\bibnamefont {Parameswaran}}, \bibinfo
  {author} {\bibfnamefont {G.}~\bibnamefont {Tasinato}},\ and\ \bibinfo
  {author} {\bibfnamefont {I.}~\bibnamefont {Zavala}},\ }\bibfield  {title}
  {\bibinfo {title} {{A new mechanism to enhance primordial tensor fluctuations
  in single field inflation}},\ }\href
  {https://doi.org/10.1088/1475-7516/2018/12/024} {\bibfield  {journal}
  {\bibinfo  {journal} {JCAP}\ }\textbf {\bibinfo {volume} {12}},\ \bibinfo
  {pages} {024}},\ \Eprint {https://arxiv.org/abs/1808.10475} {arXiv:1808.10475
  [gr-qc]} \BibitemShut {NoStop}%
\bibitem [{\citenamefont {Kawasaki}\ \emph {et~al.}(1999)\citenamefont
  {Kawasaki}, \citenamefont {Kohri},\ and\ \citenamefont
  {Sugiyama}}]{Kawasaki:1999na}%
  \BibitemOpen
  \bibfield  {author} {\bibinfo {author} {\bibfnamefont {M.}~\bibnamefont
  {Kawasaki}}, \bibinfo {author} {\bibfnamefont {K.}~\bibnamefont {Kohri}},\
  and\ \bibinfo {author} {\bibfnamefont {N.}~\bibnamefont {Sugiyama}},\
  }\bibfield  {title} {\bibinfo {title} {{Cosmological constraints on late time
  entropy production}},\ }\href {https://doi.org/10.1103/PhysRevLett.82.4168}
  {\bibfield  {journal} {\bibinfo  {journal} {Phys. Rev. Lett.}\ }\textbf
  {\bibinfo {volume} {82}},\ \bibinfo {pages} {4168} (\bibinfo {year}
  {1999})},\ \Eprint {https://arxiv.org/abs/astro-ph/9811437}
  {arXiv:astro-ph/9811437} \BibitemShut {NoStop}%
\bibitem [{\citenamefont {Kawasaki}\ \emph {et~al.}(2000)\citenamefont
  {Kawasaki}, \citenamefont {Kohri},\ and\ \citenamefont
  {Sugiyama}}]{Kawasaki:2000en}%
  \BibitemOpen
  \bibfield  {author} {\bibinfo {author} {\bibfnamefont {M.}~\bibnamefont
  {Kawasaki}}, \bibinfo {author} {\bibfnamefont {K.}~\bibnamefont {Kohri}},\
  and\ \bibinfo {author} {\bibfnamefont {N.}~\bibnamefont {Sugiyama}},\
  }\bibfield  {title} {\bibinfo {title} {{MeV scale reheating temperature and
  thermalization of neutrino background}},\ }\href
  {https://doi.org/10.1103/PhysRevD.62.023506} {\bibfield  {journal} {\bibinfo
  {journal} {Phys. Rev. D}\ }\textbf {\bibinfo {volume} {62}},\ \bibinfo
  {pages} {023506} (\bibinfo {year} {2000})},\ \Eprint
  {https://arxiv.org/abs/astro-ph/0002127} {arXiv:astro-ph/0002127}
  \BibitemShut {NoStop}%
\bibitem [{\citenamefont {Hannestad}(2004)}]{Hannestad:2004px}%
  \BibitemOpen
  \bibfield  {author} {\bibinfo {author} {\bibfnamefont {S.}~\bibnamefont
  {Hannestad}},\ }\bibfield  {title} {\bibinfo {title} {{What is the lowest
  possible reheating temperature?}},\ }\href
  {https://doi.org/10.1103/PhysRevD.70.043506} {\bibfield  {journal} {\bibinfo
  {journal} {Phys. Rev. D}\ }\textbf {\bibinfo {volume} {70}},\ \bibinfo
  {pages} {043506} (\bibinfo {year} {2004})},\ \Eprint
  {https://arxiv.org/abs/astro-ph/0403291} {arXiv:astro-ph/0403291}
  \BibitemShut {NoStop}%
\bibitem [{\citenamefont {Hasegawa}\ \emph {et~al.}(2019)\citenamefont
  {Hasegawa}, \citenamefont {Hiroshima}, \citenamefont {Kohri}, \citenamefont
  {Hansen}, \citenamefont {Tram},\ and\ \citenamefont
  {Hannestad}}]{Hasegawa:2019jsa}%
  \BibitemOpen
  \bibfield  {author} {\bibinfo {author} {\bibfnamefont {T.}~\bibnamefont
  {Hasegawa}}, \bibinfo {author} {\bibfnamefont {N.}~\bibnamefont {Hiroshima}},
  \bibinfo {author} {\bibfnamefont {K.}~\bibnamefont {Kohri}}, \bibinfo
  {author} {\bibfnamefont {R.~S.~L.}\ \bibnamefont {Hansen}}, \bibinfo {author}
  {\bibfnamefont {T.}~\bibnamefont {Tram}},\ and\ \bibinfo {author}
  {\bibfnamefont {S.}~\bibnamefont {Hannestad}},\ }\bibfield  {title} {\bibinfo
  {title} {{MeV-scale reheating temperature and thermalization of oscillating
  neutrinos by radiative and hadronic decays of massive particles}},\ }\href
  {https://doi.org/10.1088/1475-7516/2019/12/012} {\bibfield  {journal}
  {\bibinfo  {journal} {JCAP}\ }\textbf {\bibinfo {volume} {12}},\ \bibinfo
  {pages} {012}},\ \Eprint {https://arxiv.org/abs/1908.10189} {arXiv:1908.10189
  [hep-ph]} \BibitemShut {NoStop}%
\bibitem [{\citenamefont {Tomita}(1967)}]{Tomita}%
  \BibitemOpen
  \bibfield  {author} {\bibinfo {author} {\bibfnamefont {K.}~\bibnamefont
  {Tomita}},\ }\bibfield  {title} {\bibinfo {title} {{Non-Linear Theory of
  Gravitational Instability in the Expanding Universe}},\ }\href
  {https://doi.org/10.1143/PTP.37.831} {\bibfield  {journal} {\bibinfo
  {journal} {Progress of Theoretical Physics}\ }\textbf {\bibinfo {volume}
  {37}},\ \bibinfo {pages} {831} (\bibinfo {year} {1967})},\ \Eprint
  {https://arxiv.org/abs/https://academic.oup.com/ptp/article-pdf/37/5/831/5234391/37-5-831.pdf}
  {https://academic.oup.com/ptp/article-pdf/37/5/831/5234391/37-5-831.pdf}
  \BibitemShut {NoStop}%
\bibitem [{\citenamefont {Matarrese}\ \emph {et~al.}(1993)\citenamefont
  {Matarrese}, \citenamefont {Pantano},\ and\ \citenamefont
  {Saez}}]{Matarrese:1992rp}%
  \BibitemOpen
  \bibfield  {author} {\bibinfo {author} {\bibfnamefont {S.}~\bibnamefont
  {Matarrese}}, \bibinfo {author} {\bibfnamefont {O.}~\bibnamefont {Pantano}},\
  and\ \bibinfo {author} {\bibfnamefont {D.}~\bibnamefont {Saez}},\ }\bibfield
  {title} {\bibinfo {title} {{A General relativistic approach to the nonlinear
  evolution of collisionless matter}},\ }\href
  {https://doi.org/10.1103/PhysRevD.47.1311} {\bibfield  {journal} {\bibinfo
  {journal} {Phys. Rev. D}\ }\textbf {\bibinfo {volume} {47}},\ \bibinfo
  {pages} {1311} (\bibinfo {year} {1993})}\BibitemShut {NoStop}%
\bibitem [{\citenamefont {Matarrese}\ \emph {et~al.}(1994)\citenamefont
  {Matarrese}, \citenamefont {Pantano},\ and\ \citenamefont
  {Saez}}]{Matarrese:1993zf}%
  \BibitemOpen
  \bibfield  {author} {\bibinfo {author} {\bibfnamefont {S.}~\bibnamefont
  {Matarrese}}, \bibinfo {author} {\bibfnamefont {O.}~\bibnamefont {Pantano}},\
  and\ \bibinfo {author} {\bibfnamefont {D.}~\bibnamefont {Saez}},\ }\bibfield
  {title} {\bibinfo {title} {{General relativistic dynamics of irrotational
  dust: Cosmological implications}},\ }\href
  {https://doi.org/10.1103/PhysRevLett.72.320} {\bibfield  {journal} {\bibinfo
  {journal} {Phys. Rev. Lett.}\ }\textbf {\bibinfo {volume} {72}},\ \bibinfo
  {pages} {320} (\bibinfo {year} {1994})},\ \Eprint
  {https://arxiv.org/abs/astro-ph/9310036} {arXiv:astro-ph/9310036}
  \BibitemShut {NoStop}%
\bibitem [{\citenamefont {Matarrese}\ \emph {et~al.}(1998)\citenamefont
  {Matarrese}, \citenamefont {Mollerach},\ and\ \citenamefont
  {Bruni}}]{Matarrese:1997ay}%
  \BibitemOpen
  \bibfield  {author} {\bibinfo {author} {\bibfnamefont {S.}~\bibnamefont
  {Matarrese}}, \bibinfo {author} {\bibfnamefont {S.}~\bibnamefont
  {Mollerach}},\ and\ \bibinfo {author} {\bibfnamefont {M.}~\bibnamefont
  {Bruni}},\ }\bibfield  {title} {\bibinfo {title} {{Second order perturbations
  of the Einstein-de Sitter universe}},\ }\href
  {https://doi.org/10.1103/PhysRevD.58.043504} {\bibfield  {journal} {\bibinfo
  {journal} {Phys. Rev. D}\ }\textbf {\bibinfo {volume} {58}},\ \bibinfo
  {pages} {043504} (\bibinfo {year} {1998})},\ \Eprint
  {https://arxiv.org/abs/astro-ph/9707278} {arXiv:astro-ph/9707278}
  \BibitemShut {NoStop}%
\bibitem [{\citenamefont {Lin}\ and\ \citenamefont
  {Sasaki}(2016)}]{Lin:2015nda}%
  \BibitemOpen
  \bibfield  {author} {\bibinfo {author} {\bibfnamefont {C.}~\bibnamefont
  {Lin}}\ and\ \bibinfo {author} {\bibfnamefont {M.}~\bibnamefont {Sasaki}},\
  }\bibfield  {title} {\bibinfo {title} {{Resonant Primordial Gravitational
  Waves Amplification}},\ }\href
  {https://doi.org/10.1016/j.physletb.2015.11.021} {\bibfield  {journal}
  {\bibinfo  {journal} {Phys. Lett. B}\ }\textbf {\bibinfo {volume} {752}},\
  \bibinfo {pages} {84} (\bibinfo {year} {2016})},\ \Eprint
  {https://arxiv.org/abs/1504.01373} {arXiv:1504.01373 [astro-ph.CO]}
  \BibitemShut {NoStop}%
\bibitem [{\citenamefont {Kuroyanagi}\ \emph {et~al.}(2018)\citenamefont
  {Kuroyanagi}, \citenamefont {Lin}, \citenamefont {Sasaki},\ and\
  \citenamefont {Tsujikawa}}]{Kuroyanagi:2017kfx}%
  \BibitemOpen
  \bibfield  {author} {\bibinfo {author} {\bibfnamefont {S.}~\bibnamefont
  {Kuroyanagi}}, \bibinfo {author} {\bibfnamefont {C.}~\bibnamefont {Lin}},
  \bibinfo {author} {\bibfnamefont {M.}~\bibnamefont {Sasaki}},\ and\ \bibinfo
  {author} {\bibfnamefont {S.}~\bibnamefont {Tsujikawa}},\ }\bibfield  {title}
  {\bibinfo {title} {{Observational signatures of the parametric amplification
  of gravitational waves during reheating after inflation}},\ }\href
  {https://doi.org/10.1103/PhysRevD.97.023516} {\bibfield  {journal} {\bibinfo
  {journal} {Phys. Rev. D}\ }\textbf {\bibinfo {volume} {97}},\ \bibinfo
  {pages} {023516} (\bibinfo {year} {2018})},\ \Eprint
  {https://arxiv.org/abs/1710.06789} {arXiv:1710.06789 [gr-qc]} \BibitemShut
  {NoStop}%
\bibitem [{\citenamefont {Cai}\ \emph {et~al.}(2021)\citenamefont {Cai},
  \citenamefont {Lin}, \citenamefont {Wang},\ and\ \citenamefont
  {Yan}}]{Cai:2020ovp}%
  \BibitemOpen
  \bibfield  {author} {\bibinfo {author} {\bibfnamefont {Y.-F.}\ \bibnamefont
  {Cai}}, \bibinfo {author} {\bibfnamefont {C.}~\bibnamefont {Lin}}, \bibinfo
  {author} {\bibfnamefont {B.}~\bibnamefont {Wang}},\ and\ \bibinfo {author}
  {\bibfnamefont {S.-F.}\ \bibnamefont {Yan}},\ }\bibfield  {title} {\bibinfo
  {title} {{Sound speed resonance of the stochastic gravitational wave
  background}},\ }\href {https://doi.org/10.1103/PhysRevLett.126.071303}
  {\bibfield  {journal} {\bibinfo  {journal} {Phys. Rev. Lett.}\ }\textbf
  {\bibinfo {volume} {126}},\ \bibinfo {pages} {071303} (\bibinfo {year}
  {2021})},\ \Eprint {https://arxiv.org/abs/2009.09833} {arXiv:2009.09833
  [gr-qc]} \BibitemShut {NoStop}%
\bibitem [{\citenamefont {Ye}\ and\ \citenamefont
  {Silvestri}(2024)}]{Ye:2023xyr}%
  \BibitemOpen
  \bibfield  {author} {\bibinfo {author} {\bibfnamefont {G.}~\bibnamefont
  {Ye}}\ and\ \bibinfo {author} {\bibfnamefont {A.}~\bibnamefont {Silvestri}},\
  }\bibfield  {title} {\bibinfo {title} {{Can the Gravitational Wave Background
  Feel Wiggles in Spacetime?}},\ }\href
  {https://doi.org/10.3847/2041-8213/ad2851} {\bibfield  {journal} {\bibinfo
  {journal} {Astrophys. J. Lett.}\ }\textbf {\bibinfo {volume} {963}},\
  \bibinfo {pages} {L15} (\bibinfo {year} {2024})},\ \Eprint
  {https://arxiv.org/abs/2307.05455} {arXiv:2307.05455 [astro-ph.CO]}
  \BibitemShut {NoStop}%
\bibitem [{\citenamefont {Cai}\ \emph {et~al.}(2023{\natexlab{a}})\citenamefont
  {Cai}, \citenamefont {Dom\`enech}, \citenamefont {Ganz}, \citenamefont
  {Jiang}, \citenamefont {Lin},\ and\ \citenamefont {Wang}}]{Cai:2023ykr}%
  \BibitemOpen
  \bibfield  {author} {\bibinfo {author} {\bibfnamefont {Y.-F.}\ \bibnamefont
  {Cai}}, \bibinfo {author} {\bibfnamefont {G.}~\bibnamefont {Dom\`enech}},
  \bibinfo {author} {\bibfnamefont {A.}~\bibnamefont {Ganz}}, \bibinfo {author}
  {\bibfnamefont {J.}~\bibnamefont {Jiang}}, \bibinfo {author} {\bibfnamefont
  {C.}~\bibnamefont {Lin}},\ and\ \bibinfo {author} {\bibfnamefont
  {B.}~\bibnamefont {Wang}},\ }\bibfield  {title} {\bibinfo {title}
  {{Parametric resonance of gravitational waves in general scalar-tensor
  theories}},\ }\href@noop {} {\  (\bibinfo {year} {2023}{\natexlab{a}})},\
  \Eprint {https://arxiv.org/abs/2311.18546} {arXiv:2311.18546 [gr-qc]}
  \BibitemShut {NoStop}%
\bibitem [{\citenamefont {Ananda}\ \emph {et~al.}(2007)\citenamefont {Ananda},
  \citenamefont {Clarkson},\ and\ \citenamefont {Wands}}]{Ananda:2006af}%
  \BibitemOpen
  \bibfield  {author} {\bibinfo {author} {\bibfnamefont {K.~N.}\ \bibnamefont
  {Ananda}}, \bibinfo {author} {\bibfnamefont {C.}~\bibnamefont {Clarkson}},\
  and\ \bibinfo {author} {\bibfnamefont {D.}~\bibnamefont {Wands}},\ }\bibfield
   {title} {\bibinfo {title} {{The Cosmological gravitational wave background
  from primordial density perturbations}},\ }\href
  {https://doi.org/10.1103/PhysRevD.75.123518} {\bibfield  {journal} {\bibinfo
  {journal} {Phys. Rev. D}\ }\textbf {\bibinfo {volume} {75}},\ \bibinfo
  {pages} {123518} (\bibinfo {year} {2007})},\ \Eprint
  {https://arxiv.org/abs/gr-qc/0612013} {arXiv:gr-qc/0612013} \BibitemShut
  {NoStop}%
\bibitem [{\citenamefont {Baumann}\ \emph {et~al.}(2007)\citenamefont
  {Baumann}, \citenamefont {Steinhardt}, \citenamefont {Takahashi},\ and\
  \citenamefont {Ichiki}}]{Baumann:2007zm}%
  \BibitemOpen
  \bibfield  {author} {\bibinfo {author} {\bibfnamefont {D.}~\bibnamefont
  {Baumann}}, \bibinfo {author} {\bibfnamefont {P.~J.}\ \bibnamefont
  {Steinhardt}}, \bibinfo {author} {\bibfnamefont {K.}~\bibnamefont
  {Takahashi}},\ and\ \bibinfo {author} {\bibfnamefont {K.}~\bibnamefont
  {Ichiki}},\ }\bibfield  {title} {\bibinfo {title} {{Gravitational Wave
  Spectrum Induced by Primordial Scalar Perturbations}},\ }\href
  {https://doi.org/10.1103/PhysRevD.76.084019} {\bibfield  {journal} {\bibinfo
  {journal} {Phys. Rev. D}\ }\textbf {\bibinfo {volume} {76}},\ \bibinfo
  {pages} {084019} (\bibinfo {year} {2007})},\ \Eprint
  {https://arxiv.org/abs/hep-th/0703290} {arXiv:hep-th/0703290} \BibitemShut
  {NoStop}%
\bibitem [{\citenamefont {Assadullahi}\ and\ \citenamefont
  {Wands}(2009)}]{Assadullahi:2009nf}%
  \BibitemOpen
  \bibfield  {author} {\bibinfo {author} {\bibfnamefont {H.}~\bibnamefont
  {Assadullahi}}\ and\ \bibinfo {author} {\bibfnamefont {D.}~\bibnamefont
  {Wands}},\ }\bibfield  {title} {\bibinfo {title} {{Gravitational waves from
  an early matter era}},\ }\href {https://doi.org/10.1103/PhysRevD.79.083511}
  {\bibfield  {journal} {\bibinfo  {journal} {Phys. Rev. D}\ }\textbf {\bibinfo
  {volume} {79}},\ \bibinfo {pages} {083511} (\bibinfo {year} {2009})},\
  \Eprint {https://arxiv.org/abs/0901.0989} {arXiv:0901.0989 [astro-ph.CO]}
  \BibitemShut {NoStop}%
\bibitem [{\citenamefont {Alabidi}\ \emph {et~al.}(2013)\citenamefont
  {Alabidi}, \citenamefont {Kohri}, \citenamefont {Sasaki},\ and\ \citenamefont
  {Sendouda}}]{Alabidi:2013lya}%
  \BibitemOpen
  \bibfield  {author} {\bibinfo {author} {\bibfnamefont {L.}~\bibnamefont
  {Alabidi}}, \bibinfo {author} {\bibfnamefont {K.}~\bibnamefont {Kohri}},
  \bibinfo {author} {\bibfnamefont {M.}~\bibnamefont {Sasaki}},\ and\ \bibinfo
  {author} {\bibfnamefont {Y.}~\bibnamefont {Sendouda}},\ }\bibfield  {title}
  {\bibinfo {title} {{Observable induced gravitational waves from an early
  matter phase}},\ }\href {https://doi.org/10.1088/1475-7516/2013/05/033}
  {\bibfield  {journal} {\bibinfo  {journal} {JCAP}\ }\textbf {\bibinfo
  {volume} {05}},\ \bibinfo {pages} {033}},\ \Eprint
  {https://arxiv.org/abs/1303.4519} {arXiv:1303.4519 [astro-ph.CO]}
  \BibitemShut {NoStop}%
\bibitem [{\citenamefont {Saito}\ and\ \citenamefont
  {Yokoyama}(2009)}]{Saito:2008jc}%
  \BibitemOpen
  \bibfield  {author} {\bibinfo {author} {\bibfnamefont {R.}~\bibnamefont
  {Saito}}\ and\ \bibinfo {author} {\bibfnamefont {J.}~\bibnamefont
  {Yokoyama}},\ }\bibfield  {title} {\bibinfo {title} {{Gravitational wave
  background as a probe of the primordial black hole abundance}},\ }\href
  {https://doi.org/10.1103/PhysRevLett.102.161101} {\bibfield  {journal}
  {\bibinfo  {journal} {Phys. Rev. Lett.}\ }\textbf {\bibinfo {volume} {102}},\
  \bibinfo {pages} {161101} (\bibinfo {year} {2009})},\ \bibinfo {note}
  {[Erratum: Phys.Rev.Lett. 107, 069901 (2011)]},\ \Eprint
  {https://arxiv.org/abs/0812.4339} {arXiv:0812.4339 [astro-ph]} \BibitemShut
  {NoStop}%
\bibitem [{\citenamefont {Saito}\ and\ \citenamefont
  {Yokoyama}(2010)}]{Saito:2009jt}%
  \BibitemOpen
  \bibfield  {author} {\bibinfo {author} {\bibfnamefont {R.}~\bibnamefont
  {Saito}}\ and\ \bibinfo {author} {\bibfnamefont {J.}~\bibnamefont
  {Yokoyama}},\ }\bibfield  {title} {\bibinfo {title} {{Gravitational-Wave
  Constraints on the Abundance of Primordial Black Holes}},\ }\href
  {https://doi.org/10.1143/PTP.126.351} {\bibfield  {journal} {\bibinfo
  {journal} {Prog. Theor. Phys.}\ }\textbf {\bibinfo {volume} {123}},\ \bibinfo
  {pages} {867} (\bibinfo {year} {2010})},\ \bibinfo {note} {[Erratum:
  Prog.Theor.Phys. 126, 351--352 (2011)]},\ \Eprint
  {https://arxiv.org/abs/0912.5317} {arXiv:0912.5317 [astro-ph.CO]}
  \BibitemShut {NoStop}%
\bibitem [{\citenamefont {Bugaev}\ and\ \citenamefont
  {Klimai}(2010{\natexlab{a}})}]{Bugaev:2009zh}%
  \BibitemOpen
  \bibfield  {author} {\bibinfo {author} {\bibfnamefont {E.}~\bibnamefont
  {Bugaev}}\ and\ \bibinfo {author} {\bibfnamefont {P.}~\bibnamefont
  {Klimai}},\ }\bibfield  {title} {\bibinfo {title} {{Induced gravitational
  wave background and primordial black holes}},\ }\href
  {https://doi.org/10.1103/PhysRevD.81.023517} {\bibfield  {journal} {\bibinfo
  {journal} {Phys. Rev. D}\ }\textbf {\bibinfo {volume} {81}},\ \bibinfo
  {pages} {023517} (\bibinfo {year} {2010}{\natexlab{a}})},\ \Eprint
  {https://arxiv.org/abs/0908.0664} {arXiv:0908.0664 [astro-ph.CO]}
  \BibitemShut {NoStop}%
\bibitem [{\citenamefont {Bugaev}\ and\ \citenamefont
  {Klimai}(2010{\natexlab{b}})}]{Bugaev:2009kq}%
  \BibitemOpen
  \bibfield  {author} {\bibinfo {author} {\bibfnamefont {E.~V.}\ \bibnamefont
  {Bugaev}}\ and\ \bibinfo {author} {\bibfnamefont {P.~A.}\ \bibnamefont
  {Klimai}},\ }\bibfield  {title} {\bibinfo {title} {{Bound on induced
  gravitational wave background from primordial black holes}},\ }\href
  {https://doi.org/10.1134/S0021364010010017} {\bibfield  {journal} {\bibinfo
  {journal} {JETP Lett.}\ }\textbf {\bibinfo {volume} {91}},\ \bibinfo {pages}
  {1} (\bibinfo {year} {2010}{\natexlab{b}})},\ \Eprint
  {https://arxiv.org/abs/0911.0611} {arXiv:0911.0611 [astro-ph.CO]}
  \BibitemShut {NoStop}%
\bibitem [{\citenamefont {Bugaev}\ and\ \citenamefont
  {Klimai}(2011)}]{Bugaev:2010bb}%
  \BibitemOpen
  \bibfield  {author} {\bibinfo {author} {\bibfnamefont {E.}~\bibnamefont
  {Bugaev}}\ and\ \bibinfo {author} {\bibfnamefont {P.}~\bibnamefont
  {Klimai}},\ }\bibfield  {title} {\bibinfo {title} {{Constraints on the
  induced gravitational wave background from primordial black holes}},\ }\href
  {https://doi.org/10.1103/PhysRevD.83.083521} {\bibfield  {journal} {\bibinfo
  {journal} {Phys. Rev. D}\ }\textbf {\bibinfo {volume} {83}},\ \bibinfo
  {pages} {083521} (\bibinfo {year} {2011})},\ \Eprint
  {https://arxiv.org/abs/1012.4697} {arXiv:1012.4697 [astro-ph.CO]}
  \BibitemShut {NoStop}%
\bibitem [{\citenamefont {Khlopov}(2010)}]{Khlopov:2008qy}%
  \BibitemOpen
  \bibfield  {author} {\bibinfo {author} {\bibfnamefont {M.~Y.}\ \bibnamefont
  {Khlopov}},\ }\bibfield  {title} {\bibinfo {title} {{Primordial Black
  Holes}},\ }\href {https://doi.org/10.1088/1674-4527/10/6/001} {\bibfield
  {journal} {\bibinfo  {journal} {Res. Astron. Astrophys.}\ }\textbf {\bibinfo
  {volume} {10}},\ \bibinfo {pages} {495} (\bibinfo {year} {2010})},\ \Eprint
  {https://arxiv.org/abs/0801.0116} {arXiv:0801.0116 [astro-ph]} \BibitemShut
  {NoStop}%
\bibitem [{\citenamefont {Sasaki}\ \emph {et~al.}(2018)\citenamefont {Sasaki},
  \citenamefont {Suyama}, \citenamefont {Tanaka},\ and\ \citenamefont
  {Yokoyama}}]{Sasaki:2018dmp}%
  \BibitemOpen
  \bibfield  {author} {\bibinfo {author} {\bibfnamefont {M.}~\bibnamefont
  {Sasaki}}, \bibinfo {author} {\bibfnamefont {T.}~\bibnamefont {Suyama}},
  \bibinfo {author} {\bibfnamefont {T.}~\bibnamefont {Tanaka}},\ and\ \bibinfo
  {author} {\bibfnamefont {S.}~\bibnamefont {Yokoyama}},\ }\bibfield  {title}
  {\bibinfo {title} {{Primordial black holes\textemdash{}perspectives in
  gravitational wave astronomy}},\ }\href
  {https://doi.org/10.1088/1361-6382/aaa7b4} {\bibfield  {journal} {\bibinfo
  {journal} {Class. Quant. Grav.}\ }\textbf {\bibinfo {volume} {35}},\ \bibinfo
  {pages} {063001} (\bibinfo {year} {2018})},\ \Eprint
  {https://arxiv.org/abs/1801.05235} {arXiv:1801.05235 [astro-ph.CO]}
  \BibitemShut {NoStop}%
\bibitem [{\citenamefont {Green}\ and\ \citenamefont
  {Kavanagh}(2021)}]{Green:2020jor}%
  \BibitemOpen
  \bibfield  {author} {\bibinfo {author} {\bibfnamefont {A.~M.}\ \bibnamefont
  {Green}}\ and\ \bibinfo {author} {\bibfnamefont {B.~J.}\ \bibnamefont
  {Kavanagh}},\ }\bibfield  {title} {\bibinfo {title} {{Primordial Black Holes
  as a dark matter candidate}},\ }\href
  {https://doi.org/10.1088/1361-6471/abc534} {\bibfield  {journal} {\bibinfo
  {journal} {J. Phys. G}\ }\textbf {\bibinfo {volume} {48}},\ \bibinfo {pages}
  {4} (\bibinfo {year} {2021})},\ \Eprint {https://arxiv.org/abs/2007.10722}
  {arXiv:2007.10722 [astro-ph.CO]} \BibitemShut {NoStop}%
\bibitem [{\citenamefont {Carr}\ \emph {et~al.}(2021)\citenamefont {Carr},
  \citenamefont {Kohri}, \citenamefont {Sendouda},\ and\ \citenamefont
  {Yokoyama}}]{Carr:2020gox}%
  \BibitemOpen
  \bibfield  {author} {\bibinfo {author} {\bibfnamefont {B.}~\bibnamefont
  {Carr}}, \bibinfo {author} {\bibfnamefont {K.}~\bibnamefont {Kohri}},
  \bibinfo {author} {\bibfnamefont {Y.}~\bibnamefont {Sendouda}},\ and\
  \bibinfo {author} {\bibfnamefont {J.}~\bibnamefont {Yokoyama}},\ }\bibfield
  {title} {\bibinfo {title} {{Constraints on primordial black holes}},\ }\href
  {https://doi.org/10.1088/1361-6633/ac1e31} {\bibfield  {journal} {\bibinfo
  {journal} {Rept. Prog. Phys.}\ }\textbf {\bibinfo {volume} {84}},\ \bibinfo
  {pages} {116902} (\bibinfo {year} {2021})},\ \Eprint
  {https://arxiv.org/abs/2002.12778} {arXiv:2002.12778 [astro-ph.CO]}
  \BibitemShut {NoStop}%
\bibitem [{\citenamefont {Escriv\`a}\ \emph {et~al.}(2022)\citenamefont
  {Escriv\`a}, \citenamefont {Kuhnel},\ and\ \citenamefont
  {Tada}}]{Escriva:2022duf}%
  \BibitemOpen
  \bibfield  {author} {\bibinfo {author} {\bibfnamefont {A.}~\bibnamefont
  {Escriv\`a}}, \bibinfo {author} {\bibfnamefont {F.}~\bibnamefont {Kuhnel}},\
  and\ \bibinfo {author} {\bibfnamefont {Y.}~\bibnamefont {Tada}},\ }\bibfield
  {title} {\bibinfo {title} {{Primordial Black Holes}},\ }\href@noop {} {\
  (\bibinfo {year} {2022})},\ \Eprint {https://arxiv.org/abs/2211.05767}
  {arXiv:2211.05767 [astro-ph.CO]} \BibitemShut {NoStop}%
\bibitem [{\citenamefont {Dandoy}\ \emph {et~al.}(2023)\citenamefont {Dandoy},
  \citenamefont {Domcke},\ and\ \citenamefont {Rompineve}}]{Dandoy:2023jot}%
  \BibitemOpen
  \bibfield  {author} {\bibinfo {author} {\bibfnamefont {V.}~\bibnamefont
  {Dandoy}}, \bibinfo {author} {\bibfnamefont {V.}~\bibnamefont {Domcke}},\
  and\ \bibinfo {author} {\bibfnamefont {F.}~\bibnamefont {Rompineve}},\
  }\bibfield  {title} {\bibinfo {title} {{Search for scalar induced
  gravitational waves in the international pulsar timing array data release 2
  and NANOgrav 12.5 years datasets}},\ }\href
  {https://doi.org/10.21468/SciPostPhysCore.6.3.060} {\bibfield  {journal}
  {\bibinfo  {journal} {SciPost Phys. Core}\ }\textbf {\bibinfo {volume} {6}},\
  \bibinfo {pages} {060} (\bibinfo {year} {2023})},\ \Eprint
  {https://arxiv.org/abs/2302.07901} {arXiv:2302.07901 [astro-ph.CO]}
  \BibitemShut {NoStop}%
\bibitem [{\citenamefont {Franciolini}\ \emph
  {et~al.}(2023{\natexlab{a}})\citenamefont {Franciolini}, \citenamefont
  {Iovino}, \citenamefont {Vaskonen},\ and\ \citenamefont
  {Veermae}}]{Franciolini:2023pbf}%
  \BibitemOpen
  \bibfield  {author} {\bibinfo {author} {\bibfnamefont {G.}~\bibnamefont
  {Franciolini}}, \bibinfo {author} {\bibfnamefont {A.}~\bibnamefont {Iovino},
  \bibfnamefont {Junior.}}, \bibinfo {author} {\bibfnamefont {V.}~\bibnamefont
  {Vaskonen}},\ and\ \bibinfo {author} {\bibfnamefont {H.}~\bibnamefont
  {Veermae}},\ }\bibfield  {title} {\bibinfo {title} {{The recent gravitational
  wave observation by pulsar timing arrays and primordial black holes: the
  importance of non-gaussianities}},\ }\href@noop {} {\  (\bibinfo {year}
  {2023}{\natexlab{a}})},\ \Eprint {https://arxiv.org/abs/2306.17149}
  {arXiv:2306.17149 [astro-ph.CO]} \BibitemShut {NoStop}%
\bibitem [{\citenamefont {Franciolini}\ \emph
  {et~al.}(2023{\natexlab{b}})\citenamefont {Franciolini}, \citenamefont
  {Racco},\ and\ \citenamefont {Rompineve}}]{Franciolini:2023wjm}%
  \BibitemOpen
  \bibfield  {author} {\bibinfo {author} {\bibfnamefont {G.}~\bibnamefont
  {Franciolini}}, \bibinfo {author} {\bibfnamefont {D.}~\bibnamefont {Racco}},\
  and\ \bibinfo {author} {\bibfnamefont {F.}~\bibnamefont {Rompineve}},\
  }\bibfield  {title} {\bibinfo {title} {{Footprints of the QCD Crossover on
  Cosmological Gravitational Waves at Pulsar Timing Arrays}},\ }\href@noop {}
  {\  (\bibinfo {year} {2023}{\natexlab{b}})},\ \Eprint
  {https://arxiv.org/abs/2306.17136} {arXiv:2306.17136 [astro-ph.CO]}
  \BibitemShut {NoStop}%
\bibitem [{\citenamefont {Inomata}\ \emph
  {et~al.}(2023{\natexlab{a}})\citenamefont {Inomata}, \citenamefont {Kohri},\
  and\ \citenamefont {Terada}}]{Inomata:2023zup}%
  \BibitemOpen
  \bibfield  {author} {\bibinfo {author} {\bibfnamefont {K.}~\bibnamefont
  {Inomata}}, \bibinfo {author} {\bibfnamefont {K.}~\bibnamefont {Kohri}},\
  and\ \bibinfo {author} {\bibfnamefont {T.}~\bibnamefont {Terada}},\
  }\bibfield  {title} {\bibinfo {title} {{The Detected Stochastic Gravitational
  Waves and Sub-Solar Primordial Black Holes}},\ }\href@noop {} {\  (\bibinfo
  {year} {2023}{\natexlab{a}})},\ \Eprint {https://arxiv.org/abs/2306.17834}
  {arXiv:2306.17834 [astro-ph.CO]} \BibitemShut {NoStop}%
\bibitem [{\citenamefont {Cai}\ \emph {et~al.}(2023{\natexlab{b}})\citenamefont
  {Cai}, \citenamefont {He}, \citenamefont {Ma}, \citenamefont {Yan},\ and\
  \citenamefont {Yuan}}]{Cai:2023dls}%
  \BibitemOpen
  \bibfield  {author} {\bibinfo {author} {\bibfnamefont {Y.-F.}\ \bibnamefont
  {Cai}}, \bibinfo {author} {\bibfnamefont {X.-C.}\ \bibnamefont {He}},
  \bibinfo {author} {\bibfnamefont {X.}~\bibnamefont {Ma}}, \bibinfo {author}
  {\bibfnamefont {S.-F.}\ \bibnamefont {Yan}},\ and\ \bibinfo {author}
  {\bibfnamefont {G.-W.}\ \bibnamefont {Yuan}},\ }\bibfield  {title} {\bibinfo
  {title} {{Limits on scalar-induced gravitational waves from the stochastic
  background by pulsar timing array observations}},\ }\href@noop {} {\
  (\bibinfo {year} {2023}{\natexlab{b}})},\ \Eprint
  {https://arxiv.org/abs/2306.17822} {arXiv:2306.17822 [gr-qc]} \BibitemShut
  {NoStop}%
\bibitem [{\citenamefont {Wang}\ \emph {et~al.}(2023)\citenamefont {Wang},
  \citenamefont {Zhao}, \citenamefont {Li},\ and\ \citenamefont
  {Zhu}}]{Wang:2023ost}%
  \BibitemOpen
  \bibfield  {author} {\bibinfo {author} {\bibfnamefont {S.}~\bibnamefont
  {Wang}}, \bibinfo {author} {\bibfnamefont {Z.-C.}\ \bibnamefont {Zhao}},
  \bibinfo {author} {\bibfnamefont {J.-P.}\ \bibnamefont {Li}},\ and\ \bibinfo
  {author} {\bibfnamefont {Q.-H.}\ \bibnamefont {Zhu}},\ }\bibfield  {title}
  {\bibinfo {title} {{Exploring the Implications of 2023 Pulsar Timing Array
  Datasets for Scalar-Induced Gravitational Waves and Primordial Black
  Holes}},\ }\href@noop {} {\  (\bibinfo {year} {2023})},\ \Eprint
  {https://arxiv.org/abs/2307.00572} {arXiv:2307.00572 [astro-ph.CO]}
  \BibitemShut {NoStop}%
\bibitem [{\citenamefont {Liu}\ \emph {et~al.}(2023{\natexlab{a}})\citenamefont
  {Liu}, \citenamefont {Chen},\ and\ \citenamefont {Huang}}]{Liu:2023ymk}%
  \BibitemOpen
  \bibfield  {author} {\bibinfo {author} {\bibfnamefont {L.}~\bibnamefont
  {Liu}}, \bibinfo {author} {\bibfnamefont {Z.-C.}\ \bibnamefont {Chen}},\ and\
  \bibinfo {author} {\bibfnamefont {Q.-G.}\ \bibnamefont {Huang}},\ }\bibfield
  {title} {\bibinfo {title} {{Implications for the non-Gaussianity of curvature
  perturbation from pulsar timing arrays}},\ }\href@noop {} {\  (\bibinfo
  {year} {2023}{\natexlab{a}})},\ \Eprint {https://arxiv.org/abs/2307.01102}
  {arXiv:2307.01102 [astro-ph.CO]} \BibitemShut {NoStop}%
\bibitem [{\citenamefont {Unal}\ \emph {et~al.}(2023)\citenamefont {Unal},
  \citenamefont {Papageorgiou},\ and\ \citenamefont {Obata}}]{Unal:2023srk}%
  \BibitemOpen
  \bibfield  {author} {\bibinfo {author} {\bibfnamefont {C.}~\bibnamefont
  {Unal}}, \bibinfo {author} {\bibfnamefont {A.}~\bibnamefont {Papageorgiou}},\
  and\ \bibinfo {author} {\bibfnamefont {I.}~\bibnamefont {Obata}},\ }\bibfield
   {title} {\bibinfo {title} {{Axion-Gauge Dynamics During Inflation as the
  Origin of Pulsar Timing Array Signals and Primordial Black Holes}},\
  }\href@noop {} {\  (\bibinfo {year} {2023})},\ \Eprint
  {https://arxiv.org/abs/2307.02322} {arXiv:2307.02322 [astro-ph.CO]}
  \BibitemShut {NoStop}%
\bibitem [{\citenamefont {Figueroa}\ \emph {et~al.}(2023)\citenamefont
  {Figueroa}, \citenamefont {Pieroni}, \citenamefont {Ricciardone},\ and\
  \citenamefont {Simakachorn}}]{Figueroa:2023zhu}%
  \BibitemOpen
  \bibfield  {author} {\bibinfo {author} {\bibfnamefont {D.~G.}\ \bibnamefont
  {Figueroa}}, \bibinfo {author} {\bibfnamefont {M.}~\bibnamefont {Pieroni}},
  \bibinfo {author} {\bibfnamefont {A.}~\bibnamefont {Ricciardone}},\ and\
  \bibinfo {author} {\bibfnamefont {P.}~\bibnamefont {Simakachorn}},\
  }\bibfield  {title} {\bibinfo {title} {{Cosmological Background
  Interpretation of Pulsar Timing Array Data}},\ }\href@noop {} {\  (\bibinfo
  {year} {2023})},\ \Eprint {https://arxiv.org/abs/2307.02399}
  {arXiv:2307.02399 [astro-ph.CO]} \BibitemShut {NoStop}%
\bibitem [{\citenamefont {Yi}\ \emph {et~al.}(2023{\natexlab{a}})\citenamefont
  {Yi}, \citenamefont {Gao}, \citenamefont {Gong}, \citenamefont {Wang},\ and\
  \citenamefont {Zhang}}]{Yi:2023mbm}%
  \BibitemOpen
  \bibfield  {author} {\bibinfo {author} {\bibfnamefont {Z.}~\bibnamefont
  {Yi}}, \bibinfo {author} {\bibfnamefont {Q.}~\bibnamefont {Gao}}, \bibinfo
  {author} {\bibfnamefont {Y.}~\bibnamefont {Gong}}, \bibinfo {author}
  {\bibfnamefont {Y.}~\bibnamefont {Wang}},\ and\ \bibinfo {author}
  {\bibfnamefont {F.}~\bibnamefont {Zhang}},\ }\bibfield  {title} {\bibinfo
  {title} {{The waveform of the scalar induced gravitational waves in light of
  Pulsar Timing Array data}},\ }\href@noop {} {\  (\bibinfo {year}
  {2023}{\natexlab{a}})},\ \Eprint {https://arxiv.org/abs/2307.02467}
  {arXiv:2307.02467 [gr-qc]} \BibitemShut {NoStop}%
\bibitem [{\citenamefont {Zhu}\ \emph {et~al.}(2023)\citenamefont {Zhu},
  \citenamefont {Zhao},\ and\ \citenamefont {Wang}}]{Zhu:2023faa}%
  \BibitemOpen
  \bibfield  {author} {\bibinfo {author} {\bibfnamefont {Q.-H.}\ \bibnamefont
  {Zhu}}, \bibinfo {author} {\bibfnamefont {Z.-C.}\ \bibnamefont {Zhao}},\ and\
  \bibinfo {author} {\bibfnamefont {S.}~\bibnamefont {Wang}},\ }\bibfield
  {title} {\bibinfo {title} {{Joint implications of BBN, CMB, and PTA Datasets
  for Scalar-Induced Gravitational Waves of Second and Third orders}},\
  }\href@noop {} {\  (\bibinfo {year} {2023})},\ \Eprint
  {https://arxiv.org/abs/2307.03095} {arXiv:2307.03095 [astro-ph.CO]}
  \BibitemShut {NoStop}%
\bibitem [{\citenamefont {Firouzjahi}\ and\ \citenamefont
  {Talebian}(2023)}]{Firouzjahi:2023lzg}%
  \BibitemOpen
  \bibfield  {author} {\bibinfo {author} {\bibfnamefont {H.}~\bibnamefont
  {Firouzjahi}}\ and\ \bibinfo {author} {\bibfnamefont {A.}~\bibnamefont
  {Talebian}},\ }\bibfield  {title} {\bibinfo {title} {{Induced Gravitational
  Waves from Ultra Slow-Roll Inflation and Pulsar Timing Arrays
  Observations}},\ }\href@noop {} {\  (\bibinfo {year} {2023})},\ \Eprint
  {https://arxiv.org/abs/2307.03164} {arXiv:2307.03164 [gr-qc]} \BibitemShut
  {NoStop}%
\bibitem [{\citenamefont {Li}\ \emph {et~al.}(2023{\natexlab{a}})\citenamefont
  {Li}, \citenamefont {Wang}, \citenamefont {Zhao},\ and\ \citenamefont
  {Kohri}}]{Li:2023qua}%
  \BibitemOpen
  \bibfield  {author} {\bibinfo {author} {\bibfnamefont {J.-P.}\ \bibnamefont
  {Li}}, \bibinfo {author} {\bibfnamefont {S.}~\bibnamefont {Wang}}, \bibinfo
  {author} {\bibfnamefont {Z.-C.}\ \bibnamefont {Zhao}},\ and\ \bibinfo
  {author} {\bibfnamefont {K.}~\bibnamefont {Kohri}},\ }\bibfield  {title}
  {\bibinfo {title} {{Primordial Non-Gaussianity and Anisotropies in
  Gravitational Waves induced by Scalar Perturbations}},\ }\href@noop {} {\
  (\bibinfo {year} {2023}{\natexlab{a}})},\ \Eprint
  {https://arxiv.org/abs/2305.19950} {arXiv:2305.19950 [astro-ph.CO]}
  \BibitemShut {NoStop}%
\bibitem [{\citenamefont {You}\ \emph {et~al.}(2023)\citenamefont {You},
  \citenamefont {Yi},\ and\ \citenamefont {Wu}}]{You:2023rmn}%
  \BibitemOpen
  \bibfield  {author} {\bibinfo {author} {\bibfnamefont {Z.-Q.}\ \bibnamefont
  {You}}, \bibinfo {author} {\bibfnamefont {Z.}~\bibnamefont {Yi}},\ and\
  \bibinfo {author} {\bibfnamefont {Y.}~\bibnamefont {Wu}},\ }\bibfield
  {title} {\bibinfo {title} {{Constraints on primordial curvature power
  spectrum with pulsar timing arrays}},\ }\href@noop {} {\  (\bibinfo {year}
  {2023})},\ \Eprint {https://arxiv.org/abs/2307.04419} {arXiv:2307.04419
  [gr-qc]} \BibitemShut {NoStop}%
\bibitem [{\citenamefont {Balaji}\ \emph {et~al.}(2023)\citenamefont {Balaji},
  \citenamefont {Dom\`enech},\ and\ \citenamefont
  {Franciolini}}]{Balaji:2023ehk}%
  \BibitemOpen
  \bibfield  {author} {\bibinfo {author} {\bibfnamefont {S.}~\bibnamefont
  {Balaji}}, \bibinfo {author} {\bibfnamefont {G.}~\bibnamefont {Dom\`enech}},\
  and\ \bibinfo {author} {\bibfnamefont {G.}~\bibnamefont {Franciolini}},\
  }\bibfield  {title} {\bibinfo {title} {{Scalar-induced gravitational wave
  interpretation of PTA data: the role of scalar fluctuation propagation
  speed}},\ }\href@noop {} {\  (\bibinfo {year} {2023})},\ \Eprint
  {https://arxiv.org/abs/2307.08552} {arXiv:2307.08552 [gr-qc]} \BibitemShut
  {NoStop}%
\bibitem [{\citenamefont {Hosseini~Mansoori}\ \emph {et~al.}(2023)\citenamefont
  {Hosseini~Mansoori}, \citenamefont {Felegray}, \citenamefont {Talebian},\
  and\ \citenamefont {Sami}}]{HosseiniMansoori:2023mqh}%
  \BibitemOpen
  \bibfield  {author} {\bibinfo {author} {\bibfnamefont {S.~A.}\ \bibnamefont
  {Hosseini~Mansoori}}, \bibinfo {author} {\bibfnamefont {F.}~\bibnamefont
  {Felegray}}, \bibinfo {author} {\bibfnamefont {A.}~\bibnamefont {Talebian}},\
  and\ \bibinfo {author} {\bibfnamefont {M.}~\bibnamefont {Sami}},\ }\bibfield
  {title} {\bibinfo {title} {{PBHs and GWs from
  \ensuremath{\mathbb{T}}$^{2}$-inflation and NANOGrav 15-year data}},\ }\href
  {https://doi.org/10.1088/1475-7516/2023/08/067} {\bibfield  {journal}
  {\bibinfo  {journal} {JCAP}\ }\textbf {\bibinfo {volume} {08}},\ \bibinfo
  {pages} {067}},\ \Eprint {https://arxiv.org/abs/2307.06757} {arXiv:2307.06757
  [astro-ph.CO]} \BibitemShut {NoStop}%
\bibitem [{\citenamefont {Zhao}\ \emph {et~al.}(2023)\citenamefont {Zhao},
  \citenamefont {Zhu}, \citenamefont {Wang},\ and\ \citenamefont
  {Zhang}}]{Zhao:2023joc}%
  \BibitemOpen
  \bibfield  {author} {\bibinfo {author} {\bibfnamefont {Z.-C.}\ \bibnamefont
  {Zhao}}, \bibinfo {author} {\bibfnamefont {Q.-H.}\ \bibnamefont {Zhu}},
  \bibinfo {author} {\bibfnamefont {S.}~\bibnamefont {Wang}},\ and\ \bibinfo
  {author} {\bibfnamefont {X.}~\bibnamefont {Zhang}},\ }\bibfield  {title}
  {\bibinfo {title} {{Exploring the Equation of State of the Early Universe:
  Insights from BBN, CMB, and PTA Observations}},\ }\href@noop {} {\  (\bibinfo
  {year} {2023})},\ \Eprint {https://arxiv.org/abs/2307.13574}
  {arXiv:2307.13574 [astro-ph.CO]} \BibitemShut {NoStop}%
\bibitem [{\citenamefont {Liu}\ \emph {et~al.}(2023{\natexlab{b}})\citenamefont
  {Liu}, \citenamefont {Chen},\ and\ \citenamefont {Huang}}]{Liu:2023pau}%
  \BibitemOpen
  \bibfield  {author} {\bibinfo {author} {\bibfnamefont {L.}~\bibnamefont
  {Liu}}, \bibinfo {author} {\bibfnamefont {Z.-C.}\ \bibnamefont {Chen}},\ and\
  \bibinfo {author} {\bibfnamefont {Q.-G.}\ \bibnamefont {Huang}},\ }\bibfield
  {title} {\bibinfo {title} {{Probing the equation of state of the early
  Universe with pulsar timing arrays}},\ }\href@noop {} {\  (\bibinfo {year}
  {2023}{\natexlab{b}})},\ \Eprint {https://arxiv.org/abs/2307.14911}
  {arXiv:2307.14911 [astro-ph.CO]} \BibitemShut {NoStop}%
\bibitem [{\citenamefont {Yi}\ \emph {et~al.}(2023{\natexlab{b}})\citenamefont
  {Yi}, \citenamefont {You},\ and\ \citenamefont {Wu}}]{Yi:2023tdk}%
  \BibitemOpen
  \bibfield  {author} {\bibinfo {author} {\bibfnamefont {Z.}~\bibnamefont
  {Yi}}, \bibinfo {author} {\bibfnamefont {Z.-Q.}\ \bibnamefont {You}},\ and\
  \bibinfo {author} {\bibfnamefont {Y.}~\bibnamefont {Wu}},\ }\bibfield
  {title} {\bibinfo {title} {{Model-independent reconstruction of the
  primordial curvature power spectrum from PTA data}},\ }\href@noop {} {\
  (\bibinfo {year} {2023}{\natexlab{b}})},\ \Eprint
  {https://arxiv.org/abs/2308.05632} {arXiv:2308.05632 [astro-ph.CO]}
  \BibitemShut {NoStop}%
\bibitem [{\citenamefont {Bhaumik}\ \emph {et~al.}(2023)\citenamefont
  {Bhaumik}, \citenamefont {Jain},\ and\ \citenamefont
  {Lewicki}}]{Bhaumik:2023wmw}%
  \BibitemOpen
  \bibfield  {author} {\bibinfo {author} {\bibfnamefont {N.}~\bibnamefont
  {Bhaumik}}, \bibinfo {author} {\bibfnamefont {R.~K.}\ \bibnamefont {Jain}},\
  and\ \bibinfo {author} {\bibfnamefont {M.}~\bibnamefont {Lewicki}},\
  }\bibfield  {title} {\bibinfo {title} {{Ultra-low mass PBHs in the early
  universe can explain the PTA signal}},\ }\href@noop {} {\  (\bibinfo {year}
  {2023})},\ \Eprint {https://arxiv.org/abs/2308.07912} {arXiv:2308.07912
  [astro-ph.CO]} \BibitemShut {NoStop}%
\bibitem [{\citenamefont {Choudhury}\ \emph
  {et~al.}(2023{\natexlab{a}})\citenamefont {Choudhury}, \citenamefont {Karde},
  \citenamefont {Panda},\ and\ \citenamefont {Sami}}]{Choudhury:2023hfm}%
  \BibitemOpen
  \bibfield  {author} {\bibinfo {author} {\bibfnamefont {S.}~\bibnamefont
  {Choudhury}}, \bibinfo {author} {\bibfnamefont {A.}~\bibnamefont {Karde}},
  \bibinfo {author} {\bibfnamefont {S.}~\bibnamefont {Panda}},\ and\ \bibinfo
  {author} {\bibfnamefont {M.}~\bibnamefont {Sami}},\ }\bibfield  {title}
  {\bibinfo {title} {{Scalar induced gravity waves from ultra slow-roll
  Galileon inflation}},\ }\href@noop {} {\  (\bibinfo {year}
  {2023}{\natexlab{a}})},\ \Eprint {https://arxiv.org/abs/2308.09273}
  {arXiv:2308.09273 [astro-ph.CO]} \BibitemShut {NoStop}%
\bibitem [{\citenamefont {Yi}\ \emph {et~al.}(2023{\natexlab{c}})\citenamefont
  {Yi}, \citenamefont {You}, \citenamefont {Wu}, \citenamefont {Chen},\ and\
  \citenamefont {Liu}}]{Yi:2023npi}%
  \BibitemOpen
  \bibfield  {author} {\bibinfo {author} {\bibfnamefont {Z.}~\bibnamefont
  {Yi}}, \bibinfo {author} {\bibfnamefont {Z.-Q.}\ \bibnamefont {You}},
  \bibinfo {author} {\bibfnamefont {Y.}~\bibnamefont {Wu}}, \bibinfo {author}
  {\bibfnamefont {Z.-C.}\ \bibnamefont {Chen}},\ and\ \bibinfo {author}
  {\bibfnamefont {L.}~\bibnamefont {Liu}},\ }\bibfield  {title} {\bibinfo
  {title} {{Exploring the NANOGrav Signal and Planet-mass Primordial Black
  Holes through Higgs Inflation}},\ }\href@noop {} {\  (\bibinfo {year}
  {2023}{\natexlab{c}})},\ \Eprint {https://arxiv.org/abs/2308.14688}
  {arXiv:2308.14688 [astro-ph.CO]} \BibitemShut {NoStop}%
\bibitem [{\citenamefont {Harigaya}\ \emph {et~al.}(2023)\citenamefont
  {Harigaya}, \citenamefont {Inomata},\ and\ \citenamefont
  {Terada}}]{Harigaya:2023pmw}%
  \BibitemOpen
  \bibfield  {author} {\bibinfo {author} {\bibfnamefont {K.}~\bibnamefont
  {Harigaya}}, \bibinfo {author} {\bibfnamefont {K.}~\bibnamefont {Inomata}},\
  and\ \bibinfo {author} {\bibfnamefont {T.}~\bibnamefont {Terada}},\
  }\bibfield  {title} {\bibinfo {title} {{Induced Gravitational Waves with
  Kination Era for Recent Pulsar Timing Array Signals}},\ }\href@noop {} {\
  (\bibinfo {year} {2023})},\ \Eprint {https://arxiv.org/abs/2309.00228}
  {arXiv:2309.00228 [astro-ph.CO]} \BibitemShut {NoStop}%
\bibitem [{\citenamefont {Basilakos}\ \emph
  {et~al.}(2023{\natexlab{a}})\citenamefont {Basilakos}, \citenamefont
  {Nanopoulos}, \citenamefont {Papanikolaou}, \citenamefont {Saridakis},\ and\
  \citenamefont {Tzerefos}}]{Basilakos:2023xof}%
  \BibitemOpen
  \bibfield  {author} {\bibinfo {author} {\bibfnamefont {S.}~\bibnamefont
  {Basilakos}}, \bibinfo {author} {\bibfnamefont {D.~V.}\ \bibnamefont
  {Nanopoulos}}, \bibinfo {author} {\bibfnamefont {T.}~\bibnamefont
  {Papanikolaou}}, \bibinfo {author} {\bibfnamefont {E.~N.}\ \bibnamefont
  {Saridakis}},\ and\ \bibinfo {author} {\bibfnamefont {C.}~\bibnamefont
  {Tzerefos}},\ }\bibfield  {title} {\bibinfo {title} {{Gravitational wave
  signatures of no-scale Supergravity in NANOGrav and beyond}},\ }\href@noop {}
  {\  (\bibinfo {year} {2023}{\natexlab{a}})},\ \Eprint
  {https://arxiv.org/abs/2307.08601} {arXiv:2307.08601 [hep-th]} \BibitemShut
  {NoStop}%
\bibitem [{\citenamefont {Jin}\ \emph {et~al.}(2023)\citenamefont {Jin},
  \citenamefont {Chen}, \citenamefont {Yi}, \citenamefont {You}, \citenamefont
  {Liu},\ and\ \citenamefont {Wu}}]{Jin:2023wri}%
  \BibitemOpen
  \bibfield  {author} {\bibinfo {author} {\bibfnamefont {J.-H.}\ \bibnamefont
  {Jin}}, \bibinfo {author} {\bibfnamefont {Z.-C.}\ \bibnamefont {Chen}},
  \bibinfo {author} {\bibfnamefont {Z.}~\bibnamefont {Yi}}, \bibinfo {author}
  {\bibfnamefont {Z.-Q.}\ \bibnamefont {You}}, \bibinfo {author} {\bibfnamefont
  {L.}~\bibnamefont {Liu}},\ and\ \bibinfo {author} {\bibfnamefont
  {Y.}~\bibnamefont {Wu}},\ }\bibfield  {title} {\bibinfo {title} {{Confronting
  sound speed resonance with pulsar timing arrays}},\ }\href
  {https://doi.org/10.1088/1475-7516/2023/09/016} {\bibfield  {journal}
  {\bibinfo  {journal} {JCAP}\ }\textbf {\bibinfo {volume} {09}},\ \bibinfo
  {pages} {016}},\ \Eprint {https://arxiv.org/abs/2307.08687} {arXiv:2307.08687
  [astro-ph.CO]} \BibitemShut {NoStop}%
\bibitem [{\citenamefont {Cannizzaro}\ \emph {et~al.}(2023)\citenamefont
  {Cannizzaro}, \citenamefont {Franciolini},\ and\ \citenamefont
  {Pani}}]{Cannizzaro:2023mgc}%
  \BibitemOpen
  \bibfield  {author} {\bibinfo {author} {\bibfnamefont {E.}~\bibnamefont
  {Cannizzaro}}, \bibinfo {author} {\bibfnamefont {G.}~\bibnamefont
  {Franciolini}},\ and\ \bibinfo {author} {\bibfnamefont {P.}~\bibnamefont
  {Pani}},\ }\bibfield  {title} {\bibinfo {title} {{Novel tests of gravity
  using nano-Hertz stochastic gravitational-wave background signals}},\
  }\href@noop {} {\  (\bibinfo {year} {2023})},\ \Eprint
  {https://arxiv.org/abs/2307.11665} {arXiv:2307.11665 [gr-qc]} \BibitemShut
  {NoStop}%
\bibitem [{\citenamefont {Zhang}\ \emph {et~al.}(2023)\citenamefont {Zhang},
  \citenamefont {Cai}, \citenamefont {Su}, \citenamefont {Wang}, \citenamefont
  {Yu},\ and\ \citenamefont {Zhang}}]{Zhang:2023nrs}%
  \BibitemOpen
  \bibfield  {author} {\bibinfo {author} {\bibfnamefont {Z.}~\bibnamefont
  {Zhang}}, \bibinfo {author} {\bibfnamefont {C.}~\bibnamefont {Cai}}, \bibinfo
  {author} {\bibfnamefont {Y.-H.}\ \bibnamefont {Su}}, \bibinfo {author}
  {\bibfnamefont {S.}~\bibnamefont {Wang}}, \bibinfo {author} {\bibfnamefont
  {Z.-H.}\ \bibnamefont {Yu}},\ and\ \bibinfo {author} {\bibfnamefont {H.-H.}\
  \bibnamefont {Zhang}},\ }\bibfield  {title} {\bibinfo {title} {{Nano-Hertz
  gravitational waves from collapsing domain walls associated with freeze-in
  dark matter in light of pulsar timing array observations}},\ }\href
  {https://doi.org/10.1103/PhysRevD.108.095037} {\bibfield  {journal} {\bibinfo
   {journal} {Phys. Rev. D}\ }\textbf {\bibinfo {volume} {108}},\ \bibinfo
  {pages} {095037} (\bibinfo {year} {2023})},\ \Eprint
  {https://arxiv.org/abs/2307.11495} {arXiv:2307.11495 [hep-ph]} \BibitemShut
  {NoStop}%
\bibitem [{\citenamefont {Liu}\ \emph {et~al.}(2023{\natexlab{c}})\citenamefont
  {Liu}, \citenamefont {Wu},\ and\ \citenamefont {Chen}}]{Liu:2023hpw}%
  \BibitemOpen
  \bibfield  {author} {\bibinfo {author} {\bibfnamefont {L.}~\bibnamefont
  {Liu}}, \bibinfo {author} {\bibfnamefont {Y.}~\bibnamefont {Wu}},\ and\
  \bibinfo {author} {\bibfnamefont {Z.-C.}\ \bibnamefont {Chen}},\ }\bibfield
  {title} {\bibinfo {title} {{Simultaneously probing the sound speed and
  equation of state of the early Universe with pulsar timing arrays}},\
  }\href@noop {} {\  (\bibinfo {year} {2023}{\natexlab{c}})},\ \Eprint
  {https://arxiv.org/abs/2310.16500} {arXiv:2310.16500 [astro-ph.CO]}
  \BibitemShut {NoStop}%
\bibitem [{\citenamefont {Choudhury}\ \emph
  {et~al.}(2023{\natexlab{b}})\citenamefont {Choudhury}, \citenamefont {Dey},
  \citenamefont {Karde}, \citenamefont {Panda},\ and\ \citenamefont
  {Sami}}]{Choudhury:2023fwk}%
  \BibitemOpen
  \bibfield  {author} {\bibinfo {author} {\bibfnamefont {S.}~\bibnamefont
  {Choudhury}}, \bibinfo {author} {\bibfnamefont {K.}~\bibnamefont {Dey}},
  \bibinfo {author} {\bibfnamefont {A.}~\bibnamefont {Karde}}, \bibinfo
  {author} {\bibfnamefont {S.}~\bibnamefont {Panda}},\ and\ \bibinfo {author}
  {\bibfnamefont {M.}~\bibnamefont {Sami}},\ }\bibfield  {title} {\bibinfo
  {title} {{Primordial non-Gaussianity as a saviour for PBH overproduction in
  SIGWs generated by Pulsar Timing Arrays for Galileon inflation}},\
  }\href@noop {} {\  (\bibinfo {year} {2023}{\natexlab{b}})},\ \Eprint
  {https://arxiv.org/abs/2310.11034} {arXiv:2310.11034 [astro-ph.CO]}
  \BibitemShut {NoStop}%
\bibitem [{\citenamefont {Tagliazucchi}\ \emph {et~al.}(2023)\citenamefont
  {Tagliazucchi}, \citenamefont {Braglia}, \citenamefont {Finelli},\ and\
  \citenamefont {Pieroni}}]{Tagliazucchi:2023dai}%
  \BibitemOpen
  \bibfield  {author} {\bibinfo {author} {\bibfnamefont {M.}~\bibnamefont
  {Tagliazucchi}}, \bibinfo {author} {\bibfnamefont {M.}~\bibnamefont
  {Braglia}}, \bibinfo {author} {\bibfnamefont {F.}~\bibnamefont {Finelli}},\
  and\ \bibinfo {author} {\bibfnamefont {M.}~\bibnamefont {Pieroni}},\
  }\bibfield  {title} {\bibinfo {title} {{The quest of CMB spectral distortions
  to probe the scalar-induced gravitational wave background interpretation in
  PTA data}},\ }\href@noop {} {\  (\bibinfo {year} {2023})},\ \Eprint
  {https://arxiv.org/abs/2310.08527} {arXiv:2310.08527 [astro-ph.CO]}
  \BibitemShut {NoStop}%
\bibitem [{\citenamefont {Basilakos}\ \emph
  {et~al.}(2023{\natexlab{b}})\citenamefont {Basilakos}, \citenamefont
  {Nanopoulos}, \citenamefont {Papanikolaou}, \citenamefont {Saridakis},\ and\
  \citenamefont {Tzerefos}}]{Basilakos:2023jvp}%
  \BibitemOpen
  \bibfield  {author} {\bibinfo {author} {\bibfnamefont {S.}~\bibnamefont
  {Basilakos}}, \bibinfo {author} {\bibfnamefont {D.~V.}\ \bibnamefont
  {Nanopoulos}}, \bibinfo {author} {\bibfnamefont {T.}~\bibnamefont
  {Papanikolaou}}, \bibinfo {author} {\bibfnamefont {E.~N.}\ \bibnamefont
  {Saridakis}},\ and\ \bibinfo {author} {\bibfnamefont {C.}~\bibnamefont
  {Tzerefos}},\ }\bibfield  {title} {\bibinfo {title} {{Induced gravitational
  waves from flipped SU(5) superstring theory at $\mathrm{nHz}$}},\ }\href@noop
  {} {\  (\bibinfo {year} {2023}{\natexlab{b}})},\ \Eprint
  {https://arxiv.org/abs/2309.15820} {arXiv:2309.15820 [astro-ph.CO]}
  \BibitemShut {NoStop}%
\bibitem [{\citenamefont {Inomata}\ \emph
  {et~al.}(2023{\natexlab{b}})\citenamefont {Inomata}, \citenamefont
  {Kawasaki}, \citenamefont {Mukaida},\ and\ \citenamefont
  {Yanagida}}]{Inomata:2023drn}%
  \BibitemOpen
  \bibfield  {author} {\bibinfo {author} {\bibfnamefont {K.}~\bibnamefont
  {Inomata}}, \bibinfo {author} {\bibfnamefont {M.}~\bibnamefont {Kawasaki}},
  \bibinfo {author} {\bibfnamefont {K.}~\bibnamefont {Mukaida}},\ and\ \bibinfo
  {author} {\bibfnamefont {T.~T.}\ \bibnamefont {Yanagida}},\ }\bibfield
  {title} {\bibinfo {title} {{Axion Curvaton Model for the Gravitational Waves
  Observed by Pulsar Timing Arrays}},\ }\href@noop {} {\  (\bibinfo {year}
  {2023}{\natexlab{b}})},\ \Eprint {https://arxiv.org/abs/2309.11398}
  {arXiv:2309.11398 [astro-ph.CO]} \BibitemShut {NoStop}%
\bibitem [{\citenamefont {Li}\ \emph {et~al.}(2023{\natexlab{b}})\citenamefont
  {Li}, \citenamefont {Wang}, \citenamefont {Zhao},\ and\ \citenamefont
  {Kohri}}]{Li:2023xtl}%
  \BibitemOpen
  \bibfield  {author} {\bibinfo {author} {\bibfnamefont {J.-P.}\ \bibnamefont
  {Li}}, \bibinfo {author} {\bibfnamefont {S.}~\bibnamefont {Wang}}, \bibinfo
  {author} {\bibfnamefont {Z.-C.}\ \bibnamefont {Zhao}},\ and\ \bibinfo
  {author} {\bibfnamefont {K.}~\bibnamefont {Kohri}},\ }\bibfield  {title}
  {\bibinfo {title} {{Complete Analysis of Scalar-Induced Gravitational Waves
  and Primordial Non-Gaussianities $f_{\mathrm{NL}}$ and $g_{\mathrm{NL}}$}},\
  }\href@noop {} {\  (\bibinfo {year} {2023}{\natexlab{b}})},\ \Eprint
  {https://arxiv.org/abs/2309.07792} {arXiv:2309.07792 [astro-ph.CO]}
  \BibitemShut {NoStop}%
\bibitem [{\citenamefont {Dom\`enech}\ \emph {et~al.}(2023)\citenamefont
  {Dom\`enech}, \citenamefont {Vargas},\ and\ \citenamefont
  {Vargas}}]{Domenech:2023dxx}%
  \BibitemOpen
  \bibfield  {author} {\bibinfo {author} {\bibfnamefont {G.}~\bibnamefont
  {Dom\`enech}}, \bibinfo {author} {\bibfnamefont {G.}~\bibnamefont {Vargas}},\
  and\ \bibinfo {author} {\bibfnamefont {T.}~\bibnamefont {Vargas}},\
  }\bibfield  {title} {\bibinfo {title} {{An exact model for
  enhancing/suppressing primordial fluctuations}},\ }\href@noop {} {\
  (\bibinfo {year} {2023})},\ \Eprint {https://arxiv.org/abs/2309.05750}
  {arXiv:2309.05750 [astro-ph.CO]} \BibitemShut {NoStop}%
\bibitem [{\citenamefont {Gangopadhyay}\ \emph {et~al.}(2023)\citenamefont
  {Gangopadhyay}, \citenamefont {Godithi}, \citenamefont {Ichiki},
  \citenamefont {Inui}, \citenamefont {Kajino}, \citenamefont {Manusankar},
  \citenamefont {Mathews},\ and\ \citenamefont
  {Yogesh}}]{Gangopadhyay:2023qjr}%
  \BibitemOpen
  \bibfield  {author} {\bibinfo {author} {\bibfnamefont {M.~R.}\ \bibnamefont
  {Gangopadhyay}}, \bibinfo {author} {\bibfnamefont {V.~V.}\ \bibnamefont
  {Godithi}}, \bibinfo {author} {\bibfnamefont {K.}~\bibnamefont {Ichiki}},
  \bibinfo {author} {\bibfnamefont {R.}~\bibnamefont {Inui}}, \bibinfo {author}
  {\bibfnamefont {T.}~\bibnamefont {Kajino}}, \bibinfo {author} {\bibfnamefont
  {A.}~\bibnamefont {Manusankar}}, \bibinfo {author} {\bibfnamefont {G.~J.}\
  \bibnamefont {Mathews}},\ and\ \bibinfo {author} {\bibnamefont {Yogesh}},\
  }\bibfield  {title} {\bibinfo {title} {{Is the NANOGrav detection evidence of
  resonant particle creation during inflation?}},\ }\href@noop {} {\  (\bibinfo
  {year} {2023})},\ \Eprint {https://arxiv.org/abs/2309.03101}
  {arXiv:2309.03101 [astro-ph.CO]} \BibitemShut {NoStop}%
\bibitem [{\citenamefont {Cyr}\ \emph {et~al.}(2023)\citenamefont {Cyr},
  \citenamefont {Kite}, \citenamefont {Chluba}, \citenamefont {Hill},
  \citenamefont {Jeong}, \citenamefont {Acharya}, \citenamefont {Bolliet},\
  and\ \citenamefont {Patil}}]{Cyr:2023pgw}%
  \BibitemOpen
  \bibfield  {author} {\bibinfo {author} {\bibfnamefont {B.}~\bibnamefont
  {Cyr}}, \bibinfo {author} {\bibfnamefont {T.}~\bibnamefont {Kite}}, \bibinfo
  {author} {\bibfnamefont {J.}~\bibnamefont {Chluba}}, \bibinfo {author}
  {\bibfnamefont {J.~C.}\ \bibnamefont {Hill}}, \bibinfo {author}
  {\bibfnamefont {D.}~\bibnamefont {Jeong}}, \bibinfo {author} {\bibfnamefont
  {S.~K.}\ \bibnamefont {Acharya}}, \bibinfo {author} {\bibfnamefont
  {B.}~\bibnamefont {Bolliet}},\ and\ \bibinfo {author} {\bibfnamefont {S.~P.}\
  \bibnamefont {Patil}},\ }\bibfield  {title} {\bibinfo {title} {{Disentangling
  the primordial nature of stochastic gravitational wave backgrounds with CMB
  spectral distortions}},\ }\href@noop {} {\  (\bibinfo {year} {2023})},\
  \Eprint {https://arxiv.org/abs/2309.02366} {arXiv:2309.02366 [astro-ph.CO]}
  \BibitemShut {NoStop}%
\bibitem [{\citenamefont {Lozanov}\ \emph {et~al.}(2023)\citenamefont
  {Lozanov}, \citenamefont {Pi}, \citenamefont {Sasaki}, \citenamefont
  {Takhistov},\ and\ \citenamefont {Wang}}]{Lozanov:2023rcd}%
  \BibitemOpen
  \bibfield  {author} {\bibinfo {author} {\bibfnamefont {K.~D.}\ \bibnamefont
  {Lozanov}}, \bibinfo {author} {\bibfnamefont {S.}~\bibnamefont {Pi}},
  \bibinfo {author} {\bibfnamefont {M.}~\bibnamefont {Sasaki}}, \bibinfo
  {author} {\bibfnamefont {V.}~\bibnamefont {Takhistov}},\ and\ \bibinfo
  {author} {\bibfnamefont {A.}~\bibnamefont {Wang}},\ }\bibfield  {title}
  {\bibinfo {title} {{Axion Universal Gravitational Wave Interpretation of
  Pulsar Timing Array Data}},\ }\href@noop {} {\  (\bibinfo {year} {2023})},\
  \Eprint {https://arxiv.org/abs/2310.03594} {arXiv:2310.03594 [astro-ph.CO]}
  \BibitemShut {NoStop}%
\bibitem [{\citenamefont {Madge}\ \emph {et~al.}(2023)\citenamefont {Madge},
  \citenamefont {Morgante}, \citenamefont {Puchades-Ib\'a\~nez}, \citenamefont
  {Ramberg}, \citenamefont {Ratzinger}, \citenamefont {Schenk},\ and\
  \citenamefont {Schwaller}}]{Madge:2023dxc}%
  \BibitemOpen
  \bibfield  {author} {\bibinfo {author} {\bibfnamefont {E.}~\bibnamefont
  {Madge}}, \bibinfo {author} {\bibfnamefont {E.}~\bibnamefont {Morgante}},
  \bibinfo {author} {\bibfnamefont {C.}~\bibnamefont {Puchades-Ib\'a\~nez}},
  \bibinfo {author} {\bibfnamefont {N.}~\bibnamefont {Ramberg}}, \bibinfo
  {author} {\bibfnamefont {W.}~\bibnamefont {Ratzinger}}, \bibinfo {author}
  {\bibfnamefont {S.}~\bibnamefont {Schenk}},\ and\ \bibinfo {author}
  {\bibfnamefont {P.}~\bibnamefont {Schwaller}},\ }\bibfield  {title} {\bibinfo
  {title} {{Primordial gravitational waves in the nano-Hertz regime and PTA
  data \textemdash{} towards solving the GW inverse problem}},\ }\href
  {https://doi.org/10.1007/JHEP10(2023)171} {\bibfield  {journal} {\bibinfo
  {journal} {JHEP}\ }\textbf {\bibinfo {volume} {10}},\ \bibinfo {pages}
  {171}},\ \Eprint {https://arxiv.org/abs/2306.14856} {arXiv:2306.14856
  [hep-ph]} \BibitemShut {NoStop}%
\bibitem [{\citenamefont {Dom\`enech}\ \emph {et~al.}(2024)\citenamefont
  {Dom\`enech}, \citenamefont {Pi}, \citenamefont {Wang},\ and\ \citenamefont
  {Wang}}]{Domenech:2024rks}%
  \BibitemOpen
  \bibfield  {author} {\bibinfo {author} {\bibfnamefont {G.}~\bibnamefont
  {Dom\`enech}}, \bibinfo {author} {\bibfnamefont {S.}~\bibnamefont {Pi}},
  \bibinfo {author} {\bibfnamefont {A.}~\bibnamefont {Wang}},\ and\ \bibinfo
  {author} {\bibfnamefont {J.}~\bibnamefont {Wang}},\ }\bibfield  {title}
  {\bibinfo {title} {{Induced Gravitational Wave interpretation of PTA data: a
  complete study for general equation of state}},\ }\href@noop {} {\  (\bibinfo
  {year} {2024})},\ \Eprint {https://arxiv.org/abs/2402.18965}
  {arXiv:2402.18965 [astro-ph.CO]} \BibitemShut {NoStop}%
\bibitem [{\citenamefont {Antoniadis}\ \emph
  {et~al.}(2023{\natexlab{a}})\citenamefont {Antoniadis} \emph
  {et~al.}}]{EPTA:2023fyk}%
  \BibitemOpen
  \bibfield  {author} {\bibinfo {author} {\bibfnamefont {J.}~\bibnamefont
  {Antoniadis}} \emph {et~al.} (\bibinfo {collaboration} {EPTA, InPTA:}),\
  }\bibfield  {title} {\bibinfo {title} {{The second data release from the
  European Pulsar Timing Array - III. Search for gravitational wave signals}},\
  }\href {https://doi.org/10.1051/0004-6361/202346844} {\bibfield  {journal}
  {\bibinfo  {journal} {Astron. Astrophys.}\ }\textbf {\bibinfo {volume}
  {678}},\ \bibinfo {pages} {A50} (\bibinfo {year} {2023}{\natexlab{a}})},\
  \Eprint {https://arxiv.org/abs/2306.16214} {arXiv:2306.16214 [astro-ph.HE]}
  \BibitemShut {NoStop}%
\bibitem [{\citenamefont {Antoniadis}\ \emph
  {et~al.}(2023{\natexlab{b}})\citenamefont {Antoniadis} \emph
  {et~al.}}]{EPTA:2023sfo}%
  \BibitemOpen
  \bibfield  {author} {\bibinfo {author} {\bibfnamefont {J.}~\bibnamefont
  {Antoniadis}} \emph {et~al.} (\bibinfo {collaboration} {EPTA}),\ }\bibfield
  {title} {\bibinfo {title} {{The second data release from the European Pulsar
  Timing Array - I. The dataset and timing analysis}},\ }\href
  {https://doi.org/10.1051/0004-6361/202346841} {\bibfield  {journal} {\bibinfo
   {journal} {Astron. Astrophys.}\ }\textbf {\bibinfo {volume} {678}},\
  \bibinfo {pages} {A48} (\bibinfo {year} {2023}{\natexlab{b}})},\ \Eprint
  {https://arxiv.org/abs/2306.16224} {arXiv:2306.16224 [astro-ph.HE]}
  \BibitemShut {NoStop}%
\bibitem [{\citenamefont {Antoniadis}\ \emph
  {et~al.}(2023{\natexlab{c}})\citenamefont {Antoniadis} \emph
  {et~al.}}]{EPTA:2023xxk}%
  \BibitemOpen
  \bibfield  {author} {\bibinfo {author} {\bibfnamefont {J.}~\bibnamefont
  {Antoniadis}} \emph {et~al.} (\bibinfo {collaboration} {EPTA}),\ }\bibfield
  {title} {\bibinfo {title} {{The second data release from the European Pulsar
  Timing Array: V. Implications for massive black holes, dark matter and the
  early Universe}},\ }\href@noop {} {\  (\bibinfo {year}
  {2023}{\natexlab{c}})},\ \Eprint {https://arxiv.org/abs/2306.16227}
  {arXiv:2306.16227 [astro-ph.CO]} \BibitemShut {NoStop}%
\bibitem [{\citenamefont {Zic}\ \emph {et~al.}(2023)\citenamefont {Zic} \emph
  {et~al.}}]{Zic:2023gta}%
  \BibitemOpen
  \bibfield  {author} {\bibinfo {author} {\bibfnamefont {A.}~\bibnamefont
  {Zic}} \emph {et~al.},\ }\bibfield  {title} {\bibinfo {title} {{The Parkes
  Pulsar Timing Array Third Data Release}},\ }\href@noop {} {\  (\bibinfo
  {year} {2023})},\ \Eprint {https://arxiv.org/abs/2306.16230}
  {arXiv:2306.16230 [astro-ph.HE]} \BibitemShut {NoStop}%
\bibitem [{\citenamefont {Reardon}\ \emph
  {et~al.}(2023{\natexlab{a}})\citenamefont {Reardon} \emph
  {et~al.}}]{Reardon:2023gzh}%
  \BibitemOpen
  \bibfield  {author} {\bibinfo {author} {\bibfnamefont {D.~J.}\ \bibnamefont
  {Reardon}} \emph {et~al.},\ }\bibfield  {title} {\bibinfo {title} {{Search
  for an Isotropic Gravitational-wave Background with the Parkes Pulsar Timing
  Array}},\ }\href {https://doi.org/10.3847/2041-8213/acdd02} {\bibfield
  {journal} {\bibinfo  {journal} {Astrophys. J. Lett.}\ }\textbf {\bibinfo
  {volume} {951}},\ \bibinfo {pages} {L6} (\bibinfo {year}
  {2023}{\natexlab{a}})},\ \Eprint {https://arxiv.org/abs/2306.16215}
  {arXiv:2306.16215 [astro-ph.HE]} \BibitemShut {NoStop}%
\bibitem [{\citenamefont {Reardon}\ \emph
  {et~al.}(2023{\natexlab{b}})\citenamefont {Reardon} \emph
  {et~al.}}]{Reardon:2023zen}%
  \BibitemOpen
  \bibfield  {author} {\bibinfo {author} {\bibfnamefont {D.~J.}\ \bibnamefont
  {Reardon}} \emph {et~al.},\ }\bibfield  {title} {\bibinfo {title} {{The
  Gravitational-wave Background Null Hypothesis: Characterizing Noise in
  Millisecond Pulsar Arrival Times with the Parkes Pulsar Timing Array}},\
  }\href {https://doi.org/10.3847/2041-8213/acdd03} {\bibfield  {journal}
  {\bibinfo  {journal} {Astrophys. J. Lett.}\ }\textbf {\bibinfo {volume}
  {951}},\ \bibinfo {pages} {L7} (\bibinfo {year} {2023}{\natexlab{b}})},\
  \Eprint {https://arxiv.org/abs/2306.16229} {arXiv:2306.16229 [astro-ph.HE]}
  \BibitemShut {NoStop}%
\bibitem [{\citenamefont {Agazie}\ \emph
  {et~al.}(2023{\natexlab{a}})\citenamefont {Agazie} \emph
  {et~al.}}]{NANOGrav:2023hde}%
  \BibitemOpen
  \bibfield  {author} {\bibinfo {author} {\bibfnamefont {G.}~\bibnamefont
  {Agazie}} \emph {et~al.} (\bibinfo {collaboration} {NANOGrav}),\ }\bibfield
  {title} {\bibinfo {title} {{The NANOGrav 15 yr Data Set: Observations and
  Timing of 68 Millisecond Pulsars}},\ }\href
  {https://doi.org/10.3847/2041-8213/acda9a} {\bibfield  {journal} {\bibinfo
  {journal} {Astrophys. J. Lett.}\ }\textbf {\bibinfo {volume} {951}},\
  \bibinfo {pages} {L9} (\bibinfo {year} {2023}{\natexlab{a}})},\ \Eprint
  {https://arxiv.org/abs/2306.16217} {arXiv:2306.16217 [astro-ph.HE]}
  \BibitemShut {NoStop}%
\bibitem [{\citenamefont {Agazie}\ \emph
  {et~al.}(2023{\natexlab{b}})\citenamefont {Agazie} \emph
  {et~al.}}]{NANOGrav:2023gor}%
  \BibitemOpen
  \bibfield  {author} {\bibinfo {author} {\bibfnamefont {G.}~\bibnamefont
  {Agazie}} \emph {et~al.} (\bibinfo {collaboration} {NANOGrav}),\ }\bibfield
  {title} {\bibinfo {title} {{The NANOGrav 15 yr Data Set: Evidence for a
  Gravitational-wave Background}},\ }\href
  {https://doi.org/10.3847/2041-8213/acdac6} {\bibfield  {journal} {\bibinfo
  {journal} {Astrophys. J. Lett.}\ }\textbf {\bibinfo {volume} {951}},\
  \bibinfo {pages} {L8} (\bibinfo {year} {2023}{\natexlab{b}})},\ \Eprint
  {https://arxiv.org/abs/2306.16213} {arXiv:2306.16213 [astro-ph.HE]}
  \BibitemShut {NoStop}%
\bibitem [{\citenamefont {Afzal}\ \emph {et~al.}(2023)\citenamefont {Afzal}
  \emph {et~al.}}]{NANOGrav:2023hvm}%
  \BibitemOpen
  \bibfield  {author} {\bibinfo {author} {\bibfnamefont {A.}~\bibnamefont
  {Afzal}} \emph {et~al.} (\bibinfo {collaboration} {NANOGrav}),\ }\bibfield
  {title} {\bibinfo {title} {{The NANOGrav 15 yr Data Set: Search for Signals
  from New Physics}},\ }\href {https://doi.org/10.3847/2041-8213/acdc91}
  {\bibfield  {journal} {\bibinfo  {journal} {Astrophys. J. Lett.}\ }\textbf
  {\bibinfo {volume} {951}},\ \bibinfo {pages} {L11} (\bibinfo {year}
  {2023})},\ \Eprint {https://arxiv.org/abs/2306.16219} {arXiv:2306.16219
  [astro-ph.HE]} \BibitemShut {NoStop}%
\bibitem [{\citenamefont {Agazie}\ \emph
  {et~al.}(2023{\natexlab{c}})\citenamefont {Agazie} \emph
  {et~al.}}]{InternationalPulsarTimingArray:2023mzf}%
  \BibitemOpen
  \bibfield  {author} {\bibinfo {author} {\bibfnamefont {G.}~\bibnamefont
  {Agazie}} \emph {et~al.} (\bibinfo {collaboration} {International Pulsar
  Timing Array}),\ }\bibfield  {title} {\bibinfo {title} {{Comparing recent PTA
  results on the nanohertz stochastic gravitational wave background}},\
  }\href@noop {} {\  (\bibinfo {year} {2023}{\natexlab{c}})},\ \Eprint
  {https://arxiv.org/abs/2309.00693} {arXiv:2309.00693 [astro-ph.HE]}
  \BibitemShut {NoStop}%
\bibitem [{\citenamefont {Xu}\ \emph {et~al.}(2023)\citenamefont {Xu} \emph
  {et~al.}}]{Xu:2023wog}%
  \BibitemOpen
  \bibfield  {author} {\bibinfo {author} {\bibfnamefont {H.}~\bibnamefont {Xu}}
  \emph {et~al.},\ }\bibfield  {title} {\bibinfo {title} {{Searching for the
  Nano-Hertz Stochastic Gravitational Wave Background with the Chinese Pulsar
  Timing Array Data Release I}},\ }\href
  {https://doi.org/10.1088/1674-4527/acdfa5} {\bibfield  {journal} {\bibinfo
  {journal} {Res. Astron. Astrophys.}\ }\textbf {\bibinfo {volume} {23}},\
  \bibinfo {pages} {075024} (\bibinfo {year} {2023})},\ \Eprint
  {https://arxiv.org/abs/2306.16216} {arXiv:2306.16216 [astro-ph.HE]}
  \BibitemShut {NoStop}%
\bibitem [{\citenamefont {Dom\`enech}(2021)}]{Domenech:2021ztg}%
  \BibitemOpen
  \bibfield  {author} {\bibinfo {author} {\bibfnamefont {G.}~\bibnamefont
  {Dom\`enech}},\ }\bibfield  {title} {\bibinfo {title} {{Scalar Induced
  Gravitational Waves Review}},\ }\href
  {https://doi.org/10.3390/universe7110398} {\bibfield  {journal} {\bibinfo
  {journal} {Universe}\ }\textbf {\bibinfo {volume} {7}},\ \bibinfo {pages}
  {398} (\bibinfo {year} {2021})},\ \Eprint {https://arxiv.org/abs/2109.01398}
  {arXiv:2109.01398 [gr-qc]} \BibitemShut {NoStop}%
\bibitem [{\citenamefont {Dom\`enech}(2024)}]{Domenech:2024kmh}%
  \BibitemOpen
  \bibfield  {author} {\bibinfo {author} {\bibfnamefont {G.}~\bibnamefont
  {Dom\`enech}},\ }\bibfield  {title} {\bibinfo {title} {{GW Backgrounds
  associated with PBHs}},\ }\href@noop {} {\  (\bibinfo {year} {2024})},\
  \Eprint {https://arxiv.org/abs/2402.17388} {arXiv:2402.17388 [gr-qc]}
  \BibitemShut {NoStop}%
\bibitem [{\citenamefont {Yuan}\ and\ \citenamefont
  {Huang}(2021)}]{Yuan:2021qgz}%
  \BibitemOpen
  \bibfield  {author} {\bibinfo {author} {\bibfnamefont {C.}~\bibnamefont
  {Yuan}}\ and\ \bibinfo {author} {\bibfnamefont {Q.-G.}\ \bibnamefont
  {Huang}},\ }\bibfield  {title} {\bibinfo {title} {{A topic review on probing
  primordial black hole dark matter with scalar induced gravitational waves}},\
  }\href@noop {} {\  (\bibinfo {year} {2021})},\ \Eprint
  {https://arxiv.org/abs/2103.04739} {arXiv:2103.04739 [astro-ph.GA]}
  \BibitemShut {NoStop}%
\bibitem [{\citenamefont {Bagui}\ \emph {et~al.}(2023)\citenamefont {Bagui}
  \emph {et~al.}}]{LISACosmologyWorkingGroup:2023njw}%
  \BibitemOpen
  \bibfield  {author} {\bibinfo {author} {\bibfnamefont {E.}~\bibnamefont
  {Bagui}} \emph {et~al.} (\bibinfo {collaboration} {LISA Cosmology Working
  Group}),\ }\bibfield  {title} {\bibinfo {title} {{Primordial black holes and
  their gravitational-wave signatures}},\ }\href@noop {} {\  (\bibinfo {year}
  {2023})},\ \Eprint {https://arxiv.org/abs/2310.19857} {arXiv:2310.19857
  [astro-ph.CO]} \BibitemShut {NoStop}%
\bibitem [{\citenamefont {Dom\`enech}(2023{\natexlab{a}})}]{Domenech:2023jve}%
  \BibitemOpen
  \bibfield  {author} {\bibinfo {author} {\bibfnamefont {G.}~\bibnamefont
  {Dom\`enech}},\ }\bibfield  {title} {\bibinfo {title} {{Cosmological
  Gravitational Waves from Isocurvature Fluctuations}},\ }\href@noop {} {\
  (\bibinfo {year} {2023}{\natexlab{a}})},\ \Eprint
  {https://arxiv.org/abs/2311.02065} {arXiv:2311.02065 [gr-qc]} \BibitemShut
  {NoStop}%
\bibitem [{\citenamefont {Dom\`enech}\ and\ \citenamefont
  {Sasaki}(2024)}]{Domenech:2024cjn}%
  \BibitemOpen
  \bibfield  {author} {\bibinfo {author} {\bibfnamefont {G.}~\bibnamefont
  {Dom\`enech}}\ and\ \bibinfo {author} {\bibfnamefont {M.}~\bibnamefont
  {Sasaki}},\ }\bibfield  {title} {\bibinfo {title} {{Probing Primordial Black
  Hole Scenarios with Terrestrial Gravitational Wave Detectors}},\ }\href@noop
  {} {\  (\bibinfo {year} {2024})},\ \Eprint {https://arxiv.org/abs/2401.07615}
  {arXiv:2401.07615 [gr-qc]} \BibitemShut {NoStop}%
\bibitem [{\citenamefont {Abbott}\ \emph {et~al.}(2017)\citenamefont {Abbott}
  \emph {et~al.}}]{LIGOScientific:2017zic}%
  \BibitemOpen
  \bibfield  {author} {\bibinfo {author} {\bibfnamefont {B.~P.}\ \bibnamefont
  {Abbott}} \emph {et~al.} (\bibinfo {collaboration} {LIGO Scientific, Virgo,
  Fermi-GBM, INTEGRAL}),\ }\bibfield  {title} {\bibinfo {title} {{Gravitational
  Waves and Gamma-rays from a Binary Neutron Star Merger: GW170817 and GRB
  170817A}},\ }\href {https://doi.org/10.3847/2041-8213/aa920c} {\bibfield
  {journal} {\bibinfo  {journal} {Astrophys. J. Lett.}\ }\textbf {\bibinfo
  {volume} {848}},\ \bibinfo {pages} {L13} (\bibinfo {year} {2017})},\ \Eprint
  {https://arxiv.org/abs/1710.05834} {arXiv:1710.05834 [astro-ph.HE]}
  \BibitemShut {NoStop}%
\bibitem [{\citenamefont {Ezquiaga}\ and\ \citenamefont
  {Zumalac\'arregui}(2017)}]{Ezquiaga:2017ekz}%
  \BibitemOpen
  \bibfield  {author} {\bibinfo {author} {\bibfnamefont {J.~M.}\ \bibnamefont
  {Ezquiaga}}\ and\ \bibinfo {author} {\bibfnamefont {M.}~\bibnamefont
  {Zumalac\'arregui}},\ }\bibfield  {title} {\bibinfo {title} {{Dark Energy
  After GW170817: Dead Ends and the Road Ahead}},\ }\href
  {https://doi.org/10.1103/PhysRevLett.119.251304} {\bibfield  {journal}
  {\bibinfo  {journal} {Phys. Rev. Lett.}\ }\textbf {\bibinfo {volume} {119}},\
  \bibinfo {pages} {251304} (\bibinfo {year} {2017})},\ \Eprint
  {https://arxiv.org/abs/1710.05901} {arXiv:1710.05901 [astro-ph.CO]}
  \BibitemShut {NoStop}%
\bibitem [{\citenamefont {Bellini}\ \emph {et~al.}(2015)\citenamefont
  {Bellini}, \citenamefont {Jimenez},\ and\ \citenamefont
  {Verde}}]{Bellini:2015wfa}%
  \BibitemOpen
  \bibfield  {author} {\bibinfo {author} {\bibfnamefont {E.}~\bibnamefont
  {Bellini}}, \bibinfo {author} {\bibfnamefont {R.}~\bibnamefont {Jimenez}},\
  and\ \bibinfo {author} {\bibfnamefont {L.}~\bibnamefont {Verde}},\ }\bibfield
   {title} {\bibinfo {title} {{Signatures of Horndeski gravity on the Dark
  Matter Bispectrum}},\ }\href {https://doi.org/10.1088/1475-7516/2015/05/057}
  {\bibfield  {journal} {\bibinfo  {journal} {JCAP}\ }\textbf {\bibinfo
  {volume} {05}},\ \bibinfo {pages} {057}},\ \Eprint
  {https://arxiv.org/abs/1504.04341} {arXiv:1504.04341 [astro-ph.CO]}
  \BibitemShut {NoStop}%
\bibitem [{\citenamefont {Thrane}\ and\ \citenamefont
  {Romano}(2013)}]{Thrane:2013oya}%
  \BibitemOpen
  \bibfield  {author} {\bibinfo {author} {\bibfnamefont {E.}~\bibnamefont
  {Thrane}}\ and\ \bibinfo {author} {\bibfnamefont {J.~D.}\ \bibnamefont
  {Romano}},\ }\bibfield  {title} {\bibinfo {title} {{Sensitivity curves for
  searches for gravitational-wave backgrounds}},\ }\href
  {https://doi.org/10.1103/PhysRevD.88.124032} {\bibfield  {journal} {\bibinfo
  {journal} {Phys. Rev.}\ }\textbf {\bibinfo {volume} {D88}},\ \bibinfo {pages}
  {124032} (\bibinfo {year} {2013})},\ \Eprint
  {https://arxiv.org/abs/1310.5300} {arXiv:1310.5300 [astro-ph.IM]}
  \BibitemShut {NoStop}%
\bibitem [{\citenamefont {Schmitz}(2021)}]{Schmitz:2020syl}%
  \BibitemOpen
  \bibfield  {author} {\bibinfo {author} {\bibfnamefont {K.}~\bibnamefont
  {Schmitz}},\ }\bibfield  {title} {\bibinfo {title} {{New Sensitivity Curves
  for Gravitational-Wave Signals from Cosmological Phase Transitions}},\ }\href
  {https://doi.org/10.1007/JHEP01(2021)097} {\bibfield  {journal} {\bibinfo
  {journal} {JHEP}\ }\textbf {\bibinfo {volume} {01}},\ \bibinfo {pages}
  {097}},\ \Eprint {https://arxiv.org/abs/2002.04615} {arXiv:2002.04615
  [hep-ph]} \BibitemShut {NoStop}%
\bibitem [{A+()}]{A+}%
  \BibitemOpen
  \href@noop {} {\bibinfo {title} {The {A+} design curve}},\ \bibinfo
  {howpublished} {\url{https://dcc.ligo.org/LIGO-T1800042/public}},\ \bibinfo
  {note} {[Online; accessed 05-May-2023]}\BibitemShut {NoStop}%
\bibitem [{\citenamefont {Branchesi}\ \emph {et~al.}(2023)\citenamefont
  {Branchesi} \emph {et~al.}}]{Branchesi:2023mws}%
  \BibitemOpen
  \bibfield  {author} {\bibinfo {author} {\bibfnamefont {M.}~\bibnamefont
  {Branchesi}} \emph {et~al.},\ }\bibfield  {title} {\bibinfo {title} {{Science
  with the Einstein Telescope: a comparison of different designs}},\ }\href
  {https://doi.org/10.1088/1475-7516/2023/07/068} {\bibfield  {journal}
  {\bibinfo  {journal} {JCAP}\ }\textbf {\bibinfo {volume} {07}},\ \bibinfo
  {pages} {068}},\ \Eprint {https://arxiv.org/abs/2303.15923} {arXiv:2303.15923
  [gr-qc]} \BibitemShut {NoStop}%
\bibitem [{ce()}]{ce}%
  \BibitemOpen
  \href@noop {} {\bibinfo {title} {Cosmic explorer sensitivity curve}},\
  \bibinfo {howpublished} {\url{https://cosmicexplorer.org/sensitivity.html}},\
  \bibinfo {note} {[Online; accessed 05-May-2023]}\BibitemShut {NoStop}%
\bibitem [{\citenamefont {Yagi}\ and\ \citenamefont
  {Seto}(2011)}]{Yagi:2011wg}%
  \BibitemOpen
  \bibfield  {author} {\bibinfo {author} {\bibfnamefont {K.}~\bibnamefont
  {Yagi}}\ and\ \bibinfo {author} {\bibfnamefont {N.}~\bibnamefont {Seto}},\
  }\bibfield  {title} {\bibinfo {title} {{Detector configuration of DECIGO/BBO
  and identification of cosmological neutron-star binaries}},\ }\href
  {https://doi.org/10.1103/PhysRevD.95.109901, 10.1103/PhysRevD.83.044011}
  {\bibfield  {journal} {\bibinfo  {journal} {Phys. Rev.}\ }\textbf {\bibinfo
  {volume} {D83}},\ \bibinfo {pages} {044011} (\bibinfo {year} {2011})},\
  \bibinfo {note} {[Erratum: Phys. Rev.D95,no.10,109901(2017)]},\ \Eprint
  {https://arxiv.org/abs/1101.3940} {arXiv:1101.3940 [astro-ph.CO]}
  \BibitemShut {NoStop}%
\bibitem [{\citenamefont {Kawamura}\ \emph {et~al.}(2020)\citenamefont
  {Kawamura} \emph {et~al.}}]{Kawamura:2020pcg}%
  \BibitemOpen
  \bibfield  {author} {\bibinfo {author} {\bibfnamefont {S.}~\bibnamefont
  {Kawamura}} \emph {et~al.},\ }\bibfield  {title} {\bibinfo {title} {{Current
  status of space gravitational wave antenna DECIGO and B-DECIGO}},\
  }\href@noop {} {\  (\bibinfo {year} {2020})},\ \Eprint
  {https://arxiv.org/abs/2006.13545} {arXiv:2006.13545 [gr-qc]} \BibitemShut
  {NoStop}%
\bibitem [{\citenamefont {Barke}\ \emph {et~al.}(2015)\citenamefont {Barke},
  \citenamefont {Wang}, \citenamefont {Esteban~Delgado}, \citenamefont
  {Tr\"obs}, \citenamefont {Heinzel},\ and\ \citenamefont
  {Danzmann}}]{Barke:2014lsa}%
  \BibitemOpen
  \bibfield  {author} {\bibinfo {author} {\bibfnamefont {S.}~\bibnamefont
  {Barke}}, \bibinfo {author} {\bibfnamefont {Y.}~\bibnamefont {Wang}},
  \bibinfo {author} {\bibfnamefont {J.~J.}\ \bibnamefont {Esteban~Delgado}},
  \bibinfo {author} {\bibfnamefont {M.}~\bibnamefont {Tr\"obs}}, \bibinfo
  {author} {\bibfnamefont {G.}~\bibnamefont {Heinzel}},\ and\ \bibinfo {author}
  {\bibfnamefont {K.}~\bibnamefont {Danzmann}},\ }\bibfield  {title} {\bibinfo
  {title} {{Towards a gravitational wave observatory designer: sensitivity
  limits of spaceborne detectors}},\ }\href
  {https://doi.org/10.1088/0264-9381/32/9/095004} {\bibfield  {journal}
  {\bibinfo  {journal} {Class. Quant. Grav.}\ }\textbf {\bibinfo {volume}
  {32}},\ \bibinfo {pages} {095004} (\bibinfo {year} {2015})},\ \Eprint
  {https://arxiv.org/abs/1411.1260} {arXiv:1411.1260 [physics.ins-det]}
  \BibitemShut {NoStop}%
\bibitem [{\citenamefont {Abbott}\ \emph {et~al.}(2021)\citenamefont {Abbott}
  \emph {et~al.}}]{KAGRA:2021kbb}%
  \BibitemOpen
  \bibfield  {author} {\bibinfo {author} {\bibfnamefont {R.}~\bibnamefont
  {Abbott}} \emph {et~al.} (\bibinfo {collaboration} {KAGRA, Virgo, LIGO
  Scientific}),\ }\bibfield  {title} {\bibinfo {title} {{Upper limits on the
  isotropic gravitational-wave background from Advanced LIGO and Advanced
  Virgo\textquoteright{}s third observing run}},\ }\href
  {https://doi.org/10.1103/PhysRevD.104.022004} {\bibfield  {journal} {\bibinfo
   {journal} {Phys. Rev. D}\ }\textbf {\bibinfo {volume} {104}},\ \bibinfo
  {pages} {022004} (\bibinfo {year} {2021})},\ \Eprint
  {https://arxiv.org/abs/2101.12130} {arXiv:2101.12130 [gr-qc]} \BibitemShut
  {NoStop}%
\bibitem [{\citenamefont {Sesana}\ \emph {et~al.}(2021)\citenamefont {Sesana}
  \emph {et~al.}}]{Sesana:2019vho}%
  \BibitemOpen
  \bibfield  {author} {\bibinfo {author} {\bibfnamefont {A.}~\bibnamefont
  {Sesana}} \emph {et~al.},\ }\bibfield  {title} {\bibinfo {title} {{Unveiling
  the gravitational universe at $\mu$-Hz frequencies}},\ }\href
  {https://doi.org/10.1007/s10686-021-09709-9} {\bibfield  {journal} {\bibinfo
  {journal} {Exper. Astron.}\ }\textbf {\bibinfo {volume} {51}},\ \bibinfo
  {pages} {1333} (\bibinfo {year} {2021})},\ \Eprint
  {https://arxiv.org/abs/1908.11391} {arXiv:1908.11391 [astro-ph.IM]}
  \BibitemShut {NoStop}%
\bibitem [{\citenamefont {Ruan}\ \emph {et~al.}(2020)\citenamefont {Ruan},
  \citenamefont {Guo}, \citenamefont {Cai},\ and\ \citenamefont
  {Zhang}}]{Ruan:2018tsw}%
  \BibitemOpen
  \bibfield  {author} {\bibinfo {author} {\bibfnamefont {W.-H.}\ \bibnamefont
  {Ruan}}, \bibinfo {author} {\bibfnamefont {Z.-K.}\ \bibnamefont {Guo}},
  \bibinfo {author} {\bibfnamefont {R.-G.}\ \bibnamefont {Cai}},\ and\ \bibinfo
  {author} {\bibfnamefont {Y.-Z.}\ \bibnamefont {Zhang}},\ }\bibfield  {title}
  {\bibinfo {title} {{Taiji program: Gravitational-wave sources}},\ }\href
  {https://doi.org/10.1142/S0217751X2050075X} {\bibfield  {journal} {\bibinfo
  {journal} {Int. J. Mod. Phys. A}\ }\textbf {\bibinfo {volume} {35}},\
  \bibinfo {pages} {2050075} (\bibinfo {year} {2020})},\ \Eprint
  {https://arxiv.org/abs/1807.09495} {arXiv:1807.09495 [gr-qc]} \BibitemShut
  {NoStop}%
\bibitem [{\citenamefont {Gong}\ \emph {et~al.}(2021)\citenamefont {Gong},
  \citenamefont {Luo},\ and\ \citenamefont {Wang}}]{Gong:2021gvw}%
  \BibitemOpen
  \bibfield  {author} {\bibinfo {author} {\bibfnamefont {Y.}~\bibnamefont
  {Gong}}, \bibinfo {author} {\bibfnamefont {J.}~\bibnamefont {Luo}},\ and\
  \bibinfo {author} {\bibfnamefont {B.}~\bibnamefont {Wang}},\ }\bibfield
  {title} {\bibinfo {title} {{Concepts and status of Chinese space
  gravitational wave detection projects}},\ }\href
  {https://doi.org/10.1038/s41550-021-01480-3} {\bibfield  {journal} {\bibinfo
  {journal} {Nature Astron.}\ }\textbf {\bibinfo {volume} {5}},\ \bibinfo
  {pages} {881} (\bibinfo {year} {2021})},\ \Eprint
  {https://arxiv.org/abs/2109.07442} {arXiv:2109.07442 [astro-ph.IM]}
  \BibitemShut {NoStop}%
\bibitem [{\citenamefont {Maggiore}\ \emph {et~al.}(2020)\citenamefont
  {Maggiore} \emph {et~al.}}]{Maggiore:2019uih}%
  \BibitemOpen
  \bibfield  {author} {\bibinfo {author} {\bibfnamefont {M.}~\bibnamefont
  {Maggiore}} \emph {et~al.},\ }\bibfield  {title} {\bibinfo {title} {{Science
  Case for the Einstein Telescope}},\ }\href
  {https://doi.org/10.1088/1475-7516/2020/03/050} {\bibfield  {journal}
  {\bibinfo  {journal} {JCAP}\ }\textbf {\bibinfo {volume} {03}},\ \bibinfo
  {pages} {050}},\ \Eprint {https://arxiv.org/abs/1912.02622} {arXiv:1912.02622
  [astro-ph.CO]} \BibitemShut {NoStop}%
\bibitem [{voy()}]{voyager}%
  \BibitemOpen
  \href@noop {} {\bibinfo {title} {Ligo unofficial sensitivity curves}},\
  \bibinfo {howpublished} {\url{https://dcc.ligo.org/LIGO-T1500293/public}},\
  \bibinfo {note} {[Online; accessed 05-May-2023]}\BibitemShut {NoStop}%
\bibitem [{\citenamefont {\"Ozsoy}\ and\ \citenamefont
  {Tasinato}(2023)}]{Ozsoy:2023ryl}%
  \BibitemOpen
  \bibfield  {author} {\bibinfo {author} {\bibfnamefont {O.}~\bibnamefont
  {\"Ozsoy}}\ and\ \bibinfo {author} {\bibfnamefont {G.}~\bibnamefont
  {Tasinato}},\ }\bibfield  {title} {\bibinfo {title} {{Inflation and
  Primordial Black Holes}},\ }\href {https://doi.org/10.3390/universe9050203}
  {\bibfield  {journal} {\bibinfo  {journal} {Universe}\ }\textbf {\bibinfo
  {volume} {9}},\ \bibinfo {pages} {203} (\bibinfo {year} {2023})},\ \Eprint
  {https://arxiv.org/abs/2301.03600} {arXiv:2301.03600 [astro-ph.CO]}
  \BibitemShut {NoStop}%
\bibitem [{\citenamefont {Kristiano}\ and\ \citenamefont
  {Yokoyama}(2024)}]{Kristiano:2024ngc}%
  \BibitemOpen
  \bibfield  {author} {\bibinfo {author} {\bibfnamefont {J.}~\bibnamefont
  {Kristiano}}\ and\ \bibinfo {author} {\bibfnamefont {J.}~\bibnamefont
  {Yokoyama}},\ }\bibfield  {title} {\bibinfo {title} {{Generating large
  primordial fluctuations in single-field inflation for PBH formation}},\
  }\href@noop {} {\  (\bibinfo {year} {2024})},\ \Eprint
  {https://arxiv.org/abs/2405.12149} {arXiv:2405.12149 [astro-ph.CO]}
  \BibitemShut {NoStop}%
\bibitem [{\citenamefont {Pujolas}\ \emph {et~al.}(2011)\citenamefont
  {Pujolas}, \citenamefont {Sawicki},\ and\ \citenamefont
  {Vikman}}]{Pujolas:2011he}%
  \BibitemOpen
  \bibfield  {author} {\bibinfo {author} {\bibfnamefont {O.}~\bibnamefont
  {Pujolas}}, \bibinfo {author} {\bibfnamefont {I.}~\bibnamefont {Sawicki}},\
  and\ \bibinfo {author} {\bibfnamefont {A.}~\bibnamefont {Vikman}},\
  }\bibfield  {title} {\bibinfo {title} {{The Imperfect Fluid behind Kinetic
  Gravity Braiding}},\ }\href {https://doi.org/10.1007/JHEP11(2011)156}
  {\bibfield  {journal} {\bibinfo  {journal} {JHEP}\ }\textbf {\bibinfo
  {volume} {11}},\ \bibinfo {pages} {156}},\ \Eprint
  {https://arxiv.org/abs/1103.5360} {arXiv:1103.5360 [hep-th]} \BibitemShut
  {NoStop}%
\bibitem [{\citenamefont {Hu}\ and\ \citenamefont {Gao}(2024)}]{Hu:2024hzo}%
  \BibitemOpen
  \bibfield  {author} {\bibinfo {author} {\bibfnamefont {Y.-M.}\ \bibnamefont
  {Hu}}\ and\ \bibinfo {author} {\bibfnamefont {X.}~\bibnamefont {Gao}},\
  }\bibfield  {title} {\bibinfo {title} {{Parity-violating scalar-tensor theory
  and the Qi-Xiu}},\ }\href@noop {} {\  (\bibinfo {year} {2024})},\ \Eprint
  {https://arxiv.org/abs/2405.20158} {arXiv:2405.20158 [hep-th]} \BibitemShut
  {NoStop}%
\bibitem [{\citenamefont {Feng}\ \emph {et~al.}(2024)\citenamefont {Feng},
  \citenamefont {Zhang},\ and\ \citenamefont {Gao}}]{Feng:2024yic}%
  \BibitemOpen
  \bibfield  {author} {\bibinfo {author} {\bibfnamefont {J.-X.}\ \bibnamefont
  {Feng}}, \bibinfo {author} {\bibfnamefont {F.}~\bibnamefont {Zhang}},\ and\
  \bibinfo {author} {\bibfnamefont {X.}~\bibnamefont {Gao}},\ }\bibfield
  {title} {\bibinfo {title} {{Scalar induced gravitational waves in chiral
  scalar-tensor theory of gravity}},\ }\href@noop {} {\  (\bibinfo {year}
  {2024})},\ \Eprint {https://arxiv.org/abs/2404.05289} {arXiv:2404.05289
  [gr-qc]} \BibitemShut {NoStop}%
\bibitem [{\citenamefont {Zhang}\ \emph {et~al.}(2024)\citenamefont {Zhang},
  \citenamefont {Feng},\ and\ \citenamefont {Gao}}]{Zhang:2024vfw}%
  \BibitemOpen
  \bibfield  {author} {\bibinfo {author} {\bibfnamefont {F.}~\bibnamefont
  {Zhang}}, \bibinfo {author} {\bibfnamefont {J.-X.}\ \bibnamefont {Feng}},\
  and\ \bibinfo {author} {\bibfnamefont {X.}~\bibnamefont {Gao}},\ }\bibfield
  {title} {\bibinfo {title} {{Scalar induced gravitational waves in metric
  teleparallel gravity with the Nieh-Yan term}},\ }\href@noop {} {\  (\bibinfo
  {year} {2024})},\ \Eprint {https://arxiv.org/abs/2404.02922}
  {arXiv:2404.02922 [gr-qc]} \BibitemShut {NoStop}%
\bibitem [{\citenamefont {Dom\`enech}(2020)}]{Domenech:2019quo}%
  \BibitemOpen
  \bibfield  {author} {\bibinfo {author} {\bibfnamefont {G.}~\bibnamefont
  {Dom\`enech}},\ }\bibfield  {title} {\bibinfo {title} {{Induced gravitational
  waves in a general cosmological background}},\ }\href
  {https://doi.org/10.1142/S0218271820500285} {\bibfield  {journal} {\bibinfo
  {journal} {Int. J. Mod. Phys. D}\ }\textbf {\bibinfo {volume} {29}},\
  \bibinfo {pages} {2050028} (\bibinfo {year} {2020})},\ \Eprint
  {https://arxiv.org/abs/1912.05583} {arXiv:1912.05583 [gr-qc]} \BibitemShut
  {NoStop}%
\bibitem [{\citenamefont {Dom\`enech}\ \emph {et~al.}(2020)\citenamefont
  {Dom\`enech}, \citenamefont {Pi},\ and\ \citenamefont
  {Sasaki}}]{Domenech:2020kqm}%
  \BibitemOpen
  \bibfield  {author} {\bibinfo {author} {\bibfnamefont {G.}~\bibnamefont
  {Dom\`enech}}, \bibinfo {author} {\bibfnamefont {S.}~\bibnamefont {Pi}},\
  and\ \bibinfo {author} {\bibfnamefont {M.}~\bibnamefont {Sasaki}},\
  }\bibfield  {title} {\bibinfo {title} {{Induced gravitational waves as a
  probe of thermal history of the universe}},\ }\href
  {https://doi.org/10.1088/1475-7516/2020/08/017} {\bibfield  {journal}
  {\bibinfo  {journal} {JCAP}\ }\textbf {\bibinfo {volume} {08}},\ \bibinfo
  {pages} {017}},\ \Eprint {https://arxiv.org/abs/2005.12314} {arXiv:2005.12314
  [gr-qc]} \BibitemShut {NoStop}%
\bibitem [{\citenamefont {Amendola}\ \emph
  {et~al.}(2018{\natexlab{a}})\citenamefont {Amendola}, \citenamefont
  {Bettoni}, \citenamefont {Dom\`enech},\ and\ \citenamefont
  {Gomes}}]{Amendola:2018ltt}%
  \BibitemOpen
  \bibfield  {author} {\bibinfo {author} {\bibfnamefont {L.}~\bibnamefont
  {Amendola}}, \bibinfo {author} {\bibfnamefont {D.}~\bibnamefont {Bettoni}},
  \bibinfo {author} {\bibfnamefont {G.}~\bibnamefont {Dom\`enech}},\ and\
  \bibinfo {author} {\bibfnamefont {A.~R.}\ \bibnamefont {Gomes}},\ }\bibfield
  {title} {\bibinfo {title} {{Doppelg\"anger dark energy: modified gravity with
  non-universal couplings after GW170817}},\ }\href
  {https://doi.org/10.1088/1475-7516/2018/06/029} {\bibfield  {journal}
  {\bibinfo  {journal} {JCAP}\ }\textbf {\bibinfo {volume} {06}},\ \bibinfo
  {pages} {029}},\ \Eprint {https://arxiv.org/abs/1803.06368} {arXiv:1803.06368
  [gr-qc]} \BibitemShut {NoStop}%
\bibitem [{\citenamefont {Amendola}(1999)}]{Amendola:1999qq}%
  \BibitemOpen
  \bibfield  {author} {\bibinfo {author} {\bibfnamefont {L.}~\bibnamefont
  {Amendola}},\ }\bibfield  {title} {\bibinfo {title} {{Scaling solutions in
  general nonminimal coupling theories}},\ }\href
  {https://doi.org/10.1103/PhysRevD.60.043501} {\bibfield  {journal} {\bibinfo
  {journal} {Phys. Rev. D}\ }\textbf {\bibinfo {volume} {60}},\ \bibinfo
  {pages} {043501} (\bibinfo {year} {1999})},\ \Eprint
  {https://arxiv.org/abs/astro-ph/9904120} {arXiv:astro-ph/9904120}
  \BibitemShut {NoStop}%
\bibitem [{\citenamefont {Amendola}\ \emph {et~al.}(2006)\citenamefont
  {Amendola}, \citenamefont {Quartin}, \citenamefont {Tsujikawa},\ and\
  \citenamefont {Waga}}]{Amendola:2006qi}%
  \BibitemOpen
  \bibfield  {author} {\bibinfo {author} {\bibfnamefont {L.}~\bibnamefont
  {Amendola}}, \bibinfo {author} {\bibfnamefont {M.}~\bibnamefont {Quartin}},
  \bibinfo {author} {\bibfnamefont {S.}~\bibnamefont {Tsujikawa}},\ and\
  \bibinfo {author} {\bibfnamefont {I.}~\bibnamefont {Waga}},\ }\bibfield
  {title} {\bibinfo {title} {{Challenges for scaling cosmologies}},\ }\href
  {https://doi.org/10.1103/PhysRevD.74.023525} {\bibfield  {journal} {\bibinfo
  {journal} {Phys. Rev. D}\ }\textbf {\bibinfo {volume} {74}},\ \bibinfo
  {pages} {023525} (\bibinfo {year} {2006})},\ \Eprint
  {https://arxiv.org/abs/astro-ph/0605488} {arXiv:astro-ph/0605488}
  \BibitemShut {NoStop}%
\bibitem [{\citenamefont {Gomes}\ and\ \citenamefont
  {Amendola}(2014)}]{Gomes:2013ema}%
  \BibitemOpen
  \bibfield  {author} {\bibinfo {author} {\bibfnamefont {A.~R.}\ \bibnamefont
  {Gomes}}\ and\ \bibinfo {author} {\bibfnamefont {L.}~\bibnamefont
  {Amendola}},\ }\bibfield  {title} {\bibinfo {title} {{Towards scaling
  cosmological solutions with full coupled Horndeski Lagrangian: the KGB
  model}},\ }\href {https://doi.org/10.1088/1475-7516/2014/03/041} {\bibfield
  {journal} {\bibinfo  {journal} {JCAP}\ }\textbf {\bibinfo {volume} {03}},\
  \bibinfo {pages} {041}},\ \Eprint {https://arxiv.org/abs/1306.3593}
  {arXiv:1306.3593 [astro-ph.CO]} \BibitemShut {NoStop}%
\bibitem [{\citenamefont {Gomes}\ and\ \citenamefont
  {Amendola}(2016)}]{Gomes:2015dhl}%
  \BibitemOpen
  \bibfield  {author} {\bibinfo {author} {\bibfnamefont {A.~R.}\ \bibnamefont
  {Gomes}}\ and\ \bibinfo {author} {\bibfnamefont {L.}~\bibnamefont
  {Amendola}},\ }\bibfield  {title} {\bibinfo {title} {{The general form of the
  coupled Horndeski Lagrangian that allows cosmological scaling solutions}},\
  }\href {https://doi.org/10.1088/1475-7516/2016/02/035} {\bibfield  {journal}
  {\bibinfo  {journal} {JCAP}\ }\textbf {\bibinfo {volume} {02}},\ \bibinfo
  {pages} {035}},\ \Eprint {https://arxiv.org/abs/1511.01004} {arXiv:1511.01004
  [gr-qc]} \BibitemShut {NoStop}%
\bibitem [{\citenamefont {Amendola}\ \emph
  {et~al.}(2018{\natexlab{b}})\citenamefont {Amendola}, \citenamefont {Rubio},\
  and\ \citenamefont {Wetterich}}]{Amendola:2017xhl}%
  \BibitemOpen
  \bibfield  {author} {\bibinfo {author} {\bibfnamefont {L.}~\bibnamefont
  {Amendola}}, \bibinfo {author} {\bibfnamefont {J.}~\bibnamefont {Rubio}},\
  and\ \bibinfo {author} {\bibfnamefont {C.}~\bibnamefont {Wetterich}},\
  }\bibfield  {title} {\bibinfo {title} {{Primordial black holes from fifth
  forces}},\ }\href {https://doi.org/10.1103/PhysRevD.97.081302} {\bibfield
  {journal} {\bibinfo  {journal} {Phys. Rev. D}\ }\textbf {\bibinfo {volume}
  {97}},\ \bibinfo {pages} {081302} (\bibinfo {year} {2018}{\natexlab{b}})},\
  \Eprint {https://arxiv.org/abs/1711.09915} {arXiv:1711.09915 [astro-ph.CO]}
  \BibitemShut {NoStop}%
\bibitem [{\citenamefont {Frusciante}\ \emph {et~al.}(2019)\citenamefont
  {Frusciante}, \citenamefont {Kase}, \citenamefont {Koyama}, \citenamefont
  {Tsujikawa},\ and\ \citenamefont {Vernieri}}]{Frusciante:2018tvu}%
  \BibitemOpen
  \bibfield  {author} {\bibinfo {author} {\bibfnamefont {N.}~\bibnamefont
  {Frusciante}}, \bibinfo {author} {\bibfnamefont {R.}~\bibnamefont {Kase}},
  \bibinfo {author} {\bibfnamefont {K.}~\bibnamefont {Koyama}}, \bibinfo
  {author} {\bibfnamefont {S.}~\bibnamefont {Tsujikawa}},\ and\ \bibinfo
  {author} {\bibfnamefont {D.}~\bibnamefont {Vernieri}},\ }\bibfield  {title}
  {\bibinfo {title} {{Tracker and scaling solutions in DHOST theories}},\
  }\href {https://doi.org/10.1016/j.physletb.2019.01.009} {\bibfield  {journal}
  {\bibinfo  {journal} {Phys. Lett. B}\ }\textbf {\bibinfo {volume} {790}},\
  \bibinfo {pages} {167} (\bibinfo {year} {2019})},\ \Eprint
  {https://arxiv.org/abs/1812.05204} {arXiv:1812.05204 [gr-qc]} \BibitemShut
  {NoStop}%
\bibitem [{\citenamefont {Frusciante}\ \emph {et~al.}(2018)\citenamefont
  {Frusciante}, \citenamefont {Kase}, \citenamefont {Nunes},\ and\
  \citenamefont {Tsujikawa}}]{Frusciante:2018aew}%
  \BibitemOpen
  \bibfield  {author} {\bibinfo {author} {\bibfnamefont {N.}~\bibnamefont
  {Frusciante}}, \bibinfo {author} {\bibfnamefont {R.}~\bibnamefont {Kase}},
  \bibinfo {author} {\bibfnamefont {N.~J.}\ \bibnamefont {Nunes}},\ and\
  \bibinfo {author} {\bibfnamefont {S.}~\bibnamefont {Tsujikawa}},\ }\bibfield
  {title} {\bibinfo {title} {{Most general cubic-order Horndeski Lagrangian
  allowing for scaling solutions and the application to dark energy}},\ }\href
  {https://doi.org/10.1103/PhysRevD.98.123517} {\bibfield  {journal} {\bibinfo
  {journal} {Phys. Rev. D}\ }\textbf {\bibinfo {volume} {98}},\ \bibinfo
  {pages} {123517} (\bibinfo {year} {2018})},\ \Eprint
  {https://arxiv.org/abs/1810.07957} {arXiv:1810.07957 [gr-qc]} \BibitemShut
  {NoStop}%
\bibitem [{\citenamefont {Vaskonen}\ and\ \citenamefont
  {Veerm\"ae}(2021)}]{Vaskonen:2020lbd}%
  \BibitemOpen
  \bibfield  {author} {\bibinfo {author} {\bibfnamefont {V.}~\bibnamefont
  {Vaskonen}}\ and\ \bibinfo {author} {\bibfnamefont {H.}~\bibnamefont
  {Veerm\"ae}},\ }\bibfield  {title} {\bibinfo {title} {{Did NANOGrav see a
  signal from primordial black hole formation?}},\ }\href
  {https://doi.org/10.1103/PhysRevLett.126.051303} {\bibfield  {journal}
  {\bibinfo  {journal} {Phys. Rev. Lett.}\ }\textbf {\bibinfo {volume} {126}},\
  \bibinfo {pages} {051303} (\bibinfo {year} {2021})},\ \Eprint
  {https://arxiv.org/abs/2009.07832} {arXiv:2009.07832 [astro-ph.CO]}
  \BibitemShut {NoStop}%
\bibitem [{\citenamefont {Chow}\ and\ \citenamefont
  {Khoury}(2009)}]{Chow:2009fm}%
  \BibitemOpen
  \bibfield  {author} {\bibinfo {author} {\bibfnamefont {N.}~\bibnamefont
  {Chow}}\ and\ \bibinfo {author} {\bibfnamefont {J.}~\bibnamefont {Khoury}},\
  }\bibfield  {title} {\bibinfo {title} {{Galileon Cosmology}},\ }\href
  {https://doi.org/10.1103/PhysRevD.80.024037} {\bibfield  {journal} {\bibinfo
  {journal} {Phys. Rev. D}\ }\textbf {\bibinfo {volume} {80}},\ \bibinfo
  {pages} {024037} (\bibinfo {year} {2009})},\ \Eprint
  {https://arxiv.org/abs/0905.1325} {arXiv:0905.1325 [hep-th]} \BibitemShut
  {NoStop}%
\bibitem [{\citenamefont {Silva}\ and\ \citenamefont
  {Koyama}(2009)}]{Silva:2009km}%
  \BibitemOpen
  \bibfield  {author} {\bibinfo {author} {\bibfnamefont {F.~P.}\ \bibnamefont
  {Silva}}\ and\ \bibinfo {author} {\bibfnamefont {K.}~\bibnamefont {Koyama}},\
  }\bibfield  {title} {\bibinfo {title} {{Self-Accelerating Universe in
  Galileon Cosmology}},\ }\href {https://doi.org/10.1103/PhysRevD.80.121301}
  {\bibfield  {journal} {\bibinfo  {journal} {Phys. Rev. D}\ }\textbf {\bibinfo
  {volume} {80}},\ \bibinfo {pages} {121301} (\bibinfo {year} {2009})},\
  \Eprint {https://arxiv.org/abs/0909.4538} {arXiv:0909.4538 [astro-ph.CO]}
  \BibitemShut {NoStop}%
\bibitem [{\citenamefont {Deffayet}\ \emph {et~al.}(2010)\citenamefont
  {Deffayet}, \citenamefont {Pujolas}, \citenamefont {Sawicki},\ and\
  \citenamefont {Vikman}}]{Deffayet:2010qz}%
  \BibitemOpen
  \bibfield  {author} {\bibinfo {author} {\bibfnamefont {C.}~\bibnamefont
  {Deffayet}}, \bibinfo {author} {\bibfnamefont {O.}~\bibnamefont {Pujolas}},
  \bibinfo {author} {\bibfnamefont {I.}~\bibnamefont {Sawicki}},\ and\ \bibinfo
  {author} {\bibfnamefont {A.}~\bibnamefont {Vikman}},\ }\bibfield  {title}
  {\bibinfo {title} {{Imperfect Dark Energy from Kinetic Gravity Braiding}},\
  }\href {https://doi.org/10.1088/1475-7516/2010/10/026} {\bibfield  {journal}
  {\bibinfo  {journal} {JCAP}\ }\textbf {\bibinfo {volume} {10}},\ \bibinfo
  {pages} {026}},\ \Eprint {https://arxiv.org/abs/1008.0048} {arXiv:1008.0048
  [hep-th]} \BibitemShut {NoStop}%
\bibitem [{\citenamefont {Kimura}\ and\ \citenamefont
  {Yamamoto}(2011)}]{Kimura:2010di}%
  \BibitemOpen
  \bibfield  {author} {\bibinfo {author} {\bibfnamefont {R.}~\bibnamefont
  {Kimura}}\ and\ \bibinfo {author} {\bibfnamefont {K.}~\bibnamefont
  {Yamamoto}},\ }\bibfield  {title} {\bibinfo {title} {{Large Scale Structures
  in Kinetic Gravity Braiding Model That Can Be Unbraided}},\ }\href
  {https://doi.org/10.1088/1475-7516/2011/04/025} {\bibfield  {journal}
  {\bibinfo  {journal} {JCAP}\ }\textbf {\bibinfo {volume} {04}},\ \bibinfo
  {pages} {025}},\ \Eprint {https://arxiv.org/abs/1011.2006} {arXiv:1011.2006
  [astro-ph.CO]} \BibitemShut {NoStop}%
\bibitem [{\citenamefont {De~Felice}\ and\ \citenamefont
  {Tsujikawa}(2010{\natexlab{b}})}]{DeFelice:2010pv}%
  \BibitemOpen
  \bibfield  {author} {\bibinfo {author} {\bibfnamefont {A.}~\bibnamefont
  {De~Felice}}\ and\ \bibinfo {author} {\bibfnamefont {S.}~\bibnamefont
  {Tsujikawa}},\ }\bibfield  {title} {\bibinfo {title} {{Cosmology of a
  covariant Galileon field}},\ }\href
  {https://doi.org/10.1103/PhysRevLett.105.111301} {\bibfield  {journal}
  {\bibinfo  {journal} {Phys. Rev. Lett.}\ }\textbf {\bibinfo {volume} {105}},\
  \bibinfo {pages} {111301} (\bibinfo {year} {2010}{\natexlab{b}})},\ \Eprint
  {https://arxiv.org/abs/1007.2700} {arXiv:1007.2700 [astro-ph.CO]}
  \BibitemShut {NoStop}%
\bibitem [{\citenamefont {De~Felice}\ and\ \citenamefont
  {Tsujikawa}(2011{\natexlab{b}})}]{DeFelice:2010nf}%
  \BibitemOpen
  \bibfield  {author} {\bibinfo {author} {\bibfnamefont {A.}~\bibnamefont
  {De~Felice}}\ and\ \bibinfo {author} {\bibfnamefont {S.}~\bibnamefont
  {Tsujikawa}},\ }\bibfield  {title} {\bibinfo {title} {{Generalized Galileon
  cosmology}},\ }\href {https://doi.org/10.1103/PhysRevD.84.124029} {\bibfield
  {journal} {\bibinfo  {journal} {Phys. Rev. D}\ }\textbf {\bibinfo {volume}
  {84}},\ \bibinfo {pages} {124029} (\bibinfo {year} {2011}{\natexlab{b}})},\
  \Eprint {https://arxiv.org/abs/1008.4236} {arXiv:1008.4236 [hep-th]}
  \BibitemShut {NoStop}%
\bibitem [{\citenamefont {Crisostomi}\ and\ \citenamefont
  {Koyama}(2018)}]{Crisostomi:2017pjs}%
  \BibitemOpen
  \bibfield  {author} {\bibinfo {author} {\bibfnamefont {M.}~\bibnamefont
  {Crisostomi}}\ and\ \bibinfo {author} {\bibfnamefont {K.}~\bibnamefont
  {Koyama}},\ }\bibfield  {title} {\bibinfo {title} {{Self-accelerating
  universe in scalar-tensor theories after GW170817}},\ }\href
  {https://doi.org/10.1103/PhysRevD.97.084004} {\bibfield  {journal} {\bibinfo
  {journal} {Phys. Rev. D}\ }\textbf {\bibinfo {volume} {97}},\ \bibinfo
  {pages} {084004} (\bibinfo {year} {2018})},\ \Eprint
  {https://arxiv.org/abs/1712.06556} {arXiv:1712.06556 [astro-ph.CO]}
  \BibitemShut {NoStop}%
\bibitem [{\citenamefont {Creminelli}\ \emph {et~al.}(2005)\citenamefont
  {Creminelli}, \citenamefont {Nicolis},\ and\ \citenamefont
  {Zaldarriaga}}]{Creminelli:2004jg}%
  \BibitemOpen
  \bibfield  {author} {\bibinfo {author} {\bibfnamefont {P.}~\bibnamefont
  {Creminelli}}, \bibinfo {author} {\bibfnamefont {A.}~\bibnamefont
  {Nicolis}},\ and\ \bibinfo {author} {\bibfnamefont {M.}~\bibnamefont
  {Zaldarriaga}},\ }\bibfield  {title} {\bibinfo {title} {{Perturbations in
  bouncing cosmologies: Dynamical attractor versus scale invariance}},\ }\href
  {https://doi.org/10.1103/PhysRevD.71.063505} {\bibfield  {journal} {\bibinfo
  {journal} {Phys. Rev. D}\ }\textbf {\bibinfo {volume} {71}},\ \bibinfo
  {pages} {063505} (\bibinfo {year} {2005})},\ \Eprint
  {https://arxiv.org/abs/hep-th/0411270} {arXiv:hep-th/0411270} \BibitemShut
  {NoStop}%
\bibitem [{\citenamefont {Khoury}\ and\ \citenamefont
  {Steinhardt}(2010)}]{Khoury:2009my}%
  \BibitemOpen
  \bibfield  {author} {\bibinfo {author} {\bibfnamefont {J.}~\bibnamefont
  {Khoury}}\ and\ \bibinfo {author} {\bibfnamefont {P.~J.}\ \bibnamefont
  {Steinhardt}},\ }\bibfield  {title} {\bibinfo {title} {{Adiabatic Ekpyrosis:
  Scale-Invariant Curvature Perturbations from a Single Scalar Field in a
  Contracting Universe}},\ }\href
  {https://doi.org/10.1103/PhysRevLett.104.091301} {\bibfield  {journal}
  {\bibinfo  {journal} {Phys. Rev. Lett.}\ }\textbf {\bibinfo {volume} {104}},\
  \bibinfo {pages} {091301} (\bibinfo {year} {2010})},\ \Eprint
  {https://arxiv.org/abs/0910.2230} {arXiv:0910.2230 [hep-th]} \BibitemShut
  {NoStop}%
\bibitem [{\citenamefont {Lucchin}\ and\ \citenamefont
  {Matarrese}(1985)}]{Lucchin:1984yf}%
  \BibitemOpen
  \bibfield  {author} {\bibinfo {author} {\bibfnamefont {F.}~\bibnamefont
  {Lucchin}}\ and\ \bibinfo {author} {\bibfnamefont {S.}~\bibnamefont
  {Matarrese}},\ }\bibfield  {title} {\bibinfo {title} {{Power Law
  Inflation}},\ }\href {https://doi.org/10.1103/PhysRevD.32.1316} {\bibfield
  {journal} {\bibinfo  {journal} {Phys. Rev. D}\ }\textbf {\bibinfo {volume}
  {32}},\ \bibinfo {pages} {1316} (\bibinfo {year} {1985})}\BibitemShut
  {NoStop}%
\bibitem [{\citenamefont {Russo}(2004)}]{Russo:2004ym}%
  \BibitemOpen
  \bibfield  {author} {\bibinfo {author} {\bibfnamefont {J.~G.}\ \bibnamefont
  {Russo}},\ }\bibfield  {title} {\bibinfo {title} {{Exact solution of scalar
  tensor cosmology with exponential potentials and transient acceleration}},\
  }\href {https://doi.org/10.1016/j.physletb.2004.09.007} {\bibfield  {journal}
  {\bibinfo  {journal} {Phys. Lett. B}\ }\textbf {\bibinfo {volume} {600}},\
  \bibinfo {pages} {185} (\bibinfo {year} {2004})},\ \Eprint
  {https://arxiv.org/abs/hep-th/0403010} {arXiv:hep-th/0403010} \BibitemShut
  {NoStop}%
\bibitem [{\citenamefont {Andrianov}\ \emph {et~al.}(2011)\citenamefont
  {Andrianov}, \citenamefont {Cannata},\ and\ \citenamefont
  {Kamenshchik}}]{Andrianov:2011fg}%
  \BibitemOpen
  \bibfield  {author} {\bibinfo {author} {\bibfnamefont {A.~A.}\ \bibnamefont
  {Andrianov}}, \bibinfo {author} {\bibfnamefont {F.}~\bibnamefont {Cannata}},\
  and\ \bibinfo {author} {\bibfnamefont {A.~Y.}\ \bibnamefont {Kamenshchik}},\
  }\bibfield  {title} {\bibinfo {title} {{General solution of scalar field
  cosmology with a (piecewise) exponential potential}},\ }\href
  {https://doi.org/10.1088/1475-7516/2011/10/004} {\bibfield  {journal}
  {\bibinfo  {journal} {JCAP}\ }\textbf {\bibinfo {volume} {10}},\ \bibinfo
  {pages} {004}},\ \Eprint {https://arxiv.org/abs/1105.4515} {arXiv:1105.4515
  [gr-qc]} \BibitemShut {NoStop}%
\bibitem [{\citenamefont {Padilla}\ and\ \citenamefont
  {Sivanesan}(2012)}]{Padilla:2012ze}%
  \BibitemOpen
  \bibfield  {author} {\bibinfo {author} {\bibfnamefont {A.}~\bibnamefont
  {Padilla}}\ and\ \bibinfo {author} {\bibfnamefont {V.}~\bibnamefont
  {Sivanesan}},\ }\bibfield  {title} {\bibinfo {title} {{Boundary Terms and
  Junction Conditions for Generalized Scalar-Tensor Theories}},\ }\href
  {https://doi.org/10.1007/JHEP08(2012)122} {\bibfield  {journal} {\bibinfo
  {journal} {JHEP}\ }\textbf {\bibinfo {volume} {08}},\ \bibinfo {pages}
  {122}},\ \Eprint {https://arxiv.org/abs/1206.1258} {arXiv:1206.1258 [gr-qc]}
  \BibitemShut {NoStop}%
\bibitem [{\citenamefont {Nishi}\ \emph {et~al.}(2014)\citenamefont {Nishi},
  \citenamefont {Kobayashi}, \citenamefont {Tanahashi},\ and\ \citenamefont
  {Yamaguchi}}]{Nishi:2014bsa}%
  \BibitemOpen
  \bibfield  {author} {\bibinfo {author} {\bibfnamefont {S.}~\bibnamefont
  {Nishi}}, \bibinfo {author} {\bibfnamefont {T.}~\bibnamefont {Kobayashi}},
  \bibinfo {author} {\bibfnamefont {N.}~\bibnamefont {Tanahashi}},\ and\
  \bibinfo {author} {\bibfnamefont {M.}~\bibnamefont {Yamaguchi}},\ }\bibfield
  {title} {\bibinfo {title} {{Cosmological matching conditionsand galilean
  genesis in Horndeski's theory}},\ }\href
  {https://doi.org/10.1088/1475-7516/2014/03/008} {\bibfield  {journal}
  {\bibinfo  {journal} {JCAP}\ }\textbf {\bibinfo {volume} {03}},\ \bibinfo
  {pages} {008}},\ \Eprint {https://arxiv.org/abs/1401.1045} {arXiv:1401.1045
  [hep-th]} \BibitemShut {NoStop}%
\bibitem [{\citenamefont {Hwang}\ \emph {et~al.}(2017)\citenamefont {Hwang},
  \citenamefont {Jeong},\ and\ \citenamefont {Noh}}]{Hwang:2017oxa}%
  \BibitemOpen
  \bibfield  {author} {\bibinfo {author} {\bibfnamefont {J.-C.}\ \bibnamefont
  {Hwang}}, \bibinfo {author} {\bibfnamefont {D.}~\bibnamefont {Jeong}},\ and\
  \bibinfo {author} {\bibfnamefont {H.}~\bibnamefont {Noh}},\ }\bibfield
  {title} {\bibinfo {title} {{Gauge dependence of gravitational waves generated
  from scalar perturbations}},\ }\href
  {https://doi.org/10.3847/1538-4357/aa74be} {\bibfield  {journal} {\bibinfo
  {journal} {Astrophys. J.}\ }\textbf {\bibinfo {volume} {842}},\ \bibinfo
  {pages} {46} (\bibinfo {year} {2017})},\ \Eprint
  {https://arxiv.org/abs/1704.03500} {arXiv:1704.03500 [astro-ph.CO]}
  \BibitemShut {NoStop}%
\bibitem [{\citenamefont {Gong}(2022)}]{Gong:2019mui}%
  \BibitemOpen
  \bibfield  {author} {\bibinfo {author} {\bibfnamefont {J.-O.}\ \bibnamefont
  {Gong}},\ }\bibfield  {title} {\bibinfo {title} {{Analytic Integral Solutions
  for Induced Gravitational Waves}},\ }\href
  {https://doi.org/10.3847/1538-4357/ac3a6c} {\bibfield  {journal} {\bibinfo
  {journal} {Astrophys. J.}\ }\textbf {\bibinfo {volume} {925}},\ \bibinfo
  {pages} {102} (\bibinfo {year} {2022})},\ \Eprint
  {https://arxiv.org/abs/1909.12708} {arXiv:1909.12708 [gr-qc]} \BibitemShut
  {NoStop}%
\bibitem [{\citenamefont {Tomikawa}\ and\ \citenamefont
  {Kobayashi}(2020)}]{Tomikawa:2019tvi}%
  \BibitemOpen
  \bibfield  {author} {\bibinfo {author} {\bibfnamefont {K.}~\bibnamefont
  {Tomikawa}}\ and\ \bibinfo {author} {\bibfnamefont {T.}~\bibnamefont
  {Kobayashi}},\ }\bibfield  {title} {\bibinfo {title} {{Gauge dependence of
  gravitational waves generated at second order from scalar perturbations}},\
  }\href {https://doi.org/10.1103/PhysRevD.101.083529} {\bibfield  {journal}
  {\bibinfo  {journal} {Phys. Rev. D}\ }\textbf {\bibinfo {volume} {101}},\
  \bibinfo {pages} {083529} (\bibinfo {year} {2020})},\ \Eprint
  {https://arxiv.org/abs/1910.01880} {arXiv:1910.01880 [gr-qc]} \BibitemShut
  {NoStop}%
\bibitem [{\citenamefont {Sipp}\ and\ \citenamefont
  {Schaefer}(2023)}]{Sipp:2022kmb}%
  \BibitemOpen
  \bibfield  {author} {\bibinfo {author} {\bibfnamefont {M.}~\bibnamefont
  {Sipp}}\ and\ \bibinfo {author} {\bibfnamefont {B.~M.}\ \bibnamefont
  {Schaefer}},\ }\bibfield  {title} {\bibinfo {title} {{Scalar-induced
  gravitational waves in a \ensuremath{\Lambda}CDM cosmology}},\ }\href
  {https://doi.org/10.1103/PhysRevD.107.063538} {\bibfield  {journal} {\bibinfo
   {journal} {Phys. Rev. D}\ }\textbf {\bibinfo {volume} {107}},\ \bibinfo
  {pages} {063538} (\bibinfo {year} {2023})},\ \Eprint
  {https://arxiv.org/abs/2212.01190} {arXiv:2212.01190 [astro-ph.CO]}
  \BibitemShut {NoStop}%
\bibitem [{\citenamefont {De~Luca}\ \emph {et~al.}(2020)\citenamefont
  {De~Luca}, \citenamefont {Franciolini}, \citenamefont {Kehagias},\ and\
  \citenamefont {Riotto}}]{DeLuca:2019ufz}%
  \BibitemOpen
  \bibfield  {author} {\bibinfo {author} {\bibfnamefont {V.}~\bibnamefont
  {De~Luca}}, \bibinfo {author} {\bibfnamefont {G.}~\bibnamefont
  {Franciolini}}, \bibinfo {author} {\bibfnamefont {A.}~\bibnamefont
  {Kehagias}},\ and\ \bibinfo {author} {\bibfnamefont {A.}~\bibnamefont
  {Riotto}},\ }\bibfield  {title} {\bibinfo {title} {{On the Gauge Invariance
  of Cosmological Gravitational Waves}},\ }\href
  {https://doi.org/10.1088/1475-7516/2020/03/014} {\bibfield  {journal}
  {\bibinfo  {journal} {JCAP}\ }\textbf {\bibinfo {volume} {03}},\ \bibinfo
  {pages} {014}},\ \Eprint {https://arxiv.org/abs/1911.09689} {arXiv:1911.09689
  [gr-qc]} \BibitemShut {NoStop}%
\bibitem [{\citenamefont {Inomata}\ and\ \citenamefont
  {Terada}(2020)}]{Inomata:2019yww}%
  \BibitemOpen
  \bibfield  {author} {\bibinfo {author} {\bibfnamefont {K.}~\bibnamefont
  {Inomata}}\ and\ \bibinfo {author} {\bibfnamefont {T.}~\bibnamefont
  {Terada}},\ }\bibfield  {title} {\bibinfo {title} {{Gauge Independence of
  Induced Gravitational Waves}},\ }\href
  {https://doi.org/10.1103/PhysRevD.101.023523} {\bibfield  {journal} {\bibinfo
   {journal} {Phys. Rev. D}\ }\textbf {\bibinfo {volume} {101}},\ \bibinfo
  {pages} {023523} (\bibinfo {year} {2020})},\ \Eprint
  {https://arxiv.org/abs/1912.00785} {arXiv:1912.00785 [gr-qc]} \BibitemShut
  {NoStop}%
\bibitem [{\citenamefont {Dom\`enech}\ and\ \citenamefont
  {Sasaki}(2021)}]{Domenech:2020xin}%
  \BibitemOpen
  \bibfield  {author} {\bibinfo {author} {\bibfnamefont {G.}~\bibnamefont
  {Dom\`enech}}\ and\ \bibinfo {author} {\bibfnamefont {M.}~\bibnamefont
  {Sasaki}},\ }\bibfield  {title} {\bibinfo {title} {{Approximate gauge
  independence of the induced gravitational wave spectrum}},\ }\href
  {https://doi.org/10.1103/PhysRevD.103.063531} {\bibfield  {journal} {\bibinfo
   {journal} {Phys. Rev. D}\ }\textbf {\bibinfo {volume} {103}},\ \bibinfo
  {pages} {063531} (\bibinfo {year} {2021})},\ \Eprint
  {https://arxiv.org/abs/2012.14016} {arXiv:2012.14016 [gr-qc]} \BibitemShut
  {NoStop}%
\bibitem [{\citenamefont {Dom\`enech}\ and\ \citenamefont
  {Sasaki}(2018)}]{Domenech:2017ems}%
  \BibitemOpen
  \bibfield  {author} {\bibinfo {author} {\bibfnamefont {G.}~\bibnamefont
  {Dom\`enech}}\ and\ \bibinfo {author} {\bibfnamefont {M.}~\bibnamefont
  {Sasaki}},\ }\bibfield  {title} {\bibinfo {title} {{Hamiltonian approach to
  second order gauge invariant cosmological perturbations}},\ }\href
  {https://doi.org/10.1103/PhysRevD.97.023521} {\bibfield  {journal} {\bibinfo
  {journal} {Phys. Rev. D}\ }\textbf {\bibinfo {volume} {97}},\ \bibinfo
  {pages} {023521} (\bibinfo {year} {2018})},\ \Eprint
  {https://arxiv.org/abs/1709.09804} {arXiv:1709.09804 [gr-qc]} \BibitemShut
  {NoStop}%
\bibitem [{\citenamefont {Chen}\ \emph {et~al.}(2024)\citenamefont {Chen},
  \citenamefont {Li}, \citenamefont {Liu},\ and\ \citenamefont
  {Yi}}]{Chen:2024fir}%
  \BibitemOpen
  \bibfield  {author} {\bibinfo {author} {\bibfnamefont {Z.-C.}\ \bibnamefont
  {Chen}}, \bibinfo {author} {\bibfnamefont {J.}~\bibnamefont {Li}}, \bibinfo
  {author} {\bibfnamefont {L.}~\bibnamefont {Liu}},\ and\ \bibinfo {author}
  {\bibfnamefont {Z.}~\bibnamefont {Yi}},\ }\bibfield  {title} {\bibinfo
  {title} {{Probing the speed of scalar-induced gravitational waves with pulsar
  timing arrays}},\ }\href {https://doi.org/10.1103/PhysRevD.109.L101302}
  {\bibfield  {journal} {\bibinfo  {journal} {Phys. Rev. D}\ }\textbf {\bibinfo
  {volume} {109}},\ \bibinfo {pages} {L101302} (\bibinfo {year} {2024})},\
  \Eprint {https://arxiv.org/abs/2401.09818} {arXiv:2401.09818 [gr-qc]}
  \BibitemShut {NoStop}%
\bibitem [{\citenamefont {Creminelli}\ \emph {et~al.}(2014)\citenamefont
  {Creminelli}, \citenamefont {Gleyzes}, \citenamefont {Nore\~na},\ and\
  \citenamefont {Vernizzi}}]{Creminelli:2014wna}%
  \BibitemOpen
  \bibfield  {author} {\bibinfo {author} {\bibfnamefont {P.}~\bibnamefont
  {Creminelli}}, \bibinfo {author} {\bibfnamefont {J.}~\bibnamefont {Gleyzes}},
  \bibinfo {author} {\bibfnamefont {J.}~\bibnamefont {Nore\~na}},\ and\
  \bibinfo {author} {\bibfnamefont {F.}~\bibnamefont {Vernizzi}},\ }\bibfield
  {title} {\bibinfo {title} {{Resilience of the standard predictions for
  primordial tensor modes}},\ }\href
  {https://doi.org/10.1103/PhysRevLett.113.231301} {\bibfield  {journal}
  {\bibinfo  {journal} {Phys. Rev. Lett.}\ }\textbf {\bibinfo {volume} {113}},\
  \bibinfo {pages} {231301} (\bibinfo {year} {2014})},\ \Eprint
  {https://arxiv.org/abs/1407.8439} {arXiv:1407.8439 [astro-ph.CO]}
  \BibitemShut {NoStop}%
\bibitem [{\citenamefont {Vainshtein}(1972)}]{Vainshtein:1972sx}%
  \BibitemOpen
  \bibfield  {author} {\bibinfo {author} {\bibfnamefont {A.~I.}\ \bibnamefont
  {Vainshtein}},\ }\bibfield  {title} {\bibinfo {title} {{To the problem of
  nonvanishing gravitation mass}},\ }\href
  {https://doi.org/10.1016/0370-2693(72)90147-5} {\bibfield  {journal}
  {\bibinfo  {journal} {Phys. Lett. B}\ }\textbf {\bibinfo {volume} {39}},\
  \bibinfo {pages} {393} (\bibinfo {year} {1972})}\BibitemShut {NoStop}%
\bibitem [{\citenamefont {Kimura}\ \emph {et~al.}(2012)\citenamefont {Kimura},
  \citenamefont {Kobayashi},\ and\ \citenamefont {Yamamoto}}]{Kimura:2011dc}%
  \BibitemOpen
  \bibfield  {author} {\bibinfo {author} {\bibfnamefont {R.}~\bibnamefont
  {Kimura}}, \bibinfo {author} {\bibfnamefont {T.}~\bibnamefont {Kobayashi}},\
  and\ \bibinfo {author} {\bibfnamefont {K.}~\bibnamefont {Yamamoto}},\
  }\bibfield  {title} {\bibinfo {title} {{Vainshtein screening in a
  cosmological background in the most general second-order scalar-tensor
  theory}},\ }\href {https://doi.org/10.1103/PhysRevD.85.024023} {\bibfield
  {journal} {\bibinfo  {journal} {Phys. Rev. D}\ }\textbf {\bibinfo {volume}
  {85}},\ \bibinfo {pages} {024023} (\bibinfo {year} {2012})},\ \Eprint
  {https://arxiv.org/abs/1111.6749} {arXiv:1111.6749 [astro-ph.CO]}
  \BibitemShut {NoStop}%
\bibitem [{\citenamefont {Koyama}\ \emph {et~al.}(2013)\citenamefont {Koyama},
  \citenamefont {Niz},\ and\ \citenamefont {Tasinato}}]{Koyama:2013paa}%
  \BibitemOpen
  \bibfield  {author} {\bibinfo {author} {\bibfnamefont {K.}~\bibnamefont
  {Koyama}}, \bibinfo {author} {\bibfnamefont {G.}~\bibnamefont {Niz}},\ and\
  \bibinfo {author} {\bibfnamefont {G.}~\bibnamefont {Tasinato}},\ }\bibfield
  {title} {\bibinfo {title} {{Effective theory for the Vainshtein mechanism
  from the Horndeski action}},\ }\href
  {https://doi.org/10.1103/PhysRevD.88.021502} {\bibfield  {journal} {\bibinfo
  {journal} {Phys. Rev. D}\ }\textbf {\bibinfo {volume} {88}},\ \bibinfo
  {pages} {021502} (\bibinfo {year} {2013})},\ \Eprint
  {https://arxiv.org/abs/1305.0279} {arXiv:1305.0279 [hep-th]} \BibitemShut
  {NoStop}%
\bibitem [{\citenamefont {Babichev}\ and\ \citenamefont
  {Deffayet}(2013)}]{Babichev:2013usa}%
  \BibitemOpen
  \bibfield  {author} {\bibinfo {author} {\bibfnamefont {E.}~\bibnamefont
  {Babichev}}\ and\ \bibinfo {author} {\bibfnamefont {C.}~\bibnamefont
  {Deffayet}},\ }\bibfield  {title} {\bibinfo {title} {{An introduction to the
  Vainshtein mechanism}},\ }\href
  {https://doi.org/10.1088/0264-9381/30/18/184001} {\bibfield  {journal}
  {\bibinfo  {journal} {Class. Quant. Grav.}\ }\textbf {\bibinfo {volume}
  {30}},\ \bibinfo {pages} {184001} (\bibinfo {year} {2013})},\ \Eprint
  {https://arxiv.org/abs/1304.7240} {arXiv:1304.7240 [gr-qc]} \BibitemShut
  {NoStop}%
\bibitem [{\citenamefont {Dom\`enech}\ \emph {et~al.}(2022)\citenamefont
  {Dom\`enech}, \citenamefont {Passaglia},\ and\ \citenamefont
  {Renaux-Petel}}]{Domenech:2021and}%
  \BibitemOpen
  \bibfield  {author} {\bibinfo {author} {\bibfnamefont {G.}~\bibnamefont
  {Dom\`enech}}, \bibinfo {author} {\bibfnamefont {S.}~\bibnamefont
  {Passaglia}},\ and\ \bibinfo {author} {\bibfnamefont {S.}~\bibnamefont
  {Renaux-Petel}},\ }\bibfield  {title} {\bibinfo {title} {{Gravitational waves
  from dark matter isocurvature}},\ }\href
  {https://doi.org/10.1088/1475-7516/2022/03/023} {\bibfield  {journal}
  {\bibinfo  {journal} {JCAP}\ }\textbf {\bibinfo {volume} {03}}\bibfield
  {number} {\bibinfo  {number} { (03)},\ \bibinfo {pages} {023}},\ }\Eprint
  {https://arxiv.org/abs/2112.10163} {arXiv:2112.10163 [astro-ph.CO]}
  \BibitemShut {NoStop}%
\bibitem [{\citenamefont {Kohri}\ and\ \citenamefont
  {Terada}(2018)}]{Kohri:2018awv}%
  \BibitemOpen
  \bibfield  {author} {\bibinfo {author} {\bibfnamefont {K.}~\bibnamefont
  {Kohri}}\ and\ \bibinfo {author} {\bibfnamefont {T.}~\bibnamefont {Terada}},\
  }\bibfield  {title} {\bibinfo {title} {{Semianalytic calculation of
  gravitational wave spectrum nonlinearly induced from primordial curvature
  perturbations}},\ }\href {https://doi.org/10.1103/PhysRevD.97.123532}
  {\bibfield  {journal} {\bibinfo  {journal} {Phys. Rev. D}\ }\textbf {\bibinfo
  {volume} {97}},\ \bibinfo {pages} {123532} (\bibinfo {year} {2018})},\
  \Eprint {https://arxiv.org/abs/1804.08577} {arXiv:1804.08577 [gr-qc]}
  \BibitemShut {NoStop}%
\bibitem [{\citenamefont {Cai}\ \emph {et~al.}(2019)\citenamefont {Cai},
  \citenamefont {Pi},\ and\ \citenamefont {Sasaki}}]{Cai:2018dig}%
  \BibitemOpen
  \bibfield  {author} {\bibinfo {author} {\bibfnamefont {R.-g.}\ \bibnamefont
  {Cai}}, \bibinfo {author} {\bibfnamefont {S.}~\bibnamefont {Pi}},\ and\
  \bibinfo {author} {\bibfnamefont {M.}~\bibnamefont {Sasaki}},\ }\bibfield
  {title} {\bibinfo {title} {{Gravitational Waves Induced by non-Gaussian
  Scalar Perturbations}},\ }\href
  {https://doi.org/10.1103/PhysRevLett.122.201101} {\bibfield  {journal}
  {\bibinfo  {journal} {Phys. Rev. Lett.}\ }\textbf {\bibinfo {volume} {122}},\
  \bibinfo {pages} {201101} (\bibinfo {year} {2019})},\ \Eprint
  {https://arxiv.org/abs/1810.11000} {arXiv:1810.11000 [astro-ph.CO]}
  \BibitemShut {NoStop}%
\bibitem [{\citenamefont {Cai}\ \emph {et~al.}(2020)\citenamefont {Cai},
  \citenamefont {Pi},\ and\ \citenamefont {Sasaki}}]{Cai:2019cdl}%
  \BibitemOpen
  \bibfield  {author} {\bibinfo {author} {\bibfnamefont {R.-G.}\ \bibnamefont
  {Cai}}, \bibinfo {author} {\bibfnamefont {S.}~\bibnamefont {Pi}},\ and\
  \bibinfo {author} {\bibfnamefont {M.}~\bibnamefont {Sasaki}},\ }\bibfield
  {title} {\bibinfo {title} {{Universal infrared scaling of gravitational wave
  background spectra}},\ }\href {https://doi.org/10.1103/PhysRevD.102.083528}
  {\bibfield  {journal} {\bibinfo  {journal} {Phys. Rev. D}\ }\textbf {\bibinfo
  {volume} {102}},\ \bibinfo {pages} {083528} (\bibinfo {year} {2020})},\
  \Eprint {https://arxiv.org/abs/1909.13728} {arXiv:1909.13728 [astro-ph.CO]}
  \BibitemShut {NoStop}%
\bibitem [{\citenamefont {Yuan}\ \emph {et~al.}(2020)\citenamefont {Yuan},
  \citenamefont {Chen},\ and\ \citenamefont {Huang}}]{Yuan:2019wwo}%
  \BibitemOpen
  \bibfield  {author} {\bibinfo {author} {\bibfnamefont {C.}~\bibnamefont
  {Yuan}}, \bibinfo {author} {\bibfnamefont {Z.-C.}\ \bibnamefont {Chen}},\
  and\ \bibinfo {author} {\bibfnamefont {Q.-G.}\ \bibnamefont {Huang}},\
  }\bibfield  {title} {\bibinfo {title} {{Log-dependent slope of scalar induced
  gravitational waves in the infrared regions}},\ }\href
  {https://doi.org/10.1103/PhysRevD.101.043019} {\bibfield  {journal} {\bibinfo
   {journal} {Phys. Rev. D}\ }\textbf {\bibinfo {volume} {101}},\ \bibinfo
  {pages} {043019} (\bibinfo {year} {2020})},\ \Eprint
  {https://arxiv.org/abs/1910.09099} {arXiv:1910.09099 [astro-ph.CO]}
  \BibitemShut {NoStop}%
\bibitem [{\citenamefont {Bellini}\ and\ \citenamefont
  {Sawicki}(2014)}]{Bellini:2014fua}%
  \BibitemOpen
  \bibfield  {author} {\bibinfo {author} {\bibfnamefont {E.}~\bibnamefont
  {Bellini}}\ and\ \bibinfo {author} {\bibfnamefont {I.}~\bibnamefont
  {Sawicki}},\ }\bibfield  {title} {\bibinfo {title} {{Maximal freedom at
  minimum cost: linear large-scale structure in general modifications of
  gravity}},\ }\href {https://doi.org/10.1088/1475-7516/2014/07/050} {\bibfield
   {journal} {\bibinfo  {journal} {JCAP}\ }\textbf {\bibinfo {volume} {07}},\
  \bibinfo {pages} {050}},\ \Eprint {https://arxiv.org/abs/1404.3713}
  {arXiv:1404.3713 [astro-ph.CO]} \BibitemShut {NoStop}%
\bibitem [{\citenamefont {Balaji}\ \emph {et~al.}(2022)\citenamefont {Balaji},
  \citenamefont {Domenech},\ and\ \citenamefont {Silk}}]{Balaji:2022dbi}%
  \BibitemOpen
  \bibfield  {author} {\bibinfo {author} {\bibfnamefont {S.}~\bibnamefont
  {Balaji}}, \bibinfo {author} {\bibfnamefont {G.}~\bibnamefont {Domenech}},\
  and\ \bibinfo {author} {\bibfnamefont {J.}~\bibnamefont {Silk}},\ }\bibfield
  {title} {\bibinfo {title} {{Induced gravitational waves from slow-roll
  inflation after an enhancing phase}},\ }\href
  {https://doi.org/10.1088/1475-7516/2022/09/016} {\bibfield  {journal}
  {\bibinfo  {journal} {JCAP}\ }\textbf {\bibinfo {volume} {09}},\ \bibinfo
  {pages} {016}},\ \Eprint {https://arxiv.org/abs/2205.01696} {arXiv:2205.01696
  [astro-ph.CO]} \BibitemShut {NoStop}%
\bibitem [{\citenamefont {Ben~Achour}\ \emph {et~al.}(2024)\citenamefont
  {Ben~Achour}, \citenamefont {Gorji},\ and\ \citenamefont
  {Roussille}}]{BenAchour:2024tqt}%
  \BibitemOpen
  \bibfield  {author} {\bibinfo {author} {\bibfnamefont {J.}~\bibnamefont
  {Ben~Achour}}, \bibinfo {author} {\bibfnamefont {M.~A.}\ \bibnamefont
  {Gorji}},\ and\ \bibinfo {author} {\bibfnamefont {H.}~\bibnamefont
  {Roussille}},\ }\bibfield  {title} {\bibinfo {title} {{Disformal
  gravitational waves}},\ }\href@noop {} {\  (\bibinfo {year} {2024})},\
  \Eprint {https://arxiv.org/abs/2402.01487} {arXiv:2402.01487 [gr-qc]}
  \BibitemShut {NoStop}%
\bibitem [{\citenamefont {Dalang}\ \emph {et~al.}(2020)\citenamefont {Dalang},
  \citenamefont {Fleury},\ and\ \citenamefont {Lombriser}}]{Dalang:2019rke}%
  \BibitemOpen
  \bibfield  {author} {\bibinfo {author} {\bibfnamefont {C.}~\bibnamefont
  {Dalang}}, \bibinfo {author} {\bibfnamefont {P.}~\bibnamefont {Fleury}},\
  and\ \bibinfo {author} {\bibfnamefont {L.}~\bibnamefont {Lombriser}},\
  }\bibfield  {title} {\bibinfo {title} {{Horndeski gravity and standard
  sirens}},\ }\href {https://doi.org/10.1103/PhysRevD.102.044036} {\bibfield
  {journal} {\bibinfo  {journal} {Phys. Rev. D}\ }\textbf {\bibinfo {volume}
  {102}},\ \bibinfo {pages} {044036} (\bibinfo {year} {2020})},\ \Eprint
  {https://arxiv.org/abs/1912.06117} {arXiv:1912.06117 [gr-qc]} \BibitemShut
  {NoStop}%
\bibitem [{\citenamefont {Dalang}\ \emph {et~al.}(2021)\citenamefont {Dalang},
  \citenamefont {Fleury},\ and\ \citenamefont {Lombriser}}]{Dalang:2020eaj}%
  \BibitemOpen
  \bibfield  {author} {\bibinfo {author} {\bibfnamefont {C.}~\bibnamefont
  {Dalang}}, \bibinfo {author} {\bibfnamefont {P.}~\bibnamefont {Fleury}},\
  and\ \bibinfo {author} {\bibfnamefont {L.}~\bibnamefont {Lombriser}},\
  }\bibfield  {title} {\bibinfo {title} {{Scalar and tensor gravitational
  waves}},\ }\href {https://doi.org/10.1103/PhysRevD.103.064075} {\bibfield
  {journal} {\bibinfo  {journal} {Phys. Rev. D}\ }\textbf {\bibinfo {volume}
  {103}},\ \bibinfo {pages} {064075} (\bibinfo {year} {2021})},\ \Eprint
  {https://arxiv.org/abs/2009.11827} {arXiv:2009.11827 [gr-qc]} \BibitemShut
  {NoStop}%
\bibitem [{\citenamefont {Garoffolo}\ \emph {et~al.}(2021)\citenamefont
  {Garoffolo}, \citenamefont {Raveri}, \citenamefont {Silvestri}, \citenamefont
  {Tasinato}, \citenamefont {Carbone}, \citenamefont {Bertacca},\ and\
  \citenamefont {Matarrese}}]{Garoffolo:2020vtd}%
  \BibitemOpen
  \bibfield  {author} {\bibinfo {author} {\bibfnamefont {A.}~\bibnamefont
  {Garoffolo}}, \bibinfo {author} {\bibfnamefont {M.}~\bibnamefont {Raveri}},
  \bibinfo {author} {\bibfnamefont {A.}~\bibnamefont {Silvestri}}, \bibinfo
  {author} {\bibfnamefont {G.}~\bibnamefont {Tasinato}}, \bibinfo {author}
  {\bibfnamefont {C.}~\bibnamefont {Carbone}}, \bibinfo {author} {\bibfnamefont
  {D.}~\bibnamefont {Bertacca}},\ and\ \bibinfo {author} {\bibfnamefont
  {S.}~\bibnamefont {Matarrese}},\ }\bibfield  {title} {\bibinfo {title}
  {{Detecting Dark Energy Fluctuations with Gravitational Waves}},\ }\href
  {https://doi.org/10.1103/PhysRevD.103.083506} {\bibfield  {journal} {\bibinfo
   {journal} {Phys. Rev. D}\ }\textbf {\bibinfo {volume} {103}},\ \bibinfo
  {pages} {083506} (\bibinfo {year} {2021})},\ \Eprint
  {https://arxiv.org/abs/2007.13722} {arXiv:2007.13722 [astro-ph.CO]}
  \BibitemShut {NoStop}%
\bibitem [{\citenamefont {Aghanim}\ \emph {et~al.}(2020)\citenamefont {Aghanim}
  \emph {et~al.}}]{Planck:2018vyg}%
  \BibitemOpen
  \bibfield  {author} {\bibinfo {author} {\bibfnamefont {N.}~\bibnamefont
  {Aghanim}} \emph {et~al.} (\bibinfo {collaboration} {Planck}),\ }\bibfield
  {title} {\bibinfo {title} {{Planck 2018 results. VI. Cosmological
  parameters}},\ }\href {https://doi.org/10.1051/0004-6361/201833910}
  {\bibfield  {journal} {\bibinfo  {journal} {Astron. Astrophys.}\ }\textbf
  {\bibinfo {volume} {641}},\ \bibinfo {pages} {A6} (\bibinfo {year} {2020})},\
  \bibinfo {note} {[Erratum: Astron.Astrophys. 652, C4 (2021)]},\ \Eprint
  {https://arxiv.org/abs/1807.06209} {arXiv:1807.06209 [astro-ph.CO]}
  \BibitemShut {NoStop}%
\bibitem [{\citenamefont {Inomata}\ \emph
  {et~al.}(2019{\natexlab{a}})\citenamefont {Inomata}, \citenamefont {Kohri},
  \citenamefont {Nakama},\ and\ \citenamefont {Terada}}]{Inomata:2019ivs}%
  \BibitemOpen
  \bibfield  {author} {\bibinfo {author} {\bibfnamefont {K.}~\bibnamefont
  {Inomata}}, \bibinfo {author} {\bibfnamefont {K.}~\bibnamefont {Kohri}},
  \bibinfo {author} {\bibfnamefont {T.}~\bibnamefont {Nakama}},\ and\ \bibinfo
  {author} {\bibfnamefont {T.}~\bibnamefont {Terada}},\ }\bibfield  {title}
  {\bibinfo {title} {{Enhancement of Gravitational Waves Induced by Scalar
  Perturbations due to a Sudden Transition from an Early Matter Era to the
  Radiation Era}},\ }\href {https://doi.org/10.1103/PhysRevD.100.043532}
  {\bibfield  {journal} {\bibinfo  {journal} {Phys. Rev. D}\ }\textbf {\bibinfo
  {volume} {100}},\ \bibinfo {pages} {043532} (\bibinfo {year}
  {2019}{\natexlab{a}})},\ \Eprint {https://arxiv.org/abs/1904.12879}
  {arXiv:1904.12879 [astro-ph.CO]} \BibitemShut {NoStop}%
\bibitem [{\citenamefont {Inomata}\ \emph
  {et~al.}(2019{\natexlab{b}})\citenamefont {Inomata}, \citenamefont {Kohri},
  \citenamefont {Nakama},\ and\ \citenamefont {Terada}}]{Inomata:2019zqy}%
  \BibitemOpen
  \bibfield  {author} {\bibinfo {author} {\bibfnamefont {K.}~\bibnamefont
  {Inomata}}, \bibinfo {author} {\bibfnamefont {K.}~\bibnamefont {Kohri}},
  \bibinfo {author} {\bibfnamefont {T.}~\bibnamefont {Nakama}},\ and\ \bibinfo
  {author} {\bibfnamefont {T.}~\bibnamefont {Terada}},\ }\bibfield  {title}
  {\bibinfo {title} {{Gravitational Waves Induced by Scalar Perturbations
  during a Gradual Transition from an Early Matter Era to the Radiation Era}},\
  }\href {https://doi.org/10.1088/1475-7516/2019/10/071} {\bibfield  {journal}
  {\bibinfo  {journal} {JCAP}\ }\textbf {\bibinfo {volume} {10}},\ \bibinfo
  {pages} {071}},\ \Eprint {https://arxiv.org/abs/1904.12878} {arXiv:1904.12878
  [astro-ph.CO]} \BibitemShut {NoStop}%
\bibitem [{\citenamefont {Inomata}\ \emph {et~al.}(2017)\citenamefont
  {Inomata}, \citenamefont {Kawasaki}, \citenamefont {Mukaida}, \citenamefont
  {Tada},\ and\ \citenamefont {Yanagida}}]{Inomata:2016rbd}%
  \BibitemOpen
  \bibfield  {author} {\bibinfo {author} {\bibfnamefont {K.}~\bibnamefont
  {Inomata}}, \bibinfo {author} {\bibfnamefont {M.}~\bibnamefont {Kawasaki}},
  \bibinfo {author} {\bibfnamefont {K.}~\bibnamefont {Mukaida}}, \bibinfo
  {author} {\bibfnamefont {Y.}~\bibnamefont {Tada}},\ and\ \bibinfo {author}
  {\bibfnamefont {T.~T.}\ \bibnamefont {Yanagida}},\ }\bibfield  {title}
  {\bibinfo {title} {{Inflationary primordial black holes for the LIGO
  gravitational wave events and pulsar timing array experiments}},\ }\href
  {https://doi.org/10.1103/PhysRevD.95.123510} {\bibfield  {journal} {\bibinfo
  {journal} {Phys. Rev. D}\ }\textbf {\bibinfo {volume} {95}},\ \bibinfo
  {pages} {123510} (\bibinfo {year} {2017})},\ \Eprint
  {https://arxiv.org/abs/1611.06130} {arXiv:1611.06130 [astro-ph.CO]}
  \BibitemShut {NoStop}%
\bibitem [{\citenamefont {Saikawa}\ and\ \citenamefont
  {Shirai}(2018)}]{Saikawa:2018rcs}%
  \BibitemOpen
  \bibfield  {author} {\bibinfo {author} {\bibfnamefont {K.}~\bibnamefont
  {Saikawa}}\ and\ \bibinfo {author} {\bibfnamefont {S.}~\bibnamefont
  {Shirai}},\ }\bibfield  {title} {\bibinfo {title} {{Primordial gravitational
  waves, precisely: The role of thermodynamics in the Standard Model}},\ }\href
  {https://doi.org/10.1088/1475-7516/2018/05/035} {\bibfield  {journal}
  {\bibinfo  {journal} {JCAP}\ }\textbf {\bibinfo {volume} {05}},\ \bibinfo
  {pages} {035}},\ \Eprint {https://arxiv.org/abs/1803.01038} {arXiv:1803.01038
  [hep-ph]} \BibitemShut {NoStop}%
\bibitem [{\citenamefont {Peebles}\ and\ \citenamefont
  {Vilenkin}(1999)}]{Peebles:1998qn}%
  \BibitemOpen
  \bibfield  {author} {\bibinfo {author} {\bibfnamefont {P.~J.~E.}\
  \bibnamefont {Peebles}}\ and\ \bibinfo {author} {\bibfnamefont
  {A.}~\bibnamefont {Vilenkin}},\ }\bibfield  {title} {\bibinfo {title}
  {{Quintessential inflation}},\ }\href
  {https://doi.org/10.1103/PhysRevD.59.063505} {\bibfield  {journal} {\bibinfo
  {journal} {Phys. Rev. D}\ }\textbf {\bibinfo {volume} {59}},\ \bibinfo
  {pages} {063505} (\bibinfo {year} {1999})},\ \Eprint
  {https://arxiv.org/abs/astro-ph/9810509} {arXiv:astro-ph/9810509}
  \BibitemShut {NoStop}%
\bibitem [{\citenamefont {Hossain}\ \emph {et~al.}(2014)\citenamefont
  {Hossain}, \citenamefont {Myrzakulov}, \citenamefont {Sami},\ and\
  \citenamefont {Saridakis}}]{Hossain:2014xha}%
  \BibitemOpen
  \bibfield  {author} {\bibinfo {author} {\bibfnamefont {M.~W.}\ \bibnamefont
  {Hossain}}, \bibinfo {author} {\bibfnamefont {R.}~\bibnamefont {Myrzakulov}},
  \bibinfo {author} {\bibfnamefont {M.}~\bibnamefont {Sami}},\ and\ \bibinfo
  {author} {\bibfnamefont {E.~N.}\ \bibnamefont {Saridakis}},\ }\bibfield
  {title} {\bibinfo {title} {{Variable gravity: A suitable framework for
  quintessential inflation}},\ }\href
  {https://doi.org/10.1103/PhysRevD.90.023512} {\bibfield  {journal} {\bibinfo
  {journal} {Phys. Rev. D}\ }\textbf {\bibinfo {volume} {90}},\ \bibinfo
  {pages} {023512} (\bibinfo {year} {2014})},\ \Eprint
  {https://arxiv.org/abs/1402.6661} {arXiv:1402.6661 [gr-qc]} \BibitemShut
  {NoStop}%
\bibitem [{\citenamefont {Guzzetti}\ \emph {et~al.}(2016)\citenamefont
  {Guzzetti}, \citenamefont {Bartolo}, \citenamefont {Liguori},\ and\
  \citenamefont {Matarrese}}]{Guzzetti:2016mkm}%
  \BibitemOpen
  \bibfield  {author} {\bibinfo {author} {\bibfnamefont {M.~C.}\ \bibnamefont
  {Guzzetti}}, \bibinfo {author} {\bibfnamefont {N.}~\bibnamefont {Bartolo}},
  \bibinfo {author} {\bibfnamefont {M.}~\bibnamefont {Liguori}},\ and\ \bibinfo
  {author} {\bibfnamefont {S.}~\bibnamefont {Matarrese}},\ }\bibfield  {title}
  {\bibinfo {title} {{Gravitational waves from inflation}},\ }\href
  {https://doi.org/10.1393/ncr/i2016-10127-1} {\bibfield  {journal} {\bibinfo
  {journal} {Riv. Nuovo Cim.}\ }\textbf {\bibinfo {volume} {39}},\ \bibinfo
  {pages} {399} (\bibinfo {year} {2016})},\ \Eprint
  {https://arxiv.org/abs/1605.01615} {arXiv:1605.01615 [astro-ph.CO]}
  \BibitemShut {NoStop}%
\bibitem [{\citenamefont {Caprini}\ and\ \citenamefont
  {Figueroa}(2018)}]{Caprini:2018mtu}%
  \BibitemOpen
  \bibfield  {author} {\bibinfo {author} {\bibfnamefont {C.}~\bibnamefont
  {Caprini}}\ and\ \bibinfo {author} {\bibfnamefont {D.~G.}\ \bibnamefont
  {Figueroa}},\ }\bibfield  {title} {\bibinfo {title} {{Cosmological
  Backgrounds of Gravitational Waves}},\ }\href
  {https://doi.org/10.1088/1361-6382/aac608} {\bibfield  {journal} {\bibinfo
  {journal} {Class. Quant. Grav.}\ }\textbf {\bibinfo {volume} {35}},\ \bibinfo
  {pages} {163001} (\bibinfo {year} {2018})},\ \Eprint
  {https://arxiv.org/abs/1801.04268} {arXiv:1801.04268 [astro-ph.CO]}
  \BibitemShut {NoStop}%
\bibitem [{\citenamefont {Creminelli}\ \emph {et~al.}(2020)\citenamefont
  {Creminelli}, \citenamefont {Tambalo}, \citenamefont {Vernizzi},\ and\
  \citenamefont {Yingcharoenrat}}]{Creminelli:2019kjy}%
  \BibitemOpen
  \bibfield  {author} {\bibinfo {author} {\bibfnamefont {P.}~\bibnamefont
  {Creminelli}}, \bibinfo {author} {\bibfnamefont {G.}~\bibnamefont {Tambalo}},
  \bibinfo {author} {\bibfnamefont {F.}~\bibnamefont {Vernizzi}},\ and\
  \bibinfo {author} {\bibfnamefont {V.}~\bibnamefont {Yingcharoenrat}},\
  }\bibfield  {title} {\bibinfo {title} {{Dark-Energy Instabilities induced by
  Gravitational Waves}},\ }\href
  {https://doi.org/10.1088/1475-7516/2020/05/002} {\bibfield  {journal}
  {\bibinfo  {journal} {JCAP}\ }\textbf {\bibinfo {volume} {05}},\ \bibinfo
  {pages} {002}},\ \Eprint {https://arxiv.org/abs/1910.14035} {arXiv:1910.14035
  [gr-qc]} \BibitemShut {NoStop}%
\bibitem [{\citenamefont {Creminelli}\ \emph {et~al.}(2019)\citenamefont
  {Creminelli}, \citenamefont {Tambalo}, \citenamefont {Vernizzi},\ and\
  \citenamefont {Yingcharoenrat}}]{Creminelli:2019nok}%
  \BibitemOpen
  \bibfield  {author} {\bibinfo {author} {\bibfnamefont {P.}~\bibnamefont
  {Creminelli}}, \bibinfo {author} {\bibfnamefont {G.}~\bibnamefont {Tambalo}},
  \bibinfo {author} {\bibfnamefont {F.}~\bibnamefont {Vernizzi}},\ and\
  \bibinfo {author} {\bibfnamefont {V.}~\bibnamefont {Yingcharoenrat}},\
  }\bibfield  {title} {\bibinfo {title} {{Resonant Decay of Gravitational Waves
  into Dark Energy}},\ }\href {https://doi.org/10.1088/1475-7516/2019/10/072}
  {\bibfield  {journal} {\bibinfo  {journal} {JCAP}\ }\textbf {\bibinfo
  {volume} {10}},\ \bibinfo {pages} {072}},\ \Eprint
  {https://arxiv.org/abs/1906.07015} {arXiv:1906.07015 [gr-qc]} \BibitemShut
  {NoStop}%
\bibitem [{\citenamefont {Chang}\ \emph {et~al.}(2023)\citenamefont {Chang},
  \citenamefont {Zhang},\ and\ \citenamefont {Zhou}}]{Chang:2022vlv}%
  \BibitemOpen
  \bibfield  {author} {\bibinfo {author} {\bibfnamefont {Z.}~\bibnamefont
  {Chang}}, \bibinfo {author} {\bibfnamefont {X.}~\bibnamefont {Zhang}},\ and\
  \bibinfo {author} {\bibfnamefont {J.-Z.}\ \bibnamefont {Zhou}},\ }\bibfield
  {title} {\bibinfo {title} {{Gravitational waves from primordial scalar and
  tensor perturbations}},\ }\href {https://doi.org/10.1103/PhysRevD.107.063510}
  {\bibfield  {journal} {\bibinfo  {journal} {Phys. Rev. D}\ }\textbf {\bibinfo
  {volume} {107}},\ \bibinfo {pages} {063510} (\bibinfo {year} {2023})},\
  \Eprint {https://arxiv.org/abs/2209.07693} {arXiv:2209.07693 [astro-ph.CO]}
  \BibitemShut {NoStop}%
\bibitem [{\citenamefont {Yu}\ and\ \citenamefont {Wang}(2024)}]{Yu:2023lmo}%
  \BibitemOpen
  \bibfield  {author} {\bibinfo {author} {\bibfnamefont {Y.-H.}\ \bibnamefont
  {Yu}}\ and\ \bibinfo {author} {\bibfnamefont {S.}~\bibnamefont {Wang}},\
  }\bibfield  {title} {\bibinfo {title} {{Primordial gravitational waves
  assisted by cosmological scalar perturbations}},\ }\href
  {https://doi.org/10.1140/epjc/s10052-024-12937-w} {\bibfield  {journal}
  {\bibinfo  {journal} {Eur. Phys. J. C}\ }\textbf {\bibinfo {volume} {84}},\
  \bibinfo {pages} {555} (\bibinfo {year} {2024})},\ \Eprint
  {https://arxiv.org/abs/2303.03897} {arXiv:2303.03897 [astro-ph.CO]}
  \BibitemShut {NoStop}%
\bibitem [{\citenamefont {Bari}\ \emph {et~al.}(2023)\citenamefont {Bari},
  \citenamefont {Bartolo}, \citenamefont {Dom\`enech},\ and\ \citenamefont
  {Matarrese}}]{Bari:2023rcw}%
  \BibitemOpen
  \bibfield  {author} {\bibinfo {author} {\bibfnamefont {P.}~\bibnamefont
  {Bari}}, \bibinfo {author} {\bibfnamefont {N.}~\bibnamefont {Bartolo}},
  \bibinfo {author} {\bibfnamefont {G.}~\bibnamefont {Dom\`enech}},\ and\
  \bibinfo {author} {\bibfnamefont {S.}~\bibnamefont {Matarrese}},\ }\bibfield
  {title} {\bibinfo {title} {{Gravitational waves induced by scalar-tensor
  mixing}},\ }\href@noop {} {\  (\bibinfo {year} {2023})},\ \Eprint
  {https://arxiv.org/abs/2307.05404} {arXiv:2307.05404 [astro-ph.CO]}
  \BibitemShut {NoStop}%
\bibitem [{\citenamefont {Picard}\ and\ \citenamefont
  {Malik}(2023)}]{Picard:2023sbz}%
  \BibitemOpen
  \bibfield  {author} {\bibinfo {author} {\bibfnamefont {R.}~\bibnamefont
  {Picard}}\ and\ \bibinfo {author} {\bibfnamefont {K.~A.}\ \bibnamefont
  {Malik}},\ }\bibfield  {title} {\bibinfo {title} {{Induced gravitational
  waves: the effect of first order tensor perturbations}},\ }\href@noop {} {\
  (\bibinfo {year} {2023})},\ \Eprint {https://arxiv.org/abs/2311.14513}
  {arXiv:2311.14513 [astro-ph.CO]} \BibitemShut {NoStop}%
\bibitem [{\citenamefont {Bari}\ \emph {et~al.}(2022)\citenamefont {Bari},
  \citenamefont {Ricciardone}, \citenamefont {Bartolo}, \citenamefont
  {Bertacca},\ and\ \citenamefont {Matarrese}}]{Bari:2021xvf}%
  \BibitemOpen
  \bibfield  {author} {\bibinfo {author} {\bibfnamefont {P.}~\bibnamefont
  {Bari}}, \bibinfo {author} {\bibfnamefont {A.}~\bibnamefont {Ricciardone}},
  \bibinfo {author} {\bibfnamefont {N.}~\bibnamefont {Bartolo}}, \bibinfo
  {author} {\bibfnamefont {D.}~\bibnamefont {Bertacca}},\ and\ \bibinfo
  {author} {\bibfnamefont {S.}~\bibnamefont {Matarrese}},\ }\bibfield  {title}
  {\bibinfo {title} {{Signatures of Primordial Gravitational Waves on the
  Large-Scale Structure of the Universe}},\ }\href
  {https://doi.org/10.1103/PhysRevLett.129.091301} {\bibfield  {journal}
  {\bibinfo  {journal} {Phys. Rev. Lett.}\ }\textbf {\bibinfo {volume} {129}},\
  \bibinfo {pages} {091301} (\bibinfo {year} {2022})},\ \Eprint
  {https://arxiv.org/abs/2111.06884} {arXiv:2111.06884 [astro-ph.CO]}
  \BibitemShut {NoStop}%
\bibitem [{\citenamefont {Gao}\ and\ \citenamefont {Steer}(2011)}]{Gao:2011qe}%
  \BibitemOpen
  \bibfield  {author} {\bibinfo {author} {\bibfnamefont {X.}~\bibnamefont
  {Gao}}\ and\ \bibinfo {author} {\bibfnamefont {D.~A.}\ \bibnamefont
  {Steer}},\ }\bibfield  {title} {\bibinfo {title} {{Inflation and primordial
  non-Gaussianities of 'generalized Galileons'}},\ }\href
  {https://doi.org/10.1088/1475-7516/2011/12/019} {\bibfield  {journal}
  {\bibinfo  {journal} {JCAP}\ }\textbf {\bibinfo {volume} {12}},\ \bibinfo
  {pages} {019}},\ \Eprint {https://arxiv.org/abs/1107.2642} {arXiv:1107.2642
  [astro-ph.CO]} \BibitemShut {NoStop}%
\bibitem [{\citenamefont {Malik}\ and\ \citenamefont
  {Wands}(2009)}]{Malik_2009}%
  \BibitemOpen
  \bibfield  {author} {\bibinfo {author} {\bibfnamefont {K.~A.}\ \bibnamefont
  {Malik}}\ and\ \bibinfo {author} {\bibfnamefont {D.}~\bibnamefont {Wands}},\
  }\bibfield  {title} {\bibinfo {title} {Cosmological perturbations},\ }\href
  {https://doi.org/10.1016/j.physrep.2009.03.001} {\bibfield  {journal}
  {\bibinfo  {journal} {Physics Reports}\ }\textbf {\bibinfo {volume} {475}},\
  \bibinfo {pages} {1–51} (\bibinfo {year} {2009})}\BibitemShut {NoStop}%
\bibitem [{\citenamefont {Dom\`enech}(2023{\natexlab{b}})}]{Domenech:2023fuz}%
  \BibitemOpen
  \bibfield  {author} {\bibinfo {author} {\bibfnamefont {G.}~\bibnamefont
  {Dom\`enech}},\ }\bibfield  {title} {\bibinfo {title} {{Lectures on
  Gravitational Wave Signatures of Primordial Black Holes}}\ }(\bibinfo {year}
  {2023})\ \Eprint {https://arxiv.org/abs/2307.06964} {arXiv:2307.06964
  [gr-qc]} \BibitemShut {NoStop}%
\bibitem [{\citenamefont {Maldacena}(2003)}]{Maldacena:2002vr}%
  \BibitemOpen
  \bibfield  {author} {\bibinfo {author} {\bibfnamefont {J.~M.}\ \bibnamefont
  {Maldacena}},\ }\bibfield  {title} {\bibinfo {title} {{Non-Gaussian features
  of primordial fluctuations in single field inflationary models}},\ }\href
  {https://doi.org/10.1088/1126-6708/2003/05/013} {\bibfield  {journal}
  {\bibinfo  {journal} {JHEP}\ }\textbf {\bibinfo {volume} {05}},\ \bibinfo
  {pages} {013}},\ \Eprint {https://arxiv.org/abs/astro-ph/0210603}
  {arXiv:astro-ph/0210603} \BibitemShut {NoStop}%
\bibitem [{\citenamefont {Dom\`enech}\ \emph {et~al.}(2015)\citenamefont
  {Dom\`enech}, \citenamefont {Naruko},\ and\ \citenamefont
  {Sasaki}}]{Domenech:2015hka}%
  \BibitemOpen
  \bibfield  {author} {\bibinfo {author} {\bibfnamefont {G.}~\bibnamefont
  {Dom\`enech}}, \bibinfo {author} {\bibfnamefont {A.}~\bibnamefont {Naruko}},\
  and\ \bibinfo {author} {\bibfnamefont {M.}~\bibnamefont {Sasaki}},\
  }\bibfield  {title} {\bibinfo {title} {{Cosmological disformal invariance}},\
  }\href {https://doi.org/10.1088/1475-7516/2015/10/067} {\bibfield  {journal}
  {\bibinfo  {journal} {JCAP}\ }\textbf {\bibinfo {volume} {10}},\ \bibinfo
  {pages} {067}},\ \Eprint {https://arxiv.org/abs/1505.00174} {arXiv:1505.00174
  [gr-qc]} \BibitemShut {NoStop}%
\bibitem [{\citenamefont {Minamitsuji}(2014)}]{Minamitsuji:2014waa}%
  \BibitemOpen
  \bibfield  {author} {\bibinfo {author} {\bibfnamefont {M.}~\bibnamefont
  {Minamitsuji}},\ }\bibfield  {title} {\bibinfo {title} {{Disformal
  transformation of cosmological perturbations}},\ }\href
  {https://doi.org/10.1016/j.physletb.2014.08.037} {\bibfield  {journal}
  {\bibinfo  {journal} {Phys. Lett. B}\ }\textbf {\bibinfo {volume} {737}},\
  \bibinfo {pages} {139} (\bibinfo {year} {2014})},\ \Eprint
  {https://arxiv.org/abs/1409.1566} {arXiv:1409.1566 [astro-ph.CO]}
  \BibitemShut {NoStop}%
\bibitem [{\citenamefont {Motohashi}\ and\ \citenamefont
  {White}(2016)}]{Motohashi:2015pra}%
  \BibitemOpen
  \bibfield  {author} {\bibinfo {author} {\bibfnamefont {H.}~\bibnamefont
  {Motohashi}}\ and\ \bibinfo {author} {\bibfnamefont {J.}~\bibnamefont
  {White}},\ }\bibfield  {title} {\bibinfo {title} {{Disformal invariance of
  curvature perturbation}},\ }\href
  {https://doi.org/10.1088/1475-7516/2016/02/065} {\bibfield  {journal}
  {\bibinfo  {journal} {JCAP}\ }\textbf {\bibinfo {volume} {02}},\ \bibinfo
  {pages} {065}},\ \Eprint {https://arxiv.org/abs/1504.00846} {arXiv:1504.00846
  [gr-qc]} \BibitemShut {NoStop}%
\end{thebibliography}%

\end{document}